\documentclass[preprint2]{aastex631}

\usepackage{amsmath}
\usepackage{multirow}
\usepackage{xspace}

\newcommand{\aij}{{\tt AIJ}\xspace}
\newcommand{\celerite}{{\tt celerite}\xspace}
\newcommand{\celeritetwo}{{\tt celerite2}\xspace}
\newcommand{\emcee}{{\tt emcee}\xspace}
\newcommand{\exofast}{{\tt EXOFASTv2}\xspace}
\newcommand{\exostriker}{{\tt Exo-Striker}\xspace}
\newcommand{\megno}{{\tt MEGNO}\xspace}
\newcommand{\nbody}{{\tt Nbody}\xspace}
\newcommand{\rebound}{{\tt rebound}\xspace}
\newcommand{\spock}{{\tt SPOCK}\xspace}

\newcommand{\mstar}{M$_{\star}$\xspace}
\newcommand{\rstar}{R$_{\star}$\xspace}
\newcommand{\lstar}{L$_{\star}$\xspace}
\newcommand{\msun}{M$_{\sun}$\xspace}
\newcommand{\rsun}{R$_{\sun}$\xspace}
\newcommand{\lsun}{L$_{\sun}$\xspace}
\newcommand{\mplan}{M$_{\rm p}$\xspace}
\newcommand{\rplan}{R$_{\rm p}$\xspace}
\newcommand{\mjup}{M$_{\rm J}$\xspace}
\newcommand{\rjup}{R$_{\rm J}$\xspace}
\newcommand{\mear}{M$_{\earth}$\xspace}
\newcommand{\rear}{R$_{\earth}$\xspace}
\newcommand{\teff}{T$_{\rm eff}$\xspace}
\newcommand{\prot}{P$_{\rm rot}$\xspace}
\newcommand{\porb}{P$_{\rm orb}$\xspace}
\newcommand{\tc}{T$_{\rm C}$\xspace}
\newcommand{\zs}{z$_{\rm s}$\xspace}
\newcommand{\reduced}{$\chi^{2}_{\rm red}$\xspace}
\newcommand{\logmstar}{log$_{10}\left(\frac{\rm M_{\star}}{\rm M_{\sun}}\right)$}
\newcommand{\logfracpd}{log$_{10}\left(\frac{\rm Period}{\rm days}\right)$}
\newcommand{\logg}{log$_{10}\left({\rm g~cm^{-3}}\right)$}
\newcommand{\ergfrac}{${{\rm erg~s^{-1}~cm}^{-2}}$}
\newcommand{\radroot}{$\sqrt{\left(1-\frac{{\rm R}_{\rm p}}{{\rm R}_{\star}}\right)^{2}-\mbox{b}^{2}}$}

\begin{document}

\title{Validating AU Microscopii d with Transit Timing Variations}

\author[0000-0002-7424-9891]{Justin M. Wittrock}
\affiliation{Department of Physics \& Astronomy, George Mason University, 4400 University Drive MS 3F3, Fairfax, VA 22030, USA}
\email{jwittroc@gmu.edu}

\author[0000-0002-8864-1667]{Peter P. Plavchan}
\affiliation{Department of Physics \& Astronomy, George Mason University, 4400 University Drive MS 3F3, Fairfax, VA 22030, USA}

\author[0000-0002-2078-6536]{Bryson L. Cale}
\affiliation{NASA JPL, 4800 Oak Grove Drive, Pasadena, CA 91109, USA}
\affiliation{IPAC, 770 South Wilson Avenue, Pasadena, CA 91125, USA}

\author[0000-0001-7139-2724]{Thomas Barclay}
\affiliation{University of Maryland, Baltimore County, 1000 Hilltop Circle, Baltimore, MD 21250, USA}
\affiliation{NASA Goddard Space Flight Center, 8800 Greenbelt Road, Greenbelt, MD 20771, USA}

\author[0000-0002-8382-0236]{Mathis R. Ludwig}
\affiliation{Institut f\"ur Astrophysik, Georg-August-Universit\"at, Friedrich-Hund-Platz 1, 37077 G\"ottingen, Germany}

\author[0000-0001-8227-1020]{Richard P. Schwarz}
\affiliation{Center for Astrophysics \textbar \ Harvard \& Smithsonian, 60 Garden Street, Cambridge, MA 02138, USA}

\author[0000-0001-5000-7292]{Djamel M\'ekarnia}
\affiliation{Universit\'e C\^ote d’Azur, Observatoire de la C\^ote d’Azur, CNRS, Laboratoire Lagrange, Bd de l’Observatoire, CS 34229, F-06304 Nice cedex 4, France}

\author[0000-0002-5510-8751]{Amaury H. M. J. Triaud}
\affiliation{School of Physics \& Astronomy, University of Birmingham, Edgbaston, Birmingham B15 2TT, United Kingdom}

\author[0000-0002-0856-4527]{Lyu Abe}
\affiliation{Universit\'e C\^ote d’Azur, Observatoire de la C\^ote d’Azur, CNRS, Laboratoire Lagrange, Bd de l’Observatoire, CS 34229, F-06304 Nice cedex 4, France}

\author[0000-0002-3503-3617]{Olga Suarez}
\affiliation{Universit\'e C\^ote d’Azur, Observatoire de la C\^ote d’Azur, CNRS, Laboratoire Lagrange, Bd de l’Observatoire, CS 34229, F-06304 Nice cedex 4, France}

\author[0000-0002-7188-8428]{Tristan Guillot}
\affiliation{Universit\'e C\^ote d’Azur, Observatoire de la C\^ote d’Azur, CNRS, Laboratoire Lagrange, Bd de l’Observatoire, CS 34229, F-06304 Nice cedex 4, France}

\author[0000-0003-2239-0567]{Dennis M. Conti}
\affiliation{American Association of Variable Star Observers, 185 Alewife Brooke Parkway, Suite 410, Cambridge, MA 02138 USA}

\author[0000-0001-6588-9574]{Karen A. Collins}
\affiliation{Center for Astrophysics \textbar \ Harvard \& Smithsonian, 60 Garden Street, Cambridge, MA 02138, USA}

\author[0000-0002-3249-3538]{Ian A. Waite}
\affiliation{Centre for Astrophysics, University of Southern Queensland, Toowoomba, QLD, 4350, Australia}

\author[0000-0003-0497-2651]{John F. Kielkopf}
\affiliation{Department of Physics and Astronomy, University of Louisville, Louisville, KY 40292, USA}


\author[0000-0003-2781-3207]{Kevin I. Collins}
\affiliation{Department of Physics \& Astronomy, George Mason University, 4400 University Drive MS 3F3, Fairfax, VA 22030, USA}

\author[0000-0001-6187-5941]{Stefan Dreizler}
\affiliation{Institut f\"ur Astrophysik, Georg-August-Universit\"at, Friedrich-Hund-Platz 1, 37077 G\"ottingen, Germany}

\author[0000-0001-8364-2903]{Mohammed El Mufti}
\affiliation{Department of Physics \& Astronomy, George Mason University, 4400 University Drive MS 3F3, Fairfax, VA 22030, USA}

\author[0000-0002-2457-7889]{Dax L. Feliz}
\affiliation{Department of Astrophysics, American Museum of Natural History, New York, NY 10024, USA}

\author[0000-0002-5258-6846]{Eric Gaidos}
\affiliation{Department of Earth Sciences, University of Hawai`i at M$\bar{a}$noa, 1680 East-West Road, Honolulu, HI 96822, USA}

\author[0000-0001-9596-8820]{Claire S. Geneser}
\affiliation{Mississippi State University, 75 B. S. Hood Road, Mississippi State, MS 39762, USA}

\author[0000-0003-1728-0304]{Keith D. Horne}
\affiliation{SUPA Physics and Astronomy, University of St. Andrews, Fife, KY16 9SS Scotland, UK}

\author[0000-0002-7084-0529]{Stephen R. Kane}
\affiliation{University of California, Riverside, 900 University Ave., Riverside, CA 92521, USA}

\author[0000-0001-8014-0270]{Patrick J. Lowrance}
\affiliation{IPAC, California Institute of Technology, MC 314-6, 1200 E. California Blvd., Pasadena, California 91125, USA}

\author[0000-0002-5084-168X]{Eder Martioli}
\affiliation{Laborat\'orio Nacional de Astrof\'isica, Rua Estados Unidos
154, Itajub\'a, MG 37504-364, Brazil}
\affiliation{Sorbonne Universit\'e, CNRS, UMR 7095, Institut
d’Astrophysique de Paris, 98 bis bd Arago, 75014 Paris, France}

\author[0000-0002-3940-2360]{Don J. Radford}
\affiliation{Brierfield Observatory, Bowral, New South Wales, Australia}

\author[0000-0003-4701-8497]{Michael A. Reefe}
\affiliation{Kavli Institute for Astrophysics and Space Research, Massachusetts Institute of Technology, 77 Massachusetts Avenue, Cambridge, MA 02139, USA}

\author[0000-0002-4650-594X]{Veronica Roccatagliata}
\affiliation{INAF-Osservatorio Astrofisico di Arcetri, Largo E. Fermi 5, 50125 Firenze, Italy}
\affiliation{INFN, Sezione di Pisa, Largo Bruno Pontecorvo 3, 56127 Pisa, Italy}
\affiliation{Department of Physics ``E. Fermi'', University of Pisa, Largo Bruno Pontecorvo 3, 56127 Pisa, Italy}

\author[0000-0002-1836-3120]{Avi Shporer}
\affiliation{Department of Physics and Kavli Institute for Astrophysics and Space Research, Massachusetts Institute of Technology, Cambridge, MA 02139, USA}

\author[0000-0002-3481-9052]{Keivan G. Stassun}
\affiliation{Department of Physics \& Astronomy, Vanderbilt University, 6301 Stevenson Center Ln., Nashville, TN 37235, USA}

\author[0000-0003-2163-1437]{Christopher Stockdale}
\affiliation{Hazelwood Observatory, Hazelwood South, Victoria, Australia}

\author[0000-0001-5603-6895]{Thiam-Guan Tan}
\affiliation{Perth Exoplanet Survey Telescope, Perth, Western Australia, Australia}
\affiliation{Curtin Institute of Radio Astronomy, Curtin University, Bentley, Western Australia 6102}

\author[0000-0002-2903-2140]{Angelle M. Tanner}
\affiliation{Mississippi State University, 75 B. S. Hood Road, Mississippi State, MS 39762, USA}

\author[0000-0002-5928-2685]{Laura D. Vega}
\affiliation{Department of Astronomy, University of Maryland, College Park, MD 20742, USA}
\affiliation{NASA Goddard Space Flight Center, 8800 Greenbelt Road, Greenbelt, MD 20771, USA}
\affiliation{Center for Research and Exploration in Space Science \& Technology, NASA/GSFC, Greenbelt, MD 20771, USA}

\begin{abstract}
{AU Mic is a young (22 Myr) nearby exoplanetary system that exhibits excess TTVs that cannot be accounted for by the two known transiting planets nor stellar activity.}
{We present the statistical ``validation'' of the tentative planet AU Mic d (even though there are examples of ``confirmed'' planets with ambiguous orbital periods).}
{We add 18 new transits and nine midpoint times in an updated TTV analysis to prior work. We perform the joint modeling of transit light curves using {\tt EXOFASTv2} and extract the transit midpoint times. Next, we construct an O--C diagram and use {\tt Exo-Striker} to model the TTVs. We generate TTV log-likelihood periodograms to explore possible solutions for d's period, then follow those up with detailed TTV and RV MCMC modeling and stability tests.}
{We find several candidate periods for AU Mic d, all of which are near resonances with AU Mic b and c of varying order. Based on our model comparisons, the most-favored orbital period of AU Mic d is 12.73596$\pm$0.00793 days ($T_{C,d}$=2458340.55781$\pm$0.11641 BJD), which puts the three planets near 4:6:9 MMR. The mass for d is 1.053$\pm$0.511 \mear, making this planet Earth-like in mass.}
{If confirmed, AU Mic d would be the first known Earth-mass planet orbiting a young star and would provide a valuable opportunity in probing a young terrestrial planet's atmosphere. Additional TTV observation of AU Mic system are needed to further constrain the planetary masses, search for possible transits of AU Mic d, and detect possible additional planets beyond AU Mic c.}
\end{abstract}

\section{Introduction}

AU Microscopii (TOI-2221, TIC 441420236, HD 197481, GJ 803) is a fundamental system proven to be quite viable for study of planetary formation and orbital dynamics of young systems given its youthfulness \citep[$22 \pm 3$ Myr,][]{mamajek2014}, proximity \citep[9.71 pc;][]{gaia2021}, and relative brightness (m$_{\rm V}$=8.81 mag). It is highly active \citep{butler1981, kundu1987, cully1993, tsikoudi2000, gilbert2022, feinstein2022} and reported to have the largest number of flare events among the Kepler and TESS targets to date \citep{ilin2022}. \citet{ilin2022} examined TESS data for any flares caused by star-to-planet interaction (SPI) between AU Mic b and the host star; they did not find any and concluded that most flares were attributed to stellar activity, but they have not ruled this phenomenon out yet. The aforementioned heightened stellar activity of AU Mic had made it a challenging target for radial velocity (RV) and transit observations \citep{addison2021, cale2021, gilbert2022, zicher2022, wittrock2022}.

AU Mic hosts two transiting planets \citep{plavchan2020, martioli2021, gilbert2022} and a large dust disk \citep{fajardo-acosta2000, zuckerman2001, song2002, kalas2004, liu2004, plavchan2005, strubbe2006, macgregor2013, grady2020, arnold2022, olofsson2022, vizgan2022}. \citet{gallenne2022} searched for additional companions in the inner disk region (0.4 -- 2.4 au) with high-angular resolution observations through VLT/SPHERE and with combined data from VLT/NACO, VLTI/PIONIER, and VLTI/GRAVITY but did not find any brighter than $K_{s}\approx$ 11.2 mag, which in turn caps the upper planetary mass limit at 12.3 $\pm$ 0.5 \mjup. \citet{szabo2021, szabo2022} probed the AU Mic system with CHEOPS and did a joint TESS + CHEOPS TTV analysis, which found that the timing of the transits during summer of 2022 to be 30 -- 85 min later than predicted using the linear ephemeris available at the time. \citet{wittrock2022} did a TTV and photodynamical analysis of the AU Mic systems and detected the TTV excess that cannot be accounted with the presence of both planets b and c and the stellar activities of AU Mic; thus, a non-transiting hypothetical planet d between AU Mic b and c was proposed. This would have made AU Mic among the few systems that have a non-transiting planet between its adjacent transiting planets; other systems include HD 3167 d \citep{christiansen2017}, Kepler-20 g \citep{buchhave2016}, Kepler-411 e \citep{sun2019}, and TOI-431 c \citep{osborn2021}. \citet{kane2022} explored the orbital dynamics of the AU Mic system by injecting the hypothetical planet d; they found that it lies at the very edge of instability, and that d's eccentricity will vary between 0.0 and 0.3 even on short timescales.

This paper is a continuation of the work done in \citet{wittrock2022}, with an emphasis now placed on validating the tentative planet AU Mic d through the TTV method. In $\S$\ref{sec:dataobs}, we highlight the new light curves in addition to the old data sets we include for TTV analyses. Next, we summarize the steps taken in modeling the observed transits using the \exofast package \citep{eastman2019} in $\S$\ref{sec:exofastmod}. $\S$\ref{sec:exostrikermod} presents our modeling of the extracted TTVs using the \exostriker package \citep{trifonov2019} and presents a novel technique called the TTV log-likelihood periodogram, which searches for parameters that maximize the log-likelihood. In $\S$\ref{sec:rv_analysis}, we perform the RV vetting of our TTV analysis to check for consistency between the RV results and those of the TTV analysis. We discuss the results in $\S$\ref{sec:results} and present our conclusion in $\S$\ref{sec:conclude}.

\section{Data from Observations}\label{sec:dataobs}

We incorporated a total of 54 data sets from four years of AU Mic's photometric and Rossiter-McLaughlin (R-M) observations with various facilities, of which 45 of them are of AU Mic b and nine of them are of AU Mic c (Tables \ref{tab:datasets} \& \ref{tab:facilities}). The following transits had been presented in the previous works: the R-M observations \citep{martioli2020, palle2020}, the TESS and one of the Spitzer observations \citep{plavchan2020, martioli2021, gilbert2022}, the CHEOPS observations \citep{szabo2021, szabo2022}, and the Brierfield, the LCO SAAO \& SSO prior to 2021, the PEST, and two of the Spitzer observations \citep{wittrock2022}. This paper introduces 18 new and unpublished observations from ASTEP, LCOGT, and MKO CDK700 and adds nine midpoint times from CHEOPS observations. Therefore, this section will describe only the new observations that were not included in \citet{wittrock2022}.

\startlongtable
\begin{deluxetable*}{c|l|c|c|c|c|c|r|c}\label{tab:datasets}
    \tabletypesize{\scriptsize}
    \tablecaption{List of AU Mic's photometric and R-M observation data incorporated for transit and TTV analyses. All ground-based photometric observations listed here were organized via TESS Follow-up Observing Program Working Group (TFOPWG)\tablenotemark{a}.}
    \tablehead{\multirow{2}{*}{Planet} & \multirow{2}{*}{Telescope} & \multirow{2}{*}{Date (UT)} & \multirow{2}{*}{Filter} & Exposure   & No. of & Obs. Dur. & Transit  & \multirow{2}{*}{Ref} \\
                                       &                            &                            &                         & Time (sec) & Images & (min)     & Coverage &                       }
    \startdata
\multirow{3}{*}{b} & \multirow{3}{*}{ASTEP 0.4 m}                       & 2021-07-09 & R$_{\rm c}$    & 9     & 636     & 377      & full              & \multirow{3}{*}{new} \\
                   &                                                    & 2021-07-26 & R$_{\rm c}$    & 9     & 724     & 407      & full              &                      \\
                   &                                                    & 2021-08-12 & R$_{\rm c}$    & 9     & 801     & 550      & full              &                      \\ \hline
\multirow{2}{*}{c} & \multirow{2}{*}{ASTEP 0.4 m}                       & 2021-07-21 & R$_{\rm c}$    & 9     & 1\,294  & 752      & full              & \multirow{2}{*}{new} \\
                   &                                                    & 2021-08-09 & R$_{\rm c}$    & 9     & 951     & 560      & full              &                      \\ \hline
b                  & Brierfield 0.36 m                                  & 2020-08-13 & I              & 16    & 398     & 379      & full              & 1                    \\ \hline
b                  & CFHT (SPIRou)                                      & 2019-06-17 & 955-2\,515 nm  & 122.6 & 116     & 302.8    & egress            & 2                    \\ \hline
\multirow{7}{*}{b} & \multirow{7}{*}{CHEOPS}                            & 2020-07-10 & CHEOPS         & 15    & 7\,194  & 938.1    & full              & \multirow{3}{*}{3}   \\
                   &                                                    & 2020-08-21 & CHEOPS         & 15    & 8\,844  & 658.3    & full              &                      \\
                   &                                                    & 2020-09-24 & CHEOPS         & 3     & 12\,222 & 1\,066.1 & full              &                      \\
                   &                                                    & 2021-07-26 & CHEOPS         & 3     & 9\,366  & 666.9    & full              & \multirow{4}{*}{4}   \\
                   &                                                    & 2021-08-12 & CHEOPS         & 3     & 11\,746 & 687.3    & full              &                      \\
                   &                                                    & 2021-08-29 & CHEOPS         & 3     & 9\,338  & 687.3    & full              &                      \\
                   &                                                    & 2021-09-06 & CHEOPS         & 3     & 9\,002  & 687.3    & full              &                      \\ \hline
\multirow{2}{*}{c} & \multirow{2}{*}{CHEOPS}                            & 2021-08-09 & CHEOPS         & 3     & 14\,406 & 878.5    & full              & \multirow{2}{*}{4}   \\
                   &                                                    & 2021-08-28 & CHEOPS         & 3     & 12\,698 & 865.8    & full              &                      \\ \hline
b                  & IRTF (iSHELL)                                      & 2019-06-17 & 2.18-2.47 nm   & 120   & 47      & 105.2    & egress            & 2                    \\ \hline
\multirow{3}{*}{b} & \multirow{3}{*}{LCO CTIO 1.0 m}                    & 2021-06-14 & Pan-STARRS \zs & 15    & 452     & 342      & full              & \multirow{3}{*}{new} \\
                   &                                                    & 2021-07-01 & Pan-STARRS \zs & 15    & 440     & 340      & full              &                      \\
                   &                                                    & 2021-08-04 & Pan-STARRS \zs & 15    & 362     & 274      & egress            &                      \\ \hline
c                  & LCO CTIO 1.0 m                                     & 2021-10-04 & Pan-STARRS \zs & 15    & 412     & 311      & ingress           & new                  \\ \hline
\multirow{8}{*}{b} & \multirow{8}{*}{LCO SAAO 1.0 m}                    & 2020-05-20 & Pan-STARRS Y   & 35    & 99      & 262      & egress            & \multirow{5}{*}{1} \\
                   &                                                    & 2020-05-20 & Pan-STARRS \zs & 15    & 333     & 266      & egress            &                      \\
                   &                                                    & 2020-06-06 & Pan-STARRS \zs & 15    & 266     & 218      & egress            &                      \\
                   &                                                    & 2020-06-23 & Pan-STARRS \zs & 15    & 223     & 183      & egress            &                      \\
                   &                                                    & 2020-09-07 & Pan-STARRS \zs & 15    & 211     & 172      & ingress           &                      \\
                   &                                                    & 2021-07-18 & Pan-STARRS \zs & 15    & 259     & 212      & ingress           & \multirow{3}{*}{new} \\
                   &                                                    & 2021-08-03 & Pan-STARRS \zs & 15    & 317     & 252      & ingress           &                      \\
                   &                                                    & 2021-09-23 & Pan-STARRS \zs & 15    & 456     & 344      & full              &                      \\ \hline
b \& c             & LCO SAAO 1.0 m                                     & 2020-10-11 & Pan-STARRS \zs & 15    & 311     & 266      & ingress \& egress & 1 \& new             \\ \hline            
\multirow{9}{*}{b} & \multirow{9}{*}{LCO SSO 1.0 m}                     & 2020-04-25 & Pan-STARRS Y   & 35    & 40      & 104      & egress            & \multirow{5}{*}{1}   \\
                   &                                                    & 2020-04-25 & Pan-STARRS \zs & 15    & 212     & 172      & egress            &                      \\
                   &                                                    & 2020-08-13 & Pan-STARRS \zs & 15    & 379     & 312      & full              &                      \\
                   &                                                    & 2020-09-16 & Pan-STARRS \zs & 15    & 408     & 340      & full              &                      \\
                   &                                                    & 2020-10-03 & Pan-STARRS \zs & 15    & 248     & 219      & egress            &                      \\
                   &                                                    & 2021-06-22 & Pan-STARRS \zs & 15    & 351     & 281      & full              & \multirow{4}{*}{new} \\
                   &                                                    & 2021-08-12 & Pan-STARRS \zs & 15    & 406     & 343      & full              &                      \\
                   &                                                    & 2021-08-29 & Pan-STARRS \zs & 15    & 448     & 340      & full              &                      \\
                   &                                                    & 2021-09-15 & Pan-STARRS \zs & 15    & 295     & 277      & egress            &                      \\ \hline
b                  & LCO TO 1.0 m                                       & 2021-09-23 & Pan-STARRS \zs & 15    & 300     & 231      & egress            & new                  \\ \hline
c                  & MKO CDK700 0.7m                                    & 2021-08-28 & {\it r'}       & 16    & 74      & 87       & egress            & new                  \\ \hline
b                  & PEST 0.30 m                                        & 2020-07-10 & V              & 15    & 1\,143  & 556      & full              & 1                    \\ \hline
\multirow{3}{*}{b} & \multirow{3}{*}{Spitzer (IRAC)}                    & 2019-02-10 & 4.5 $\mu$m     & 0.08  & 3\,020  & 475.7    & full              & \multirow{3}{*}{1}   \\
                   &                                                    & 2019-02-27 & 4.5 $\mu$m     & 0.08  & 3\,377  & 475.7    & egress            &                      \\
                   &                                                    & 2019-09-09 & 4.5 $\mu$m     & 0.08  & 6\,002  & 990.9    & full              &                      \\ \hline
\multirow{5}{*}{b} & \multirow{5}{*}{TESS\tablenotemark{\scriptsize b}} & 2018-07-26 & TESS           & 120   & 329     & 718.0    & full              & \multirow{5}{*}{5}   \\
                   &                                                    & 2018-08-12 & TESS           & 120   & 296     & 708.0    & full              &                      \\
                   &                                                    & 2020-07-10 & TESS           & 20    & 2\,132  & 719.7    & full              &                      \\
                   &                                                    & 2020-07-19 & TESS           & 20    & 2\,137  & 719.7    & full              &                      \\
                   &                                                    & 2020-07-27 & TESS           & 20    & 2\,120  & 719.7    & full              &                      \\ \hline
\multirow{3}{*}{c} & \multirow{3}{*}{TESS\tablenotemark{\scriptsize b}} & 2018-08-11 & TESS           & 120   & 342     & 718.0    & full              & \multirow{3}{*}{5}   \\
                   &                                                    & 2020-07-09 & TESS           & 20    & 2\,138  & 719.7    & full              &                      \\
                   &                                                    & 2020-07-28 & TESS           & 20    & 2\,133  & 719.7    & full              &                      \\ \hline
b                  & VLT (ESPRESSO)                                     & 2019-08-07 & 378.2-788.7 nm & 200   & 88      & 359      & full              & 6
    \enddata
    \tablenotetext{a}{\url{https://tess.mit.edu/followup}}
    \tablenotetext{b}{$\sim$12-hour snippets of the $\sim$27-day duration TESS Cycle 1 and 3 light curves were extracted for our analysis, centered approximately on each transit.}
    \tablerefs{(1) \citet{wittrock2022}; (2) \citet{martioli2020}; (3) \citet{szabo2021}; (4) \citet{szabo2022}; (5) \citet{gilbert2022}; (6) \citet{palle2020}}
\end{deluxetable*}

\begin{deluxetable*}{l|l|l|r|c|c|c|c}\label{tab:facilities}
    \tablecaption{List of facilities utilized for photometric and Rossiter-McLaughlin follow-up observations of AU Mic.}
    \tablehead{\multirow{2}{*}{Telescope} & \multirow{2}{*}{Instrument} & \multirow{2}{*}{Location} & Aperture & Pixel Scale & Resolution & FOV      & \multirow{2}{*}{Ref} \\
                                          &                             &                           & (m)      & (arcsec)    & (pixels)   & (arcmin) &                      }
    \startdata
ASTEP 400  & FLI Proline 16800E & Concordia Research Station, Antarctica & 0.4    & 0.93  & 4\,096$\times$4\,096 & 63.5$\times$63.5 & 1   \\
Brierfield & Moravian 16803     & Bowral, New South Wales                & 0.36   & 0.732 & 4\,096$\times$4\,096 & 50.0$\times$50.0 & 2   \\
CFHT       & SPIRou             & Maunakea, Hawai`i                      & 3.58   & ...   & ...                  & ...              & 3   \\
CHEOPS     & ...                & ...                                    & 0.32   & 1.11  & 1\,024$\times$1\,024 & 19$\times$19     & 4   \\
IRTF       & iSHELL             & Maunakea, Hawai`i                      & 3.2    & ...   & ...                  & ...              & 5   \\
LCO CTIO   & Sinistro           & Cerro Tololo, Chile                    & 1.0    & 0.389 & 4\,096$\times$4\,096 & 26.5$\times$26.5 & 6   \\
LCO SAAO   & Sinistro           & Sutherland, South Africa               & 1.0    & 0.389 & 4\,096$\times$4\,096 & 26.5$\times$26.5 & 6   \\
LCO SSO    & Sinistro           & Mount Woorut, New South Wales          & 1.0    & 0.389 & 4\,096$\times$4\,096 & 26.5$\times$26.5 & 6   \\
LCO TO     & Sinistro           & Mount Teide, Tenerife, Canary Islands  & 1.0    & 0.389 & 4\,096$\times$4\,096 & 26.5$\times$26.5 & 6   \\
MKO CDK700 & U16                & Mount Kent, Queensland                 & 0.7    & 0.401 & 4\,096$\times$4\,096 & 27.4$\times$27.4 & ... \\
PEST       & SBIG ST-8XME       & Perth, Western Australia               & 0.3048 & 1.23  & 1\,530$\times$1\,020 & 31$\times$21     & 7   \\
Spitzer    & IRAC               & ...                                    & 0.85   & 1.22  & 256$\times$256       & 5.2$\times$5.2   & 8   \\
VLT        & ESPRESSO           & Cerro Paranal, Chile                   & 8.2    & ...   & ...                  & ...              & 9
    \enddata
    \tablerefs{(1) \url{https://astep.oca.eu}; (2) \url{https://www.brierfieldobservatory.com}; (3) \url{https://www.cfht.hawaii.edu}; (4) \url{https://cheops.unibe.ch}; (5) \url{http://irtfweb.ifa.hawaii.edu}; (6) \url{https://lco.global/observatory}; (7) \url{http://pestobservatory.com}; (8) \url{https://www.spitzer.caltech.edu}; (9) \url{https://www.eso.org/public/teles-instr/paranal-observatory/vlt}}
\end{deluxetable*}

\subsection{CHEOPS Photometry}

The CHaracterising ExOPlanet Satellite (CHEOPS) is a space-based telescope whose mission is to search for transits of known exoplanets and recover their radii more accurately, which will then place constraints on atmospheric and interior modeling and formation process \citep{rando2020, benz2021}. \citet{szabo2021, szabo2022} observed seven transits for AU Mic b and two transits for AU Mic c. The CHEOPS light curves have been processed and modeled separately as described in \citet{szabo2021, szabo2022}, and this paper only incorporates the transit midpoint times from those works.

\subsection{Ground-Based Photometry}

All of the ground-based observations listed in Table \ref{tab:datasets} have been coordinated through the TESS Follow-up Observing Program (TFOP) Working Group (WG)\footnote{\url{https://tess.mit.edu/followup}}. Along with the ones mentioned in \citet{wittrock2022}, we added 13 new follow-up photometric transit observations, including one from ASTEP, four from LCO CTIO 1.0 m, three from LCO SAAO 1.0 m, four from LCO SSO 1.0 m, and one from LCO TO 1.0 m. These light curves are available on ExoFOP-TESS\footnote{\url{https://exofop.ipac.caltech.edu/tess}} \citep{akeson2013}. The follow-up observation schedules were conducted with the online version of the {\tt TAPIR} package \citep{jensen2013}. {\tt AstroImageJ} \citep[\aij,][]{collins2017} had been utilized to process the ground-based light curves and then create a subset table containing only BJD\_TDB, normalized detrended flux, flux uncertainty, and detrending parameter columns from the ground-based light curves (e.g. airmass, position centroid, FWHM, etc.). We use these detrending parameter columns for \exofast modeling and extraction of midpoint times to assess the impact of systematic trends in the ground-based light curves on the modeled transit midpoint time posterior distributions ($\S$\ref{sec:exofastmod}). The choice of detrending parameters were modified for some ground transit observations from \citet{wittrock2022} to improve signal RMS and minimize the uncertainty in the timing of the transits; see Table \ref{tab:detrend} for a complete list of detrending parameters applied to each transit observation for this paper.

\subsubsection{ASTEP (FLI Proline 16800E) Photometry}

ASTEP 400, part of the Antarctic Search for Transiting ExoPlanets (ASTEP) program and located at the Concordia Research Station, Antarctica, is a 0.4 m telescope that has been utilized for transiting exoplanet search \citep{guillot2015, mekarnia2016}. The data collected with ASTEP 400 was processed on-site with an IDL-based aperture photometry pipeline \citep{abe2013, mekarnia2016}. Although five AU Mic transit observations were made with ASTEP 400, the photometric conditions during four of those nights were of suboptimal quality, so only the first transit observation of AU Mic b is included in this paper.

\subsubsection{LCOGT (Sinistro) Photometry}

The Las Cumbres Observatory Global Telescope network \citep[LCOGT,][]{brown2013}\footnote{\url{https://lco.global/observatory}} partipated in collecting the transits of the AU Mic systems through four different 1.0 m LCO Ritchey-Chretien Cassegrain telescopes equipped with Sinistro; Pan-STARRS \zs was used with exposure time of 15 seconds for all new LCO observations.

The third night from LCO CTIO was impacted by poor sky conditions. The third transit observation from LCO SAAO was affected by intermittent clouds. Additionally, the 2020 October 11 night from LCO SAAO fortuitously observed the egress of planet c while intending to observe the transit of planet b; this transit was not included in the previous analysis of \citet{wittrock2022} but is now included in this work. The first night from LCO SSO was impacted by poor sky conditions and tracking. All LCOGT light curves have been reduced and detrended with \aij, then a subset table was generated from each light curve.

\subsubsection{MKO CDK700 (U16) Photometry}

The light curve from MKO CDK700 has been reduced and detrended with \aij, then a subset table was generated from it. However, this observation was found to have missed the AU Mic c's predicted transit calculated by the \exofast package. Since we are employing the {\tt rejectflatmodel} option to improve the convergence of our transit models, this set is dropped from the analysis (see $\S$\ref{sec:exofastmod} regarding the use of {\tt rejectflatmodel}).

\section{\exofast Transit Modeling}\label{sec:exofastmod}

We model the 33 photometric transits of AU Mic b and five photometric transits of AU Mic c using the \exofast package \citep{eastman2013, eastman2019}; CHEOPS and R-M observations are not included in this transit modeling step. \exofast computes the Markov chain Monte Carlo (MCMC) to estimate the posterior probabilities and determine the statistical significance of our detections and the confident intervals in their corresponding transit midpoint times. Since \exofast is written in {\tt IDL}, it uses the differential evolution MCMC algorithm \citep{terbraak2006} instead of \emcee \citep{foreman-mackey2013} for sampling purposes \citep{eastman2013}. The detrending parameters flare (Spitzer), sky (Spitzer \& PEST), Sky/Pixel\_T1 (LCOGT), \& SKY (ASTEP) are treated as additive while the remaining detrending parameters are treated as multiplicative. See Table \ref{tab:detrend} for a full list of nights included for \exofast analysis and their corresponding detrending parameters. Since both Pan-STARRS Y and Pan-STARRS \zs are not available among the filters in \exofast, y and z' (Sloan z) were used as respective approximate substitutes.

\startlongtable
\begin{deluxetable*}{l|c|c|c|c}\label{tab:detrend}
    \tablecaption{Detrending parameters incorporated into \exofast modeling of AU Mic b transits. The flare (Spitzer), sky (Spitzer \& PEST), Sky/Pixel\_T1 (LCOGT), \& SKY (ASTEP) were implemented as additives; the remaining detrending parameters were implemented as multiplicative. See $\S$\ref{sec:dataobs} for details on detrending parameters used for each observation. Since both Pan-STARRS Y and Pan-STARRS \zs are not available among the filters in \exofast, y and z' (Sloan z) were used as respective approximate substitutes.}
    \tablehead{Telescope & Date (UT) & Filter & Detrending Parameter(s) & Note}
    \startdata
TESS       & 2018-07-26 & TESS       & ...                                                                   & a \\
TESS       & 2018-08-11 & TESS       & ...                                                                   & a \\
TESS       & 2018-08-12 & TESS       & ...                                                                   & a \\
Spitzer    & 2019-02-10 & 4.5 $\mu$m & x, y, noise/pixel, FWHM\_x, FWHM\_y, sky, linear, quadratic           & b \\
Spitzer    & 2019-02-27 & 4.5 $\mu$m & x, y, noise/pixel, FWHM\_x, FWHM\_y, sky, linear, quadratic, flare    & b \\
Spitzer    & 2019-09-09 & 4.5 $\mu$m & x, y, noise/pixel, FWHM\_x, FWHM\_y, sky, linear, quadratic, Gaussian & b \\
LCO SSO    & 2020-04-25 & z'         & AIRMASS                                                               & c \\
LCO SSO    & 2020-04-25 & y          & AIRMASS                                                               & c \\
LCO SAAO   & 2020-05-20 & z'         & AIRMASS                                                               & c \\
LCO SAAO   & 2020-05-20 & y          & AIRMASS                                                               & c \\
LCO SAAO   & 2020-06-06 & z'         & AIRMASS                                                               & c \\
LCO SAAO   & 2020-06-23 & z'         & AIRMASS, Width\_T1                                                    & c \\
TESS       & 2020-07-09 & TESS       & ...                                                                   & a \\
TESS       & 2020-07-10 & TESS       & ...                                                                   & a \\
PEST       & 2020-07-10 & V          & comp\_flux, dist\_center, fwhm, airmass, sky                          & d \\
TESS       & 2020-07-19 & TESS       & ...                                                                   & a \\
TESS       & 2020-07-27 & TESS       & ...                                                                   & a \\
TESS       & 2020-07-28 & TESS       & ...                                                                   & a \\
Brierfield & 2020-08-13 & I          & Meridian\_Flip, tot\_C\_cnts, X(FITS)\_T1, Y(FITS)\_T1                & c \\
LCO SSO    & 2020-08-13 & z'         & AIRMASS, Width\_T1                                                    & c \\
LCO SAAO   & 2020-09-07 & z'         & Sky/Pixel\_T1, Width\_T1                                              & c \\
LCO SSO    & 2020-09-16 & z'         & AIRMASS, Width\_T1                                                    & c \\
LCO SSO    & 2020-10-03 & z'         & Sky/Pixel\_T1, Width\_T1                                              & c \\
LCO SAAO   & 2020-10-11 & z'         & AIRMASS                                                               & c \\
LCO CTIO   & 2021-06-14 & z'         & tot\_C\_cnts, Sky/Pixel\_T1, Width\_T1                                & c \\
LCO SSO    & 2021-06-22 & z'         & tot\_C\_cnts, Y(FITS)\_T1, Width\_T1                                  & c \\
LCO CTIO   & 2021-07-01 & z'         & tot\_C\_cnts, Sky/Pixel\_T1, Width\_T1                                & c \\
ASTEP      & 2021-07-09 & R          & SKY                                                                   & e \\
LCO SAAO   & 2021-07-18 & z'         & AIRMASS, Sky/Pixel\_T1                                                & c \\
LCO SAAO   & 2021-08-03 & z'         & X(FITS)\_T1, Y(FITS)\_T1, Sky/Pixel\_T1                               & c \\
LCO CTIO   & 2021-08-04 & z'         & AIRMASS                                                               & c \\
LCO SSO    & 2021-08-12 & z'         & Y(FITS)\_T1, Sky/Pixel\_T1, Width\_T1                                 & c \\
LCO SSO    & 2021-08-29 & z'         & tot\_C\_cnts, X(FITS)\_T1, Width\_T1                                  & c \\
LCO SSO    & 2021-09-15 & z'         & AIRMASS, X(FITS)\_T1, Y(FITS)\_T1                                     & c \\
LCO SAAO   & 2021-09-23 & z'         & Sky/Pixel\_T1, Width\_T1                                              & c \\
LCO TO     & 2021-09-23 & z'         & tot\_C\_cnts, Width\_T1                                               & c \\
LCO CTIO   & 2021-10-04 & z'         & AIRMASS, X(FITS)\_T1, Width\_T1                                       & c
    \enddata
    \tablecomments{(a) See \citet{gilbert2022} for details on the detrending parameters applied to TESS data. (b) See \citet{wittrock2022} for details on the detrending parameters applied to Spitzer data. (c) Detrending parameters generated from \aij \citep{collins2017}. (d) Detrending parameters generated from {\tt PEST} pipeline (\url{http://pestobservatory.com/the-pest-pipeline}). (e) Detrending parameters generated from {\tt IDL}-based aperture photometry pipeline \citep{abe2013, mekarnia2016}.}
\end{deluxetable*}

\begin{deluxetable*}{l|c|cc|c}\label{tab:exofastv2priors}
    \tablecaption{Stellar, planetary, and transit priors for \exofast modeling.}
    \tablehead{\multirow{2}{*}{Prior} & \multirow{2}{*}{Unit} & \multicolumn{2}{c|}{Input} & \multirow{2}{*}{Ref} \\
                                      &                       & AU Mic b & AU Mic c        &                      }
    \startdata
\logmstar     & ...      & \multicolumn{2}{c|}{$\mathcal{N}$(-0.301, 0.026)}                           & 1   \\
\rstar        & \rsun    & \multicolumn{2}{c|}{$\mathcal{N}$(0.75, 0.03)}                              & 2   \\
\teff         & K        & \multicolumn{2}{c|}{$\mathcal{N}$(3700, 100)}                               & 3   \\
Age           & Gyr      & \multicolumn{2}{c|}{$\mathcal{N}$(0.022, 0.003)}                            & 4   \\ \hline
\tc           & BJD\_TDB & $\mathcal{N}$(2458330.39080, 0.00058) & $\mathcal{N}$(2458342.2239, 0.0019) & 5   \\
\logfracpd    & ...      & $\mathcal{N}$(0.92752436, 0.00000031) & $\mathcal{N}$(1.2755182, 0.0000012) & 5   \\
\rplan/\rstar & ...      & $\mathcal{N}$(0.0512, 0.0020)         & $\mathcal{N}$(0.0340, 0.0034)       & 5   \\ \hline
TTV Offset    & days     & \multicolumn{2}{c|}{$\mathcal{U}$(-0.02, 0.02)}                             & ... \\
Depth Offset  & ...      & \multicolumn{2}{c|}{$\mathcal{U}$(-0.01, 0.01)}                             & ...
    \enddata
    \tablerefs{(1) \citet{plavchan2020}; (2) \citet{white2019}; (3) \citet{plavchan2009}; (4) \citet{mamajek2014}; (5) \citet{gilbert2022}}
    \tablecomments{$\mathcal{N}$ denotes the Gaussian priors, and $\mathcal{U}$ denotes the uniform priors. TTV and depth offsets are arbitrary and applied as constraints to all transits. The logarithmic version of stellar mass and orbital period were used because they are the fitted priors in \exofast. Equivalent Evolutionary Point (EEP) was set to 1 but is allowed to float freely, so it is not included in the prior table above.}
\end{deluxetable*}

\begin{figure*}
    \centering
    \includegraphics[width=0.31\textwidth]{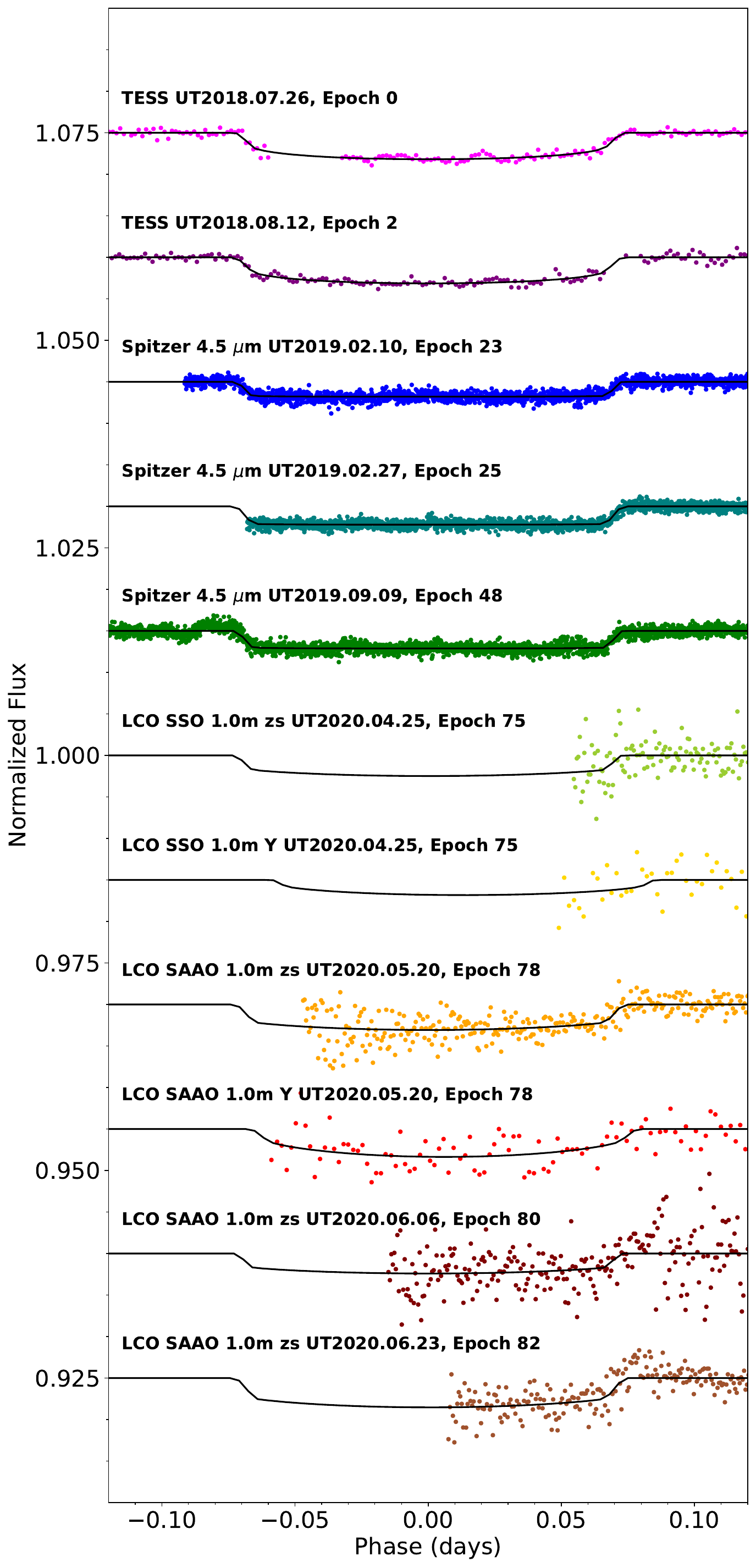}
    \includegraphics[width=0.31\textwidth]{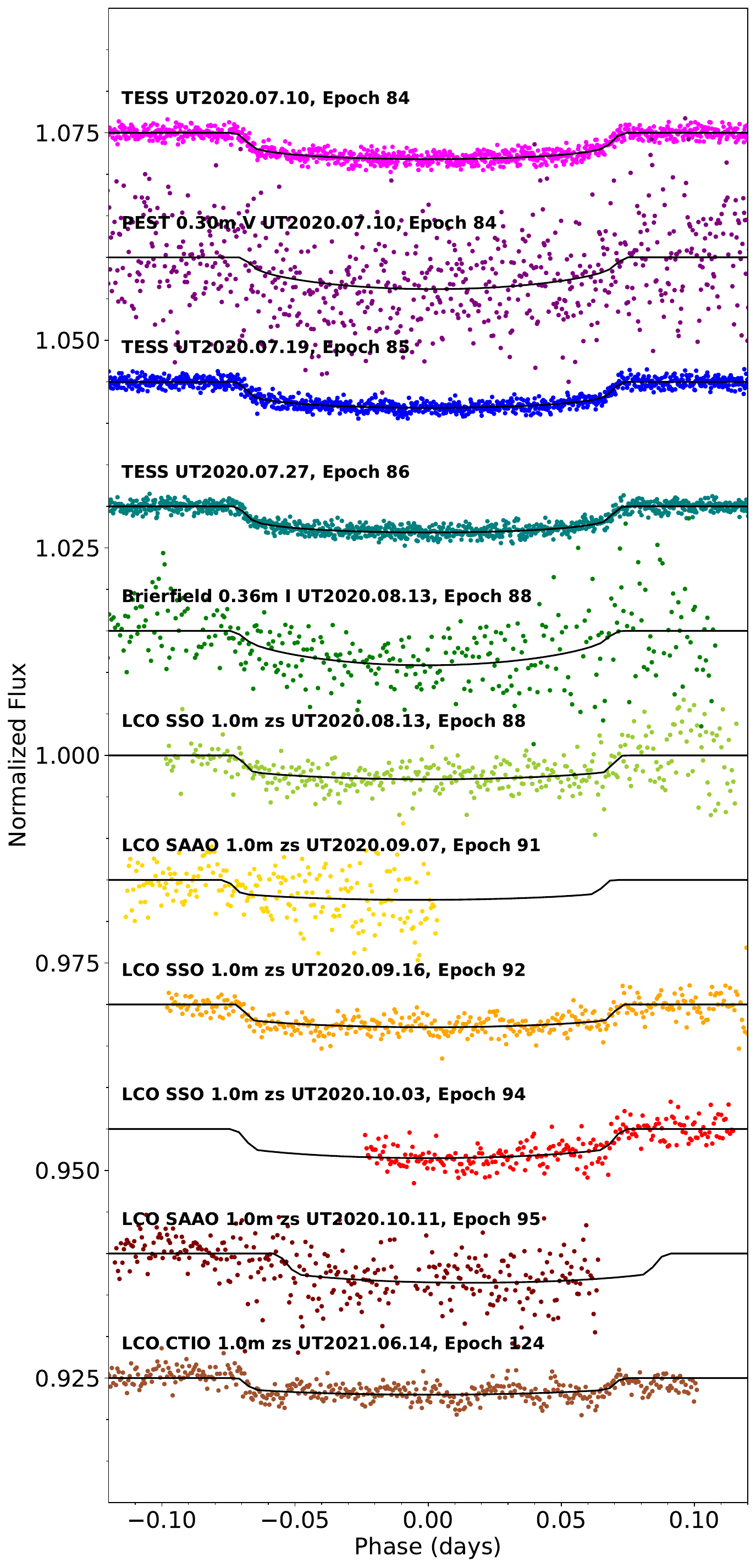}
    \includegraphics[width=0.31\textwidth]{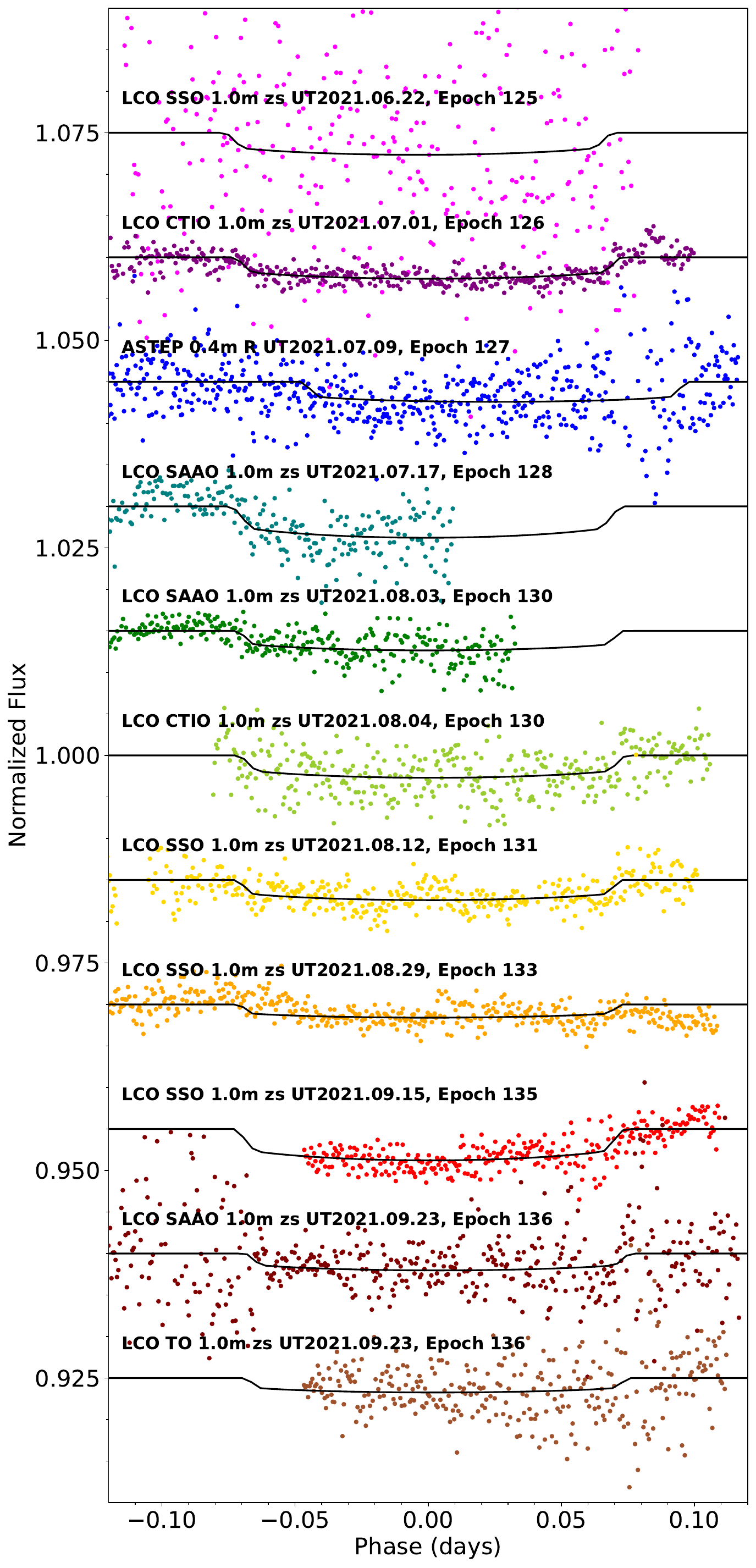} \\
    \includegraphics[width=0.31\textwidth]{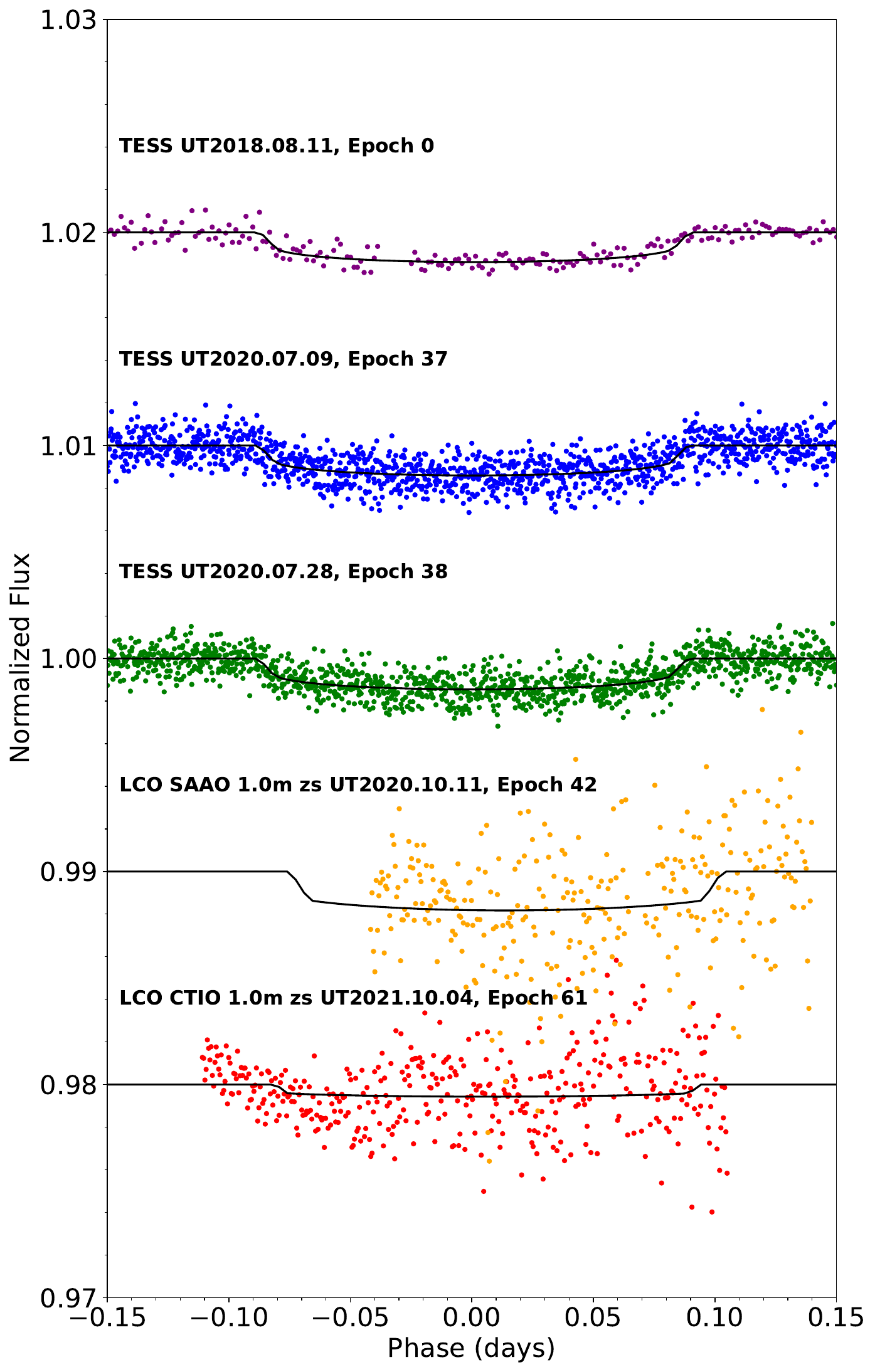}
    \caption{Comparison between ground-based + Spitzer + TESS transits (multi-colors) and \exofast's best fit model (black) for AU Mic b (first three columns in the top row) and c (last column in the bottom row). Each transit is labeled with the name of telescope, the date of observation in UT, and the epoch, which refers to the number of transits since the first transit of b and c, respectively.}
    \label{fig:exofastv2models}
\end{figure*}

We use MIST for evolutionary models \citep{choi2016, dotter2016} and have \exofast ignore the Claret \& Bloemen limb darkening tables \citep{claret2011} since AU Mic is a low-mass red dwarf. Table \ref{tab:exofastv2priors} provides a list of priors for \exofast. The logarithmic version of stellar mass and orbital period were used because they are the fitted priors in \exofast. The purpose of TTV and depth offset priors are to place constraints on the variation of transit timing and depth of all light curves; any transit depth variability was not explored for this paper. We set {\tt MAXSTEPS} = 7\,500 and {\tt NTHIN} = 25 and include the {\tt rejectflatmodel} option for all light curves with {\tt NTEMPS} = 8 to aid in faster convergence. The reason for allowing the {\tt rejectflatmodel} option to become active for this work is that both AU Mic b and c are confirmed transiting planets with well-established orbital periods, so we ``reject'' any flat models to help narrow the vast range of possible outcomes. After \exofast completes the transit modeling, it generates the transit models (Figure \ref{fig:exofastv2models}), median posteriors (Tables \ref{tab:exofastv2median1}, \ref{tab:exofastv2median2}, \ref{tab:exofastv2median3}, \& \ref{tab:exofastv2median4}), and midpoint times (Table \ref{tab:ttvpriors}).

\startlongtable
\begin{deluxetable*}{l|c|c|cc}\label{tab:exofastv2median1}
    \tablecaption{\exofast-generated median values and 68\% confidence interval for AU Mic system.}
    \tablehead{\multirow{2}{*}{Posterior} & \multirow{2}{*}{Description} & \multirow{2}{*}{Unit} & \multicolumn{2}{c}{Output} \\
                                          &                              &                       & AU Mic b & AU Mic c          }
    \startdata
\mstar                     & Stellar Mass                                                      & \msun             & \multicolumn{2}{c}{$0.510^{+0.028}_{-0.027}$}                               \\
\rstar                     & Stellar Radius                                                    & \rsun             & \multicolumn{2}{c}{$0.744^{+0.023}_{-0.021}$}                               \\
\lstar                     & Stellar Luminosity                                                & \lsun             & \multicolumn{2}{c}{$0.0916^{+0.011}_{-0.0098}$}                             \\
$\rho_{\star}$             & Stellar Density                                                   & g cm$^{-3}$       & \multicolumn{2}{c}{$1.75^{+0.14}_{-0.16}$}                                  \\
$\log{g}$                  & Surface Gravity                                                   & \logg             & \multicolumn{2}{c}{$4.404^{+0.026}_{-0.031}$}                               \\
\teff                      & Effective Temperature                                             & K                 & \multicolumn{2}{c}{$3678^{+90}_{-88}$}                                      \\
$[{\rm Fe/H}]$             & Metallicity                                                       & dex               & \multicolumn{2}{c}{$0.23^{+0.24}_{-0.30}$}                                  \\
$[{\rm Fe/H}]_{0}$         & Initial Metallicity\tablenotemark{\scriptsize a}                  & dex               & \multicolumn{2}{c}{$0.17^{+0.22}_{-0.28}$}                                  \\
Age                        & ...                                                               & Gyr               & \multicolumn{2}{c}{$0.0201^{+0.0025}_{-0.0024}$}                            \\
EEP                        & Equal Evolutionary Phase\tablenotemark{\scriptsize b}             & ...               & \multicolumn{2}{c}{$162.0 \pm 2.9$}                                         \\ \hline
\porb                      & Orbital Period                                                    & days              & $8.4630177^{+0.0000052}_{-0.0000050}$ & $18.858970^{+0.000051}_{-0.000052}$ \\
\mplan                     & Planetary Mass\tablenotemark{\scriptsize c}                       & \mjup             & $0.053^{+0.019}_{-0.012}$             & $0.0247^{+0.010}_{-0.0065}$         \\
\rplan                     & Planetary Radius                                                  & \rjup             & $0.353 \pm 0.013$                     & $0.225 \pm 0.022$                   \\
\tc                        & Time of Conjunction\tablenotemark{\scriptsize d}                  & BJD\_TDB          & $2458330.39168 \pm 0.00053$           & $2458342.2240^{+0.0019}_{-0.0018}$  \\
T$_{\rm T}$                & Time of Minimum Projected Separation\tablenotemark{\scriptsize e} & BJD\_TDB          & $2458330.39169 \pm 0.00053$           & $2458342.2240^{+0.0019}_{-0.0018}$  \\
T$_{0}$                    & Optimal Conjunction Time\tablenotemark{\scriptsize f}             & BJD\_TDB          & $2458525.04109^{+0.00052}_{-0.00051}$ & $2458342.2240^{+0.0019}_{-0.0018}$  \\
a                          & Semi-Major Axis                                                   & au                & $0.0649 \pm 0.0012$                   & $0.1108 \pm 0.0020$                 \\
e                          & Eccentricity                                                      & ...               & $0.081^{+0.17}_{-0.058}$              & $0.101^{+0.11}_{-0.066}$            \\
i                          & Inclination                                                       & deg               & $89.57^{+0.28}_{-0.31}$               & $89.43^{+0.35}_{-0.23}$             \\
$\omega$                   & Argument of Periastron                                            & deg               & $-202^{+44}_{-120}$                   & $115^{+66}_{-95}$                   \\
T$_{\rm eq}$               & Equilibrium Temperature\tablenotemark{\scriptsize g}              & K                 & $600^{+17}_{-16}$                     & $459^{+13}_{-12}$                   \\
$\tau_{\rm circ}$          & Tidal Circularization Timescale{\scriptsize h}                    & Gyr               & $161 \pm 76$                          & $22800^{+15000}_{-9500}$            \\
K                          & RV Semi-Amplitude\tablenotemark{\scriptsize c}                    & m s$^{-1}$        & $8.4^{+3.0}_{-1.9}$                   & $2.99^{+1.2}_{-0.78}$               \\
\rplan/\rstar              & ...                                                               & ...               & $0.0488 \pm 0.0010$                   & $0.0311 \pm 0.0028$                 \\
a/\rstar                   & ...                                                               & ...               & $18.79^{+0.50}_{-0.59}$               & $32.05^{+0.86}_{-1.0}$              \\
$\delta$                   & Transit Depth (\rplan/\rstar)$^2$                                 & ...               & $0.002379^{+0.000100}_{-0.000099}$    & $0.00097^{+0.00018}_{-0.00017}$     \\
$\delta_{\rm I}$           & Transit Depth in I                                                & ...               & $0.0039^{+0.0025}_{-0.0011}$          & $0.00156^{+0.00099}_{-0.00047}$     \\
$\delta_{\rm R}$           & Transit Depth in R                                                & ...               & $0.0039^{+0.0031}_{-0.0012}$          & $0.00154^{+0.0012}_{-0.00049}$      \\
$\delta_{\rm z'}$          & Transit Depth in z'                                               & ...               & $0.00310 \pm 0.00023$                 & $0.00123^{+0.00025}_{-0.00022}$     \\
$\delta_{\rm 4.5\mu m}$    & Transit Depth in 4.5 $\mu$m                                       & ...               & $0.00239\pm0.00010$                   & $0.00097^{+0.00019}_{-0.00017}$     \\
$\delta_{\rm TESS}$        & Transit Depth in TESS                                             & ...               & $0.00308\pm0.00017$                   & $0.00123^{+0.00024}_{-0.00022}$     \\
$\delta_{\rm V}$           & Transit Depth in V                                                & ...               & $0.00344^{+0.0012}_{-0.00057}$        & $0.00133^{+0.00057}_{-0.00033}$     \\
$\delta_{\rm y}$           & Transit Depth in y                                                & ...               & $0.0038^{+0.0027}_{-0.0010}$          & $0.00151^{+0.0011}_{-0.00043}$      \\
$\tau$                     & Ingress/Egress Transit Duration                                   & days              & $0.00691^{+0.00028}_{-0.00020}$       & $0.00589^{+0.00086}_{-0.00069}$     \\
T$_{14}$                   & Total Transit Duration                                            & days              & $0.14553^{+0.00031}_{-0.00028}$       & $0.1765 \pm 0.0012$                 \\
T$_{\rm FWHM}$             & FWHM Transit Duration                                             & days              & $0.13859 \pm 0.00019$                 & $0.17053^{+0.00098}_{-0.00097}$     \\
b                          & Transit Impact Parameter                                          & ...               & $0.134^{+0.096}_{-0.089}$             & $0.30^{+0.13}_{-0.19}$              \\
b$_{\rm S}$                & Eclipse Impact Parameter                                          & ...               & $0.136^{+0.095}_{-0.090}$             & $0.32^{+0.11}_{-0.19}$              \\
$\tau_{\rm S}$             & Ingress/Egress Eclipse Duration                                   & days              & $0.00711^{+0.00052}_{-0.00054}$       & $0.00631^{+0.00076}_{-0.00072}$     \\
T$_{\rm S,14}$             & Total Eclipse Duration                                            & days              & $0.1486^{+0.0099}_{-0.010}$           & $0.186^{+0.018}_{-0.012}$           \\
T$_{\rm S,FWHM}$           & FWHM Eclipse Duration                                             & days              & $0.1415^{+0.0095}_{-0.0097}$          & $0.180^{+0.018}_{-0.011}$           \\
$\delta_{\rm S,2.5 \mu m}$ & Blackbody Eclipse Depth at 2.5 $\mu$m                             & ppm               & $0.62^{+0.15}_{-0.13}$                & $0.0134^{+0.0055}_{-0.0040}$        \\
$\delta_{\rm S,5.0 \mu m}$ & Blackbody Eclipse Depth at 5.0 $\mu$m                             & ppm               & $23.6^{+2.8}_{-2.5}$                  & $2.19^{+0.56}_{-0.46}$              \\
$\delta_{\rm S,7.5 \mu m}$ & Blackbody Eclipse Depth at 7.5 $\mu$m                             & ppm               & $69.6^{+5.6}_{-5.2}$                  & $10.4^{+2.3}_{-2.0}$                \\
$\rho_{\rm p}$             & Planetary Density\tablenotemark{\scriptsize c}                    & g cm$^{-3}$       & $1.48^{+0.52}_{-0.34}$                & $2.70^{+1.0}_{-0.67}$               \\
$\log{g_{\rm p}}$          & Surface Gravity\tablenotemark{\scriptsize c}                      & ...               & $3.02^{+0.13}_{-0.11}$                & $3.08^{+0.13}_{-0.11}$              \\
$\Theta$                   & Safronov Number                                                   & ...               & $0.0380^{+0.013}_{-0.0085}$           & $0.047^{+0.017}_{-0.011}$           \\
$\langle F \rangle$        & Incident Flux                                                     & 10$^{9}$ \ergfrac & $0.0288^{+0.0034}_{-0.0033}$          & $0.0099^{+0.0012}_{-0.0011}$        \\
T$_{\rm P}$                & Time of Periastron                                                & BJD\_TDB          & $2458322.77^{+0.83}_{-2.3}$           & $2458324.0^{+2.9}_{-3.7}$           \\
T$_{\rm S}$                & Time of Eclipse                                                   & BJD\_TDB          & $2458334.56^{+0.52}_{-0.93}$          & $2458351.6^{+1.5}_{-1.6}$           \\
T$_{\rm A}$                & Time of Ascending Node                                            & BJD\_TDB          & $2458328.28^{+0.29}_{-0.47}$          & $2458337.67^{+0.81}_{-0.75}$        \\
T$_{\rm D}$                & Time of Descending Node                                           & BJD\_TDB          & $2458332.42^{+0.26}_{-0.44}$          & $2458346.73^{+0.71}_{-0.83}$        \\
V$_{\rm c}$/V$_{\rm e}$    & {\it i}                                                           & ...               & $0.980^{+0.022}_{-0.033}$             & $0.962^{+0.037}_{-0.044}$           \\
\radroot                   & Transit Chord                                                     & ...               & $1.0401^{+0.0074}_{-0.017}$           & $0.985^{+0.039}_{-0.050}$           \\
sign                       & {\it j}                                                           & ...               & $0.51^{+0.41}_{-0.26}$                & $0.51^{+0.39}_{-0.27}$              \\
e$\cos{\omega}$            & ...                                                               & ...               & $-0.011^{+0.097}_{-0.17}$             & $-0.00^{+0.12}_{-0.13}$             \\
e$\sin{\omega}$            & ...                                                               & ...               & $0.010^{+0.033}_{-0.036}$             & $0.030^{+0.047}_{-0.037}$           \\
\mplan$\sin{i}$            & Minimum Mass\tablenotemark{\scriptsize c}                         & \mjup             & $0.053^{+0.019}_{-0.012}$             & $0.0247^{+0.010}_{-0.0065}$         \\
\mplan/\mstar              & Mass Ratio\tablenotemark{\scriptsize c}                           & ...               & $0.000099^{+0.000036}_{-0.000023}$    & $0.000046^{+0.000019}_{-0.000012}$  \\
d/\rstar                   & Separation at Mid Transit                                         & ...               & $18.19^{+0.96}_{-1.3}$                & $30.5^{+1.8}_{-2.0}$                \\
P$_{\rm T}$                & A Priori Non-grazing Transit Probability                          & ...               & $0.0523^{+0.0042}_{-0.0026}$          & $0.0317^{+0.0023}_{-0.0018}$        \\
P$_{\rm T,G}$              & A Priori Transit Probability                                      & ...               & $0.0577^{+0.0046}_{-0.0029}$          & $0.0338^{+0.0024}_{-0.0019}$        \\
P$_{\rm S}$                & A Priori Non-grazing Eclipse Probability                          & ...               & $0.05024^{+0.0045}_{-0.00099}$        & $0.0297^{+0.0021}_{-0.0014}$        \\
P$_{\rm S,G}$              & A Priori Eclipse Probability                                      & ...               & $0.0554^{+0.0050}_{-0.0011}$          & $0.0316^{+0.0023}_{-0.0015}$        \\ \hline
u$_{\rm 1,I}$              & Linear Limb-darkening Coefficient in I                            & ...               & \multicolumn{2}{c}{$0.80^{+0.47}_{-0.48}$}                                  \\
u$_{\rm 1,R}$              & Linear Limb-darkening Coefficient in R                            & ...               & \multicolumn{2}{c}{$0.80^{+0.53}_{-0.56}$}                                  \\
u$_{\rm 1,z'}$             & Linear Limb-darkening Coefficient in z'                           & ...               & \multicolumn{2}{c}{$0.477^{+0.094}_{-0.10}$}                                \\
u$_{\rm 1,4.5\mu m}$       & Linear Limb-darkening Coefficient in 4.5 $\mu$m                   & ...               & \multicolumn{2}{c}{$0.0076^{+0.012}_{-0.0057}$}                             \\
u$_{\rm 1,TESS}$           & Linear Limb-darkening Coefficient in TESS                         & ...               & \multicolumn{2}{c}{$0.467^{+0.057}_{-0.058}$}                               \\
u$_{\rm 1,V}$              & Linear Limb-darkening Coefficient in V                            & ...               & \multicolumn{2}{c}{$0.63^{+0.46}_{-0.42}$}                                  \\
u$_{\rm 1,y}$              & Linear Limb-darkening Coefficient in y                            & ...               & \multicolumn{2}{c}{$0.76^{+0.52}_{-0.47}$}                                  \\
u$_{\rm 2,I}$              & Quadratic Limb-darkening Coefficient in I                         & ...               & \multicolumn{2}{c}{$-0.10^{+0.49}_{-0.40}$}                                 \\
u$_{\rm 2,R}$              & Quadratic Limb-darkening Coefficient in R                         & ...               & \multicolumn{2}{c}{$-0.13^{+0.41}_{-0.40}$}                                 \\
u$_{\rm 2,z'}$             & Quadratic Limb-darkening Coefficient in z'                        & ...               & \multicolumn{2}{c}{$-0.174^{+0.12}_{-0.078}$}                               \\
u$_{\rm 2,4.5\mu m}$       & Quadratic Limb-darkening Coefficient in 4.5 $\mu$m                & ...               & \multicolumn{2}{c}{$0.067 \pm 0.038$}                                       \\
u$_{\rm 2,TESS}$           & Quadratic Limb-darkening Coefficient in TESS                      & ...               & \multicolumn{2}{c}{$0.186^{+0.092}_{-0.091}$}                               \\
u$_{\rm 2,V}$              & Quadratic Limb-darkening Coefficient in V                         & ...               & \multicolumn{2}{c}{$0.15^{+0.45}_{-0.47}$}                                  \\
u$_{\rm 2,y}$              & Quadratic Limb-darkening Coefficient in y                         & ...               & \multicolumn{2}{c}{$-0.08^{+0.47}_{-0.41}$}
    \enddata
\tablecomments{See Table 3 in \citet{eastman2019} for a detailed description of all parameters. Since both Pan-STARRS Y and Pan-STARRS \zs are not available among the filters in \exofast, y and z' (Sloan z) were used as respective approximate substitutes. Additionally, the Claret \& Bloemen limb darkening tables \citep{claret2011} default option has been disabled since AU Mic is a low-mass red dwarf.}
\tablenotetext{a}{The metallicity of the star at birth.}
\tablenotetext{b}{Corresponds to static points in a star's evolutionary history. See $\S$2 in \citet{dotter2016}.}
\tablenotetext{c}{Uses measured radius and estimated mass from \citet{chen2017}.}
\tablenotetext{d}{Time of conjunction is commonly reported as the ``transit time''.}
\tablenotetext{e}{Time of minimum projected separation is a more correct ``transit time''.}
\tablenotetext{f}{Optimal time of conjunction minimizes the covariance between \tc and Period.}
\tablenotetext{g}{Assumes no albedo and perfect redistribution.}
\tablenotetext{h}{Depends on the tidal Q factor.}
\tablenotetext{i}{The velocity at \tc of an assumed circular orbit divided by the velocity of the modeled eccentric orbit.}
\tablenotetext{j}{The sign of the solution to the quadratic mapping from V$_{\rm c}$/V$_{\rm e}$ to e.}
\end{deluxetable*}

\begin{longrotatetable}
\begin{deluxetable*}{c|l|c|c|c|c|c|c}\label{tab:exofastv2median2}
    \tablecaption{\exofast-generated median values and 68\% confidence interval for follow-up observations of AU Mic transits (Part I).}
    \tablehead{Planet & Telescope & Date (UT) & Filter & $\sigma^{2}$ (Added Variance) & TTV\tablenotemark{a} (days) & T$\delta$ V\tablenotemark{b} & F$_{0}$\tablenotemark{c}}
    \startdata
b      & TESS       & 2018-07-26 & TESS       & $0.000000010^{+0.000000012}_{-0.000000011}$    & $-0.00242^{+0.00065}_{-0.00066}$ & $0.0020 \pm 0.0011$              & $0.999994 \pm 0.000030$            \\
c      & TESS       & 2018-08-11 & TESS       & $0.000000006^{+0.000000012}_{-0.000000010}$    & $-0.0007 \pm 0.0020$             & $0.0030 \pm 0.0029$              & $0.999992 \pm 0.000030$            \\
b      & TESS       & 2018-08-12 & TESS       & $0.000000003^{+0.000000017}_{-0.000000015}$    & $-0.00063 \pm 0.00075$           & $0.0021 \pm 0.0011$              & $1.000012 \pm 0.000039$            \\
b      & Spitzer    & 2019-02-10 & 4.5 $\mu$m & $0.0000000700^{+0.0000000056}_{-0.0000000054}$ & $0.00336^{+0.00053}_{-0.00054}$  & $-0.0063 \pm 0.0012$             & $1.000028^{+0.000034}_{-0.000033}$ \\
b      & Spitzer    & 2019-02-27 & 4.5 $\mu$m & $0.0000000384^{+0.0000000043}_{-0.0000000041}$ & $0.00413^{+0.00052}_{-0.00054}$  & $-0.0024 \pm 0.0011$             & $1.000312 \pm 0.000027$            \\
b      & Spitzer    & 2019-09-09 & 4.5 $\mu$m & $0.0000001237^{+0.0000000061}_{-0.0000000058}$ & $0.00082^{+0.00054}_{-0.00055}$  & $-0.0025 \pm 0.0012$             & $0.999765 \pm 0.000027$            \\
b      & LCO SSO    & 2020-04-25 & z'         & $0.00000275^{+0.00000033}_{-0.00000029}$       & $-0.0047^{+0.0016}_{-0.0014}$    & $-0.0033^{+0.0054}_{-0.0044}$    & $0.99999 \pm 0.00014$              \\
b      & LCO SSO    & 2020-04-25 & y          & $0.00000066^{+0.0000013}_{-0.00000093}$        & $0.003 \pm 0.010$                & $0.0005^{+0.0065}_{-0.0070}$     & $1.00009^{+0.00048}_{-0.00044}$    \\
b      & LCO SAAO   & 2020-05-20 & z'         & $0.00000188^{+0.00000017}_{-0.00000016}$       & $-0.0048^{+0.0018}_{-0.0016}$    & $0.0040^{+0.0023}_{-0.0024}$     & $1.00034 \pm 0.00017$              \\
b      & LCO SAAO   & 2020-05-20 & y          & $0.00000141^{+0.00000054}_{-0.00000043}$       & $0.0088^{+0.0065}_{-0.0041}$     & $0.0005^{+0.0051}_{-0.0053}$     & $0.99984^{+0.00038}_{-0.00037}$    \\
b      & LCO SAAO   & 2020-06-06 & z'         & $0.00000673^{+0.00000069}_{-0.00000062}$       & $-0.0079 \pm 0.0032$             & $0.0000^{+0.0053}_{-0.0055}$     & $0.99995^{+0.00035}_{-0.00034}$    \\
b      & LCO SAAO   & 2020-06-23 & z'         & $0.00000189^{+0.00000024}_{-0.00000021}$       & $-0.0019 \pm 0.0014$             & $0.0075^{+0.0018}_{-0.0029}$     & $0.99982^{+0.00016}_{-0.00019}$    \\
c      & TESS       & 2020-07-09 & TESS       & $-0.000000003^{+0.000000016}_{-0.000000015}$   & $-0.0011^{+0.0029}_{-0.0026}$    & $0.0034 \pm 0.0029$              & $1.000009 \pm 0.000024$            \\
b      & TESS       & 2020-07-10 & TESS       & $0.000000005^{+0.000000014}_{-0.000000013}$    & $-0.00393^{+0.00063}_{-0.00067}$ & $0.0020 \pm 0.0011$              & $1.000002 \pm 0.000021$            \\
b      & PEST       & 2020-07-10 & V          & $0.0000163^{+0.0000013}_{-0.0000012}$          & $-0.0116^{+0.0033}_{-0.0034}$    & $0.00879^{+0.00090}_{-0.0020}$   & $0.99931^{+0.00020}_{-0.00021}$    \\
b      & TESS       & 2020-07-19 & TESS       & $-0.000000005^{+0.000000012}_{-0.000000011}$   & $-0.00326^{+0.00064}_{-0.00066}$ & $0.0020 \pm 0.0011$              & $0.999998 \pm 0.000020$            \\
b      & TESS       & 2020-07-27 & TESS       & $-0.000000001^{+0.000000012}_{-0.000000011}$   & $-0.00217^{+0.00063}_{-0.00066}$ & $0.0019^{+0.0011}_{-0.0010}$     & $0.999994 \pm 0.000020$            \\
c      & TESS       & 2020-07-28 & TESS       & $0.000000010^{+0.000000014}_{-0.000000013}$    & $0.0008^{+0.0029}_{-0.0026}$     & $0.0033 \pm 0.0029$              & $1.000004 \pm 0.000022$            \\
b      & Brierfield & 2020-08-13 & I          & $0.0000171^{+0.0000015}_{-0.0000013}$          & $0.0074^{+0.0039}_{-0.0030}$     & $0.0047^{+0.0035}_{-0.0051}$     & $0.99996^{+0.00031}_{-0.00034}$    \\
b      & LCO SSO    & 2020-08-13 & z'         & $0.00000406^{+0.00000035}_{-0.00000032}$       & $-0.0035^{+0.0018}_{-0.0021}$    & $0.0013^{+0.0034}_{-0.0035}$     & $1.00003^{+0.00026}_{-0.00025}$    \\
b      & LCO SAAO   & 2020-09-07 & z'         & $0.00000688^{+0.00000077}_{-0.00000067}$       & $0.0060^{+0.0090}_{-0.0072}$     & $-0.0038^{+0.0049}_{-0.0040}$    & $0.99972^{+0.00043}_{-0.00034}$    \\
b      & LCO SSO    & 2020-09-16 & z'         & $0.000001191^{+0.00000011}_{-0.000000100}$     & $-0.00314^{+0.00093}_{-0.00097}$ & $-0.0009 \pm 0.0021$             & $1.00000 \pm 0.00013$              \\
b      & LCO SSO    & 2020-10-03 & z'         & $0.00000125^{+0.00000016}_{-0.00000014}$       & $-0.0013 \pm 0.0011$             & $0.0071^{+0.0017}_{-0.0021}$     & $1.00225^{+0.00016}_{-0.00017}$    \\
b \& c & LCO SAAO   & 2020-10-11 & z'         & $0.00000613^{+0.00000063}_{-0.00000056}$       & $-0.002^{+0.017}_{-0.013}$       & $0.0044^{+0.0031}_{-0.0036}$     & $1.00301^{+0.00045}_{-0.00048}$    \\
b      & LCO CTIO   & 2021-06-14 & z'         & $0.000000740^{+0.000000075}_{-0.000000068}$    & $0.01230^{+0.00099}_{-0.0010}$   & $-0.0076^{+0.0018}_{-0.0015}$    & $1.000026^{+0.000092}_{-0.000086}$ \\
b      & LCO SSO    & 2021-06-22 & z'         & $0.0001177^{+0.000010}_{-0.0000095}$           & $-0.0086^{+0.013}_{-0.0070}$     & $-0.0012^{+0.0070}_{-0.0060}$    & $0.99832^{+0.00076}_{-0.00074}$    \\
b      & LCO CTIO   & 2021-07-01 & z'         & $0.000000980^{+0.000000089}_{-0.000000082}$    & $0.0139^{+0.0011}_{-0.0010}$     & $0.0009 \pm 0.0019$              & $1.00003^{+0.00012}_{-0.00011}$    \\
b      & ASTEP      & 2021-07-09 & R          & $0.00001316^{+0.00000087}_{-0.00000081}$       & $-0.0130^{+0.020}_{-0.0045}$     & $-0.0050^{+0.0036}_{-0.0030}$    & $0.99998^{+0.00021}_{-0.00020}$    \\
b      & LCO SAAO   & 2021-07-17 & z'         & $0.00000535^{+0.00000054}_{-0.00000049}$       & $0.0131^{+0.0022}_{-0.0018}$     & $0.0087^{+0.0010}_{-0.0020}$     & $0.99904^{+0.00020}_{-0.00021}$    \\
b      & LCO SAAO   & 2021-08-03 & z'         & $0.00000215^{+0.00000022}_{-0.00000020}$       & $0.0179^{+0.0014}_{-0.0018}$     & $-0.0039^{+0.0045}_{-0.0039}$    & $0.99990^{+0.00028}_{-0.00023}$    \\
b      & LCO CTIO   & 2021-08-04 & z'         & $0.00000574^{+0.00000047}_{-0.00000043}$       & $0.0078^{+0.0024}_{-0.0030}$     & $0.0017^{+0.0030}_{-0.0031}$     & $1.00031 \pm 0.00026$              \\
b      & LCO SSO    & 2021-08-12 & z'         & $0.00000264^{+0.00000024}_{-0.00000022}$       & $0.0132^{+0.0015}_{-0.0016}$     & $-0.0029^{+0.0032}_{-0.0033}$    & $0.99992^{+0.00021}_{-0.00020}$    \\
b      & LCO SSO    & 2021-08-29 & z'         & $0.00000197^{+0.00000016}_{-0.00000015}$       & $0.01952^{+0.00036}_{-0.00072}$  & $-0.00973^{+0.00042}_{-0.00020}$ & $0.999574^{+0.000088}_{-0.000087}$ \\
b      & LCO SSO    & 2021-09-15 & z'         & $0.00000194^{+0.00000020}_{-0.00000018}$       & $0.01954^{+0.00034}_{-0.00069}$  & $0.00914^{+0.00063}_{-0.0012}$   & $0.99952 \pm 0.00015$              \\
b      & LCO SAAO   & 2021-09-23 & z'         & $0.0000181^{+0.0000014}_{-0.0000013}$          & $0.0124^{+0.0030}_{-0.011}$      & $-0.0040^{+0.0044}_{-0.0038}$    & $1.00029^{+0.00031}_{-0.00029}$    \\
b      & LCO TO     & 2021-09-23 & z'         & $0.0000106^{+0.0000012}_{-0.0000010}$          & $0.0169^{+0.0022}_{-0.0035}$     & $0.0001^{+0.0061}_{-0.0064}$     & $0.99959^{+0.00053}_{-0.00048}$    \\
c      & LCO CTIO   & 2021-10-04 & z'         & $0.00000273^{+0.00000022}_{-0.00000020}$       & $-0.0105^{+0.015}_{-0.0066}$     & $-0.0077^{+0.0031}_{-0.0017}$    & $1.00047^{+0.00018}_{-0.00015}$
    \enddata
\tablecomments{See Table 3 in \citet{eastman2019} for a detailed description of all parameters. Since both Pan-STARRS Y and Pan-STARRS \zs are not available among the filters in \exofast, y and z' (Sloan z) were used as respective approximate substitutes.}
\tablenotetext{a}{Transit Timing Variation}
\tablenotetext{b}{Transit Depth Variation}
\tablenotetext{c}{Baseline Flux}
\end{deluxetable*}
\end{longrotatetable}

\begin{longrotatetable}
\begin{deluxetable*}{c|l|c|c|c|c|c|c|c}\label{tab:exofastv2median3}
    \tablecaption{\exofast-generated median values and 68\% confidence interval for follow-up observations of AU Mic b (Part II).}
    \tablehead{Planet & Telescope & Date (UT) & Filter & C$_{0}$\tablenotemark{a} & C$_{1}$\tablenotemark{a} & M$_{0}$\tablenotemark{b} & M$_{1}$\tablenotemark{b} & M$_{2}$\tablenotemark{b}}
    \startdata
b      & TESS       & 2018-07-26 & TESS       & ...                              & ...                     & ...                              & ...                              & ...                              \\
c      & TESS       & 2018-08-11 & TESS       & ...                              & ...                     & ...                              & ...                              & ...                              \\
b      & TESS       & 2018-08-12 & TESS       & ...                              & ...                     & ...                              & ...                              & ...                              \\
b      & Spitzer    & 2019-02-10 & 4.5 $\mu$m & $-0.000148 \pm 0.000046$         & ...                     & $-0.00097 \pm 0.00036$           & $-0.0105 \pm 0.0014$             & $0.0121 \pm 0.0014$              \\
b      & Spitzer    & 2019-02-27 & 4.5 $\mu$m & $-0.000271 \pm 0.000046$         & $0.000077 \pm 0.000031$ & $0.00021^{+0.00040}_{-0.00041}$  & $0.00037 \pm 0.00044$            & $-0.00208^{+0.00092}_{-0.00089}$ \\
b      & Spitzer    & 2019-09-09 & 4.5 $\mu$m & $-0.00051 \pm 0.00014$           & ...                     & $-0.00129 \pm 0.00032$           & $0.00280 \pm 0.00027$            & $0.0020 \pm 0.0017$              \\
b      & LCO SSO    & 2020-04-25 & z'         & ...                              & ...                     & $-0.00009^{+0.00038}_{-0.00035}$ & ...                              & ...                              \\
b      & LCO SSO    & 2020-04-25 & y          & ...                              & ...                     & $-0.00005^{+0.0010}_{-0.00099}$  & ...                              & ...                              \\
b      & LCO SAAO   & 2020-05-20 & z'         & ...                              & ...                     & $0.00104 \pm 0.00031$            & ...                              & ...                              \\
b      & LCO SAAO   & 2020-05-20 & y          & ...                              & ...                     & $-0.00034^{+0.00070}_{-0.00073}$ & ...                              & ...                              \\
b      & LCO SAAO   & 2020-06-06 & z'         & ...                              & ...                     & $0.00000^{+0.00070}_{-0.00065}$  & ...                              & ...                              \\
b      & LCO SAAO   & 2020-06-23 & z'         & ...                              & ...                     & $-0.00032^{+0.00049}_{-0.00052}$ & $-0.00007^{+0.00056}_{-0.00055}$ & ...                              \\
c      & TESS       & 2020-07-09 & TESS       & ...                              & ...                     & ...                              & ...                              & ...                              \\
b      & TESS       & 2020-07-10 & TESS       & ...                              & ...                     & ...                              & ...                              & ...                              \\
b      & PEST       & 2020-07-10 & V          & $-0.0043 \pm 0.0025$             & ...                     & $0.00017^{+0.00076}_{-0.00074}$  & $0.00026 \pm 0.00080$            & $0.0007^{+0.0012}_{-0.0011}$     \\
b      & TESS       & 2020-07-19 & TESS       & ...                              & ...                     & ...                              & ...                              & ...                              \\
b      & TESS       & 2020-07-27 & TESS       & ...                              & ...                     & ...                              & ...                              & ...                              \\
c      & TESS       & 2020-07-28 & TESS       & ...                              & ...                     & ...                              & ...                              & ...                              \\
b      & Brierfield & 2020-08-13 & I          & ...                              & ...                     & $0.0036 \pm 0.0055$              & $-0.0021 \pm 0.0018$             & $0.0032^{+0.0050}_{-0.0051}$     \\
b      & LCO SSO    & 2020-08-13 & z'         & ...                              & ...                     & $-0.00003 \pm 0.00039$           & $0.00001^{+0.00027}_{-0.00026}$  & ...                              \\
b      & LCO SAAO   & 2020-09-07 & z'         & $-0.00021 \pm 0.00060$           & ...                     & $-0.00056^{+0.00097}_{-0.00089}$ & ...                              & ...                              \\
b      & LCO SSO    & 2020-09-16 & z'         & ...                              & ...                     & $0.00002 \pm 0.00035$            & $0.00001^{+0.00032}_{-0.00033}$  & ...                              \\
b      & LCO SSO    & 2020-10-03 & z'         & $-0.00153 \pm 0.00019$           & ...                     & $-0.00134 \pm 0.00026$           & ...                              & ...                              \\
b \& c & LCO SAAO   & 2020-10-11 & z'         & ...                              & ...                     & $0.00073^{+0.00082}_{-0.0012}$   & ...                              & ...                              \\
b      & LCO CTIO   & 2021-06-14 & z'         & $-0.00003 \pm 0.00017$           & ...                     & $0.00002 \pm 0.00037$            & $-0.00001 \pm 0.00022$           & ...                              \\
b      & LCO SSO    & 2021-06-22 & z'         & ...                              & ...                     & $0.0014^{+0.0020}_{-0.0019}$     & $0.0016 \pm 0.0016$              & $-0.0006^{+0.0021}_{-0.0022}$    \\
b      & LCO CTIO   & 2021-07-01 & z'         & $-0.00004 \pm 0.00018$           & ...                     & $0.00002 \pm 0.00049$            & $0.00001 \pm 0.00017$            & ...                              \\
b      & ASTEP      & 2021-07-09 & R          & $-0.00001 \pm 0.00044$           & ...                     & ...                              & ...                              & ...                              \\
b      & LCO SAAO   & 2021-07-17 & z'         & $-0.00088^{+0.00065}_{-0.00067}$ & ...                     & $-0.00100^{+0.00068}_{-0.00067}$ & ...                              & ...                              \\
b      & LCO SAAO   & 2021-08-03 & z'         & $-0.00002^{+0.00048}_{-0.00043}$ & ...                     & $-0.00033^{+0.00057}_{-0.00056}$ & $0.00004 \pm 0.00035$            & ...                              \\
b      & LCO CTIO   & 2021-08-04 & z'         & ...                              & ...                     & $-0.00022^{+0.00046}_{-0.00045}$ & ...                              & ...                              \\
b      & LCO SSO    & 2021-08-12 & z'         & $0.00032^{+0.00056}_{-0.00057}$  & ...                     & $-0.00000 \pm 0.00023$           & $-0.00020^{+0.00030}_{-0.00029}$ & ...                              \\
b      & LCO SSO    & 2021-08-29 & z'         & ...                              & ...                     & $-0.00301 \pm 0.00050$           & $0.00131 \pm 0.00022$            & $-0.00131 \pm 0.00048$           \\
b      & LCO SSO    & 2021-09-15 & z'         & ...                              & ...                     & $0.00081 \pm 0.00033$            & $-0.00047 \pm 0.00047$           & $-0.00001^{+0.00037}_{-0.00038}$ \\
b      & LCO SAAO   & 2021-09-23 & z'         & $-0.00038^{+0.00061}_{-0.00066}$ & ...                     & $0.00095^{+0.00077}_{-0.00080}$  & ...                              & ...                              \\
b      & LCO TO     & 2021-09-23 & z'         & ...                              & ...                     & $0.00052^{+0.00059}_{-0.00060}$  & $0.00002^{+0.00071}_{-0.00074}$  & ...                              \\
c      & LCO CTIO   & 2021-10-04 & z'         & ...                              & ...                     & $-0.00054^{+0.00070}_{-0.00071}$ & $0.00009 \pm 0.00065$            & $-0.00070^{+0.00062}_{-0.00064}$
    \enddata
\tablecomments{See Table 3 in \citet{eastman2019} for a detailed description of all parameters. Since both Pan-STARRS Y and Pan-STARRS \zs are not available among the filters in \exofast, y and z' (Sloan z) were used as respective approximate substitutes. Additionally, the detrending parameters flare (Spitzer), sky (Spitzer \& PEST), Sky/Pixel\_T1 (LCOGT), \& SKY (ASTEP) were set as additive while the remaining detrending parameters were set as multiplicative.}
\tablenotetext{a}{Additive Detrending Coefficient}
\tablenotetext{b}{Multiplicative Detrending Coefficient}
\end{deluxetable*}
\end{longrotatetable}

\begin{longrotatetable}
\begin{deluxetable*}{c|l|c|c|c|c|c|c|c}\label{tab:exofastv2median4}
    \tablecaption{\exofast-generated median values and 68\% confidence interval for follow-up observations of AU Mic b (Part III).}
    \tablehead{Planet & Telescope & Date (UT) & Filter & M$_{3}$\tablenotemark{a} & M$_{4}$\tablenotemark{a} & M$_{5}$\tablenotemark{a} & M$_{6}$\tablenotemark{a} & M$_{7}$\tablenotemark{a}}
    \startdata
b      & TESS       & 2018-07-26 & TESS       & ...                             & ...                             & ...                                & ...                                 & ...                                 \\
c      & TESS       & 2018-08-11 & TESS       & ...                             & ...                             & ...                                & ...                                 & ...                                 \\
b      & TESS       & 2018-08-12 & TESS       & ...                             & ...                             & ...                                & ...                                 & ...                                 \\
b      & Spitzer    & 2019-02-10 & 4.5 $\mu$m & $-0.00618 \pm 0.00092$          & $0.00376 \pm 0.00064$           & $-0.000281 \pm 0.000045$           & $-0.000316^{+0.000062}_{-0.000061}$ & ...                                 \\
b      & Spitzer    & 2019-02-27 & 4.5 $\mu$m & $0.00226^{+0.00088}_{-0.00090}$ & $0.00061^{+0.00071}_{-0.00072}$ & $0.00065 \pm 0.00022$              & $-0.00065 \pm 0.00024$              & ...                                 \\
b      & Spitzer    & 2019-09-09 & 4.5 $\mu$m & $-0.00183 \pm 0.00091$          & $-0.0039 \pm 0.0013$            & $0.000004^{+0.000031}_{-0.000032}$ & $-0.000486^{+0.000095}_{-0.000094}$ & $-0.000761^{+0.000088}_{-0.000086}$ \\
b      & LCO SSO    & 2020-04-25 & z'         & ...                             & ...                             & ...                                & ...                                 & ...                                 \\
b      & LCO SSO    & 2020-04-25 & y          & ...                             & ...                             & ...                                & ...                                 & ...                                 \\
b      & LCO SAAO   & 2020-05-20 & z'         & ...                             & ...                             & ...                                & ...                                 & ...                                 \\
b      & LCO SAAO   & 2020-05-20 & y          & ...                             & ...                             & ...                                & ...                                 & ...                                 \\
b      & LCO SAAO   & 2020-06-06 & z'         & ...                             & ...                             & ...                                & ...                                 & ...                                 \\
b      & LCO SAAO   & 2020-06-23 & z'         & ...                             & ...                             & ...                                & ...                                 & ...                                 \\
c      & TESS       & 2020-07-09 & TESS       & ...                             & ...                             & ...                                & ...                                 & ...                                 \\
b      & TESS       & 2020-07-10 & TESS       & ...                             & ...                             & ...                                & ...                                 & ...                                 \\
b      & PEST       & 2020-07-10 & V          & $0.0013 \pm 0.0014$             & ...                             & ...                                & ...                                 & ...                                 \\
b      & TESS       & 2020-07-19 & TESS       & ...                             & ...                             & ...                                & ...                                 & ...                                 \\
b      & TESS       & 2020-07-27 & TESS       & ...                             & ...                             & ...                                & ...                                 & ...                                 \\
c      & TESS       & 2020-07-28 & TESS       & ...                             & ...                             & ...                                & ...                                 & ...                                 \\
b      & Brierfield & 2020-08-13 & I          & $0.0013 \pm 0.0024$             & ...                             & ...                                & ...                                 & ...                                 \\
b      & LCO SSO    & 2020-08-13 & z'         & ...                             & ...                             & ...                                & ...                                 & ...                                 \\
b      & LCO SAAO   & 2020-09-07 & z'         & ...                             & ...                             & ...                                & ...                                 & ...                                 \\
b      & LCO SSO    & 2020-09-16 & z'         & ...                             & ...                             & ...                                & ...                                 & ...                                 \\
b      & LCO SSO    & 2020-10-03 & z'         & ...                             & ...                             & ...                                & ...                                 & ...                                 \\
b \& c & LCO SAAO   & 2020-10-11 & z'         & ...                             & ...                             & ...                                & ...                                 & ...                                 \\
b      & LCO CTIO   & 2021-06-14 & z'         & ...                             & ...                             & ...                                & ...                                 & ...                                 \\
b      & LCO SSO    & 2021-06-22 & z'         & ...                             & ...                             & ...                                & ...                                 & ...                                 \\
b      & LCO CTIO   & 2021-07-01 & z'         & ...                             & ...                             & ...                                & ...                                 & ...                                 \\
b      & ASTEP      & 2021-07-09 & R          & ...                             & ...                             & ...                                & ...                                 & ...                                 \\
b      & LCO SAAO   & 2021-07-17 & z'         & ...                             & ...                             & ...                                & ...                                 & ...                                 \\
b      & LCO SAAO   & 2021-08-03 & z'         & ...                             & ...                             & ...                                & ...                                 & ...                                 \\
b      & LCO CTIO   & 2021-08-04 & z'         & ...                             & ...                             & ...                                & ...                                 & ...                                 \\
b      & LCO SSO    & 2021-08-12 & z'         & ...                             & ...                             & ...                                & ...                                 & ...                                 \\
b      & LCO SSO    & 2021-08-29 & z'         & ...                             & ...                             & ...                                & ...                                 & ...                                 \\
b      & LCO SSO    & 2021-09-15 & z'         & ...                             & ...                             & ...                                & ...                                 & ...                                 \\
b      & LCO SAAO   & 2021-09-23 & z'         & ...                             & ...                             & ...                                & ...                                 & ...                                 \\
b      & LCO TO     & 2021-09-23 & z'         & ...                             & ...                             & ...                                & ...                                 & ...                                 \\
c      & LCO CTIO   & 2021-10-04 & z'         & ...                             & ...                             & ...                                & ...                                 & ...
    \enddata
\tablecomments{See Table 3 in \citet{eastman2019} for a detailed description of all parameters. Since both Pan-STARRS Y and Pan-STARRS \zs are not available among the filters in \exofast, y and z' (Sloan z) were used as respective approximate substitutes. Additionally, the detrending parameters flare (Spitzer), sky (Spitzer \& PEST), Sky/Pixel\_T1 (LCOGT), \& SKY (ASTEP) were set as additive while the remaining detrending parameters were set as multiplicative.}
\tablenotetext{a}{Multiplicative Detrending Coefficient}
\end{deluxetable*}
\end{longrotatetable}

\section{\exostriker TTV Modeling}\label{sec:exostrikermod}

In this section, we present our O--C diagram from \exofast, the super-period of AU Mic b's TTVs, and the TTV dynamical modeling with \exostriker. We explore three different scenarios: a system with two planets b and c, a system with three planets where planet d is interior to b, and a system with three planets where planet d is between b and c. Additionally, we develop the novel technique TTV log-likelihood periodogram, which aids us in exploring the TTV parameters that maximize the log-likelihood.

\subsection{O--C Diagram and TTV Super-Period}\label{sec:ocdiagram}

\begin{figure*}
    \centering
    \includegraphics[width=\textwidth]{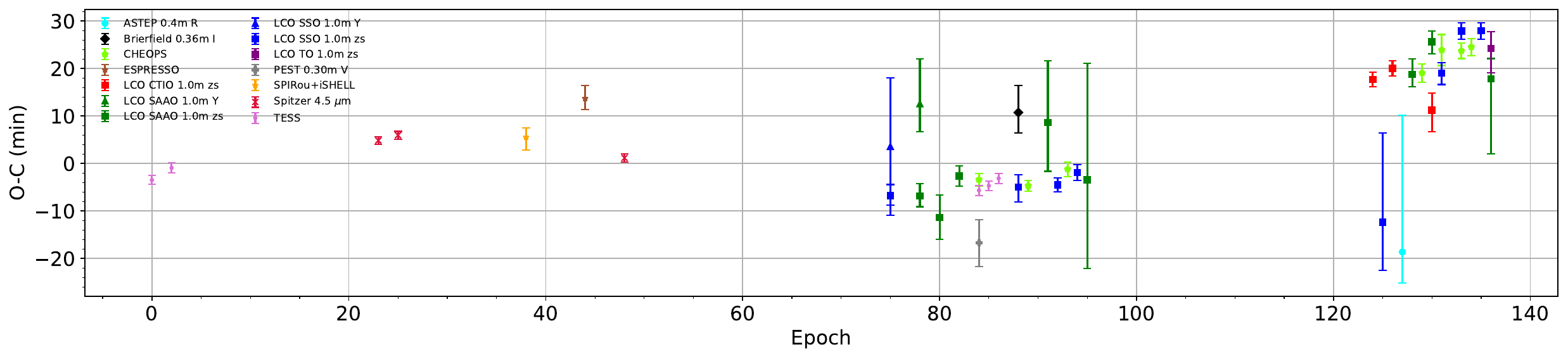}
    \includegraphics[width=\textwidth]{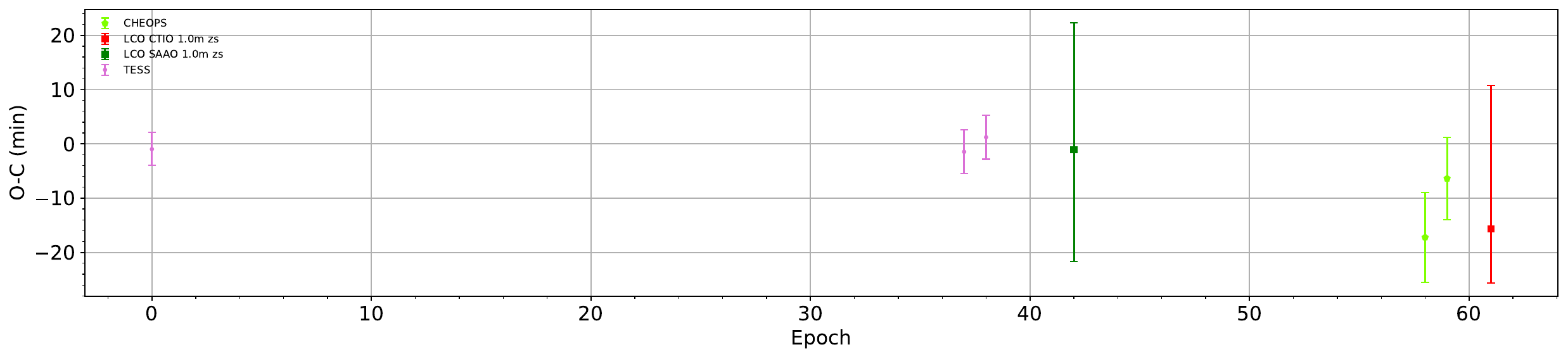}
    \caption{O--C Diagram of AU Mic b ({\it top}) and AU Mic c ({\it bottom}), using the \exofast-generated measured midpoint times (Table \ref{tab:ttvpriors}) and the calculated expected midpoint times for all transit data sets from Table \ref{tab:datasets}. The planets' period and \tc from \exofast posteriors were used for the calculation of expected midpoint times. The epoch refers to the number of transits since the first transit of b and c, respectively.}
    \label{fig:ocdiagram}
\end{figure*}

In the previous section, we ran the \exofast package to model the transits and generate the observed midpoint times (Table \ref{tab:ttvpriors}). CHEOPS and R-M data were not included in transit modeling, but CHEOPS midpoint times from \citet{szabo2021, szabo2022} and R-M midpoints times from \citet{martioli2020} and \citet{palle2020} are now added to the list of midpoint times generated from \exofast. We calculate the expected midpoint times using the planets' \exofast-generated periods and \tc. Then, we use both the calculated and the observed midpoint times to create the O--C diagram of both AU Mic b and c (Figure \ref{fig:ocdiagram}).

Relative to the space-based transit midpoint times, most ground-based photometric transits have larger timing uncertainties and some larger scatter. Some of the notable outliers in transit midpoint time of AU Mic b (regardless of timing uncertainty) include ESPRESSO, PEST 0.3 m, Brierfield 0.36 m, ASTEP 0.4 m, and at least one each of LCO CTIO, SAAO, \& SSO 1.0 m. Many of AU Mic b's transits observed in 2021 are considerably later ($\sim$20 min) than those observed from 2020 given the previously measured ephemerides.

\begin{figure*}
    \centering
    \includegraphics[width=\textwidth]{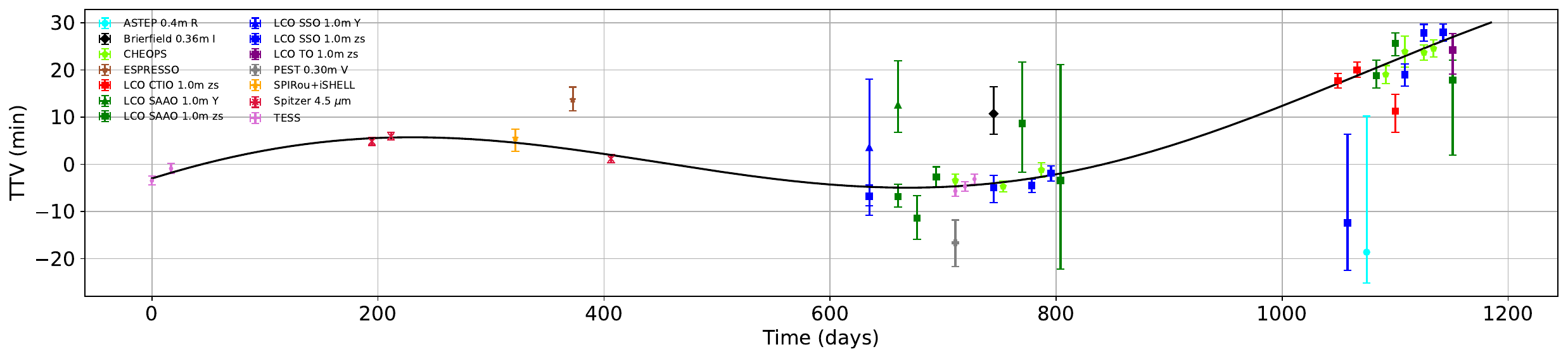}
    \caption{O--C Diagram of AU Mic b with the super-period model overlaid (black). The super-period model was generated using the coefficients from Table \ref{tab:ttvsin} and equation \ref{eqn:sinfunction}, and the O--C diagram was generated using the \exofast-generated measured midpoint times (Table \ref{tab:ttvpriors}) and the calculated expected midpoint times for all transit data sets from Table \ref{tab:datasets}. The planet's period and \tc from \exofast posteriors were used for the calculation of expected midpoint times. The time is with respect to the first transit of b.}
    \label{fig:ocsin}
\end{figure*}

AU Mic b's TTVs appear to have a quasi-sinusoidal pattern. This indicates that there is potentially a ``super-period'' embedded in the TTV observations. To model the super-period, we construct a sinusoidal model which includes a linear trend to account for the apparent drift in the TTVs:
\begin{equation}\label{eqn:sinfunction}
    y(t) = A\sin(Bt + C) + Dt + E
\end{equation}
where $t$ is time since the first transit of b and $A$, $B$, $C$, $D$, \& $E$ are the unknown coefficients. The coefficients are optimized using the \texttt{scipy.optimize.curve\_fit}\footnote{\url{https://docs.scipy.org/doc/scipy/reference/generated/scipy.optimize.curve_fit.html}} (Table \ref{tab:ttvsin}), and we use equation \ref{eqn:sinfunction} to model the super-period onto AU Mic b's O--C diagram (Figure \ref{fig:ocsin}).

\begin{deluxetable}{c|c|c}\label{tab:ttvsin}
    \tablecaption{Coefficients from Equation \ref{eqn:sinfunction} as part of modeling the super-period of AU Mic b's TTVs (Figure \ref{fig:ocsin}). The coefficients were solved using the \texttt{scipy.optimize.curve\_fit}\tablenotemark{a}.}
    \tablehead{Coefficient & Unit & Output}
    \startdata
A & min            & 13.46394 $\pm$ 3.06443  \\
B & day$^{-1}$     & 0.00504 $\pm$ 0.00070   \\
C & ...            & 0.87455 $\pm$ 0.28102   \\
D & min day$^{-1}$ & 0.03052 $\pm$ 0.00964   \\
E & min            & -13.31819 $\pm$ 4.52818
    \enddata
    \tablenotetext{a}{\url{https://docs.scipy.org/doc/scipy/reference/generated/scipy.optimize.curve\_fit.html}}
\end{deluxetable}

We can use that super-period to estimate the period of AU Mic d using equation 5 from \citet{lithwick2012},
\begin{equation}\label{eqn:superp}
    P^{j} \equiv \frac{1}{\left| j/P' - (j-1)/P \right|}
\end{equation}
where $j$ is an integer that represents the $j$:$j-1$ mean-motion orbital resonant (MMR) chain, $P$ is the orbital period of the inner planet, $P'$ is the orbital period of the outer planet, and $P^{j}$ is the super-period of the TTVs. Since the 3:2 period ratio is the most common pairing for the MMR \citep{fabrycky2014}, we assume $j$=3 and will not consider other j values for this problem. From the coefficient B, we obtain the super-period $P^{3}$ = 1186.44931 $\pm$ 153.24946 days. From equation \ref{eqn:superp}, we end up with the estimated periods $P_{\rm d}$ = 5.62927 $\pm$ 0.00177 days or 5.65482 $\pm$ 0.00178 days (assuming AU Mic d is orbiting interior to AU Mic b), and $P_{\rm d}$ = 12.65156 $\pm$ 0.00594 days or 12.73779 $\pm$ 0.00603 days (assuming AU Mic d is orbiting between AU Mic b and c); the absolute value brackets in the denominator of Equation \ref{eqn:superp} allow us to have two potential solutions for both the inner d and middle d scenarios. Additionally, we note a statistically significant (3.2$\sigma$) non-zero linear drift of 0.03052 $\pm$ 0.00964 min day$^{-1}$ in the TTVs of AU Mic b.

This approach in using AU Mic b's TTV super-period is meant to provide a starting point for estimating the orbital period of AU Mic d that could drive the observed excess TTVs of AU Mic b as reported by \citet{wittrock2022}. Additionally, numerous compact systems have planets in near-MMR chains -- e.g., HD 158259 \citep{hara2020}, TRAPPIST-1 \citep{gillon2016, gillon2017, luger2017}, V1298 Tau \citep{david2019b}, and several Kepler systems \citep{lissauer2011, fabrycky2014} -- and most Kepler systems with measured TTVs have planets in near-MMR chains (\citealt{lithwick2012}; also \citealt{steffen2013} for example), so it is not unreasonable or unwarranted to assume that AU Mic d might be near MMR pairing with b and with c.

\citet{wittrock2022} generated the massless no-TTV 2-planet model with \exostriker as a control test on the presence and statistical significance of TTVs, but it also serves as a useful snapshot on whether AU Mic b's TTVs are behaving linearly over time. The massless planets O-C diagram (Figure 5 from \citealt{wittrock2022}) and the statistical comparison between the massless 2-planet model and the non-massless 2-planet model -- e.g., reduced chi-square \reduced = 8.7$\times$10$^{8}$ vs 38 respectively and log-likelihood $\ln\mathcal{L}$ = -6.1$\times$10$^{9}$ vs -75 respectively \citep{wittrock2022} -- show very clear indications that a linear trend does not fit very well with AU Mic b’s observed TTVs.

In the meanwhile, given the relatively sparse observations of AU Mic c, we cannot draw any meaningful interpretation of what c's TTV super-period could be. Thus, the potential super-period in AU Mic c's observed TTVs is not being explored in this paper.

\subsection{\exostriker Dynamical Modeling Preparations and Processes}

The O--C diagram from \citet{wittrock2022} displayed the apparent deviation of the TTVs from a linear ephemeris; this was believed to be attributed to the yet-to-be-confirmed planet AU Mic d even after accounting for the impacts from stellar activity, such as flaring and spot crossings. With the new data from $\S$\ref{sec:dataobs} added to this work, we model the TTVs of AU Mic b and c using the \exostriker package \citep{trifonov2019}, with our focus now on validating AU Mic d. \exostriker is capable of applying the Markov chain Monte Carlo (MCMC) via \emcee \citep{foreman-mackey2013} to determine the statistical significance of our TTV measurements and the confidence in their corresponding dynamical model posteriors.

\begin{deluxetable*}{l|c|ccc}\label{tab:exostriker_priors}
    \tablecaption{AU Mic's stellar and planetary priors for \exostriker best fit and MCMC modeling.}
    \tablehead{\multirow{2}{*}{Prior} & \multirow{2}{*}{Unit} & \multicolumn{3}{c}{Input}       \\
                                      &                       & AU Mic b & AU Mic c & AU Mic d  }
    \startdata
Mass       & \msun       & \multicolumn{3}{c}{$\mathcal{N}$(0.510, 0.028)}                                           \\
Radius     & \rsun       & \multicolumn{3}{c}{$\mathcal{N}$(0.744, 0.023)}                                           \\
Luminosity & \lsun       & \multicolumn{3}{c}{$\mathcal{N}$(0.0916, 0.011)}                                          \\
\teff      & K           & \multicolumn{3}{c}{$\mathcal{N}$(3678, 90)}                                               \\
v sin i    & km s$^{-1}$ & \multicolumn{3}{c}{$\mathcal{N}$(8.7, 0.2)}                                               \\ \hline
K          & m s$^{-1}$  & $\mathcal{U}$(0.0, 10000.0) & $\mathcal{U}$(0.0, 10000.0)  & $\mathcal{U}$(0.0, 10000.0)  \\
\porb      & day         & $\mathcal{N}$(8.463, 0.001) & $\mathcal{N}$(18.859, 0.001) & $\mathcal{U}$(0.0, 100000.0) \\
e          & ...         & $\mathcal{U}$(0.000, 0.999) & $\mathcal{U}$(0.000, 0.999)  & $\mathcal{U}$(0.000, 0.999)  \\
$\omega$   & deg         & $\mathcal{U}$(0.0, 360.0)   & $\mathcal{U}$(0.0, 360.0)    & $\mathcal{U}$(0.0, 360.0)    \\
M$_{0}$    & deg         & $\mathcal{U}$(0.0, 360.0)   & $\mathcal{U}$(0.0, 360.0)    & $\mathcal{U}$(0.0, 360.0)    \\
i          & deg         & $\mathcal{N}$(89.57, 0.31)  & $\mathcal{N}$(89.43, 0.35)   & $\mathcal{U}$(0.0, 180.0)    \\
$\Omega$   & deg         & $\mathcal{U}$(0.0, 360.0)   & $\mathcal{U}$(0.0, 360.0)    & $\mathcal{U}$(0.0, 360.0)
    \enddata
    \tablecomments{The stellar, orbital period, and inclination priors are taken from \exofast posteriors. AU Mic d's priors apply to 3-planet modeling only. M$_{\rm 0}$ $\equiv$ mean anomaly, and $\Omega$ $\equiv$ longitude of ascending node.}
\end{deluxetable*}

For our \exostriker modeling, we incorporate the priors for the host star and the planets from Table \ref{tab:exostriker_priors} and the midpoint time priors from Table \ref{tab:ttvpriors}. \exostriker uses only the {\tt Simplex} algorithm for any TTV models. We use {\tt N-body} algorithm for all model fittings and MCMC runs, and we set the dynamical model time steps to 0.01 days. For our model fittings, we use the following {\tt scipy} minimizer algorithms: truncated Newton algorithm (TNC)\footnote{\url{https://docs.scipy.org/doc/scipy/reference/optimize.minimize-tnc.html}} as a primary minimizer and Nelder-Mead algorithm\footnote{\url{https://docs.scipy.org/doc/scipy/reference/optimize.minimize-neldermead.html}} as a secondary minimizer, with the configurations of both minimizers set at default, including one consecutive integration and 5\,000 integration steps. We manually fit the model to the data using the previously mentioned minimizer algorithms, then each time we find a possible best fit model, we proceed to perform MCMC computations by adopting MCMC parameters as best $\ln\mathcal{L}$ and with 1\,000 burn-in steps, 10\,000 main steps for the 2-planet dynamical models and 8\,000 main steps for the 3-planet dynamical models, and 196 walkers for the 2-planet dynamical models and 441 walkers for the 3-planet dynamical models. The three aforementioned scenarios are explored and presented in the following sections.

\begin{deluxetable*}{c|l|c|c|c|c}\label{tab:ttvpriors}
    \tablecaption{AU Mic planets' midpoint time priors for \exostriker models. These midpoint times, with the exception of those from CHEOPS \citep{szabo2021, szabo2022} and R-M observations \citep{martioli2020, palle2020}, were generated from \exofast transit modeling.}
    \tablehead{Planet & Telescope & Transit N & Date (UT) & Band & T$_{0}$ (BJD)}
    \startdata
\multirow{42}{*}{b} & TESS            & 1   & 2018-07-26 & TESS           & 2458330.38927 $\pm$ 0.00039 \\
                    & TESS            & 3   & 2018-08-12 & TESS           & 2458347.31710 $\pm$ 0.00053 \\
                    & Spitzer         & 24  & 2019-02-10 & 4.5 $\mu$m     & 2458525.04446 $\pm$ 0.00016 \\
                    & Spitzer         & 26  & 2019-02-27 & 4.5 $\mu$m     & 2458541.97126 $\pm$ 0.00014 \\
                    & SPIRou + iSHELL & 39  & 2019-06-17 & (a)            & 2458651.99020 $\pm$ 0.00180 \\
                    & ESPRESSO        & 45  & 2019-08-07 & 378.2-788.7 nm & 2458702.77397 $\pm$ 0.00178 \\
                    & Spitzer         & 49  & 2019-09-09 & 4.5 $\mu$m     & 2458736.61735 $\pm$ 0.00014 \\
                    & LCO SSO         & 76  & 2020-04-25 & Pan-STARRS Y   & 2458965.11330 $\pm$ 0.00150 \\
                    & LCO SSO         & 76  & 2020-04-25 & Pan-STARRS \zs & 2458965.12050 $\pm$ 0.01000 \\
                    & LCO SAAO        & 79  & 2020-05-20 & Pan-STARRS \zs & 2458990.50230 $\pm$ 0.00170 \\
                    & LCO SAAO        & 79  & 2020-05-20 & Pan-STARRS Y   & 2458990.51580 $\pm$ 0.00650 \\
                    & LCO SAAO        & 81  & 2020-06-06 & Pan-STARRS \zs & 2459007.42520 $\pm$ 0.00320 \\
                    & LCO SAAO        & 83  & 2020-06-23 & Pan-STARRS \zs & 2459024.35730 $\pm$ 0.00130 \\
                    & TESS            & 85  & 2020-07-10 & TESS           & 2459041.28124 $\pm$ 0.00026 \\
                    & PEST            & 85  & 2020-07-10 & V              & 2459041.27360 $\pm$ 0.00340 \\
                    & CHEOPS          & 85  & 2020-07-10 & CHEOPS         & 2459041.28280 $\pm$ 0.00060 \\
                    & TESS            & 86  & 2020-07-19 & TESS           & 2459049.74493 $\pm$ 0.00024 \\
                    & TESS            & 87  & 2020-07-27 & TESS           & 2459058.20903 $\pm$ 0.00024 \\
                    & Brierfield      & 89  & 2020-08-13 & I              & 2459075.14470 $\pm$ 0.00390 \\
                    & LCO SSO         & 89  & 2020-08-13 & Pan-STARRS \zs & 2459075.13380 $\pm$ 0.00210 \\
                    & CHEOPS          & 90  & 2020-08-21 & CHEOPS         & 2459083.59700 $\pm$ 0.00040 \\
                    & LCO SAAO        & 92  & 2020-09-07 & Pan-STARRS \zs & 2459100.53230 $\pm$ 0.00900 \\
                    & LCO SSO         & 93  & 2020-09-16 & Pan-STARRS \zs & 2459108.98618 $\pm$ 0.00072 \\
                    & CHEOPS          & 94  & 2020-09-24 & CHEOPS         & 2459117.45150 $\pm$ 0.00080 \\
                    & LCO SSO         & 95  & 2020-10-03 & Pan-STARRS \zs & 2459125.91403 $\pm$ 0.00090 \\
                    & LCO SAAO        & 96  & 2020-10-11 & Pan-STARRS \zs & 2459134.37600 $\pm$ 0.01700 \\
                    & LCO CTIO        & 125 & 2021-06-14 & Pan-STARRS \zs & 2459379.81818 $\pm$ 0.00069 \\
                    & LCO SSO         & 126 & 2021-06-22 & Pan-STARRS \zs & 2459388.26030 $\pm$ 0.01300 \\
                    & LCO CTIO        & 127 & 2021-07-01 & Pan-STARRS \zs & 2459396.74582 $\pm$ 0.00079 \\
                    & ASTEP           & 128 & 2021-07-09 & R              & 2459405.18200 $\pm$ 0.02000 \\
                    & LCO SAAO        & 129 & 2021-07-17 & Pan-STARRS \zs & 2459413.67100 $\pm$ 0.00210 \\
                    & CHEOPS          & 130 & 2021-07-26 & CHEOPS         & 2459422.13420 $\pm$ 0.00100 \\
                    & LCO SAAO        & 131 & 2021-08-03 & Pan-STARRS \zs & 2459430.60180 $\pm$ 0.00160 \\
                    & LCO CTIO        & 131 & 2021-08-04 & Pan-STARRS \zs & 2459430.59180 $\pm$ 0.00300 \\
                    & LCO SSO         & 132 & 2021-08-12 & Pan-STARRS \zs & 2459439.06020 $\pm$ 0.00140 \\
                    & CHEOPS          & 132 & 2021-08-12 & CHEOPS         & 2459439.06360 $\pm$ 0.00210 \\
                    & LCO SSO         & 134 & 2021-08-29 & Pan-STARRS \zs & 2459455.99240 $\pm$ 0.00089 \\
                    & CHEOPS          & 134 & 2021-08-29 & CHEOPS         & 2459455.98950 $\pm$ 0.00070 \\
                    & CHEOPS          & 135 & 2021-09-06 & CHEOPS         & 2459464.45310 $\pm$ 0.00090 \\
                    & LCO SSO         & 136 & 2021-09-15 & Pan-STARRS \zs & 2459472.91850 $\pm$ 0.00088 \\
                    & LCO SAAO        & 137 & 2021-09-23 & Pan-STARRS \zs & 2459481.37450 $\pm$ 0.01100 \\
                    & LCO TO          & 137 & 2021-09-23 & Pan-STARRS \zs & 2459481.37890 $\pm$ 0.00340 \\ \hline
\multirow{7}{*}{c}  & TESS            & 1   & 2018-08-11 & TESS           & 2458342.22333 $\pm$ 0.00093 \\
                    & TESS            & 38  & 2020-07-09 & TESS           & 2459040.00487 $\pm$ 0.00072 \\
                    & TESS            & 39  & 2020-07-28 & TESS           & 2459058.86571 $\pm$ 0.00066 \\
                    & LCO SAAO        & 43  & 2020-10-11 & Pan-STARRS \zs & 2459134.30000 $\pm$ 0.01600 \\
                    & CHEOPS          & 59  & 2021-08-09 & CHEOPS         & 2459436.03230 $\pm$ 0.00450 \\
                    & CHEOPS          & 60  & 2021-08-28 & CHEOPS         & 2459454.89880 $\pm$ 0.00380 \\
                    & LCO CTIO        & 62  & 2021-10-04 & Pan-STARRS \zs & 2459492.61030 $\pm$ 0.01800
    \enddata
    \tablecomments{(a) 955-2\,515 nm (SPIRou) and 2.18-2.47 nm (iSHELL)}
\end{deluxetable*}

\subsection{2-Planet Dynamical Modeling}\label{sec:2p_ttv_mod}

\begin{figure*}
    \centering
    \includegraphics[width=0.48\textwidth]{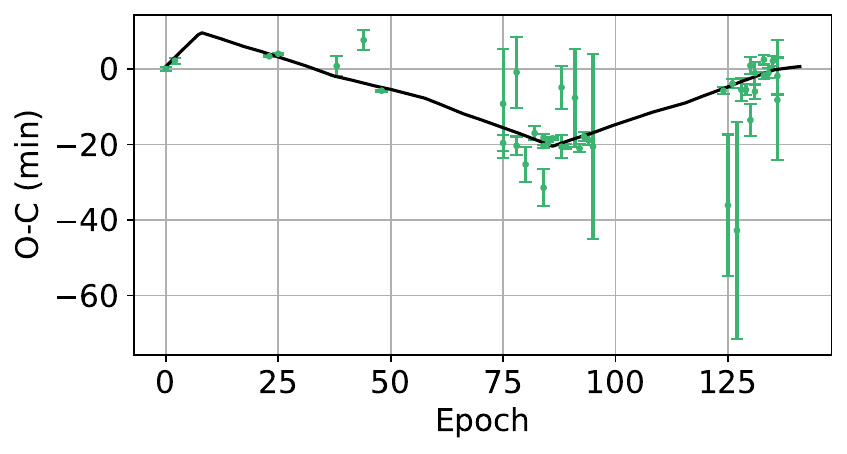}
    \includegraphics[width=0.48\textwidth]{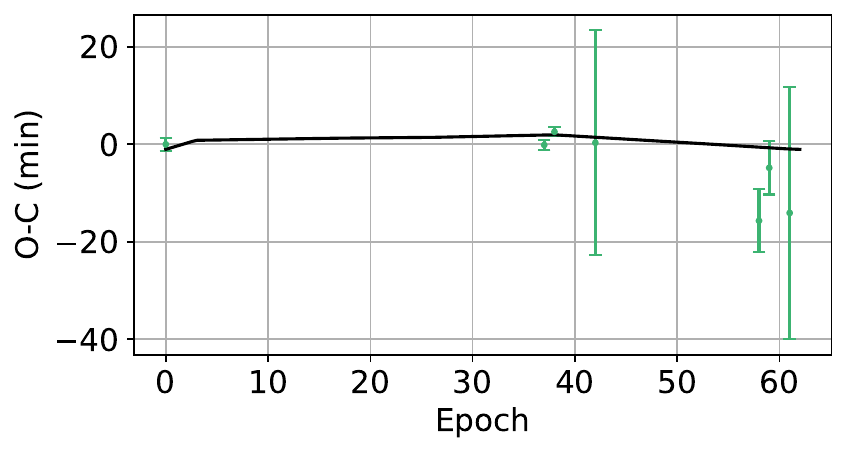}
    \caption{O--C diagram of AU Mic b ({\it left}) and AU Mic c ({\it right}), with comparison between TTVs (green) and \exostriker-generated mcmc models (black) for 2-planet high-eccentricity configuration.}
    \label{fig:2p_ext_exostriker_plots}
\end{figure*}

\begin{figure*}
    \centering
    \includegraphics[width=\textwidth]{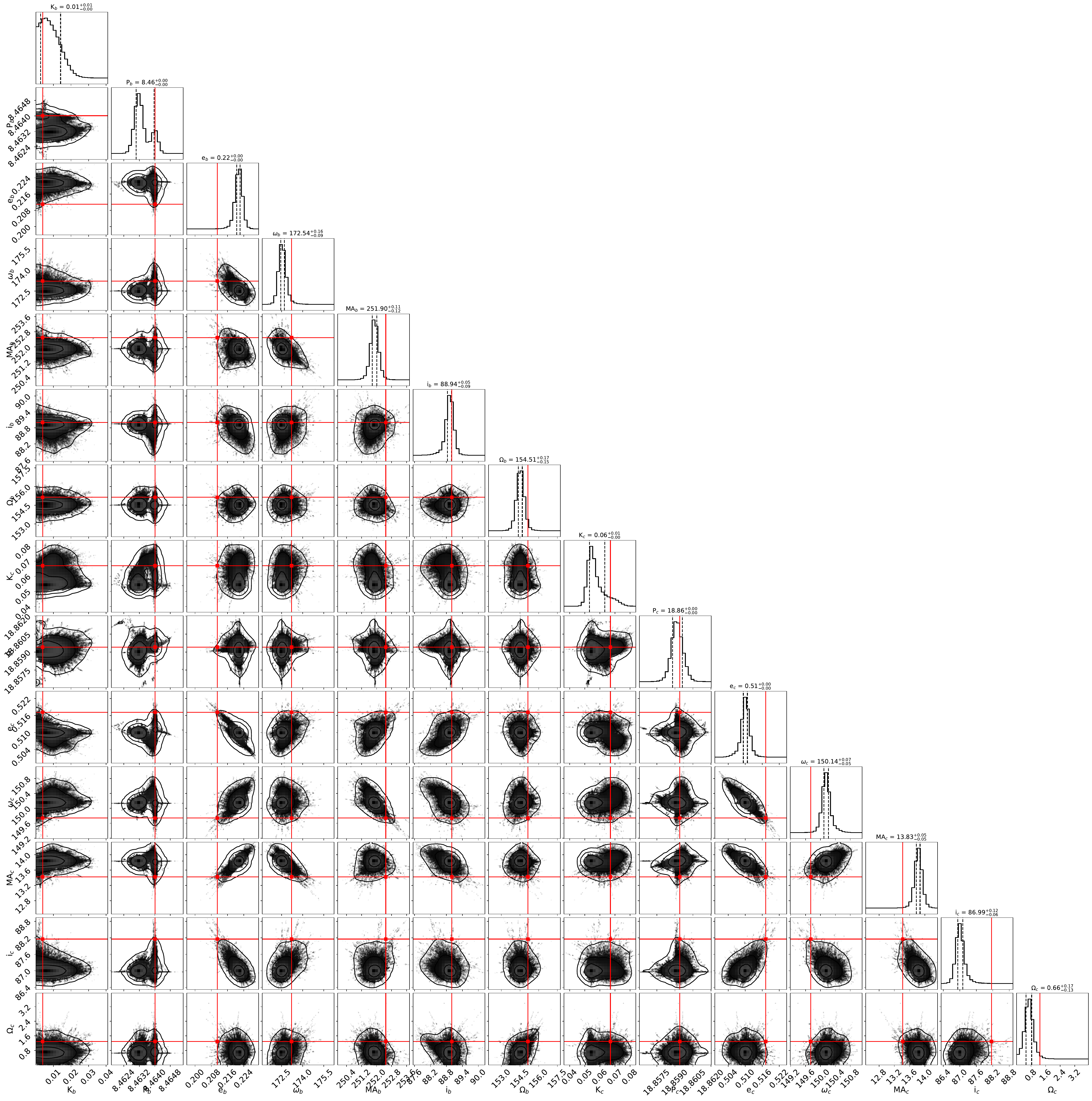}
    \caption{Corner plot of AU Mic b and c's orbital parameters from \exostriker mcmc analysis for 2-planet high-eccentricity configuration.}
    \label{fig:2p_ext_exostriker_cornerplot}
\end{figure*}

We explored a best fit scenario for a 2-planet model for the purpose of obtaining the statistical significance of TTVs and comparing the results with those of the 3-planet models. We share our maximum-likelihood 2-planet model with its MCMC O--C diagram (Figure \ref{fig:2p_ext_exostriker_plots}), posteriors (Table \ref{tab:2p_ext_exostriker_params}), and MCMC corner plot (Figure \ref{fig:2p_ext_exostriker_cornerplot}). The O--C diagram from Figure \ref{fig:2p_ext_exostriker_plots} does exhibit the super-period that was obvious in Figure \ref{fig:ocsin}. However, the planets' inferred masses are very small (K $<$ 0.07 m s$^{-1}$) and their eccentricities relatively large (e $>$ 0.2), neither of which are in agreement with those from RV literatures (e.g., \citealt{cale2021} and \citealt{donati2023}). Moreover, the Angular Momentum Deficit \citep[AMD;][]{laskar1997, laskar2000, laskar2017} criteria built within the \exostriker package indicated that this model is unstable. We followed this up by testing its stability over 20 Myr with the N-body simulator \rebound \citep{rein2012, rein2015}.

\begin{figure*}
    \centering
    \includegraphics[width=0.48\textwidth]{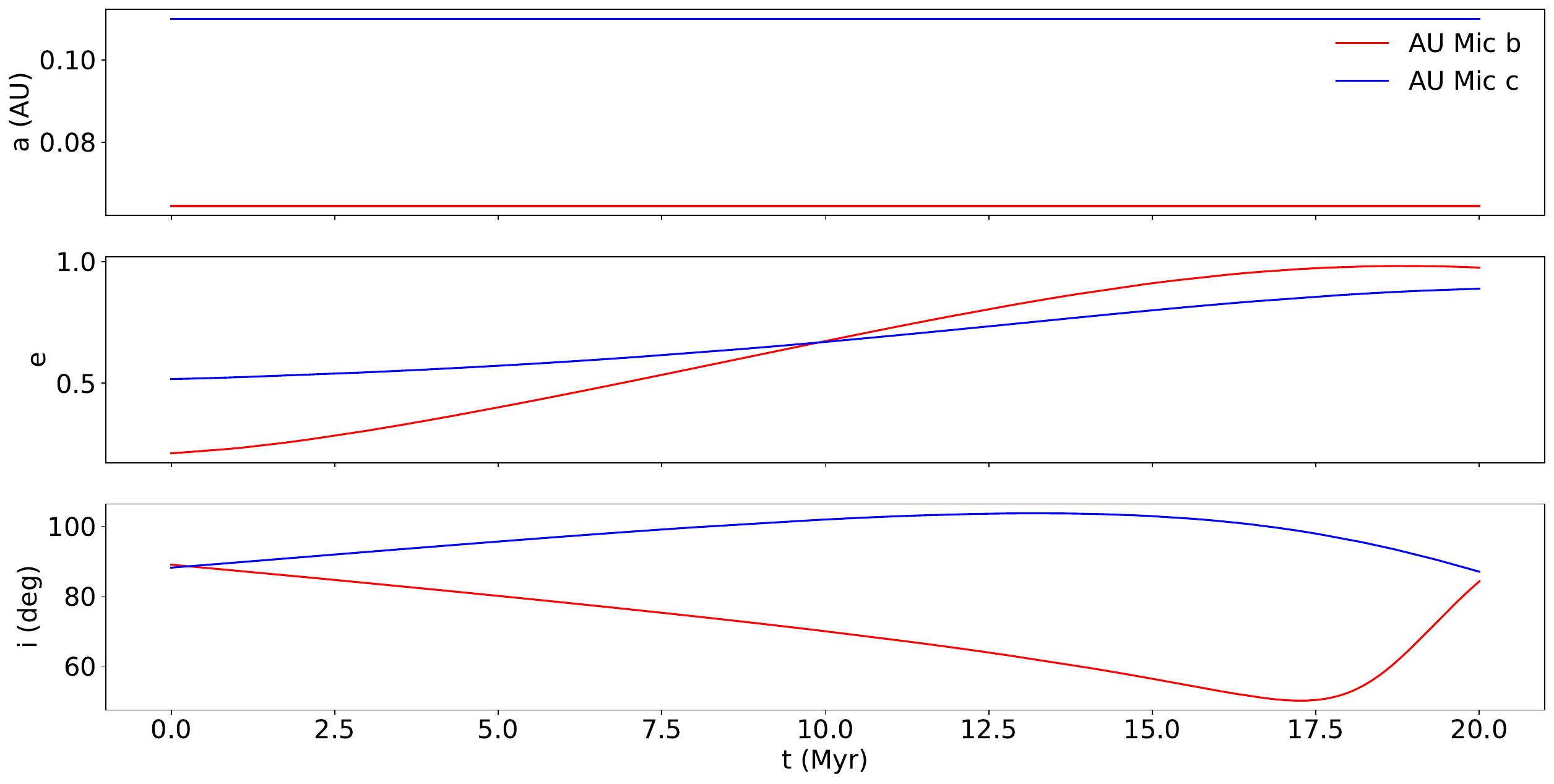}
    \includegraphics[width=0.48\textwidth]{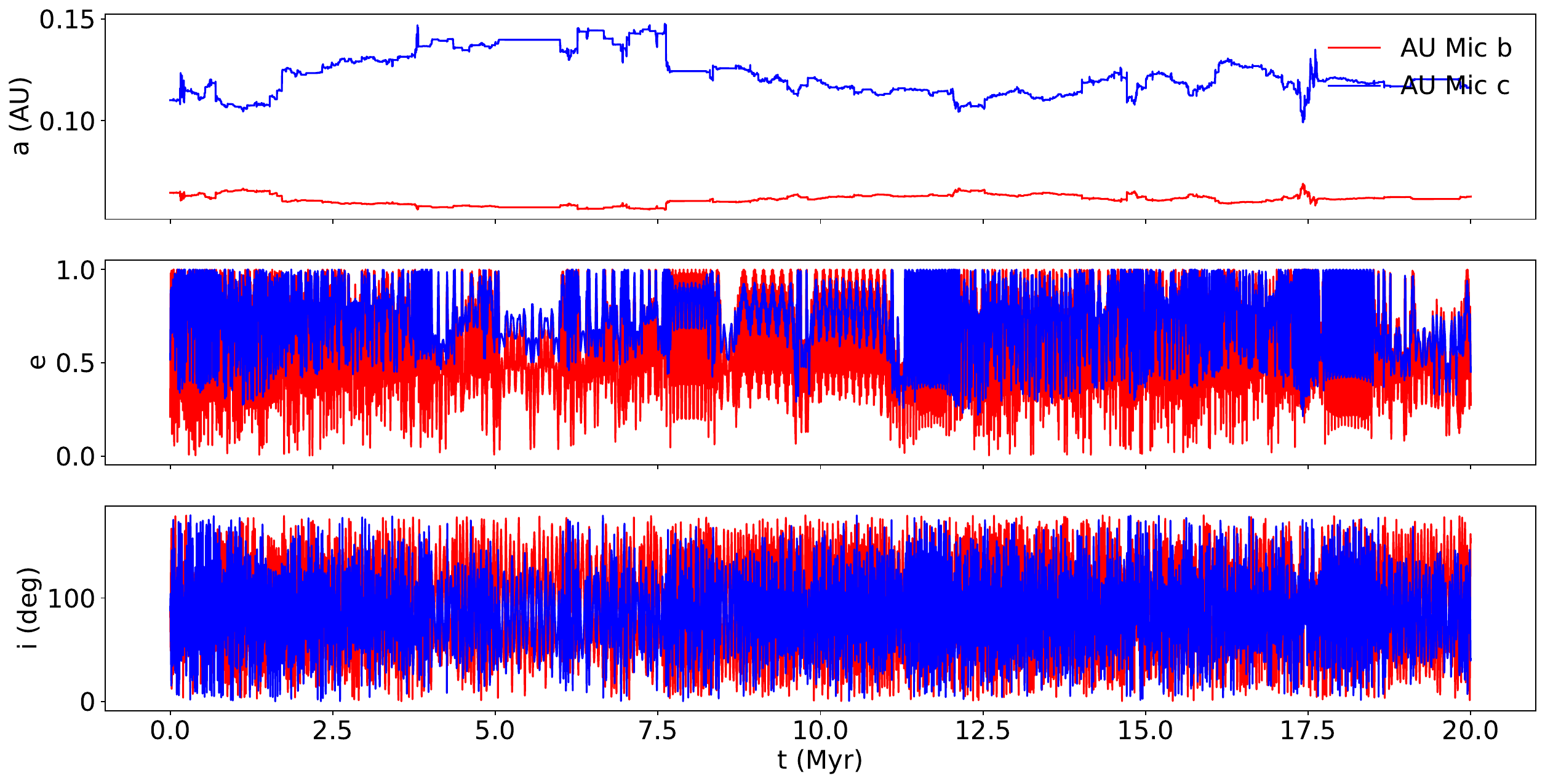}
    \caption{\rebound models of the stability of AU Mic system on timescale of 20 Myr for 2-planet high-eccentricity configuration with original K's from Table \ref{tab:2p_ext_exostriker_params} ({\it left}) and with K's from \citet{cale2021} RV models ({\it right}), due to original K's being very small ($<$0.07 m s$^{-1}$). While the original K's case appears to be stable, both planets' orbits become increasingly eccentric over time, and both planets would quickly become non-transiting due to their increasing orbital misalignment. The RV K's case exhibits signs of chaos in AU Mic system, with the planets undergoing orbital migrations but no orbital crossing, their eccentricities are extremely erratic throughout, rapidly oscillating between 0 and 1, and their highly-fluctuating inclinations would frequently put both transiting planets in non-transiting configurations.}
    \label{fig:2p_ext_rebound}
\end{figure*}

Since this model's inferred masses are very small, we performed two stability tests (Figure \ref{fig:2p_ext_rebound}): one with the planets' original K's from Table \ref{tab:2p_ext_exostriker_params}, and another with the planets' K's from \citet{cale2021} RV models. At first glance, the original K's case appears to be dynamically stable, but the planets' already highly-eccentric orbits become increasingly eccentric over time, and both planets' orbits become increasingly misaligned with the system, which would cause both planets to quickly lose their transiting status. The RV K's case exhibits signs of chaos in AU Mic system, with both planets undergoing orbital migrations, with planet c demonstrating greater wobbles in its orbital path. Also, both planets' eccentricities oscillate rapidly between 0 and 1, and their highly-fluctuating inclinations would frequently put both transiting planets in non-transiting configurations. Given that both planets' observed orbital periods are very consistent and with very small timing uncertainties \citep{plavchan2020, martioli2020, cale2021, gilbert2022, wittrock2022, donati2023}, that the RV models from \citet{cale2021} and \citet{donati2023} place both planets' orbits at much lower eccentricities, and that there have not been any known cases in which either planet ``misses'' its transit due to being outside the line of sight between the observers and the host star, this 2-planet high-eccentricity configuration appears to be nonphysical and highly implausible.

\begin{deluxetable*}{l|c|cc|cc}\label{tab:2p_ext_exostriker_params}
    \tablecaption{\exostriker-generated best fit and MCMC modeling parameters for 2-planet high-eccentricity configuration.}
    \tablehead{\multirow{2}{*}{Parameter} & \multirow{2}{*}{Unit} & \multicolumn{2}{c|}{Best Fit}  & \multicolumn{2}{c}{MCMC} \\
                                          &                       & AU Mic b & AU Mic c & AU Mic b & AU Mic c                 }
    \startdata
K                & m s$^{-1}$ & 0.00543   & 0.05368             & 0.00385 $\pm$ 0.00114   & 0.06702 $\pm$ 0.01389   \\
\porb            & day        & 8.46302   & 18.85898            & 8.46401 $\pm$ 0.00097   & 18.85941 $\pm$ 0.00060  \\
e                & ...        & 0.22188   & 0.51009             & 0.21098 $\pm$ 0.00952   & 0.51720 $\pm$ 0.00786   \\
$\omega$         & deg        & 172.51565 & 150.13429           & 173.20702 $\pm$ 0.76278 & 149.73899 $\pm$ 0.35421 \\
M$_{0}$          & deg        & 251.90049 & 13.83210            & 252.49015 $\pm$ 0.71508 & 13.42102 $\pm$ 0.35754  \\
i                & deg        & 88.94778  & 86.96414            & 89.00381 $\pm$ 0.16286  & 88.15442 $\pm$ 1.22945  \\
$\Omega$         & deg        & 154.51121 & 0.65309             & 155.13328 $\pm$ 0.77348 & 1.28668 $\pm$ 0.76259   \\ \hline
$\chi^{2}$       & ...        & 182.65010 & 8.66743             & 187.17573               & 11.10833                \\
\reduced         & ...        & 6.37725   & 38.26351            & 6.60947                 & 39.65681                \\
$\ln\mathcal{L}$ & ...        & \multicolumn{2}{c|}{176.17820}  & \multicolumn{2}{c}{176.09710}                     \\
$BIC$            & ...        & \multicolumn{2}{c|}{-297.87092} & \multicolumn{2}{c}{-297.70872}                    \\
$AIC_{c}$        & ...        & \multicolumn{2}{c|}{-312.00346} & \multicolumn{2}{c}{-311.84126}
    \enddata
    \tablecomments{The K's listed here are unconstrained, but see $\S$\ref{sec:results} for discussion regarding the planets' low K's generated by \exostriker.}
\end{deluxetable*}

\begin{figure*}
    \centering
    \includegraphics[width=0.48\textwidth]{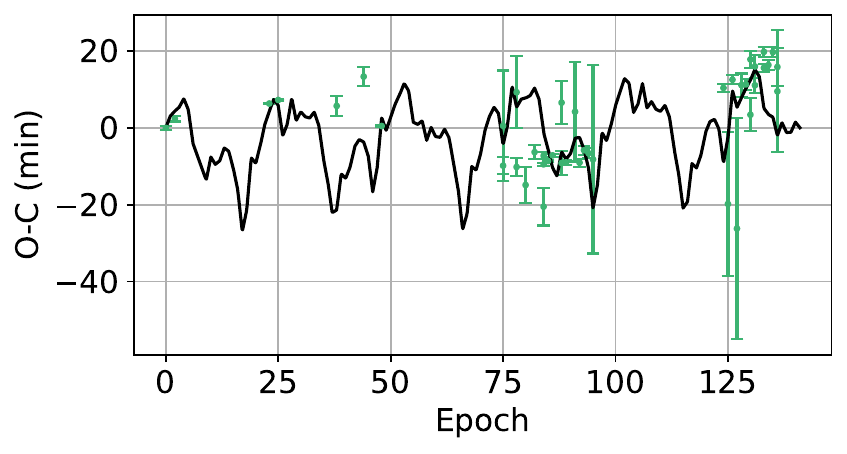}
    \includegraphics[width=0.48\textwidth]{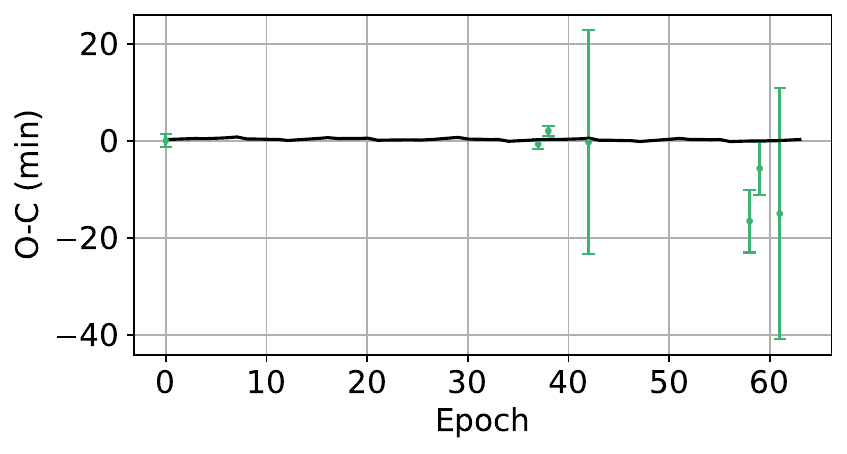}
    \caption{O--C diagram of AU Mic b ({\it left}) and AU Mic c ({\it right}), with comparison between TTVs (green) and \exostriker-generated mcmc models (black) for 2-planet low-eccentricity configuration.}
    \label{fig:2p_ord_exostriker_plots}
\end{figure*}

\begin{figure*}
    \centering
    \includegraphics[width=\textwidth]{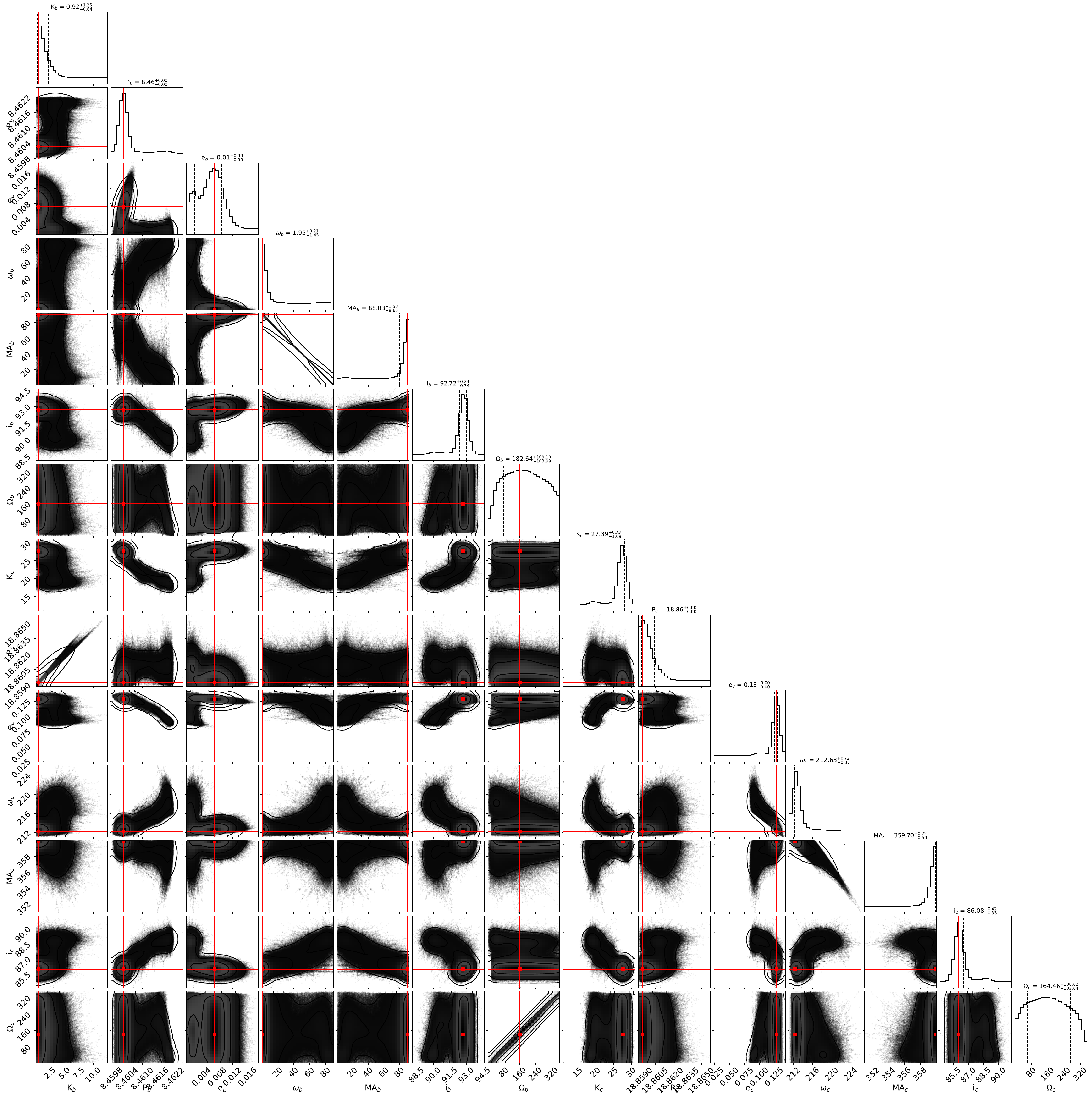}
    \caption{Corner plot of AU Mic b and c's orbital parameters from \exostriker mcmc analysis for 2-planet low-eccentricity configuration.}
    \label{fig:2p_ord_exostriker_cornerplot}
\end{figure*}

We then turned to an alternative 2-planet model with lower eccentricities. We present our low-eccentricity maximum-likelihood 2-planet model with its MCMC O--C diagram (Figure \ref{fig:2p_ord_exostriker_plots}), posteriors (Table \ref{tab:2p_ord_exostriker_params}), and MCMC corner plot (Figure \ref{fig:2p_ord_exostriker_cornerplot}). This time, the AMD criteria indicated that this 2-planet low-eccentricity model is stable. However, the inclinations from Table \ref{tab:2p_ord_exostriker_params} suggest that the planets are misaligned, which contradicts the observed transits of both AU Mic b and c and the inclinations from transit and TTV literatures (e.g., \citealt{martioli2021}, \citealt{gilbert2022}, and \citealt{wittrock2022}). Additionally, the O--C diagram from Figure \ref{fig:2p_ord_exostriker_plots} does not exhibit the super-period that was very distinctive in the O--C diagram from Figure \ref{fig:ocsin}, resulting in the model not converging very well with many observed TTVs, including those from the R-M observations, CHEOPS, and some TESS. Thus, both the 2-planet diagram and the statistics information from Table \ref{tab:2p_ord_exostriker_params} (e.g., $\chi^{2}_{\rm red,b}$ = 75.69 and $\chi^{2}_{\rm red,c}$ = 454.17) clearly show that the model is of a poor fit, suggesting that we need a third planet to account for the observed TTV excess, a conclusion also reached in \citet{wittrock2022}, but now much more obvious by eye with the additional season of 2021 transit observations.

\begin{deluxetable*}{l|c|cc|cc}\label{tab:2p_ord_exostriker_params}
    \tablecaption{\exostriker-generated best fit and MCMC modeling parameters for 2-planet low-eccentricity configuration.}
    \tablehead{\multirow{2}{*}{Parameter} & \multirow{2}{*}{Unit} & \multicolumn{2}{c|}{Best Fit}  & \multicolumn{2}{c}{MCMC} \\
                                          &                       & AU Mic b & AU Mic c & AU Mic b & AU Mic c                 }
    \startdata
K                & m s$^{-1}$ & 2.25722    & 7.65452             & 0.42086 $\pm$ 0.13722    & 27.70929 $\pm$ 1.41812   \\
\porb            & day        & 8.46265    & 18.86039            & 8.46026 $\pm$ 0.00011    & 18.85925 $\pm$ 0.00009   \\
e                & ...        & 0.00000    & 0.00000             & 0.00723 $\pm$ 0.00510    & 0.12867 $\pm$ 0.00279    \\
$\omega$         & deg        & 89.92745   & 229.20188           & 0.74540 $\pm$ 0.24389    & 212.29306 $\pm$ 0.04038  \\
M$_{\rm 0}$      & deg        & 0.00049    & 354.88367           & 90.08325 $\pm$ 10.09994  & 359.95202 $\pm$ 0.75854  \\
i                & deg        & 89.56579   & 89.43523            & 92.68980 $\pm$ 0.30871   & 85.98426 $\pm$ 0.23709   \\
$\Omega$         & deg        & 0.00814    & 0.00004             & 160.82677 $\pm$ 82.66571 & 141.76084 $\pm$ 81.43847 \\ \hline
$\chi^{2}$       & ...        & 4151.71376 & 11.59895            & 2259.05984               & 11.77892                 \\
\reduced         & ...        & 138.77709  & 832.66254           & 75.69463                 & 454.16775                \\
$\ln\mathcal{L}$ & ...        & \multicolumn{2}{c|}{-1808.30801} & \multicolumn{2}{c}{-963.74485}                      \\
$BIC$            & ...        & \multicolumn{2}{c|}{3671.10150}  & \multicolumn{2}{c}{1981.97518}                      \\
$AIC_{c}$        & ...        & \multicolumn{2}{c|}{3656.96896}  & \multicolumn{2}{c}{1967.84264}
    \enddata
    \tablecomments{The K's listed here are unconstrained, but see $\S$\ref{sec:results} for discussion regarding planet b's low K's generated by \exostriker.}
\end{deluxetable*}

\clearpage
\subsection{3-Planet Dynamical Modeling}\label{sec:3p_ttv_mod}

The 2-planet high-eccentricity case's high orbital eccentricities and very low inferred masses from Table \ref{tab:2p_ext_exostriker_params} put this configuration in strong disagreement with the RV models from \citet{cale2021} and \citet{donati2023}, and this orbital configuration exhibits potential instability (Figure \ref{fig:2p_ext_rebound}), with both planets' eccentricities and inclinations undergoing significant fluctuations; these issues suggest that the 2-planet high-eccentricity case does not provide a good fit for a 2-planet model. The 2-planet low-eccentricity case has both transiting planets being misaligned based on the inclinations from Table \ref{tab:2p_ord_exostriker_params}, and Figure \ref{fig:2p_ord_exostriker_plots} and relatively high \reduced and $AIC_{c}$ \citep[corrected Akaike information criterion;][]{sugiura1978, hurvich1989, cavanaugh1997} from Table \ref{tab:2p_ord_exostriker_params} -- e.g., $\chi^{2}_{\rm red,b}$ = 75.69, $\chi^{2}_{\rm red,c}$ = 454.17, and $AIC_{c}$ = 1967.84 -- demonstrate that the 2-planet low-eccentricity model is inadequate in describing the observed TTVs, so we consider a 3-planet model.

Both \citet{cale2021} and \citet{wittrock2022} probed the AU Mic system with RVs and TTVs respectively and modeled a tentative planet AU Mic d between the known planets AU Mic b and c with an orbital period of 12.742 and 13.466 days respectively, the former of which would put the three planets near a 4:6:9 MMR chain. Since no additional transits have been identified for AU Mic, it was proposed that a third planet might be non-transiting. A non-transiting planet between two transiting planets is unusual but not unprecedented; HD 3167 d \citep{christiansen2017}, Kepler-20 g \citep{buchhave2016}, Kepler-411 e \citep{sun2019}, and TOI-431 c \citep{osborn2021} are among the confirmed non-transiting planets orbiting between their adjacent transiting planets. We explore two possible scenarios for the 3-planet modeling: AU Mic d orbiting interior to AU Mic b, and AU Mic d orbiting between AU Mic b and c. We first investigate the interior orbit case.

\subsubsection{Planet d Interior to b}

\begin{figure}
    \centering
    \includegraphics[width=\linewidth]{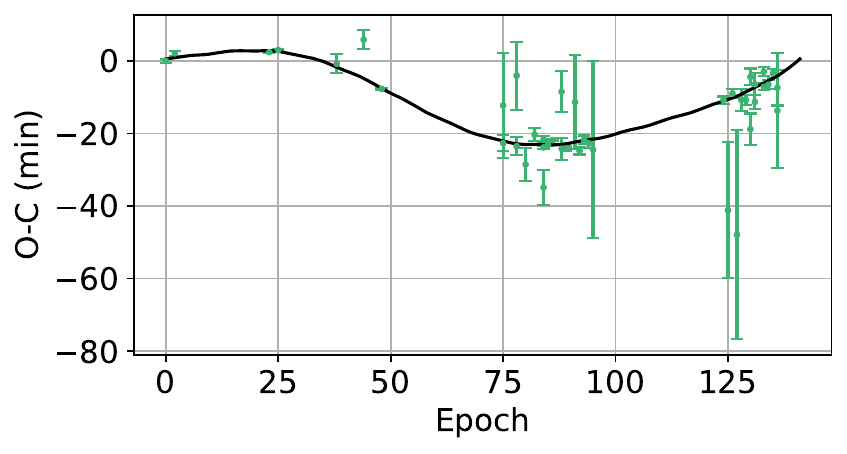}
    \includegraphics[width=\linewidth]{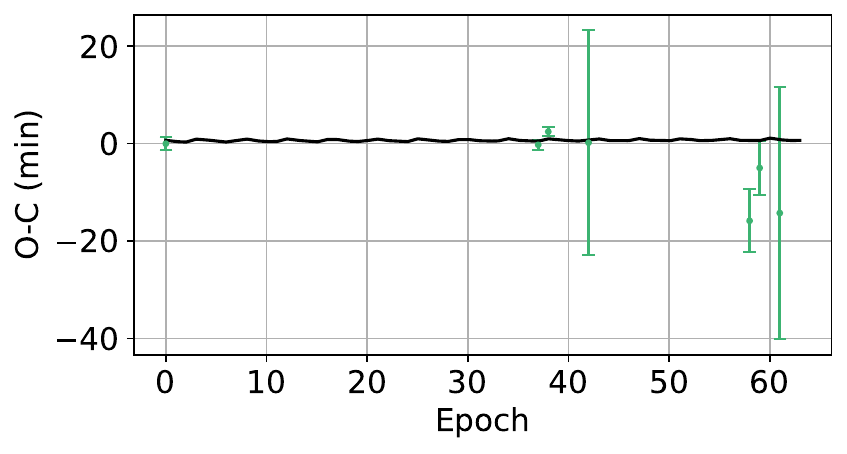}
    \caption{Initial 3-planet (with d interior to b) best fit O--C diagram of AU Mic b ({\it top}) and AU Mic c ({\it bottom}), with comparison between TTVs (green) and \exostriker-generated mcmc models (black).}
    \label{fig:3p_inner_init_exostriker_plots}
\end{figure}

As the 2:3 orbital resonance pair is the most common pairing for mature, compact multi-planet systems \citep{fabrycky2014}, and since near-resonances are necessary to produce the observed detectable TTVs, we explored the possibility that P$_{\rm d}$ = 5.629 days and modeled the best-fit 3-planet configuration using that period as a starting point (Figure \ref{fig:3p_inner_init_exostriker_plots} and Table \ref{tab:3p_inner_init_exostriker_params}). The O--C diagram from Figure \ref{fig:3p_inner_init_exostriker_plots} now displays a super-period in AU Mic b's TTVs, and the statistics from Table \ref{tab:3p_inner_init_exostriker_params} (e.g., $\chi^{2}_{\rm red,b}$ = 5.21, $\chi^{2}_{\rm red,c}$ = 31.23, and $AIC_{c}$ = -343.02) demonstrate that the 3-planet model as having significantly better fitting than both of the 2-planet models.

\begin{deluxetable}{l|c|ccc}\label{tab:3p_inner_init_exostriker_params}
    \tablecaption{\exostriker-generated initial 3-planet (with d interior to b) best fit modeling parameters.}
    \tablehead{Parameter & Unit & AU Mic b & AU Mic c & AU Mic d}
    \startdata
K                & m s$^{-1}$ & 5.03491   & 0.51952   & 0.34401   \\
\porb            & day        & 8.46334   & 18.86226  & 5.63749   \\
e                & ...        & 0.00000   & 0.00459   & 0.00019   \\
$\omega$         & deg        & 98.48619  & 223.69811 & 167.66115 \\
M$_{\rm 0}$      & deg        & 351.49848 & 0.00066   & 239.37339 \\
i                & deg        & 89.57446  & 89.43918  & 87.32379  \\
$\Omega$         & deg        & 30.38207  & 31.22049  & 9.19867   \\ \hline
$\chi^{2}$       & ...        & 113.87171 & 11.06507  & ...       \\
\reduced         & ...        & 5.20570   & 31.23419  & ...       \\
$\ln\mathcal{L}$ & ...        & \multicolumn{3}{c}{209.62069}     \\
$BIC$            & ...        & \multicolumn{3}{c}{-337.51315}    \\
$AIC_{c}$        & ...        & \multicolumn{3}{c}{-343.01916}
    \enddata
    \tablecomments{The K's listed here are unconstrained, but see $\S$\ref{sec:results} for discussion regarding the planets' low K's generated by \exostriker.}
\end{deluxetable}

\begin{figure}
    \centering
    \includegraphics[width=\linewidth]{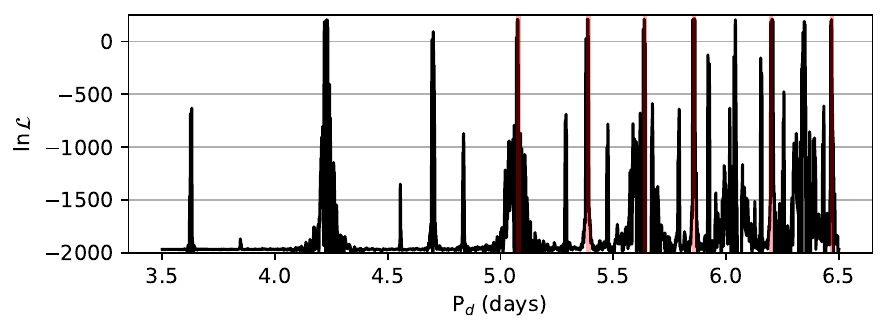}
    \includegraphics[width=\linewidth]{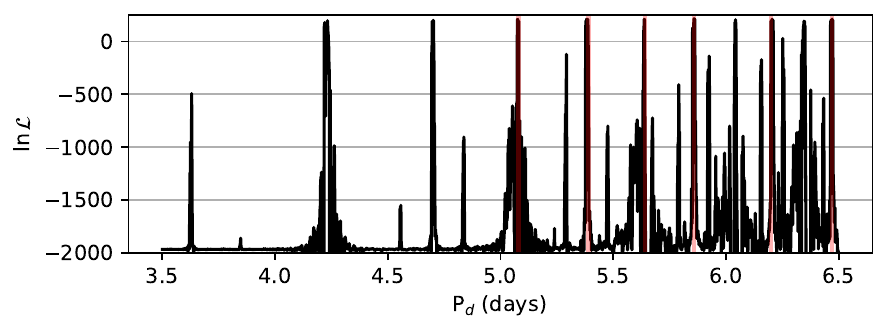}
    \caption{TTV log-likelihood periodograms of AU Mic d's orbital period using the parameters from Table \ref{tab:3p_inner_init_exostriker_params} as the starting point and initial inclinations $i_{\rm d}$ = 85$^{\circ}$ ({\it top}) and 90$^{\circ}$ ({\it bottom}). We obtained the six most favored inner d periods based on their log-likelihood: 5.08 days ($\ln\mathcal{L}$ = 208.22 \& 208.84), 5.39 days ($\ln\mathcal{L}$ = 209.99 \& 209.80), 5.64 days ($\ln\mathcal{L}$ = 209.65 \& 209.62), 5.86 days ($\ln\mathcal{L}$ = 214.36 \& 214.33), 6.20 days ($\ln\mathcal{L}$ = 209.95 \& 208.43), and 6.47 days ($\ln\mathcal{L}$ = 207.76 \& 211.40) from both respective inclination cases. The red lines denote those periods corresponding to best-fitting peaks in the periodograms.}
    \label{fig:inner_ttv_pdg}
\end{figure}

Next, we constructed a TTV log-likelihood periodogram using the optimization functions from the \exostriker library to probe a range of possible orbital periods for AU Mic d using the parameters from Table \ref{tab:3p_inner_init_exostriker_params} as the starting point; we generated a set of 4\,801 orbital periods between 3.5 and 6.5 days corresponding to a step size of 0.0005 days. Since the \exostriker modeling did not significantly deviate from its initial value during the periodogram run, we adjusted d’s initial inclination prior to each run to brute force a broader exploration of this model parameter. We performed two TTV periodograms with different initial inclinations i$_{\rm d}$ = (85$^{\circ}$, 90$^{\circ}$) respectively (Figure \ref{fig:inner_ttv_pdg}). From those runs, we came across six most favored inner d periods based on their respective log-likelihoods: 5.08 days ($\ln\mathcal{L}$ = 208.22 \& 208.84), 5.39 days ($\ln\mathcal{L}$ = 209.99 \& 209.80), 5.64 days ($\ln\mathcal{L}$ = 209.65 \& 209.62), 5.86 days ($\ln\mathcal{L}$ = 214.36 \& 214.33), 6.20 days ($\ln\mathcal{L}$ = 209.95 \& 208.43), and 6.47 days ($\ln\mathcal{L}$ = 207.76 \& 211.40).

We perform best fit modeling and MCMC calculations for each of those best inner d cases and have plotted the O--C diagrams (Figures \ref{fig:3p_inner_exostriker_plots_1} \& \ref{fig:3p_inner_exostriker_plots_2}), the output parameters (Tables \ref{tab:5-08d_exostriker_params}, \ref{tab:5-39d_exostriker_params}, \ref{tab:5-64d_exostriker_params}, \ref{tab:5-86d_exostriker_params}, \ref{tab:6-20d_exostriker_params}, \& \ref{tab:6-47d_exostriker_params}), and the corner plots (Figures \ref{fig:5-08d_exostriker_cornerplot}, \ref{fig:5-39d_exostriker_cornerplot}, \ref{fig:5-64d_exostriker_cornerplot}, \ref{fig:5-86d_exostriker_cornerplot}, \ref{fig:6-20d_exostriker_cornerplot}, \& \ref{fig:6-47d_exostriker_cornerplot}). For all inner d cases, the 3-planet models reproduces the TTV super-period of AU Mic b in O–C diagrams. The statistics criteria (e.g., \reduced and $AIC_{c}$ values from Tables \ref{tab:5-08d_exostriker_params} through \ref{tab:6-47d_exostriker_params} vs those from Tables \ref{tab:2p_ext_exostriker_params} \& \ref{tab:2p_ord_exostriker_params}) indicate that these inner d cases are strongly favored over the 2-planet cases. Notably, only P$_{\rm d}$ = 5.64 days has AU Mic d being coplanar with AU Mic system based on its inclination; the 5.86 and 6.47-day cases have the planet being moderately misaligned with the system, and the 5.08, 5.39, and 6.20-day cases have the planet being highly misaligned with the system..

The AMD criteria in \exostriker suggest that all inner d cases are unstable, so we investigate this further. We tested the stability of these 3-planet cases over 2 Myr with the N-body simulator \rebound and found that all of those configurations are stable (Figure \ref{fig:3p_others_rebound}). However, the 5.39-day case has all three planets' orbits fluctuate toward high-eccentricity configuration, and these planets, especially AU Mic b and d, become highly misaligned for much of the 2-Myr timeline. The 5.86-day case's misalignment issue appears more pronounced, with all three planets becoming quickly misaligned with the system. Although the 5.64-day case has AU Mic d initially being coplanar, \rebound has it become misaligned periodically. Lastly, the 6.47-day case has the planets exhibiting the most consistent coplanarity among the inner d cases.

\subsubsection{Planet d Between b and c}

\begin{figure}
    \centering
    \includegraphics[width=\linewidth]{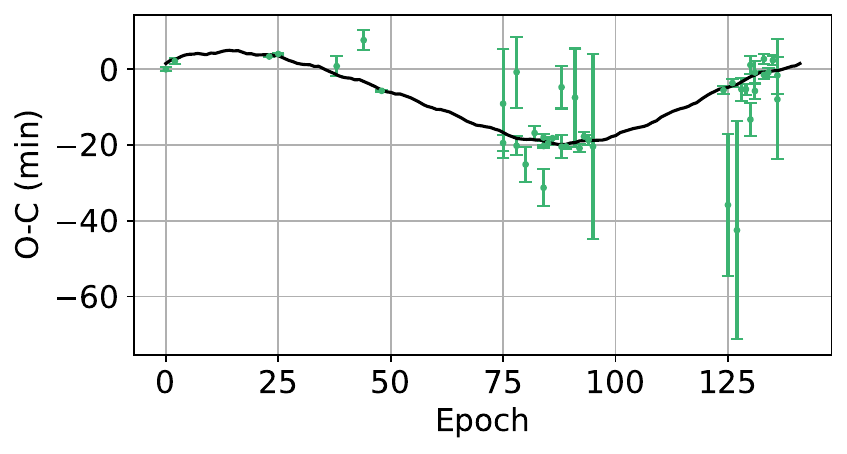}
    \includegraphics[width=\linewidth]{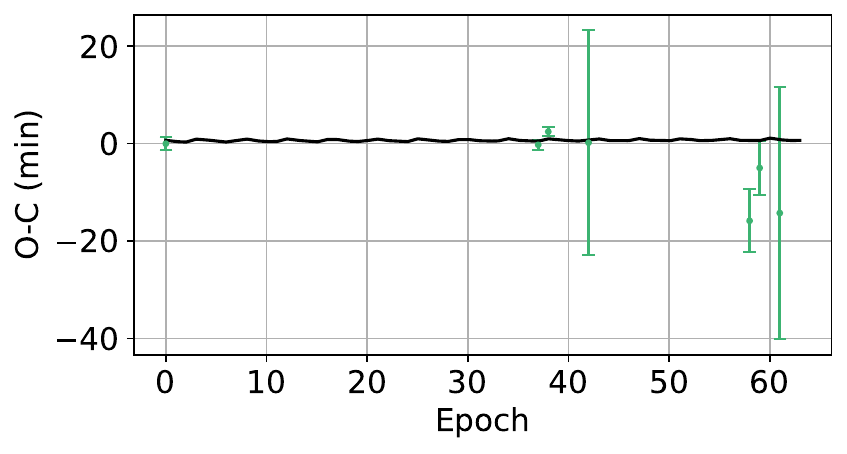}
    \caption{Initial 3-planet (with d between b and c) best fit O--C diagram of AU Mic b ({\it top}) and AU Mic c ({\it bottom}), with comparison between TTVs (green) and \exostriker-generated mcmc models (black).}
    \label{fig:3p_middle_init_exostriker_plots}
\end{figure}

We repeat the steps described in the previous section for the middle d scenario. Based on AU Mic b's TTV super-period and assuming the near-2:3 orbital resonance pair between b and d, we explored the possibility that P$_{d}$ = 12.738 days. Using this orbital period as a starting point, we obtained an initial best fit model for the 3-planet configuration (Figure \ref{fig:3p_middle_init_exostriker_plots} and Table \ref{tab:3p_middle_init_exostriker_params}). The O-C diagram from Figure \ref{fig:3p_middle_init_exostriker_plots} also displays the super-period in AU Mic b's TTVs, and the statistics from Table \ref{tab:3p_middle_init_exostriker_params} (e.g., $\chi^{2}_{\rm red,b}$ = 5.92, $\chi^{2}_{\rm red,c}$ = 35.53, and $AIC_{c}$ = -330.97) demonstrate that this 3-planet model yields a significantly better fit than the 2-planet models.

\begin{deluxetable}{l|c|ccc}\label{tab:3p_middle_init_exostriker_params}
    \tablecaption{\exostriker-generated initial 3-planet (with d between b and c) best fit modeling parameters.}
    \tablehead{\multirow{2}{*}{Parameter} & \multirow{2}{*}{Unit} & \multicolumn{3}{c}{Best Fit}   \\
                                          &                       & AU Mic b & AU Mic c & AU Mic d }
    \startdata
K                & m s$^{-1}$ & 4.85810   & 1.49734   & 0.16823   \\
\porb            & day        & 8.46325   & 18.86214  & 12.73040  \\
e                & ...        & 0.00704   & 0.00136   & 0.00283   \\
$\omega$         & deg        & 89.95582  & 223.96004 & 166.18027 \\
M$_{0}$          & deg        & 0.00002   & 0.00018   & 0.00017   \\
i                & deg        & 89.39532  & 89.41147  & 81.08949  \\
$\Omega$         & deg        & 0.00002   & 0.00000   & 0.00000   \\ \hline
$\chi^{2}$       & ...        & 131.05680 & 11.06507  & ...       \\
\reduced         & ...        & 5.92174   & 35.53047  & ...       \\
$\ln\mathcal{L}$ & ...        & \multicolumn{3}{c}{203.59374}     \\
$BIC$            & ...        & \multicolumn{3}{c}{-325.45925}    \\
$AIC_{c}$        & ...        & \multicolumn{3}{c}{-330.96526}
    \enddata
    \tablecomments{The K's listed here are unconstrained, but see $\S$\ref{sec:results} for discussion regarding the planets' low K's generated by \exostriker.}
\end{deluxetable}

\begin{figure}
    \centering
    \includegraphics[width=\linewidth]{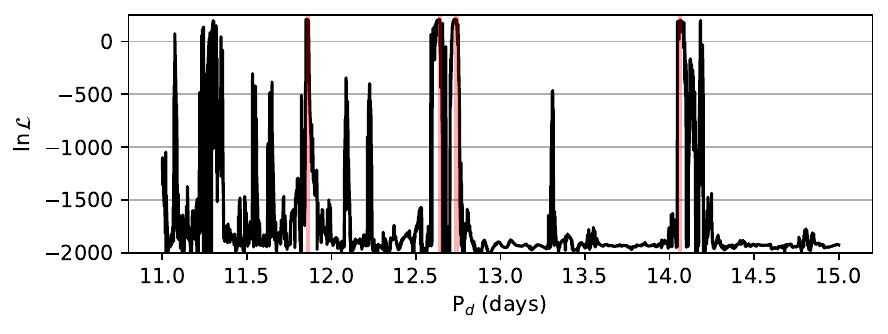}
    \includegraphics[width=\linewidth]{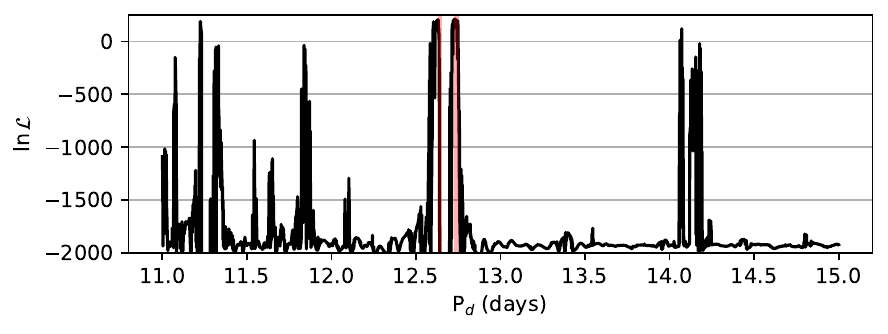}
    \caption{TTV periodograms of AU Mic d's orbital period using the parameters from Table \ref{tab:3p_middle_init_exostriker_params} as the starting point and initial inclinations $i_{\rm d}$ = 80$^{\circ}$ ({\it top}) and 90$^{\circ}$ ({\it bottom}). We obtained the four most favored periods based on their log-likelihood: 11.9 days ($\ln\mathcal{L}$ = 208.16) and 14.1 days ($\ln\mathcal{L}$ = 201.45) from initial i$_{\rm d}$ = 80$^{\circ}$ case, and 12.6 days ($\ln\mathcal{L}$ = 209.21 \& 205.70) and 12.7 days ($\ln\mathcal{L}$ = 211.25 \& 209.79) from both respective inclination cases. The red lines denote those days corresponding to best-fitting peaks in the periodograms.}
    \label{fig:middle_ttv_pdg}
\end{figure}

Next, we utilize the TTV periodogram to probe a range of possible orbital periods for AU Mic d using the parameters from Table \ref{tab:3p_middle_init_exostriker_params} as the starting point; we generated a set of 8\,001 orbital periods between 11 and 15 days with 0.0005-day interval in between. We adjusted d's initial inclination prior to each run and did two runs with different initial inclinations i$_{\rm d}$ = (80$^{\circ}$, 90$^{\circ}$) respectively (Figure \ref{fig:middle_ttv_pdg}). From those runs, we came across four most favored periods based on their log-likelihood: 11.9 days ($\ln\mathcal{L}$ = 208.16) and 14.1 days ($\ln\mathcal{L}$ = 201.45) from initial i$_{\rm d}$ = 80$^{\circ}$ case, and 12.6 days ($\ln\mathcal{L}$ = 209.21 \& 205.70) and 12.7 days ($\ln\mathcal{L}$ = 211.25 \& 209.79) from both respective inclination cases.

We perform best fit modeling and MCMC calculations for each of those four best middle d cases and have generated the O--C diagrams (Figures \ref{fig:12-7d_exostriker_plots} \& \ref{fig:3p_middle_exostriker_plots}), the output parameters (Tables \ref{tab:12-7d_exostriker_params}, \ref{tab:11-9d_exostriker_params}, \ref{tab:12-6d_exostriker_params}, \& \ref{tab:14-1d_exostriker_params}), and the corner plots (Figures \ref{fig:12-7d_exostriker_cornerplot}, \ref{fig:11-9d_exostriker_cornerplot}, \ref{fig:12-6d_exostriker_cornerplot}, \& \ref{fig:14-1d_exostriker_cornerplot}). For all of those four cases, the 3-planet models adequately account for the TTV super-period of AU Mic b in the O--C diagrams, and the statistics criteria (e.g., \reduced and $AIC_{c}$ values from Tables \ref{tab:12-7d_exostriker_params} \& \ref{tab:11-9d_exostriker_params} through \ref{tab:14-1d_exostriker_params} vs those from Tables \ref{tab:2p_ext_exostriker_params} \& \ref{tab:2p_ord_exostriker_params}) indicate that all 3-planet middle d cases are strongly favored over the 2-planet case. Curiously, planet d's inclination varies considerably among these four cases. For instance, the 12.6 and 12.7-day cases have the planet d being relatively coplanar with the AU Mic system while the 11.9 and 14.1-day cases have it being highly misaligned.

The AMD criteria in \exostriker suggest that these 3-planet cases might be unstable. As before, we utilized \rebound to test the stability of the 3-planet cases; through this package, we found that each of these 3-planet configurations is stable (Figures \ref{fig:12-7d_rebound} \& \ref{fig:3p_others_rebound}). Among the middle d cases, both the 12.6 and 12.7-day cases are the most consistent in maintaining the coplanarity of the planets.

\subsection{TTV Model Comparisons}\label{sec:ttv_mod_comp}

We perform a model comparison among both the 2-planet and 3-planet cases, as seen in Table \ref{tab:ttv_mod_comp}. Based on the TTV model information criterion, the P$_{\rm d}$ = 12.7 days configuration is the most favored overall, and the P$_{\rm d}$ = 6.47 days configuration is the most favored among the inner d cases, followed closely by the 6.20-day and 5.86-day cases. All other configurations are not as well favored based on their corresponding $\Delta BIC$ \citep[Bayesian information criterion;][]{schwarz1978, wit2012} and $\Delta AIC_{c}$, but they are not statistically ruled out. The 2-planet cases' $\Delta BIC$ and $\Delta AIC_{c}$ are much larger, so both 2-planet models can be rejected.

\begin{deluxetable*}{l|c|c|c|c|c}\label{tab:ttv_mod_comp}
    \tablecaption{TTV model information criterion for the AU Mic system.}
    \tablehead{Type & P$_{\rm d}$ (days) & Relative \reduced & $\Delta\ln{\mathcal{L}}$ & $\Delta BIC$ & $\Delta AIC_{c}$}
    \startdata
\multirow{12}{*}{MCMC}     & 12.7          & 3.95967   & 0.00000    & 0.00000    & 0.00000    \\
                           & 6.47          & 3.99117   & 1.56595    & 3.13190    & 3.13190    \\
                           & 6.20          & 3.92884   & 1.67726    & 3.35452    & 3.35452    \\
                           & 5.86          & 4.08053   & 1.75398    & 3.50796    & 3.50796    \\
                           & 14.1          & 4.12141   & 2.37983    & 4.75966    & 4.75966    \\
                           & 5.39          & 4.14006   & 2.78037    & 5.56074    & 5.56074    \\
                           & 12.6          & 4.33368   & 5.83824    & 11.67648   & 11.67648   \\
                           & 11.9          & 4.44417   & 7.12662    & 14.25324   & 14.25324   \\
                           & 5.08          & 4.45799   & 8.78461    & 17.56922   & 17.56922   \\
                           & 5.64          & 4.55658   & 9.62125    & 19.24250   & 19.24250   \\
                           & no d (high e) & 5.66526   & 52.65304   & 78.06333   & 69.43680   \\
                           & no d (low e)  & 64.88111  & 1192.49499 & 2357.74723 & 2349.12070 \\ \hline
\multirow{12}{*}{Best Fit} & 12.7          & 4.05575   & 0.00000    & 0.00000    & 0.00000    \\
                           & 5.86          & 4.08923   & 0.56826    & 1.13652    & 1.13652    \\
                           & 6.20          & 4.25992   & 2.95667    & 5.91334    & 5.91334    \\
                           & 14.1          & 4.28073   & 3.27556    & 6.55112    & 6.55112    \\
                           & 5.39          & 4.30368   & 3.52700    & 7.05400    & 7.05400    \\
                           & 6.47          & 4.30787   & 3.53291    & 7.06582    & 7.06582    \\
                           & 12.6          & 4.36285   & 4.37647    & 8.75294    & 8.75294    \\
                           & 11.9          & 4.43253   & 5.40495    & 10.80990   & 10.80990   \\
                           & 5.64          & 4.46849   & 5.87727    & 11.75454   & 11.75454   \\
                           & 5.08          & 4.47444   & 5.94329    & 11.88658   & 11.88658   \\
                           & no d (high e) & 5.46622   & 39.22908   & 51.21541   & 42.58888   \\
                           & no d (low e)  & 118.95179 & 2023.71529 & 4020.18783 & 4011.56130
    \enddata
    \tablecomments{The numbers of degrees of freedom used to calculate the relative \reduced are N$_{\rm dof}$ = 14 for a 2-planet model and N$_{\rm dof}$ = 21 for a 3-planet model.}
\end{deluxetable*}

Next, we checked the robustness of the results by including only the high-precision TTVs, namely the TESS + Spitzer + SPIRou + CHEOPS joint modeling and performing the best fit modeling and MCMC calculations, then we did a model comparison of the results (Table \ref{tab:ttv_mod_comp_hp}). This time, the 14.1-day case is the most favored overall followed closely by the 12.7-day case, and the 5.39-day case is the most favored among the inner d cases; all other cases are disfavored but not statistically ruled out. Again, the 2-planet cases are statistically ruled out due to their significantly high $\Delta BIC$ and $\Delta AIC_{c}$.

\begin{deluxetable*}{l|c|c|c|c|c}\label{tab:ttv_mod_comp_hp}
    \tablecaption{High-precision (TESS + Spitzer + SPIRou + CHEOPS) TTV model information criterion for the AU Mic system.}
    \tablehead{Type & P$_{\rm d}$ (days) & Relative \reduced & $\Delta\ln{\mathcal{L}}$ & $\Delta BIC$ & $\Delta AIC_{c}$}
    \startdata
\multirow{12}{*}{MCMC}     & 14.1          & 10.24154  & 0.00000    & 0.00000    & 0.00000    \\
                           & 12.7          & 10.88606  & 1.42079    & 2.84158    & 2.84158    \\
                           & 5.39          & 11.96840  & 2.76171    & 5.52342    & 5.52342    \\
                           & 6.20          & 12.10677  & 3.84762    & 7.69524    & 7.69524    \\
                           & 5.86          & 14.15778  & 6.27514    & 12.55028   & 12.55028   \\
                           & 12.6          & 14.71546  & 7.24787    & 14.49574   & 14.49574   \\
                           & 6.47          & 14.72007  & 7.58335    & 15.16670   & 15.16670   \\
                           & 11.9          & 16.53061  & 9.83512    & 19.67024   & 19.67024   \\
                           & 5.08          & 16.89976  & 11.34680   & 22.69360   & 22.69360   \\
                           & 5.64          & 17.35204  & 12.42087   & 24.84174   & 24.84174   \\
                           & no d (high e) & 13.55832  & 40.32199   & 59.33232   & 1060.64398 \\
                           & no d (low e)  & 164.49125 & 601.87497  & 1182.43828 & 2183.74994 \\ \hline
\multirow{12}{*}{Best Fit} & 12.7          & 10.96391  & 0.00000    & 0.00000    & 0.00000    \\
                           & 14.1          & 11.56410  & 0.83980    & 1.67960    & 1.67960    \\
                           & 6.20          & 13.89642  & 4.46176    & 8.92352    & 8.92352    \\
                           & 5.39          & 13.94522  & 4.50562    & 9.01124    & 9.01124    \\
                           & 5.86          & 14.65509  & 5.64192    & 11.28384   & 11.28384   \\
                           & 12.6          & 14.77799  & 5.72267    & 11.44534   & 11.44534   \\
                           & 6.47          & 15.66435  & 7.16597    & 14.33194   & 14.33194   \\
                           & 11.9          & 16.41644  & 8.26575    & 16.53150   & 16.53150   \\
                           & 5.08          & 16.54831  & 8.46338    & 16.92676   & 16.92676   \\
                           & 5.64          & 16.68665  & 8.69418    & 17.38836   & 17.38836   \\
                           & no d (high e) & 14.39713  & 34.30994   & 47.30822	  & 1048.61988 \\
                           & no d (low e)  & 312.57388 & 1077.13307 & 2132.95448 & 3134.26614
    \enddata
    \tablecomments{The numbers of degrees of freedom used to calculate the relative \reduced are N$_{\rm dof}$ = 14 for a 2-planet model and N$_{\rm dof}$ = 18 for a 3-planet model. N$_{\rm dof}$ = 18 is used instead of 21 due to there being 21 data observations in this high-precision model comparison.}
\end{deluxetable*}

Taking into consideration with both model comparisons, the 12.7-day, 14.1-day, 5.39-day, and 6.20-day cases are consistently favored while the 12.6-day, 11.9-day, 5.64-day, 5.08-day, and both 2-planet cases are consistently disfavored relative to the other cases. The 5.86 and 6.47-day cases are inconsistent, with both being more relatively favored in the all-TTVs model comparison and much less so in the high-precision TTVs model comparison.

We did not explore randomly removing individual TTVs to evaluate the robustness of these results and the dependence of our results on particular TTV measurements. Our data set is notably heterogeneous, and some of the earlier observations, particularly the Spitzer transit times, are critical in determining the super-period because they sparsely fill in the TTV curve. Thus, as would expected, our results are highly dependent on these space-based timing measurements. Conversely, for the 2021 season, we have much denser transit timing sampling, and thus our results will be less impacted by the removal of those data sets.

\subsection{Stability Analysis of the Ten Cases}\label{sec:stability_tests}

With both 2-planet cases statistically ruled out and therefore excluded (Tables \ref{tab:ttv_mod_comp} \& \ref{tab:ttv_mod_comp_hp}), we proceeded to test the dynamical stability of the aforementioned ten 3-planet configurations by utilizing the \spock package's {\tt FeatureClassifier} and {\tt NbodyRegressor} \citep{tamayo2020}. The {\tt FeatureClassifier}, hereafter referred to as \spock, is a trained model estimating the stability probability after 10$^{9}$ orbits of the innermost planet (for AU Mic d in inner d scenario: 13.9 Myr for 5.08 days, 14.8 Myr for 5.39 days, 15.4 Myr for 5.64 days, 16.0 Myr for 5.86 days, 17.0 Myr for 6.20 days, and 17.7 Myr for 6.47 days; for AU Mic b in middle d scenario: 23.2 Myr, which is comparable to the age of the system). Subsequently, we weight the parameters by their stability probability to sort out or give weaker weight to unstable systems that would likely have already destroyed themselves over time.

The {\tt NbodyRegressor}, hereafter referred to as \nbody, performs an \nbody simulation and checks a system for stability after an arbitrarily chosen number of orbits of the innermost planet. Due to the high numerical cost of an \nbody simulation over a large number of orbits, we chose a simulation period of $2\times10^{5}$ orbits for this part of the analysis (for AU Mic d in inner d scenario: 2.78 kyr for 5.08 days, 2.95 kyr for 5.39 days, 3.09 kyr for 5.64 days, 3.21 kyr for 5.86 days, 3.40 kyr for 6.20 days, and 3.54 kyr for 6.47 days; for AU Mic b in middle d scenario: 4.63 kyr). We utilized \nbody for two purposes: To notice any deviations that may arise between \spock and \nbody results, and to validate the results from \spock. \spock was trained with planetary configurations showing mutual inclinations $\lesssim$10$^{\circ}$. Since higher values for the mutual inclinations can be found for the cases 5.08 days, 5.39 days, 6.20 days, and 14.1 days, we considered it useful to use additional analysis methods for comparison. At this point, however, it is important to note that differences and similarities between the results of \spock and \nbody can only serve as an indication and not a proof of \spock's reliability, since the time period after which a system is tested for stability differs significantly. A system that becomes unstable after 10$^{6}$ orbits will be classified as stable by \nbody in our case, although \spock will probably assign a low probability of stability to the system. Nevertheless, we think it is useful to use \nbody as a comparison to notice significant differences and to look at the stability of the system over different timescales.

We also utilized \megno \citep[{\tt mean exponential growth factor of nearby orbits};][]{cincotta2000, cincotta2003}, which is a fast indicator for chaotic orbits, for our orbital simulations. The \megno can be used similar to \spock to estimate the orbital stability of planetary systems as done by \citet{tamayo2020}, although it is pointed out that a direct comparison is only possible to a limited extent \citep[see $\S$1.D. in][]{tamayo2020}. Since \spock uses \megno as one of ten features for internal analysis, we consider a comparison of results useful in this case as well. However, also in this case it is important to emphasize that discrepancies between \spock and \megno can only be an indication of inconsistencies of the \spock results, since orbital stability and regularity of orbits cannot be used synonymously.

We incorporated the stellar parameters from Table \ref{tab:exostriker_priors} and the planetary parameters from \exostriker-generated posteriors (Tables \ref{tab:12-7d_exostriker_params} \& \ref{tab:5-08d_exostriker_params} thru \ref{tab:14-1d_exostriker_params}). We generated 20\,000 random configurations for each of the 10 cases, drawn from the respective Gaussian distributions of the planetary parameters. Next, a simulation object is created from \rebound, to which all planetary parameters and the mass of the star of a certain random configuration are passed on. Then, this simulation object is examined through \spock for the stability probability after 10$^{9}$ orbits of the innermost planet. The same parameters were also used to estimate the stability of the system after $2\times10^{5}$ orbits using \nbody. \spock determines ten features for the classification of planetary systems, based on which it estimates the stability probability. Since \megno is one of them, it can be directly output by \spock and used for our analysis. Thus, all three analysis methods use the same orbital parameters and masses.

\begin{figure*}
    \centering
    \includegraphics[width=0.47\textwidth]{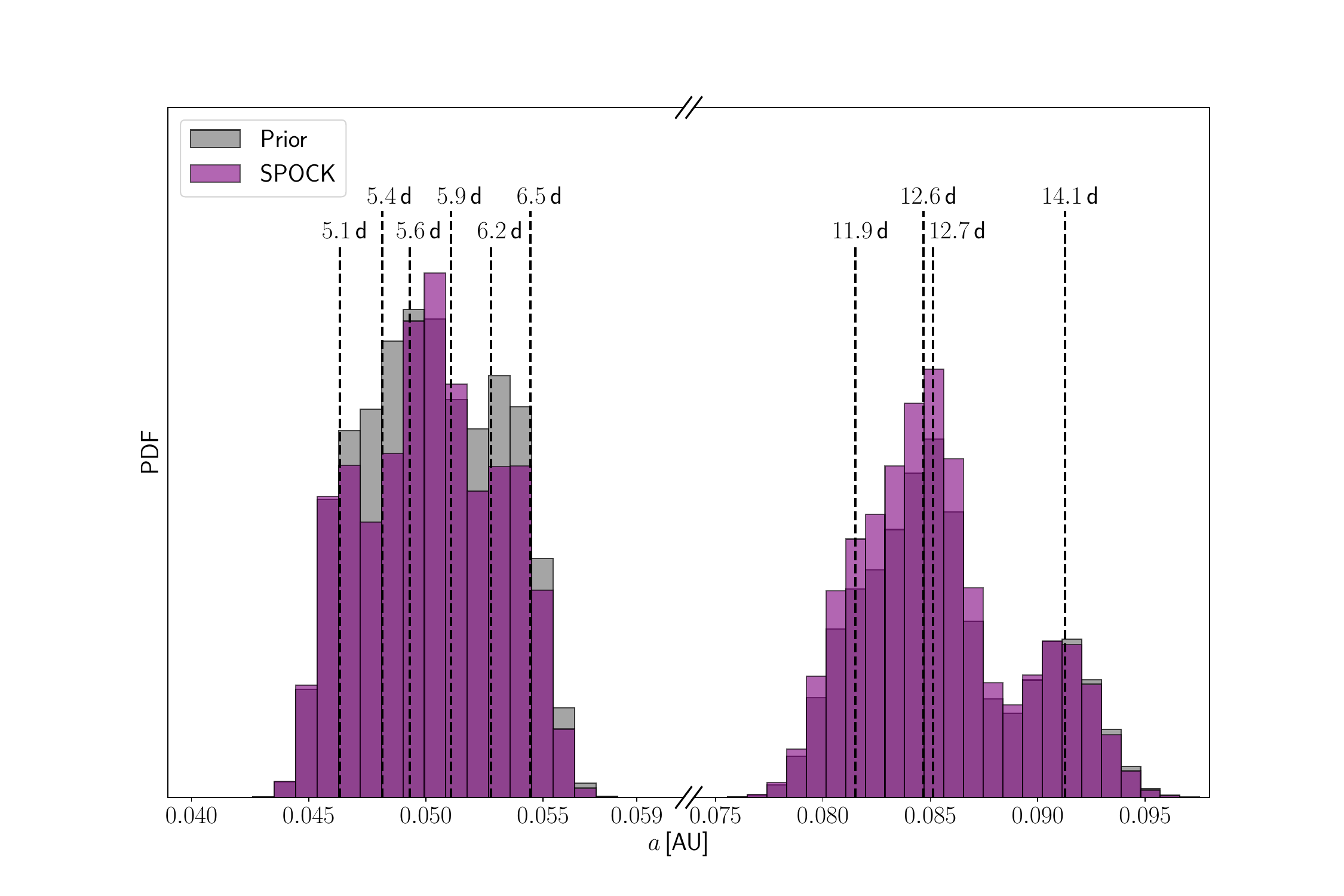}
    \includegraphics[width=0.47\textwidth]{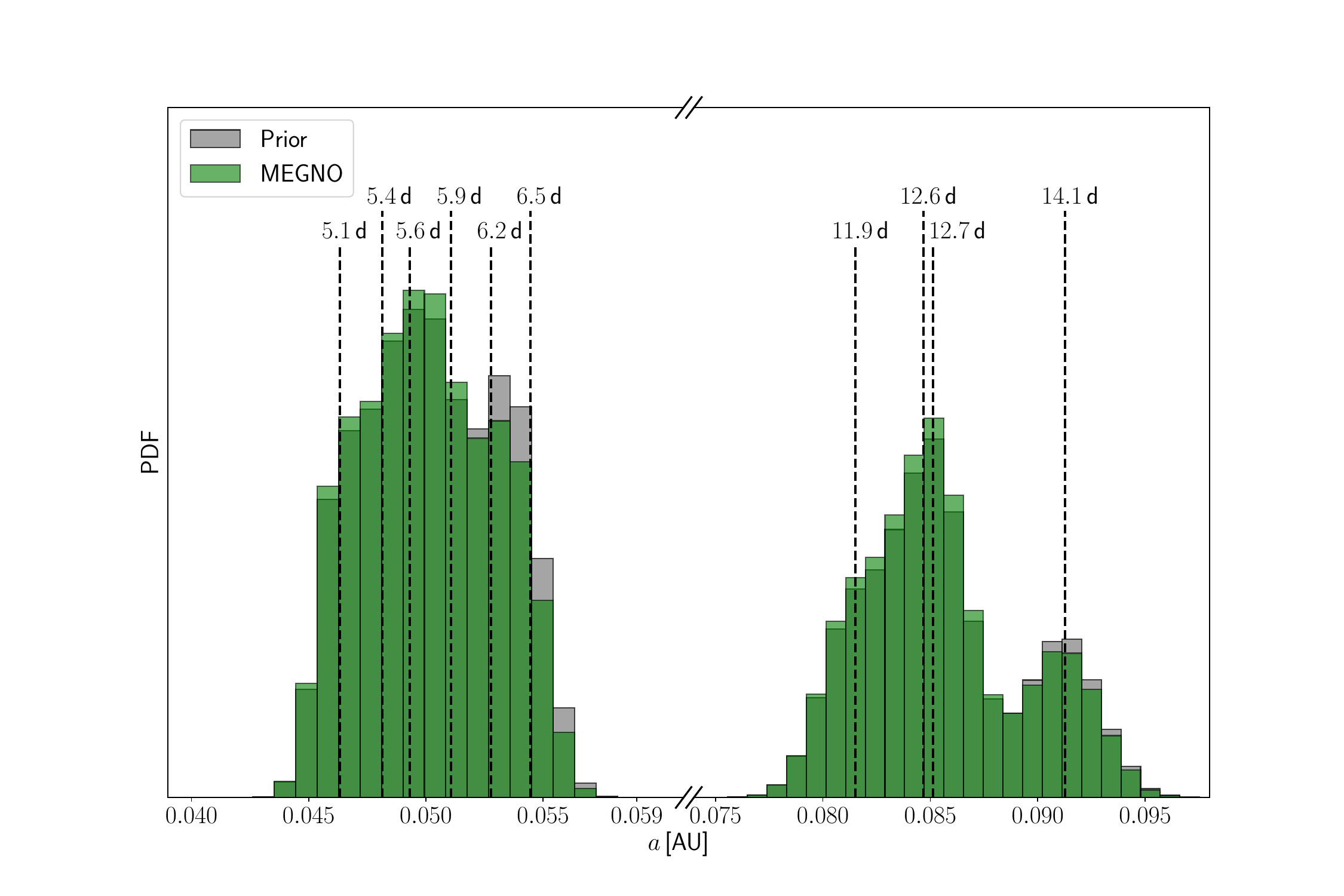}
    \includegraphics[width=0.47\textwidth]{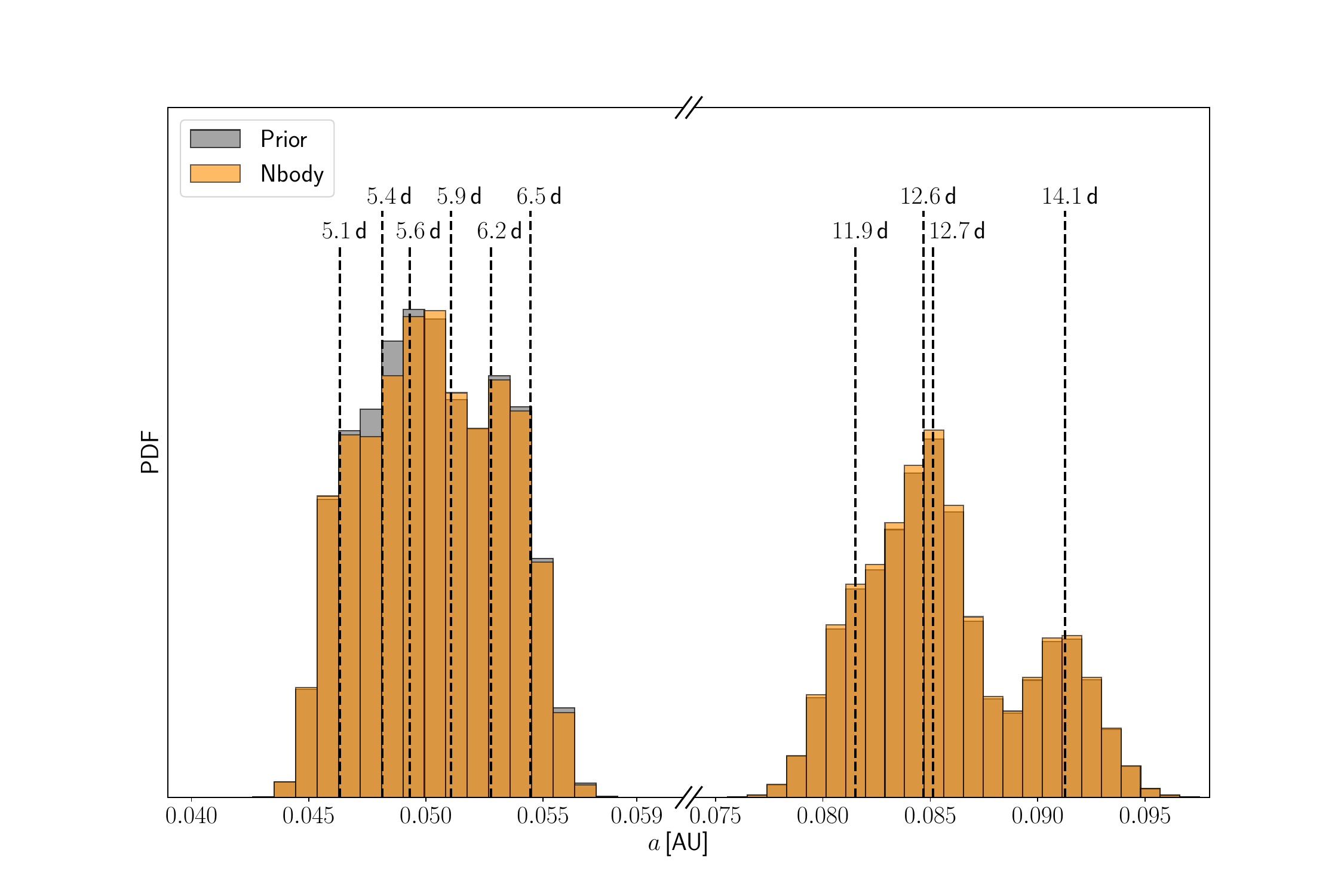}
    \caption{Histograms of the semi-major axis of AU Mic d with \spock ({\it top left}), \megno ({\it top right}), and \nbody ({\it bottom}), each marked with the corresponding orbital periods of AU Mic d. \spock and \nbody estimates the stability probability after 10$^{9}$ and 2$\times10^{5}$ orbits, respectively, of the innermost planet, while \megno is a fast indicator for chaotic orbits. For each parameter set, 20k random configurations of masses and orbital parameters were generated following the respective Gaussian distributions of planetary parameters from \exostriker; each of these random configurations was given its own stability probability by \spock, and the semi-major axis of AU Mic d is then plotted in the histogram for each of these configurations. Since the same number of random configurations was generated for each parameter set, the result is a mixture distribution in which the respective underlying distributions have the same weighting. The values were then weighted with the stability probability in the histogram, resulting in the overlapping histogram for \spock, \megno, and \nbody. Thus, each histogram contains 200k values. The \nbody uses the same aforementioned simulation object generated by \rebound to perform \nbody simulations and outputs either 0 (unstable) or 1 (stable); these outputs are then used for weighting. \nbody is more accurate than \spock but is much more computationally intensive over a large simulation time, so both the 200k-orbit integrations were performed and the pretrained classifier was used. The \megno value threshold is set to 2.1, so any configurations that lead to their \megno value exceeding this threshold are considered unstable and given a weight of 0 while all other configurations that are considered stable are given a weight of 1. By weighting with the \spock or \megno outputs, the 5.64-day case appears to be favored over other inner d cases, and both 12.6 and 12.7 days appear to be preferred among the middle d cases.}
    \label{fig:stability_hist}
\end{figure*}

In Figure \ref{fig:stability_hist}, the semi-major axis of AU Mic d was then plotted in a histogram for each of these configurations. Since the same number of random configurations was generated for each parameter set, the result is a mixture distribution in which the respective underlying distributions have the same weighting. The values are then weighted with the stability probability in the histogram, resulting in the overlapping histogram for \spock, \megno, and \nbody. Each histogram thus contains 200k values each with and without weighting by the stability probability. Unlike \spock, \nbody and \megno do not provide continuously distributed stability probabilities. \nbody provides binary stability probabilities, weighting a stable system by 1 and an unstable system by 0. For \megno, however, we had to set a limit above which we classify a system as unstable for comparison with the other methods. To reduce the number of systems incorrectly classified as unstable, we use the difference of the mean $\mu_{\megno}$ and the standard deviation $\sigma_{\megno}$ of the \megno, which can also be output by \spock, for the estimation of the orbital stability. If $\mu_{\megno} - \sigma_{\megno} >$ 2.1, the system is considered as chaotic and thus unstable and receives weight 0. On the other hand, if the value is smaller, we consider the system stable and it receives weight 1. At this point, it should be emphasized once again that a chaotic system does not necessarily have to lead to instability in the near future.

Figure \ref{fig:stability_hist} displays the results from these aforementioned runs. The \spock results appear to deviate from the \megno and \nbody results for the inner planet d cases while all three results appear to be consistent for the middle planet d case. For the inner d scenario, the 5.64-day case appears to be the most preferred, although the \spock results peak closer to the 5.86-day case. For the middle d scenario, both the 12.6 and 12.7-day cases are clearly preferred over the 11.9 and 14.1-day cases from all three results. Comparisons between the inner d orbits and the middle d orbits should be viewed with caution because of the different simulation durations due to the different innermost orbit. \megno and, in particular, \nbody show hardly any visible trends. One reason for the differences between \spock and the other methods could be that the training sample was not matched by each parameter set. However, it is more likely that the differences are due to the different number of orbits considered.

To check whether this is indeed the case, we ran 100 \nbody simulations over 10$^{7}$ orbits for each of the ten parameter sets. This is still two orders of magnitude smaller than the 10$^{9}$ orbits according to which \spock estimates the stability probability, but a smaller value was chosen because of the numerical effort. In fact, after 10$^{7}$ orbits, the systems tended to become unstable more often. The clearest difference occurred in the 5.39-day case. While 85\% of the generated configurations were still stable after $2\times10^{5}$ orbits, only 37\% were stable after 10$^{7}$ orbits. For the 6.47-day case and the 14.1-day case, the number of stable configurations fell by about 9\% from 97\% to 88\% and from 100\% to 91\%, respectively. For the other cases, only minor effects or, in some cases, no effects at all were observed. However, the results are still consistent with \spock, since the 5.39-day case and the 6.47-day case are clearly disfavored here as well. For this reason, we assume that the \spock results provide a sound basis for assessing probable parameters for AU Mic d.

To make sure that the configurations evolving instabilities over the larger timescale match the \spock results, we analyzed the fraction of equally classified configurations by \spock and \nbody. As \spock provides a stability probability instead of a binary stability classification, we have to set a limit, above which all configurations are classified as stable in order to be able to compare the results with \nbody and \megno. For a first approach, we set this limit to 0.34 because \citet{tamayo2020} used this limit to analyze the reliability of \spock as well. As \citet{tamayo2021} remark that the chosen limit can influence the false positive rate, we additionally try the limits 0.2 and 0.5 to make sure that differences in the results are not due to the selected threshold. Therefore, each configuration exceeding a stability probability of 0.34 is classified as stable, and each configuration below this limit is classified as unstable. It is important to mention that this limit is only necessary to directly compare \spock with \nbody and \megno. To weight the values in the histogram, the stability probabilities are used without a certain stability limit.

Due to the difference in the number of orbits after which the stability probability is predicted (10$^{9}$ for \spock and $2\times10^{5}$ \& 10$^{7}$ for \nbody), we cannot consider the fraction of different classified configurations as false positive or false negative rate, respectively. Nevertheless, we expect that the number of equally classified configurations should increase for the 10$^{7}$ run as the timespan, in which instabilities can occur, grows. For most parameter sets the fraction of different classified systems is clearly below 10\% for both the $2\times10^5$ and the 10$^{7}$ runs. Nevertheless, the cases 5.39 days, 6.20 days, 6.47 days, and 14.1 days are exceptions and show stronger deviating results. For 6.47 days and 14.1 days, the fraction of different classified systems decreased from 13\% and 12\% to 7\% and 5\%. For these cases, the number of considered orbits seems to be the main reason for the differences in the $2\times10^5$ run, and we assume that the \nbody results will approach closer to \spock for a longer \nbody run. However, the cases 6.20 days and 5.39 days show a different behaviour; both parameter sets show large discrepancies for the $2\times10^5$ \nbody run (41\% for 6.20 days and 46\% for 5.39 days). Despite the decrease from 41\% to 37\% for the 6.20 day case, the fraction of different classified systems is still much larger than for the other parameter sets. For the 5.39-day case, the rate is even increasing from 46\% to 53\% what corresponds to an almost random classification. To check whether the choice of the stability limit causes this effect, we compared the results again for a stability limit of 0.2 and 0.5 for \spock. For both limits, the fraction of differently classified configurations was about 50\%. Although we can not exclude the possibility that the parameter set develops instabilities on larger timescales, we would like to mention that both the 5.39-day case as well as the 6.20-day set belong to the sets for which the mutual inclinations do not match the \spock training sample. So, it is quite possible that this could be the reason for the deviations.

However, the behavior of \megno is particularly striking here. In the comparison with \nbody after 10$^{7}$ orbits, the proportion of differently classified systems for the case 5.39 days was even 64\%. For the 6.20 days the proportion is about 19\%. Since \spock uses, among other features, the \megno to estimate the stability of planetary configurations, this could be a decisive factor for the deviations. Nevertheless, for most of the parameter sets there is a very good agreement between \spock and \nbody, and even for the case of 5.39 days the trend indicates a higher number of unstable systems after a high number of orbits for both methods. This suggests that the trend shown by the \spock histogram may well be used as an indication of the stability of the system to at least partly limit the parameter space.

\subsection{Most Favored Configuration for AU Mic d}

\begin{figure*}
    \centering
    \includegraphics[width=0.48\textwidth]{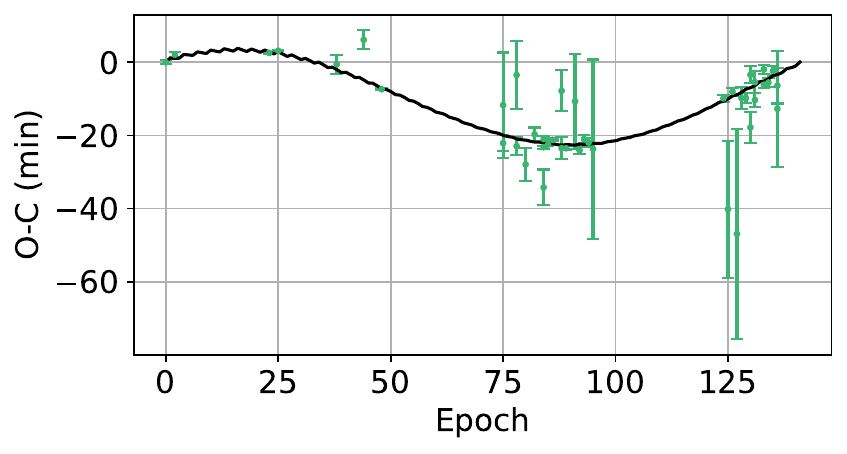}
    \includegraphics[width=0.48\textwidth]{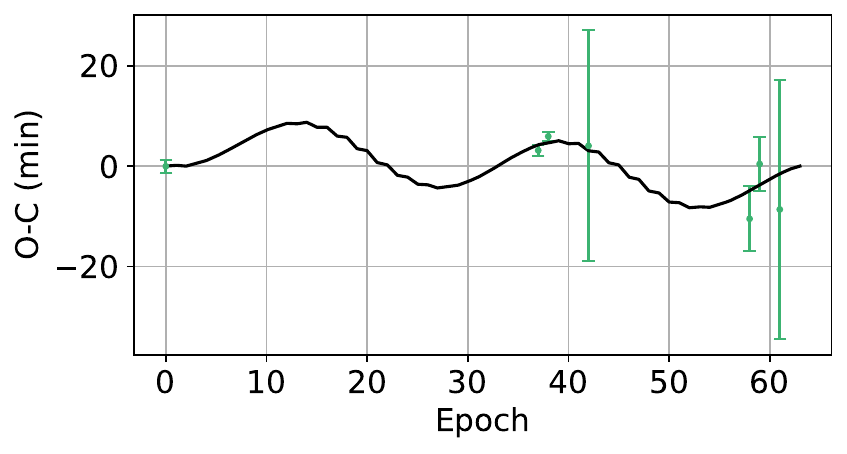}
    \caption{O--C diagram of AU Mic b ({\it left}) and AU Mic c ({\it right}), with comparison between TTVs (green) and \exostriker-generated mcmc models (black) for P$_{\rm d}$ = 12.7 days.}
    \label{fig:12-7d_exostriker_plots}
\end{figure*}

\begin{figure*}
    \centering
    \includegraphics[width=\textwidth]{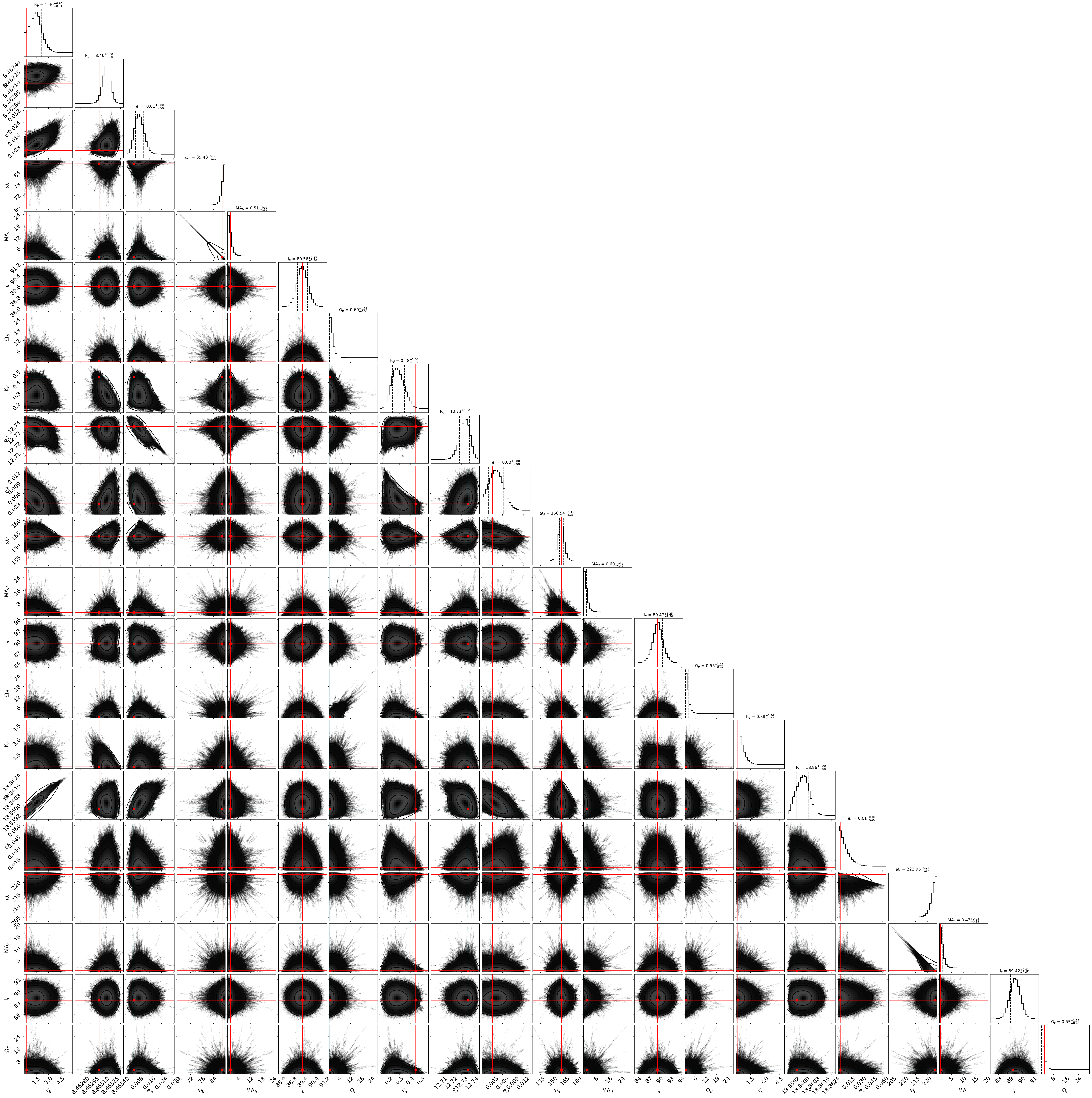}
    \caption{Corner plot of AU Mic b and c's orbital parameters from \exostriker mcmc analysis for P$_{\rm d}$ = 12.7 days.}
    \label{fig:12-7d_exostriker_cornerplot}
\end{figure*}

\begin{figure}
    \centering
    \includegraphics[width=\linewidth]{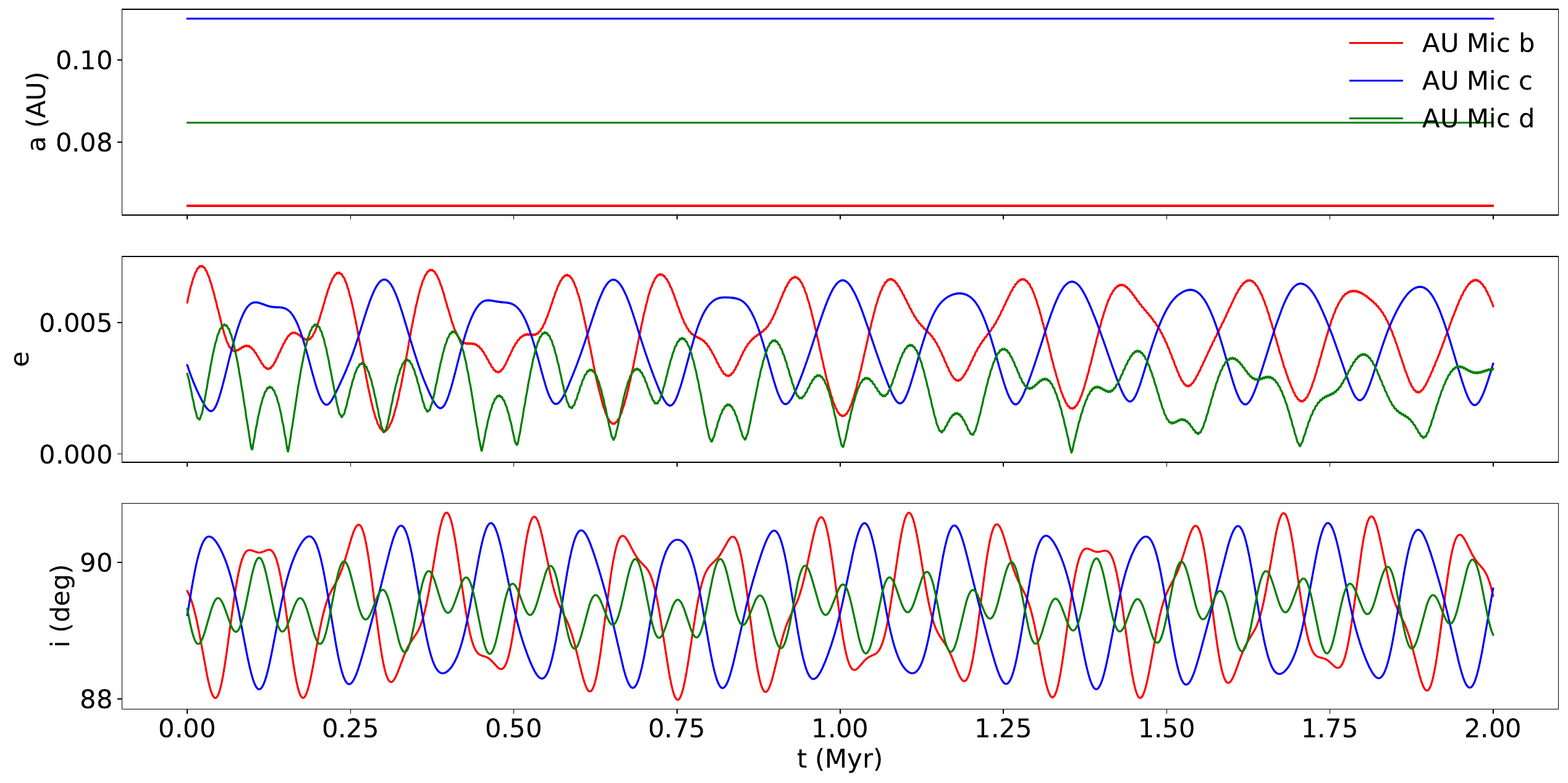}
    \caption{\rebound model of the stability of AU Mic system on timescale of 2 Myr for P$_{\rm d}$ = 12.7 days.}
    \label{fig:12-7d_rebound}
\end{figure}

Based on the calculated super-period ($\S$\ref{sec:ocdiagram}), the TTV model comparisons ($\S$\ref{sec:ttv_mod_comp}), and the stability test ($\S$\ref{sec:stability_tests}), the 12.7-day configuration is the most favored model and thus is presented in this section (Figures \ref{fig:12-7d_exostriker_plots}, \ref{fig:12-7d_exostriker_cornerplot}, \& \ref{fig:12-7d_rebound}, and Table \ref{tab:12-7d_exostriker_params}), while the other TTV cases are presented in the Appendix. More discussions regarding the results can be found in $\S$\ref{sec:results}.

\begin{deluxetable*}{l|c|ccc|ccc}\label{tab:12-7d_exostriker_params}
    \tablecaption{\exostriker-generated best fit and MCMC modeling parameters for P$_{\rm d}$ = 12.7 days.}
    \tablehead{\multirow{2}{*}{Parameter} & \multirow{2}{*}{Unit} & \multicolumn{3}{c|}{Best Fit}  & \multicolumn{3}{c}{MCMC}       \\
                                          &                       & AU Mic b & AU Mic c & AU Mic d & AU Mic b & AU Mic c & AU Mic d }
    \startdata
K	             & m s$^{-1}$ & 1.65668   & 0.45119   & 0.30658   & 0.31290 $\pm$ 0.26983  & 0.18450 $\pm$ 0.07362   & 0.45150 $\pm$ 0.21874   \\
\porb	         & day        & 8.46318   & 18.86027  & 12.73174  & 8.46308 $\pm$ 0.00006  & 18.85969 $\pm$ 0.00008  & 12.73596 $\pm$ 0.00793  \\
e	             & ...        & 0.01013   & 0.00000   & 0.00339   & 0.00577 $\pm$ 0.00101  & 0.00338 $\pm$ 0.00164   & 0.00305 $\pm$ 0.00104   \\
$\omega$	     & deg        & 89.99210  & 224.07980 & 160.53522 & 88.43038 $\pm$ 0.05783 & 223.28438 $\pm$ 1.68357 & 160.78945 $\pm$ 2.59947 \\
M$_{0}$  	     & deg        & 0.00003   & 0.00003   & 0.00013   & 1.58566 $\pm$ 1.46718  & 0.51338 $\pm$ 0.41839   & 2.11220 $\pm$ 1.97561   \\
i	             & deg        & 89.57141  & 89.42889  & 89.53818  & 89.57917 $\pm$ 0.37639 & 89.22655 $\pm$ 0.21654  & 89.31812 $\pm$ 1.15800  \\
$\Omega$	     & deg        & 0.00015   & 0.00000   & 0.00000   & 0.31487 $\pm$ 0.15366  & 2.29724 $\pm$ 2.17643   & 0.39499 $\pm$ 0.27136   \\ \hline
$\chi^{2}$       & ...        & 109.75800 & 3.80307   & ...       & 106.33254              & 4.53826                 & ...                     \\
\reduced         & ...        & 4.73171   & 28.39027  & ...       & 4.61962                & 27.71770                & ...                     \\
$\ln\mathcal{L}$ & ...        & \multicolumn{3}{c|}{215.40728}    & \multicolumn{3}{c}{228.75014}                                              \\
$BIC$            & ...        & \multicolumn{3}{c|}{-349.08633}   & \multicolumn{3}{c}{-375.77205}                                             \\
$AIC_{c}$        & ...        & \multicolumn{3}{c|}{-354.59234}   & \multicolumn{3}{c}{-381.27806}
    \enddata
    \tablecomments{The K's listed here are unconstrained, but see $\S$\ref{sec:results} for discussion regarding the planets' low K's generated by \exostriker.}
\end{deluxetable*}

\clearpage

\section{RV Vetting of TTV Analysis}\label{sec:rv_analysis}

\citet{cale2021} modeled the RVs of the AU Mic system and searched for additional candidate non-transiting planet signals. Their RV models indicated a candidate RV signal between the orbits of AU Mic b and c with an orbital period of 12.742 days but were ruled inconclusive.

\begin{figure}
    \centering
    \includegraphics[width=0.46\textwidth]{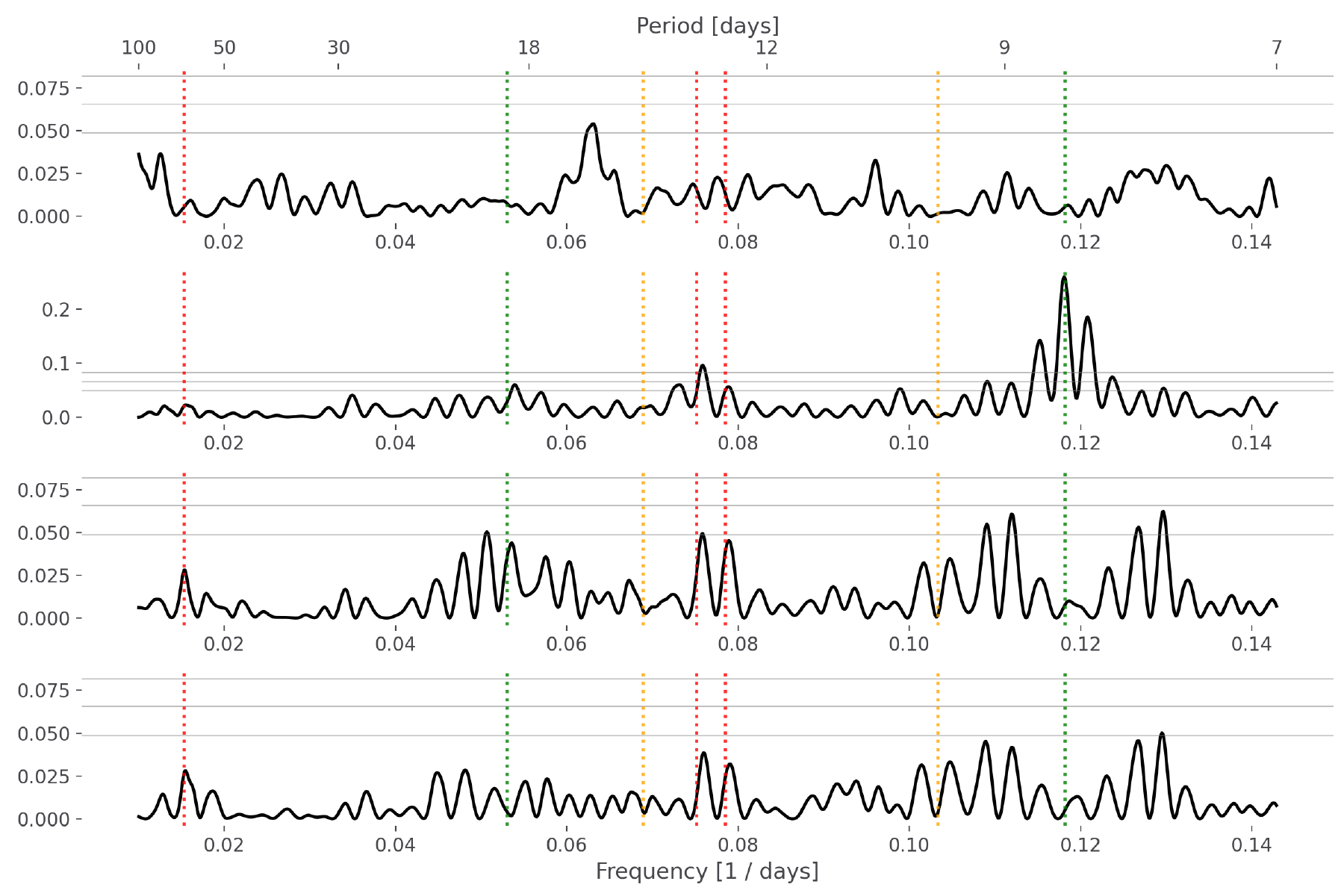}
    \includegraphics[width=0.48\textwidth]{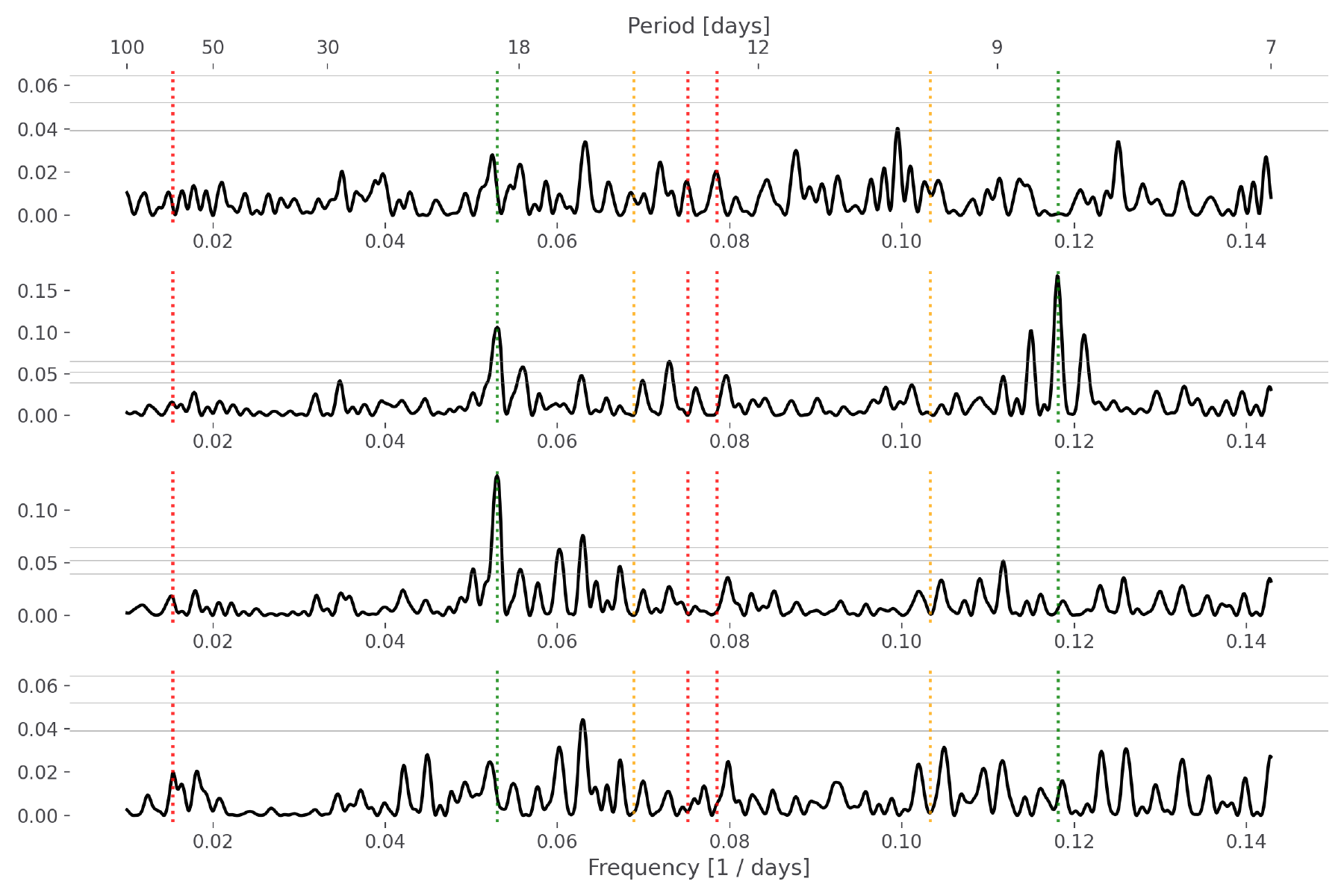}
    \caption{GLS periodograms of the AU Mic system. ({\it top four panels}): GLS generated without HARPS data. ({\it bottom four panels}): GLS generated with HARPS data. From top to bottom panels: zero point corrected, activity corrected, planet b corrected, planet c corrected. The dotted vertical lines from right to left are P$_{\rm b}$, 2*\prot, P$_{\rm d}$=12.74 days, P$_{\rm d}$=13.31 days, 3*\prot, P$_{\rm c}$, P$_{\rm candidate}$=65.22 days. AU Mic d's signal vanishes when HARPS data were added to the GLS periodogram. Also, the 16-day peak in the bottom row of the periodogram is not credible in term of orbital dynamics stability.}
    \label{fig:glsrv}
\end{figure}

\begin{figure}
    \centering
    \includegraphics[width=\linewidth]{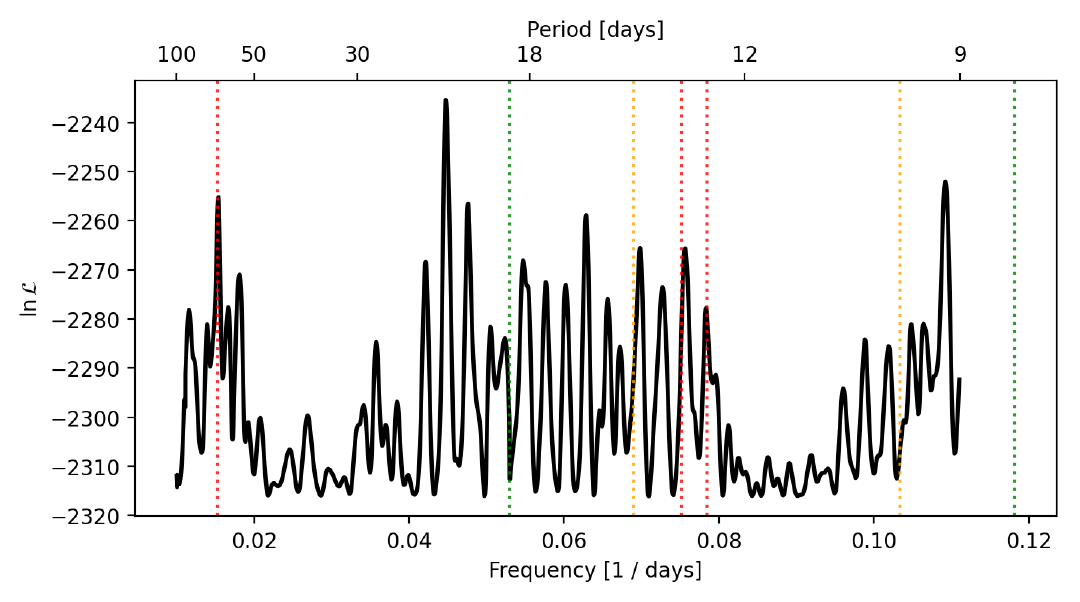}
    \caption{$\ln{\mathcal{L}}$ brute force periodogram of AU Mic's RVs, including models for both AU Mic b and c. AU Mic d's \tc was allowed to vary. The dotted vertical lines from right to left are P$_{\rm b}$, 2*\prot, P$_{\rm d}$=12.74 days, P$_{\rm d}$=13.31 days, 3*\prot, P$_{\rm c}$, P$_{\rm candidate}$=65.22 days. There appears to be no obvious peaks consistent with d's periods derived from the TTVs, but there is additional power in the periodogram, particularly near 12.75 and 13.3 days. Also note that 22-day peak would not be dynamically stable, and that there is the possibility of a 65-day planetary candidate in the RVs.}
    \label{fig:rv_pdg}
\end{figure}

We repeat the analysis of \citet{cale2021} with the addition of the SERVAL-generated \citep{zechmeister2018} HARPS RVs from \citet{zicher2022} to see if the RVs exhibit consistent behavior with the TTVs. We compute two GLS periodograms, one without HARPS RV data and another with HARPS data included (Figure \ref{fig:glsrv}). The planet d's signal was present in the periodogram without HARPS data; however, after adding HARPS data to the GLS periodogram, the signal vanishes. We also note the 16-day peak in the bottom row of the periodogram is not credible in terms of orbital dynamics stability.

We compute an $\ln{\mathcal{L}}$ brute-force periodogram  (Figure \ref{fig:rv_pdg}) over a range of possible periods for AU Mic d. We do not see any clearly identified peaks consistent with d's periods derived from the TTVs, but there is additional power in the periodogram, particularly near 12.75 (which is the exact 2:3 orbital resonance and our TTV period) and 13.3 days, indicating our RV model is incomplete and, therefore, our RV analysis is inconclusive in confirming AU Mic d. The 22-day peak would not be dynamically stable, and there is the possibility of a 65-day planetary candidate in the RVs.

We explore the RV model comparison for ten different periods of AU Mic d (Table \ref{tab:rv_model_comp}). The 3-planet configuration is favored relative to the 2-planet configuration for 6.20-day ($\Delta AIC_{c}$ = 104.83), 5.39-day ($\Delta AIC_{c}$ = 48.64), 5.64-day ($\Delta AIC_{c}$ = 36.77), 12.7-day ($\Delta AIC_{c}$ = 23.63), 6.47-day ($\Delta AIC_{c}$ = 19.12), 11.9-day ($\Delta AIC_{c}$ = 12.66), and 5.08-day ($\Delta AIC_{c}$ = 11.74) cases, while the 2-planet configuration is favored relative to the 3-planet configuration for 12.6-day ($\Delta AIC_{c}$ = 9.61), 14.1-day ($\Delta AIC_{c}$ = 46.22), and 5.86-day ($\Delta AIC_{c}$ = 69.79) cases. Based on the log-likelihood, $\Delta BIC$, and $\Delta AIC_{c}$, the 6.20-day is the most favored one overall and the 12.7-day is the most favored among the 3-planet cases. However, the TTV analysis indicated that the planet d is of very small mass compared to AU Mic b and c, so it is not surprising that the RV model struggles to detect AU Mic d with any statistical significance.

\startlongtable
\begin{deluxetable*}{c|l|c|c|c|c}\label{tab:rv_model_comp}
    \tablecaption{RV model information criterion for the AU Mic system.}
    \tablehead{P$_{\rm d}$ (days) & Planets & \reduced & $\ln{\mathcal{L}}$ & $\Delta BIC$ & $\Delta AIC_{c}$}
    \startdata
\multirow{8}{*}{5.08} & b, c, d & 4.00978 & -2248.46411 & 0.00000   & 0.00000   \\
                      & b, c    & 4.10715 & -2257.68857 & 0.85803   & 11.74349  \\
                      & b, d    & 4.16747 & -2271.93853 & 41.0852   & 44.70022  \\
                      & b       & 4.14185 & -2280.01400 & 39.64525  & 54.18589  \\
                      & c, d    & 4.15293 & -2283.78303 & 64.7742   & 68.38922  \\
                      & c       & 4.24211 & -2302.43094 & 84.47913  & 99.01977  \\
                      & d       & 4.33652 & -2320.10267 & 131.54986 & 138.79341 \\
                      & -       & 4.27910 & -2327.09888 & 127.95138 & 146.16034 \\ \hline
\multirow{8}{*}{5.39} & b, c, d & 3.95174 & -2248.37326 & 0.00000   & 0.00000   \\
                      & b, d    & 4.14993 & -2269.88140 & 37.15265  & 40.76767  \\
                      & b, c    & 4.24239 & -2276.04763 & 37.75784  & 48.64331  \\
                      & c, d    & 4.05626 & -2275.93760 & 49.26505  & 52.88007  \\
                      & c       & 4.19304 & -2287.77753 & 55.35401  & 69.89465  \\
                      & b       & 4.10217 & -2298.35680 & 76.51256  & 91.05320  \\
                      & d       & 4.19466 & -2305.18164 & 101.88949 & 109.13304 \\
                      & -       & 4.60850 & -2353.45811 & 180.85153 & 199.06049 \\ \hline
\multirow{8}{*}{5.64} & b, c, d & 3.90634 & -2223.31676 & 0.00000   & 0.00000   \\
                      & b, c    & 3.95149 & -2245.05276 & 25.88113  & 36.76659  \\
                      & c, d    & 3.95899 & -2248.70048 & 44.90381  & 48.51884  \\
                      & b, d    & 3.91049 & -2252.70267 & 52.90820  & 56.52322  \\
                      & b       & 4.18677 & -2279.81918 & 89.55033  & 104.09097 \\
                      & c       & 4.24168 & -2292.44964 & 114.81124 & 129.35188 \\
                      & -       & 4.24294 & -2311.41267 & 146.87368 & 165.08264 \\
                      & d       & 4.37061 & -2332.12947 & 205.89816 & 213.14171 \\ \hline
\multirow{8}{*}{5.86} & b, c    & 3.95261 & -2247.52212 & 0.00000   & 0.00000   \\
                      & b       & 3.99113 & -2267.43656 & 33.96527  & 37.62044  \\
                      & b, c, d & 4.24926 & -2279.06355 & 80.67377  & 69.78830  \\
                      & b, d    & 4.26988 & -2287.46508 & 91.61319  & 84.34275  \\
                      & c, d    & 4.24476 & -2291.16124 & 99.00551  & 91.73507  \\
                      & d       & 4.24737 & -2307.05552 & 124.93044 & 121.28852 \\
                      & c       & 4.27718 & -2315.40526 & 129.90265 & 133.55782 \\
                      & -       & 4.33599 & -2318.39795 & 130.02440 & 137.34790 \\ \hline
\multirow{8}{*}{6.20} & b, c, d & 3.76222 & -2223.32578 & 0.00000   & 0.00000   \\
                      & b       & 4.04026 & -2266.56250 & 63.01893  & 77.55957  \\
                      & b, c    & 4.11199 & -2279.09511 & 93.94778  & 104.83324 \\
                      & b, d    & 4.19250 & -2280.76895 & 109.02271 & 112.63774 \\
                      & c       & 4.18665 & -2289.96236 & 109.81864 & 124.35928 \\
                      & c, d    & 4.20633 & -2299.23938 & 145.96358 & 149.57860 \\
                      & -       & 4.22326 & -2311.14636 & 146.32301 & 164.53197 \\
                      & d       & 4.41166 & -2331.23825 & 204.09768 & 211.34123 \\ \hline
\multirow{8}{*}{6.47} & b, d    & 3.82012 & -2242.37068 & 0.00000   & 0.00000   \\
                      & b, c, d & 4.17966 & -2268.82978 & 58.78184  & 55.16681  \\
                      & b, c    & 4.06131 & -2281.74255 & 67.01647  & 74.28691  \\
                      & b       & 4.16875 & -2290.60045 & 78.86866  & 89.79428  \\
                      & c, d    & 4.11485 & -2289.43973 & 94.13811  & 94.13811  \\
                      & c       & 4.28375 & -2313.19199 & 124.05174 & 134.97736 \\
                      & -       & 4.27356 & -2315.79337 & 123.39086 & 137.98480 \\
                      & d       & 4.48229 & -2339.34545 & 188.08591 & 191.71444 \\ \hline
\multirow{8}{*}{11.9} & b, c, d & 4.06977 & -2256.91123 & 0.00000   & 0.00000   \\
                      & b       & 4.08732 & -2266.27497 & -4.72704  & 9.81360   \\
                      & b, c    & 4.13643 & -2266.59584 & 1.77833   & 12.66379  \\
                      & b, d    & 4.11093 & -2270.03862 & 20.39115  & 24.00618  \\
                      & c       & 4.29308 & -2306.57442 & 75.87187  & 90.41251  \\
                      & d       & 4.31650 & -2317.55744 & 109.56516 & 116.80871 \\
                      & -       & 4.39886 & -2321.13976 & 99.13891  & 117.34787 \\
                      & c, d    & 4.42181 & -2342.39152 & 165.09695 & 168.71197 \\ \hline
\multirow{8}{*}{12.6} & b, c    & 3.93956 & -2242.04753 & 0.00000   & 0.00000   \\
                      & b, c, d & 3.95070 & -2243.49828 & 20.49239  & 9.60693   \\
                      & b, d    & 4.03554 & -2253.46944 & 34.57107  & 27.30063  \\
                      & c, d    & 4.16405 & -2283.63677 & 94.90575  & 87.6353   \\
                      & c       & 4.15953 & -2288.90894 & 87.85919  & 91.51436  \\
                      & b       & 4.21603 & -2299.51198 & 109.06527 & 112.72045 \\
                      & d       & 4.37735 & -2325.24047 & 172.24951 & 168.6076  \\
                      & -       & 5.01924 & -2454.82572 & 413.82911 & 421.15261 \\ \hline
\multirow{8}{*}{12.7} & b, d    & 3.82300 & -2230.60095 & 0.00000   & 0.00000   \\
                      & b, c, d & 3.93925 & -2244.69089 & 34.04350  & 30.42848  \\
                      & b, c    & 4.03965 & -2259.85878 & 46.78839  & 54.05884  \\
                      & c, d    & 4.06094 & -2265.50468 & 69.80745  & 69.80745  \\
                      & b       & 4.06797 & -2276.96998 & 75.14716  & 86.07278  \\
                      & d       & 4.15544 & -2283.89919 & 100.73284 & 104.36137 \\
                      & -       & 4.21856 & -2316.20370 & 147.75097 & 162.34491 \\
                      & c       & 4.32722 & -2330.15944 & 181.52608 & 192.45169 \\ \hline
\multirow{8}{*}{14.1} & b, d    & 4.00062 & -2264.98707 & 0.00000   & 0.00000   \\
                      & b, c    & 4.08155 & -2278.10035 & 14.49931  & 21.76975  \\
                      & b       & 4.32125 & -2291.32638 & 35.08774  & 46.01336  \\
                      & c       & 4.25284 & -2293.92168 & 40.27834  & 51.20395  \\
                      & c, d    & 4.20466 & -2290.74089 & 51.50765  & 51.50765  \\
                      & b, c, d & 4.35643 & -2297.85791 & 71.60532  & 67.99030  \\
                      & d       & 4.42068 & -2319.25242 & 102.66709 & 106.29561 \\
                      & -       & 4.66740 & -2392.62944 & 231.83023 & 246.42417
    \enddata
\end{deluxetable*}

\section{Discussion}\label{sec:results}

We modeled the transits of AU Mic b and c to obtain the midpoint times, which we then use to model the TTVs of the AU Mic system. We attempted both the 2-planet and 3-planet dynamical models and found that the 3-planet model adequately describes the observed TTVs, including its super-period. Moreover, we generated the TTV periodograms and found ten possible solutions for the period of the non-transiting planet AU Mic d. In the next subsections, we discuss the impact of AU Mic's stellar activity on the TTVs, the convergence of transit models, the TTV and RV model comparisons, the coplanarity of AU Mic system, the planetary resonant chains, the mass of AU Mic d, the inner d orbits vs middle d orbits, and the validation vs confirmation of AU Mic d.

\subsection{Impact of AU Mic's Stellar Activity on TTVs}

\citet{wittrock2022} explored the impacts stellar flares and spot modulations may have on the TTV profile. Given the 7:4 spin-orbital commensurability between AU Mic and planet b \citep{szabo2021} and the long lifetime of AU Mic's starspots, \citet{wittrock2022} modeled the TTVs induced by AU Mic's spot crossing during AU Mic b's transit and found the effect to be relatively minimal (no more than $\sim$2 min) compared to TTVs of planet b induced by planets c and d and with no clear pattern. \citet{martioli2020}, \citet{palle2020}, \citet{gilbert2022}, \citet{wittrock2022}, and \citet{szabo2021, szabo2022} modeled the stellar flares from SPIRou, ESPRESSO, TESS, Spitzer, and CHEOPS observations, respectively, which marginalized the effect of flares on the observed TTVs. While \aij and \exofast are not yet capable of jointly modeling the flares during transits, the ground-based observations have relatively lower photometric precision and larger timing uncertainties, so the flares' effect on the TTVs are significantly down-weighted within the photometric precision timing uncertainties \citep{wittrock2022}.

\subsection{Convergence of Transit Models}

It can be challenging to get transit model convergence when using standard MCMC methods such as the \emcee package uses \citep{foreman-mackey2013}, especially with a large number of model parameters. However, as mentioned in $\S$\ref{sec:exofastmod}, \exofast is written in {\tt IDL} and uses the differential evolution MCMC algorithm instead of the \emcee package. As part of the convergence criteria, \exofast simultaneously employs two metrics, the Gelman-Rubin statistic and the number of independent draws, to help minimize their individual shortcomings of these convergence criteria \citep{gelman1992, goodman2010}. We were able to use fairly tight priors on most of the parameters (Table \ref{tab:exofastv2priors}) and have had {\tt rejectflatmodel} switched on to ease the convergence of our transit models. Numerous different walkers, while not fully independent from each other, are found to produce posteriors that are relatively consistent with one another, convincing us that the independent chains do converge. See \citet{eastman2013} and \citet{eastman2019} for more information on \exofast's use of MCMC algorithm and convergence criteria and $\S$\ref{sec:exofastmod} of this paper for details on our process with \exofast.

\subsection{Dynamically Settled Assumption of AU Mic System}\label{sec:dyn_assumption}

In $\S$\ref{sec:2p_ttv_mod}, we presented a high-eccentricity 2-planet TTV model which does exhibit the super-period (Figure \ref{fig:2p_ext_exostriker_plots}), making it seems a potentially good fit for a 2-planet system with a stable but unlikely eccentric-aligned configuration with abnormally small masses inconsistent with the RV masses from \citet{cale2021} and \citet{donati2023}. After having the very small masses being replaced with those RV masses, the high-eccentricity 2-planet model provides an example of an unstable eccentric system that does model the TTVs and does persist for 20 Myr as seen in Figure \ref{fig:2p_ext_rebound}, but it is clearly not dynamically settled given the apparent orbital migrations of AU Mic b and c and the extreme fluctuations in both planets' eccentricities and inclinations in the \rebound simulations.

We must caution that \rebound does not include tidal circularization and general relativity (GR) precession, which are relevant at these orbital periods, and highly eccentric planets do not exist in older systems. So, in theory, planet d may not exist if the system is currently dynamically unstable. However, future TTVs will clearly rule out this case, and current RVs from \citet{cale2021} and \citet{donati2023}, while challenging to constrain the eccentricity, do not give any clear indication that this system is highly eccentric. Additionally, the secondary eclipse model of AU Mic b by Kevin et al. (in prep) rules out high eccentricity and provides a vastly different argument of periastron ($\sim$90$^{\circ}$ as opposed to our 173$^{\circ}$ from Table \ref{tab:2p_ext_exostriker_params}). The dynamically unstable case does not account for the fact that the host star was larger in the recent past, so the planets would have likely collided with the star prior to the 20-Myr mark at these high eccentricities. Thus, we likely can rule out this scenario, but more work is needed on dynamical simulations that include GR precession, tidal circularization, and any disk drag from the past for this star.

\subsection{Overall Model Comparisons of AU Mic System}\label{sec:allmodelcomp}

In $\S$\ref{sec:3p_ttv_mod}, we modeled our TTVs and then generated the TTV log-likelihood periodograms to explore a set of planet d's orbital periods that would be in agreement with the observed TTVs of the AU Mic system; it was through the periodograms that we obtained ten possible orbital periods and orbital configurations for AU Mic d. Given that the high-eccentricity 2-planet configuration is ruled out for reasons provided in the previous section $\S$\ref{sec:dyn_assumption}, that the low-eccentricity 2-planet configuration has both b and c being misaligned despite their well-known transiting status \citep{plavchan2020, martioli2021, gilbert2022, wittrock2022}, and that both 2-planet models are statistically ruled out at high confidence (Tables \ref{tab:ttv_mod_comp} \& \ref{tab:ttv_mod_comp_hp}), this validates AU Mic d, and AU Mic d must possess one of these ten orbital periods and orbital configurations. Based on our TTV model comparison (Tables \ref{tab:ttv_mod_comp} \& \ref{tab:ttv_mod_comp_hp}), our calculated super-period (Figure \ref{fig:ocsin} \& Table \ref{tab:ttvsin}), our RV analysis, and our stability tests (Figure \ref{fig:stability_hist}), the 12.7-day configuration is the most favored period for AU Mic d.

In the meanwhile, the 5.64-day case is favored by the RVs and the dynamical stability tests but is strongly disfavored by the TTVs. The 5.39, 6.20, and 6.47-day cases are favored by the TTVs and RVs but are disfavored by the stability tests. The 5.08 and 11.9-day cases are favored by the RVs but are disfavored by the TTVs and stability tests. The 12.6-day case is favored by the stability tests but is disfavored by the TTVs and RVs. Lastly, the 5.86 and 14.1-day case are favored by the TTV but are disfavored by the RVs and stability tests.

Thus, while we do not statistically rule out the other nine potential periods for AU Mic d, they are all statistically disfavored to varying degrees for one reason or another. Since we do not statistically rule out these other potential orbital periods for AU Mic d, we do not consider AU Mic d to be confirmed, only statistically validated. Further, in addition to AU Mic d at its likely orbital period of 12.7 days, it's entirely plausible there could be an additional fourth planet at 5.39, 6.20, or 6.47 days interior to AU Mic b, which is favored by the TTVs and RVs and is disfavored by the dynamical stability tests. Our existing data is unable to explore this latter possibility.

\subsection{Coplanarity of AU Mic System}\label{sec:coplanar}

One curious feature among the \exostriker\ MCMC models is the inclination of AU Mic d for these ten candidate periods of AU Mic d. AU Mic d is nearly coplanar for the 5.64, 12.6, and 12.7-day cases, with i$_{\rm d}$ = 90.08$^{\circ}$ $\pm$ 4.99$^{\circ}$, 89.63$^{\circ}$ $\pm$ 1.71$^{\circ}$, and 89.32$^{\circ}$ $\pm$ 1.16$^{\circ}$, respectively, but the 5.64-day case is strongly disfavored by the TTVs and the 12.6-day case is disfavored by both the TTVs and RVs. For the 6.47-day case, AU Mic d is slightly misaligned at i$_{\rm d}$ = 93.57$^{\circ}$ $\pm$ 3.26$^{\circ}$ but is strongly disfavored from the dynamical stability tests. The remaining candidate periods required considerable misalignment with i$_{\rm d}$ = 76.88$^{\circ}$ $\pm$ 3.05$^{\circ}$, 57.19$^{\circ}$ $\pm$ 1.83$^{\circ}$, 99.59$^{\circ}$ $\pm$ 6.78$^{\circ}$, 107.16$^{\circ}$ $\pm$ 14.19$^{\circ}$, 81.89$^{\circ}$ $\pm$ 0.64$^{\circ}$, and 74.27$^{\circ}$ $\pm$ 0.40$^{\circ}$ for the 5.08, 5.39, 5.86, 6.20, 11.9, and 14.1 cases, respectively.

We do not know if the planet formation process could produce such a highly misaligned planet in this young system, especially for the cases between the two coplanar, transiting planets, but we can invoke Occam's razor, which argues that the simplest configurations are the 5.64, 12.6, and 12.74-day cases since these configuration have AU Mic d being coplanar with the system. As discussed in \citet{picogna2014}, misalignments are considered to be a particular signature of a past flyby, with a maximum value of tilting of 9$^{\circ}$, while multiple flybys can lead to higher misalignments. However, after reconstructing the flybys of debris disks using the Gaia EDR3 data, \citet{bertini2023} found that AU Mic did not experience any flyby in the last 5 Myr, so any misalignment induced by a recent close flyby appears unlikely. It is also important to point out that if a flyby had occurred, we would expect to see both AU Mic b and c become highly misaligned along with AU Mic d. Moreover, most of the Kepler compact multi-planet systems are well-aligned with a scatter of $\pm$3$^{\rm o}$ \citep{lissauer2011, fang2012, fabrycky2014}. These are yet another reasons why we favor the 12.74-day period for the validated AU Mic d, but additional TTV observations will be needed to confirm that this is the case.

\subsection{Near-Resonant Chains of AU Mic System}\label{sec:reschains}

Since AU Mic b exhibits such strong TTVs, one consequence of all ten candidate periods for AU Mic d is that they all place the AU Mic system near a resonant chain of varying order. For the inner d scenario, the near resonances between d and b are 3:5 for 5.08-day, 5:8 for 5.39-day, 2:3 for 5.64-day, 5:7 for 5.86-day, 3:4 for 6.20-day, and 7:9 for 6.47-day cases. The near orbital resonance between planets b and c for all inner d cases is 4:9, so the overall near-resonant chains are 12:20:45 for 5.08-day, 5:8:18 for 5.39-day, 8:12:27 for 5.64-day, 20:28:63 for 5.86-day, 3:4:9 for 6.20-day, and 28:36:81 for 6.47-day cases. For the 11.9-day case, the near resonances between b \& d and between d \& c are 5:7 and 5:8, or near a 25:35:56 chain overall. Both the 12.6-day and 12.7-day cases have both pairs near a 2:3 chain, or near 4:6:9 chain overall. The 14.1-day case pairs are near a 3:5 and 3:4 chains, or near a 9:15:20 chain overall. These near-MMR chains are the reason why our TTV periodograms (Figures \ref{fig:inner_ttv_pdg} \& \ref{fig:middle_ttv_pdg}) yielded ten possible solutions, each with relatively high log-likelihood.

Suggestively, as per the Occam's razor argument, the near 4:6:9 resonant chain (or near 3:2 parings between d \& b and between c \& d) is the simplest one, given that our TTV and RV models, dynamical stability tests, and the calculated TTV super-period favor the 12.7-day case (the 12.6-day case also shares this near-MMR chain but is not as favored). The 3:2 resonance is the most common among the resonant chains, the 4:3, 5:3, 7:5, and 8:5 resonances are not as common, respectively, but are not unprecedented, and the 9:4 and 9:7 resonances do not appear to be very common \citep{fabrycky2014}. Thus, the distribution of near-MMR chains among the known exoplanets gives stronger credence to the potential near 3:2 resonant chain pairs of the AU Mic system. Regardless of which of the ten candidates periods for AU Mic d is correct, with the 12.7-day case being most favored, all place the AU Mic system near a resonant chain, with the 12.7-day case being the simplest commensurability chain configuration. The presence of near-MMRs in such a young system poses significant constraints for the formation model and evolution of planetary systems. The similarly young V1298 Tau system is also near a similar resonant chain configuration, suggesting this may be a common characteristic of the formation of compact multi-planet systems \citep{david2019a, david2019b, tejada-arevalo2022}. This implies that the planetary systems can develop resonant chains very early on, particularly in compact systems, which in turns would quickly establish the stability of the dynamical systems.

\subsection{Characterizing AU Mic d}

We calculate the mass of AU Mic d (Table \ref{tab:aumic_d_char}) using the parameters generated from \exostriker\ MCMC models and equation \ref{eqn:K_to_mass} \citep{cumming1999}:
\begin{equation}\label{eqn:K_to_mass}
    K = \left(\frac{2\pi G}{P_{orb}}\right)^{1/3}\frac{M_{p}\sin{i}}{(M_{*}+M_{p})^{2/3}}\frac{1}{\sqrt{1-e^{2}}}
\end{equation}
Since AU Mic d is not known or observed to transit, its radius, and therefore its density, is unknown.

\begin{deluxetable*}{cccccc}\label{tab:aumic_d_char}
    \tablecaption{Masses, radii, and time of conjunctions of AU Mic d. The masses were calculated using the parameters from \exostriker MCMC models. AU Mic d is not known to transit, so its measured radius and density are unknown. However, we use the Chen-Kipping mass-radius relation \citep{chen2017} to estimate the radii of AU Mic d.}
    \tablehead{P$_{\rm d}$ (days) & M$_{\rm d}$ (\mear) & M$_{\rm d}$ (\mjup) & R$_{\rm d}$ (\rear) & R$_{\rm d}$ (\rjup) & $T_{\rm C,d}$ (BJD)}
    \startdata
5.08 & 0.737 $\pm$ 0.083 & 0.00232 $\pm$ 0.00026 & 0.926 $\pm$ 0.029 & 0.0826 $\pm$ 0.0002 & 2458330.88624 $\pm$ 0.16079 \\
5.39 & 0.667 $\pm$ 0.042 & 0.00210 $\pm$ 0.00013 & 0.900 $\pm$ 0.016 & 0.0803 $\pm$ 0.0001 & 2458330.79993 $\pm$ 0.29196 \\
5.64 & 0.469 $\pm$ 0.045 & 0.00148 $\pm$ 0.00014 & 0.816 $\pm$ 0.022 & 0.0728 $\pm$ 0.0001 & 2458331.07793 $\pm$ 0.30077 \\
5.86 & 1.049 $\pm$ 0.239 & 0.00330 $\pm$ 0.00075 & 1.021 $\pm$ 0.065 & 0.0911 $\pm$ 0.0005 & 2458331.46003 $\pm$ 0.28115 \\
6.20 & 0.904 $\pm$ 0.482 & 0.00284 $\pm$ 0.00152 & 0.980 $\pm$ 0.146 & 0.0874 $\pm$ 0.0011 & 2458331.66688 $\pm$ 1.08314 \\
6.47 & 0.510 $\pm$ 0.112 & 0.00161 $\pm$ 0.00035 & 0.836 $\pm$ 0.051 & 0.0745 $\pm$ 0.0003 & 2458331.03312 $\pm$ 0.33184 \\
11.9 & 0.392 $\pm$ 0.079 & 0.00123 $\pm$ 0.00025 & 0.776 $\pm$ 0.044 & 0.0693 $\pm$ 0.0003 & 2458339.91558 $\pm$ 0.08663 \\
12.6 & 0.522 $\pm$ 0.147 & 0.00164 $\pm$ 0.00046 & 0.841 $\pm$ 0.066 & 0.0750 $\pm$ 0.0004 & 2458339.99167 $\pm$ 0.18465 \\
12.7 & 1.053 $\pm$ 0.511 & 0.00331 $\pm$ 0.00161 & 1.023 $\pm$ 0.139 & 0.0912 $\pm$ 0.0011 & 2458340.55781 $\pm$ 0.11641 \\
14.1 & 0.628 $\pm$ 0.092 & 0.00198 $\pm$ 0.00029 & 0.885 $\pm$ 0.036 & 0.0790 $\pm$ 0.0003 & 2458341.05771 $\pm$ 0.20391
    \enddata
\end{deluxetable*}

The mass of AU Mic b is poorly constrained from the TTVs due to limited number of transit data from AU Mic c and that AU Mic d is is not known to be transiting, so we do not derive a TTV mass constraint for AU Mic b. AU Mic c is better constrained given the wealth of transit data from AU Mic b, but since there is no transit data from AU Mic d, we are unable to meaningfully constrain the mass of AU Mic c either. However, the mass of AU Mic d is the most constrained of the three planets due to availability of transit data from both planets b and c. The most favored model (P$_{d}$=12.7 days) has the mass M$_{\rm d}$ = 1.053 $\pm$ 0.511 \mear, which makes AU Mic d Earth-like in mass. The 5.86 and 6.20-day cases also have AU Mic d's mass being Earth-like but smaller, while the rest of the cases have AU Mic d being of even smaller mass but still not as small as that of Mars; for comparison, the mass of Mars is $M_{\rm Mars}$ = 0.107 \mear.

These relatively small masses for AU Mic d partially explain the challenges RVs face in characterizing such a planet and the divergent results between TTVs and RVs in which periods for AU Mic d were favored. Additionally, given that the host star AU Mic is highly active, it is incredibly challenging to validate AU Mic d with RVs with the current generation RV measurements and techniques for modeling and mitigating stellar activity. However, given that our RV models indicated the presence of a potential additional planet beyond AU Mic c, it will still be worthwhile to continue to intensely monitor the AU Mic system with RVs.

Given that the 12.7-day case is the most plausible one based on the super-period ($\S$\ref{sec:ocdiagram}), the stability tests ($\S$\ref{sec:stability_tests}), overall model comparisons ($\S$\ref{sec:allmodelcomp}), and the Occam's razor arguments from $\S$\ref{sec:coplanar} \& $\S$\ref{sec:reschains}, AU Mic d is most likely to have a roughly Earth-like mass. If confirmed, this would be the first known Earth-mass planet orbiting a young star. If it does transit, it will provide a crucial and valuable opportunity in probing its atmosphere and understanding the evolution of terrestrial planets' atmosphere.

We use the Chen-Kipping mass-radius relation \citep{chen2017} to estimate the radius of AU Mic d (Table \ref{tab:aumic_d_char}). For P$_{\rm d}$ = 12.7 days, AU Mic d has the radius 1.023 $\pm$ 0.139 \rear, making its size very Earth-like. The 5.86 and 6.20-day cases still have AU Mic d being Earth-like in sizes while the remaining cases have it being smaller, with the smallest size being 0.776 $\pm$ 0.044 \rear from the 11.9-day case (for comparison, Mars' radius is R$_{\rm Mars}$ = 0.532 \rear).

\begin{figure*}
    \centering
    \includegraphics[width=0.48\textwidth]{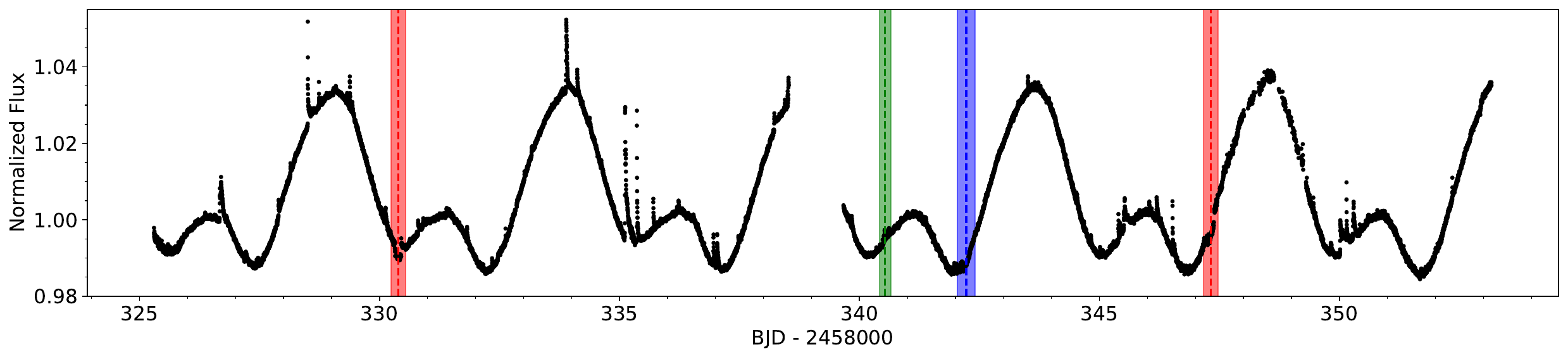}
    \includegraphics[width=0.48\textwidth]{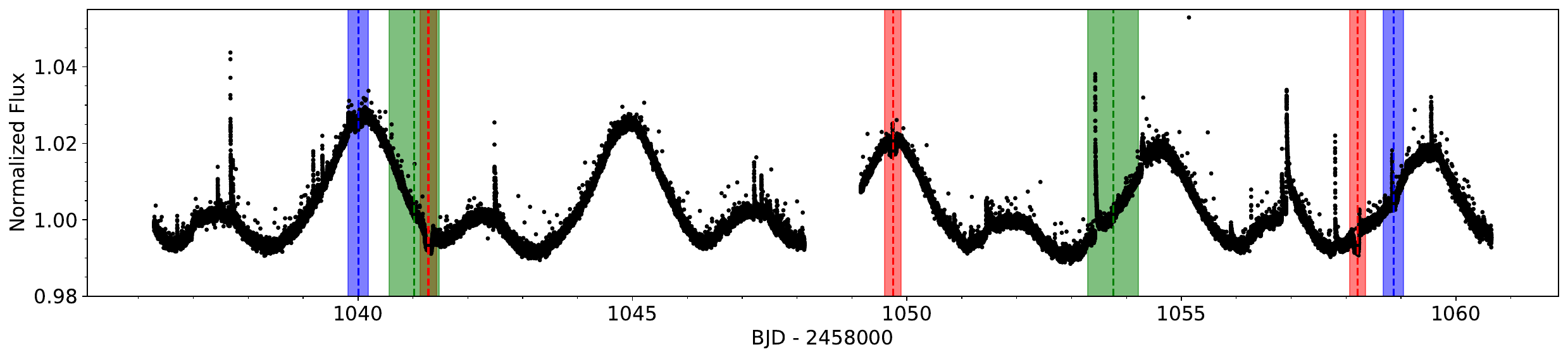}
    \caption{Raw TESS photometry of AU Mic overlaid with observed transits of AU Mic b (red) and c (blue) and range of predicted transit midpoint times for AU Mic d (green). This case is for P$_{d}$ = 12.7 days. Left panel is from cycle 1, and right panel is from cycle 3. Owing to AU Mic d's estimated 160 ppm depth and AU Mic's intense stellar activity, the potential transit signatures from AU Mic d is not readily apparent by eye, nor detectable at high statistical significance in the TESS light curve given the stellar activity and photon noise.}
    \label{fig:d_transittimes_1}
\end{figure*}

We obtain $T_{C,d}$ = 2458340.55781 $\pm$ 0.11641 BJD for P$_{\rm d}$ = 12.7 days (Table \ref{tab:aumic_d_char}). Assuming AU Mic d is coplanar with the other planets, we predict the time of transits for AU Mic d and map them onto the raw TESS photometry of AU Mic (Figures \ref{fig:d_transittimes_1} and \ref{fig:d_transittimes_2}).

Any potential transit signatures of AU Mic d are not readily apparent by eye in the TESS photometry. Based on the model of AU Mic c's depth uncertainty from \citet{gilbert2022}, we place a 3$\sigma$ upper limit on AU Mic d's transit of 460 ppm. With the estimated radius R$_{\rm d}$ = 1.023 $\pm$ 0.139 \rear for AU Mic d (Table \ref{tab:aumic_d_char}), its transit depth would have been $\sim$160 ppm. AU Mic d's small radius and very shallow transit depth could explain its lack of transit detection, especially in the presence of AU Mic's intense stellar activity. Future multi-wavelength high-precision space-based photometry could potentially confirm whether or not AU Mic d transits.

We next estimate whether or not a transit of AU Mic d could be recovered with any statistical significance. The point-to-point white noise scatter in the AU Mic TESS light curves model from \citet{gilbert2022} is 550 ppm for the 20s cadence data and 370 ppm for the 2-minute cadence data. There is also the variability in the GP component of their model, which has an amplitude of $\sim$3000 ppm and is periodic over $\sim$0.6 days; this periodic noise is, in principle, removable with the GP process, so the white noise is the remaining limiting factor. As an example, let us consider the 2-min cadence data with the 370-ppm scatter. In one hour, that noise averages down to $\sim$70 ppm. Therefore, to obtain an SNR = 7 for a 160-ppm transit, we would need approximately ten hours of transit data. The 20s cadence data with the 550-ppm scatter is less noisy per unit time; the reason for that is partially unknown but is believed to be not partially caused by the cosmic ray correction \citep{huber2022}. In that case, the noise averages down to $\sim$40 ppm in one hour; this would require approximately 4 hours of transit data to achieve a SNR = 7 for a 160-ppm transit. Therefore, accounting for the photon-noise alone, given the transit duration of AU Mic d would be less than four hours if it transited, there is marginal photon noise to recover a statistically significant detection off AU Mic d in the TESS light curves, but this statistically significance would be degraded even further after accounting for the additional uncertainty introduced from the stellar activity. This is best evidenced by the recovered depth uncertainty for AU Mic c of 1.15 $\pm$ 0.22 ppt.

\begin{figure*}
    \centering
    \includegraphics[width=0.43\textwidth]{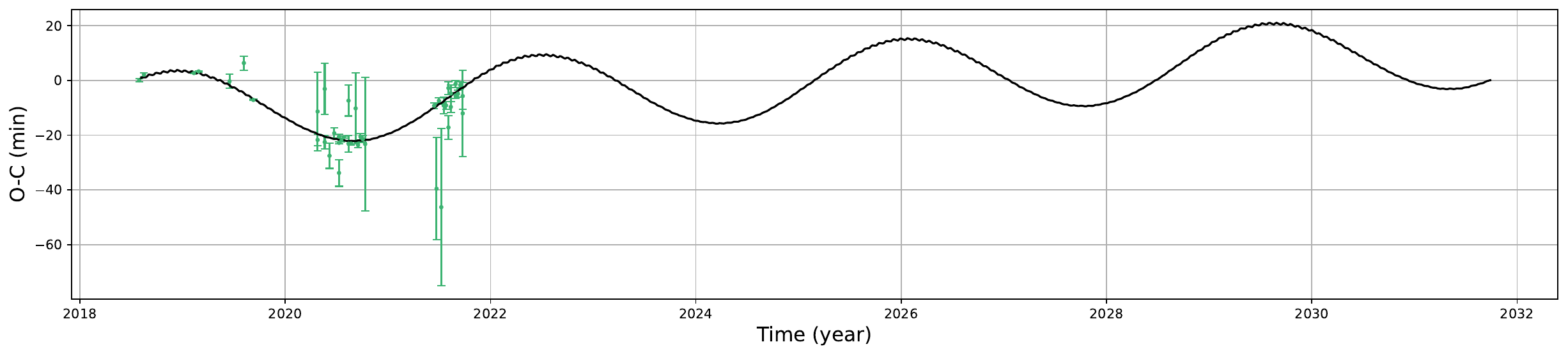}
    \includegraphics[width=0.43\textwidth]{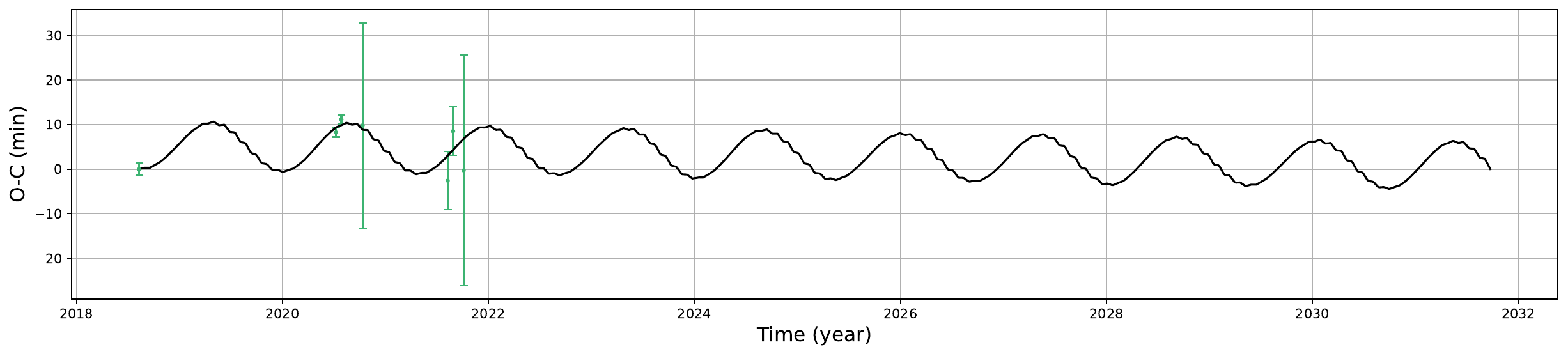}
    \caption{\exostriker-generated TTV models of AU Mic b ({\it left}) and c ({\it right}) projected over ten years since last observed transit of AU Mic system for P$_{d}$ = 12.7 days. The observed TTVs (green) is included with the model (black).}
    \label{fig:future_ttvs}
\end{figure*}

A future work is needed to involve a thorough search of all existing space-based light curves of AU Mic system for any statistically significant 160-ppm signal to confirm AU Mic d's transiting nature and to tighten the predicted timing of its transits. This will also require additional transit observations of AU Mic to minimize the photon shot noise such that AU Mic d's transits could potentially be detected. To further constrain this system, and confirm and distinguish the 12.7-day period for AU Mid d relative to the other nine candidate periods, additional precise space-based transit timing observations of AU Mic b and especially AU Mic c are needed. For this purpose, we modeled the projected TTVs of AU Mic b and c over the course of ten years since the last observed transit in AU Mic system (Figures \ref{fig:future_ttvs} \& \ref{fig:future_ttvs_other}).

\subsection{Inner d Orbits vs Middle d Orbits}

Is AU Mic d interior or exterior to AU Mic b? The TTV analysis (Tables \ref{tab:ttv_mod_comp} \& \ref{tab:ttv_mod_comp_hp}) favors AU Mic d to be between AU Mic b and c, and this is additionally supported by the Occam razor arguments covered in $\S$\ref{sec:coplanar} \& $\S$\ref{sec:reschains}, whereas the \spock, \megno, and \nbody stability analyses (Figure \ref{fig:stability_hist}) and the RV analysis (Table \ref{tab:rv_model_comp}) appear to favor AU Mic d lying interior to AU Mic b. However, this kind of comparison can only be done to a limited extent because the time periods after which the stability is estimated varies. For instance, \spock can only predict the stability probability after 10$^{9}$ orbits of the innermost planet. If AU Mic d is interior to b, this corresponds to about 14 Myr, but if AU Mic d is between b and d, this corresponds to about 23 Myr. Therefore, the results for the two ranges are not comparable 1-to-1, since the longer timescale could naturally make more systems unstable, sorting out more configurations, and making the AU Mic d between b and c case seem less likely than it is. Additionally, the current RV models, including those from \citet{cale2021} and \citet{zicher2022}, struggle in detecting low-mass planets with K$_{\rm d} <$ 1 m s$^{-1}$ like AU Mic d and most especially in the presence of heightened stellar activity such as that of AU Mic.

The determination of these cases is further complicated by the fact that AU Mic d's transiting nature is unknown. If AU Mic d does not appear to transit, it can mean that either the planet is misaligned or is so small such that its transit signature is masked by the instrumental and photon noise and/or host star's activity. Notably, almost all of the TTV orbital configurations that were explored in this paper require d to be misaligned with the other two planets, with the exception of the 5.64, 12.6, and 12.7-day cases. Additionally, AU Mic d in the inner d cases would have a higher probability of transiting AU Mic than if it were in the middle d cases (Table \ref{tab:aumic_d_transit_prob}). We also found AU Mic d to be relatively small ($\sim$1.02 \rear assuming d is Earth-like in density), so its transit signature would be very shallow ($\sim$160 ppm, which is below the 460 ppm limit for resolving transits in the AU Mic TESS light curves).

\begin{deluxetable}{ccc}\label{tab:aumic_d_transit_prob}
    \tablecaption{Probabilities of AU Mic d transiting based on its corresponding distances from the host star using the simplest estimate P$_{\rm tr}$ = \rstar/a$_{\rm d}$ \citep{borucki1984, sackett1999}.}
    \tablehead{P$_{\rm d}$ (days) & a$_{\rm d}$ (au) & P$_{\rm tr}$}
    \startdata
5.08 & 0.0462 $\pm$ 0.0008 & 0.0749 $\pm$ 0.0027 \\
5.39 & 0.0481 $\pm$ 0.0009 & 0.0720 $\pm$ 0.0026 \\
5.64 & 0.0495 $\pm$ 0.0009 & 0.0698 $\pm$ 0.0025 \\
5.86 & 0.0508 $\pm$ 0.0009 & 0.0681 $\pm$ 0.0024 \\
6.20 & 0.0528 $\pm$ 0.0010 & 0.0655 $\pm$ 0.0024 \\
6.47 & 0.0543 $\pm$ 0.0010 & 0.0637 $\pm$ 0.0023 \\
11.9 & 0.0813 $\pm$ 0.0015 & 0.0425 $\pm$ 0.0015 \\
12.6 & 0.0849 $\pm$ 0.0016 & 0.0408 $\pm$ 0.0015 \\
12.7 & 0.0853 $\pm$ 0.0016 & 0.0406 $\pm$ 0.0015 \\
14.1 & 0.0912 $\pm$ 0.0017 & 0.0380 $\pm$ 0.0014 \\
    \enddata
\end{deluxetable}

Given the complexity of the resonance orders of the different candidate periods, Occam's razor suggests that the 12.7-day period (and thus the middle d case) is the most likely scenario, given that this configuration has the simplest near-MMR chains and with planet d being coplanar with the system (as discussed in ${\S}$\ref{sec:reschains} \& and $\S$\ref{sec:coplanar}, respectively). However, we do not rule out the possibility of an interior planet to AU Mic b and the near-2:3 resonance.

With the small mass M$_{\rm d}$ = 1.053 $\pm$ 0.511 \mear of AU Mic d (based on 12.7 days being the most plausible case), this would make d the first Earth-mass planet orbiting a known young star. Thus, d will serve as a very valuable target for a test case study involving the evolution of terrestrial planets and their atmosphere, provided that AU Mid d does transit. \citet{kane2022} determined that AU Mic d presently does not lie within the habitable zone (HZ). While it is true that AU Mic is still in its pre-main sequence phase and that the HZ regions are predicted to migrate relatively quickly toward the host star until AU Mic becomes a main sequence star \citep{kane2022}, AU Mic d will still be well interior to the innermost boundary of the HZ. This implies that AU Mic d may become more like Venus instead of Earth. Additionally, the compact near-MMR orbits mentioned in $\S$\ref{sec:reschains} could lead to significant tidal heating of AU Mic planets' interior, particularly that of AU Mic d \citep[e.g.,][]{bolmont2013, vanlaerhoven2014, luger2017}, but this effect is not further explored herein.

\subsection{Is AU Mic d a candidate, validated, or confirmed planet?}

Throughout this work, we have characterized AU Mic d as a ``validated'' planet and not as a ``confirmed'' or ``candidate'' planet. We consider the threshold for ``confirmed'' planet to be that it is detected through a second technique that unequivocally rules out false positive scenarios (even if its orbital period is not well-known), for ``validated'' planet to be that the planet is statistically verified (even if its orbital period is ambiguous), and for ``candidate planet'' to be that there is a signal that may be due to a planet but is not statistically verified. Thus, by our definition, ``validated'' is more conservative than ``confirmed''. The main point of ``validation'' is that we have statistically proven AU Mic d must exist (TTV model comparisons from Tables \ref{tab:ttv_mod_comp} \& \ref{tab:ttv_mod_comp_hp} and most of RV model comparisons from Table \ref{tab:rv_model_comp} consistently ruled out the 2-planet configurations), even if there is some ambiguity in its orbital period, and \citet{wittrock2022} ruled out TTVs driven by stellar flares and spot modulations. Moreover, we have come across several published examples where the planets that are ``confirmed'' via the TTV detection technique apparently may have several possible orbital periods or do not need their orbital period well-defined or even known; such examples include Kepler-19 c \citep{ballard2011}, Kepler-82 f \citep{freudenthal2019}, Kepler-160 d \citep{heller2020}, Kepler-448 c \citep{masuda2017}, Kepler-539 c \citep{mancini2016}, and KOI-984 c \citep{sun2022}. We should also point out that many direct imaging planets are considered confirmed without precise knowledge of their orbital period.

Based on the sensitivity of RVs from \citet{cale2021} and \citet{zicher2022} and since K$_{\rm d} <$ 1 m s$^{-1}$, further characterization of AU Mic d is out of reach of the current RV sensitivity, especially in the presence of substantial stellar activity. Furthermore, this planet is too close to the host star for direct imaging at $\sim$0.1 au, which is $\sim$0.01'' maximum angular separation from the host star at 9.7 pc. Thus, TTVs and transits (if AU Mic d does transit) are the only near-term methods to further characterize this system and confirm planet d. Additional TTVs of AU Mic b and, particularly, AU Mic c are critical for the coming decade with space-based photometry and in further distinguishing the inner d and middle d scenarios. Follow-up with current and future space-based transit observing missions such as HST, CHEOPS, PLATO, and Pandora would be greatly beneficial in searching for and/or ruling out any transit signatures for AU Mic d. For that purpose, both AU Mic b's and c's predicted transit midpoint times in BJD (after accounting for the TTVs) over the next three years for each 3-planet configuration are listed in Tables \ref{tab:aumic_b_transit_3yr_1} through \ref{tab:aumic_c_transit_3yr_2}.

\section{Conclusion}\label{sec:conclude}

AU Mic is a young nearby system hosting a dusty disk and two transiting planets. AU Mic is known to be very active, which made RV and TTV observations of AU Mic system quite challenging. However, \citet{wittrock2022} determined that the impact of spot modulation on observed TTVs are minimal with respect to the observed TTV variability, and the effect of stellar flares had been mitigated in SPIRou \citep{martioli2020}, ESPRESSO \citep{palle2020}, TESS \citep{gilbert2022}, Spitzer \citep{wittrock2022}, and CHEOPS \citep{szabo2021, szabo2022} data and masked in the ground data due to larger transit timing uncertainties.

We presented the validation of the new planet AU Mic d likely orbiting between planets b and c. We collected 50 data sets from four years of observations on AU Mic system, including 18 new observations presented herein. We modeled the photometric transits with \exofast and the TTVs with \exostriker. We determined the super-period of AU Mic b's TTVs, which we then used to estimate the orbital period of AU Mic d. We modeled the high-eccentricity and low-eccentricity 2-planet configurations. The high-eccentricity configurations' high eccentricities (e $>$ 0.2) and very small inferred masses (K $<$ 0.07 m s$^{-1}$ are in strong disagreement with those from RV models \citep{cale2021, donati2023} and secondary eclipse search (Kevin et al. in prep). While this configuration does model a TTV super-period, it is clearly not dynamically settled given the apparent orbital migrations of AU Mic b and c and the extreme fluctuations in both planets' eccentricities and inclinations in the rebound simulations. Unlike the high-eccentricity configuration, the low-eccentricity 2-planet counterpart does not exhibit the super-period in AU Mic b's TTVs; this configuration also has both b and c being misaligned, which contradicts with transit models and observations \citep{plavchan2020, martioli2021, gilbert2022, wittrock2022}. Lastly, the TTV model comparisons statistically ruled out both 2-planet models. We then developed the TTV log-likelihood periodogram to explore a range of possible orbital periods for AU Mic d; through this technique, we obtained ten possible solutions, which we then follow up with \exostriker's Bayesian analysis, stability test packages including \rebound \& \spock, and RV modeling.

After performing the TTV and RV model comparisons, the stability tests, and the calculation of the super-period of AU Mic b's TTVs and taking into consideration the Occam's razor arguments regarding the near-MMR chains and coplanarity of the AU Mic system, we determined that the most favored AU Mic d's orbital period is P$_{\rm d}$ = 12.73596 $\pm$ 0.00793 days. The nine other possible periods for AU Mic d are statistically and dynamically disfavored but not ruled out. The favored solution for AU Mic d places the three planets near the 2:3 orbital resonance pairs, or near 4:6:9 orbital resonance overall. The near-3:2 resonant chain is the most common pairing among the known exoplanets \citep{fabrycky2014}. Moreover, this particular configuration is statistically the most stable one. The presence of resonant chains among the planets in a very young yet stable system is significant, as it indicates that a compact system can quickly establish resonant chains very early on and which can become dynamically stable within a few hundred short orbital periods. The other disfavored candidate periods for AU Mic d that we modeled would also establish the AU Mic system of three planets in a near-MMR chain, but in higher order and less commonly occurring resonances. Either way, the scenario with no AU Mic d is statistically ruled out through the TTV and RV model comparisons, and the system must be in a near-resonant chain.

For the period of 12.7 days, our modeling determined the mass M$_{\rm d}$ = 1.053 $\pm$ 0.511 \mear and the time of conjunction $T_{C,d}$ = 2458340.55781 $\pm$ 0.11641 BJD for AU Mic d. This will make AU Mic d the fist known Earth-mass planet to orbit a young star and which will serve as a fundamental target for young terrestrial planets' atmospheric characterization and evolution, provided that it does transit. If one assumes AU Mic d is transiting and its density is Earth-like, then its transit depth would be $\sim$160 ppm. This very shallow depth in face of AU Mic's heightened stellar activity may explain the apparent lack of transit signals from AU Mic d in the TESS light curves. Future work could involve thoroughly searching all existing space-based light curves of AU Mic system for any statistically significant 160-ppm signals. The calculated time of conjunction provided in this paper will simplify the search as it provides a constraint on the timing of the expected transits of AU Mic d.

Our RV analysis implies that there may be additional planets beyond AU Mic c, including the 65-day candidate signature from our $\ln\mathcal{L}$ brute-force periodogram. We recommend additional ground- and space-based TTV observations to further characterize AU Mic d and confirm its orbital period, and to search for additional planets beyond AU Mic c. Given the youthfulness of the AU Mic system and the numerous exciting discoveries emerging from it, AU Mic will serve as an excellent target for HST, CHEOPS, PLATO, Pandora, and future space-based transit observing missions.

PPP acknowledges support from NASA (Exoplanet Research Program Award \#80NSSC20K0251, TESS Cycle 3 Guest Investigator Program Award \#80NSSC21K0349, JPL Research and Technology Development, and Keck Observatory Data Analysis) and the NSF (Astronomy and Astrophysics Grants \#1716202 and 2006517), and the Mt Cuba Astronomical Foundation.

EG acknowledges support from the NASA Exoplanets Research Program Award \#80NSSC20K0251. The material is based upon work supported by NASA under award \#80GSFC21M0002.

LDV acknowledges funding support from the Heising-Simons Astrophysics Postdoctoral Launch Program, through a grant to Vanderbilt University.

VR acknowledges the support of the Italian National Institute of Astrophysics (INAF) through the INAF GTO Grant “ERIS \& SHARK GTO data exploitation”.

This paper includes data collected by the TESS mission, which are publicly available from the Mikulski Archive for Space Telescopes (MAST). Funding for the TESS mission is provided by NASA’s Science Mission directorate. Resources supporting this work were provided by the NASA High-End Computing (HEC) Program through the NASA Advanced Supercomputing (NAS) Division at Ames Research Center for the production of the SPOC data products.

This work is based [in part] on observations made with the Spitzer Space Telescope, which was operated by the Jet Propulsion Laboratory, California Institute of Technology under a contract with NASA. Support for this work was provided by NASA through an award issued by JPL/Caltech. This research has made use of the NASA/IPAC Infrared Science Archive, which is funded by the National Aeronautics and Space Administration and operated by the California Institute of Technology.

This work makes use of observations from the Las Cumbres Observatory global telescope network. Part of the LCOGT telescope time was granted by NOIRLab through the Mid-Scale Innovations Program (MSIP). MSIP is funded by NSF.

This research made use of the PEST photometry pipeline\footnote{\url{http://pestobservatory.com}} by Thiam-Guan Tan.

This work makes use of observations from the ASTEP telescope. ASTEP benefited from the support of the French and Italian polar agencies IPEV and PNRA in the framework of the Concordia station program, from OCA, INSU, and Idex UCAJEDI (ANR- 15-IDEX-01).

This research received funding from the European Research Council (ERC) under the European Union's Horizon 2020 research and innovation programme (grant agreement n$^\circ$ 803193/BEBOP), and from the Science and Technology Facilities Council (STFC; grant n$^\circ$ ST/S00193X/1).

This research has made use of the NASA Exoplanet Archive, which is operated by the California Institute of Technology, under contract with the National Aeronautics and Space Administration under the Exoplanet Exploration Program.

This research has made use of the Exoplanet Follow-up Observation Program (ExoFOP; DOI: 10.26134/ExoFOP5) website, which is operated by the California Institute of Technology, under contract with the National Aeronautics and Space Administration under the Exoplanet Exploration Program.

This research has made use of the SIMBAD database, operated at CDS, Strasbourg, France.

This research has made use of NASA’s Astrophysics Data System Bibliographic Services.

This research has made use of an online calculator that converts a list of Barycentric Julian Dates in Barycentric Dynamical Time (BJD\_TDB) to JD in UT \citep{eastman2010}\footnote{\url{https://astroutils.astronomy.osu.edu/time/bjd2utc.html}}.

We also give thanks to Trifon Trifonov for his assistance in the use of the \exostriker package and analysis of the AU Mic system.

\facilities{ASTEP 400:0.4 m (FLI Proline 16800E), Brierfield:0.36 m (Moravian G4-16000 KAF-16803), CFHT (SPIRou), CHEOPS, ExoFOP, Exoplanet Archive, IRSA, IRTF (iSHELL), LCOGT (CTIO:1 m, SAAO:1 m, SSO:1 m, \& TO:1 m; Sinistro), MAST, MKO CDK700:0.7 m (U16), PEST:0.30 m (SBIG ST-8XME), Spitzer (IRAC), TESS, VLT:Antu (ESPRESSO)}

\software{{\tt AstroImageJ} \citep{collins2017}, {\tt astropy} \citep{astropy2013, astropy2018}, {\tt batman} \citep{kreidberg2015}, \celerite \citep{foreman-mackey2017}, \celeritetwo \citep{foreman-mackey2017, foreman-mackey2018}, \emcee \citep{foreman-mackey2013}, {\tt EXOFAST} \citep{eastman2013}, \exofast \citep{eastman2019}, {\tt exoplanet} \citep{foreman-mackey2021}, \exostriker \citep{trifonov2019}, {\tt ipython} \citep{perez2007}, {\tt lightkurve} \citep{lightkurve2018}, {\tt matplotlib} \citep{hunter2007}, \megno \citep{cincotta2000, cincotta2003}, {\tt numpy} \citep{harris2020}, \rebound \citep{rein2012, rein2015}, {\tt scipy} \citep{virtanen2020}, \spock \citep{tamayo2020}, {\tt TAPIR} \citep{jensen2013}}

\clearpage
\appendix
\restartappendixnumbering
\section{Outputs From Alternative Cases}

The appendix lists the O--C diagrams, mcmc posteriors, corner plots, rebound plots, predicted transit times, and future (10 yrs) TTV models for P$_{d}$ = (5.08, 5.39, 5.64, 5.86, 6.20, 6.47, 11.9, 12.6, and 14.1) days. The 12.7-day case is statistically most favored and is presented in the main text sections $\S$\ref{sec:exostrikermod} and $\S$\ref{sec:results}.

\begin{figure*}[b]
    \centering
    \includegraphics[width=0.48\textwidth]{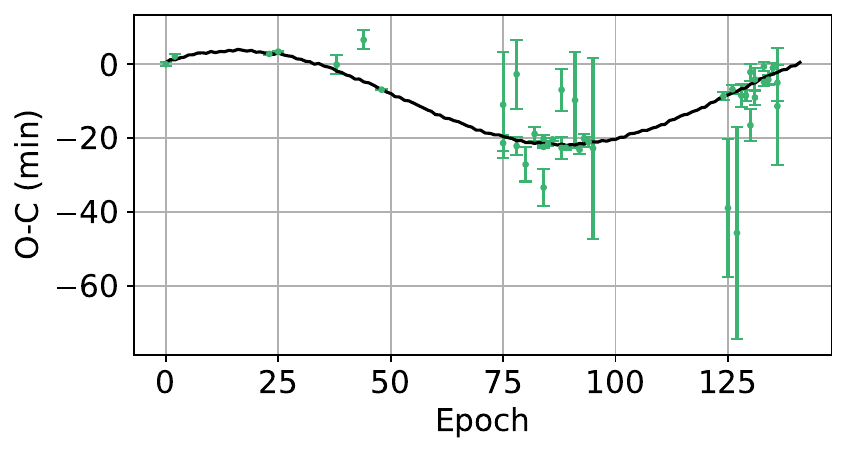}
    \includegraphics[width=0.48\textwidth]{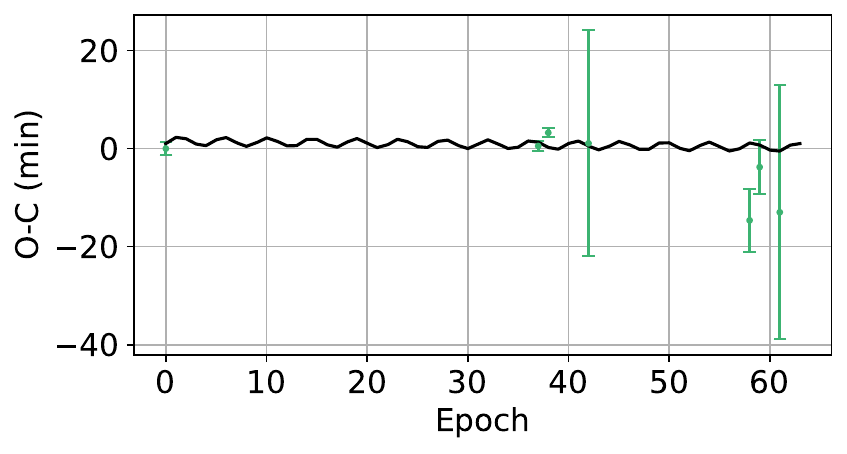}
    \includegraphics[width=0.48\textwidth]{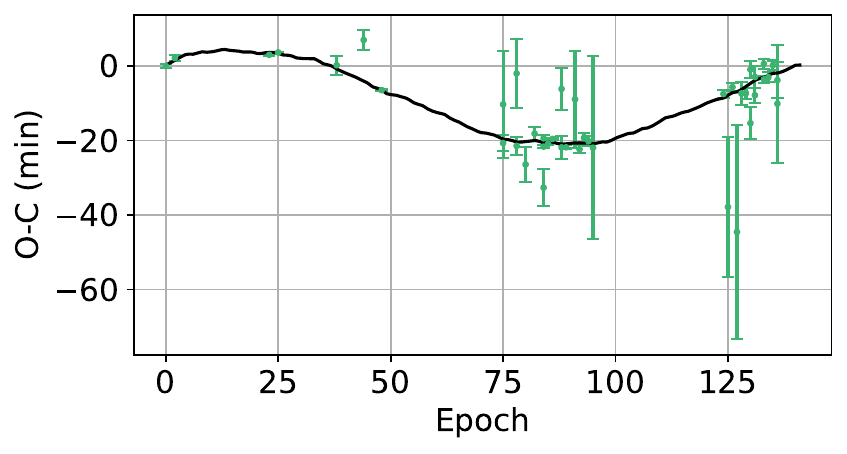}
    \includegraphics[width=0.48\textwidth]{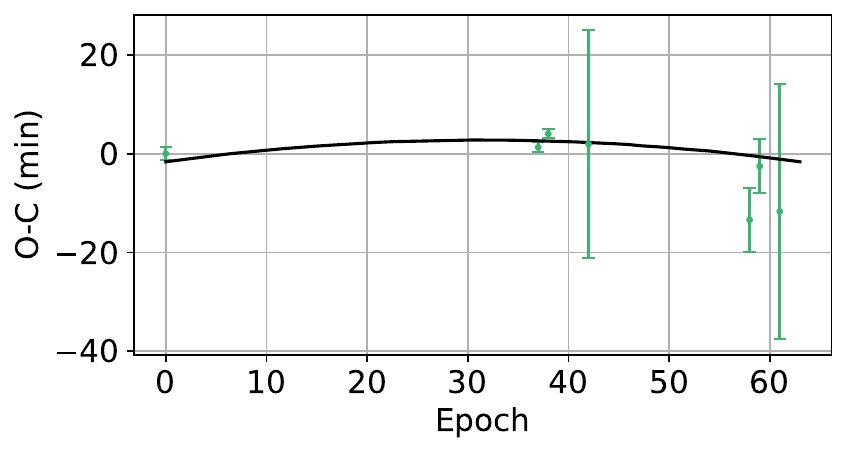}
    \includegraphics[width=0.48\textwidth]{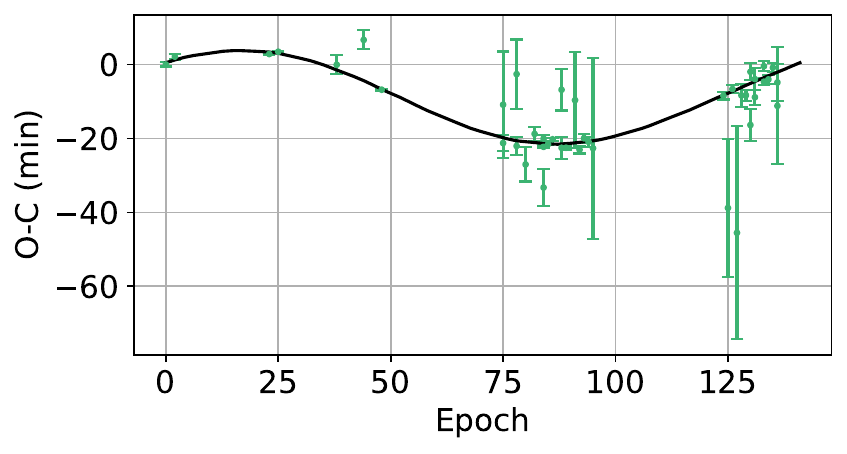}
    \includegraphics[width=0.48\textwidth]{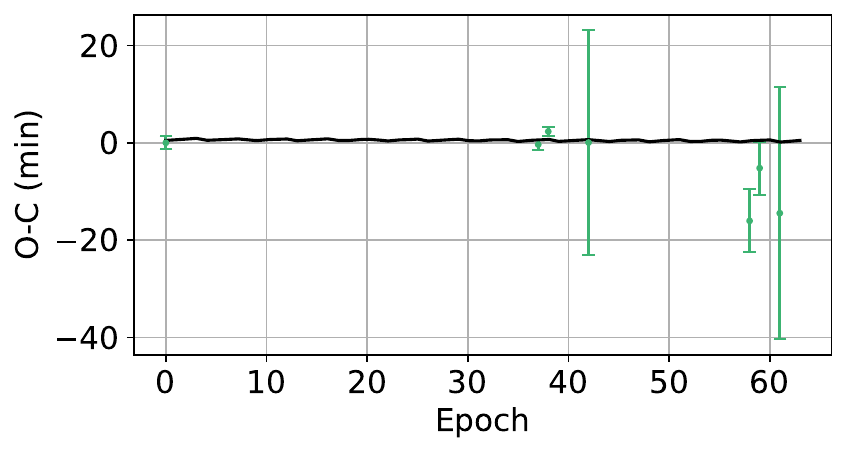}
    \caption{O--C diagram of AU Mic b ({\it left}) and AU Mic c ({\it right}), with comparison between TTVs (green) and \exostriker-generated mcmc models (black) for P$_{\rm d}$ = 5.08 days ({\it top row}), 5.39 days ({\it middle row}), \& 5.64 days ({\it bottom row}).}
    \label{fig:3p_inner_exostriker_plots_1}
\end{figure*}

\begin{figure*}
    \centering
    \includegraphics[width=0.48\textwidth]{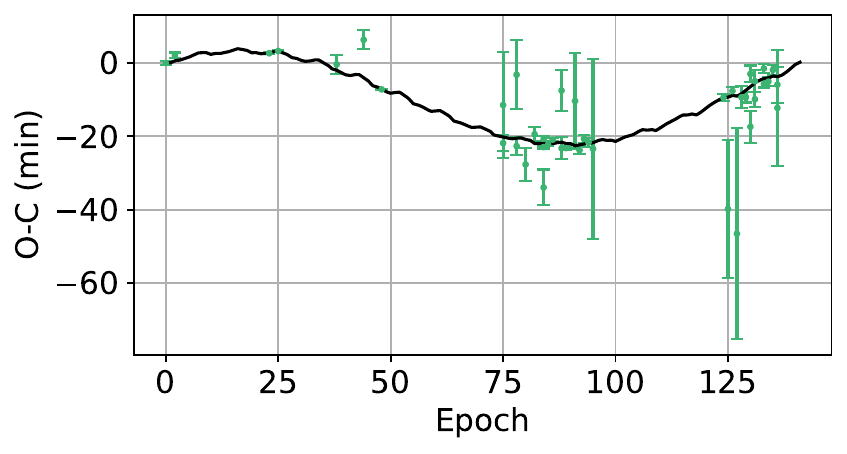}
    \includegraphics[width=0.48\textwidth]{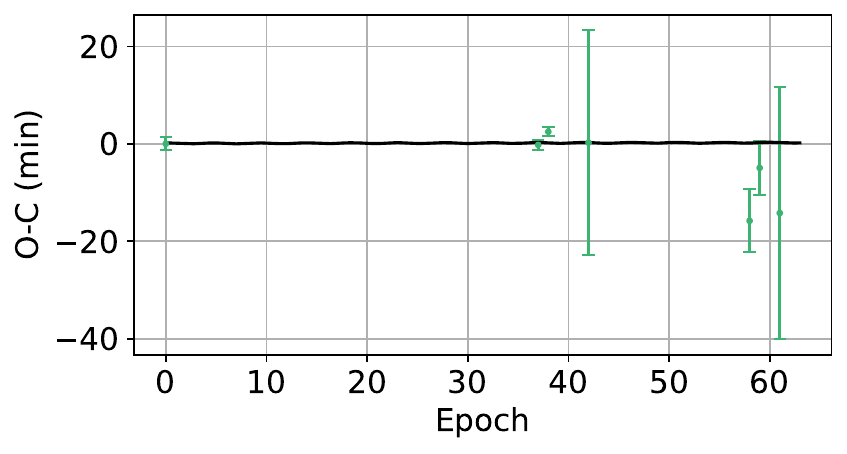}
    \includegraphics[width=0.48\textwidth]{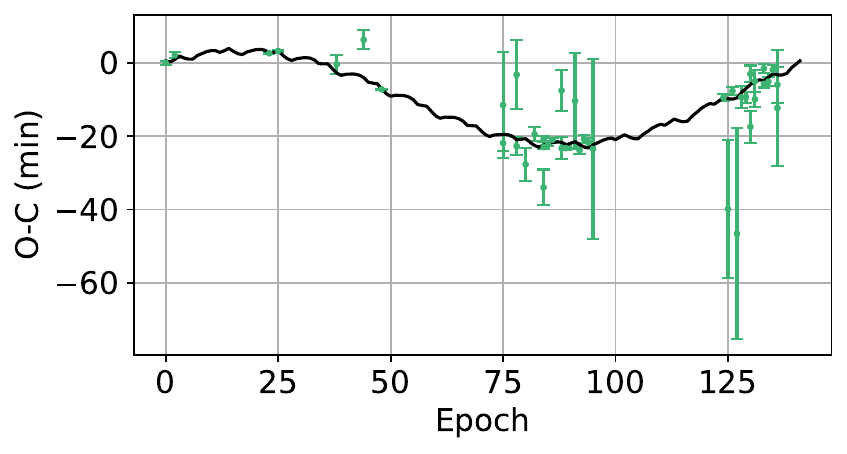}
    \includegraphics[width=0.48\textwidth]{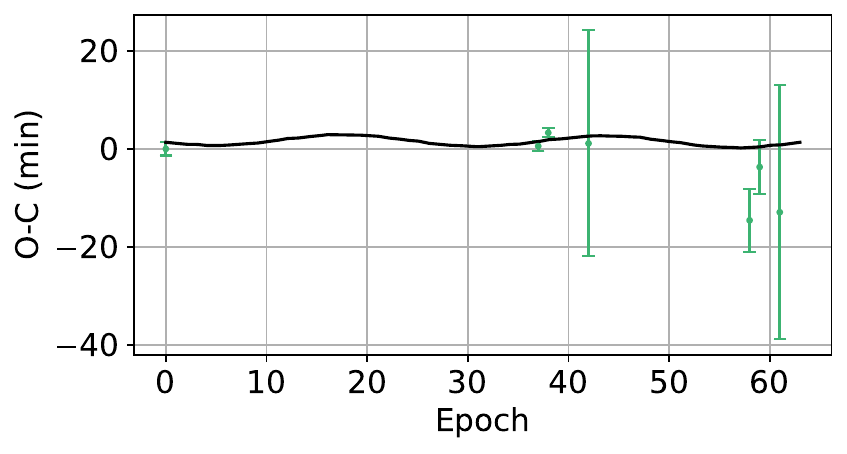}
    \includegraphics[width=0.48\textwidth]{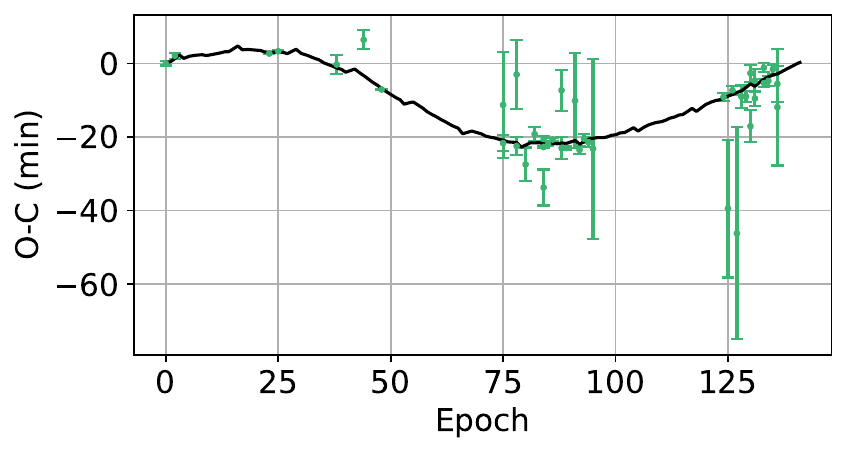}
    \includegraphics[width=0.48\textwidth]{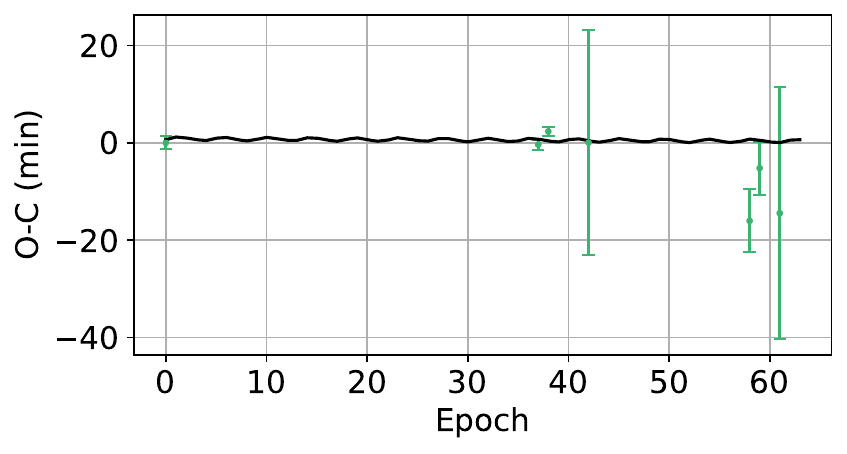}
    \caption{O--C diagram of AU Mic b ({\it left}) and AU Mic c ({\it right}), with comparison between TTVs (green) and \exostriker-generated mcmc models (black) for P$_{\rm d}$ = 5.86 days ({\it top row}), 6.20 days ({\it middle row}), \& 6.47 days ({\it bottom row}).}
    \label{fig:3p_inner_exostriker_plots_2}
\end{figure*}

\begin{figure*}
    \centering
    \includegraphics[width=0.48\textwidth]{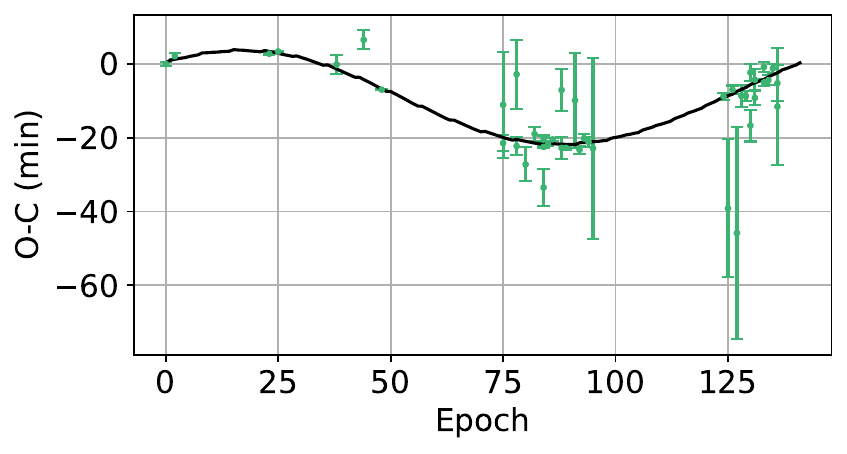}
    \includegraphics[width=0.48\textwidth]{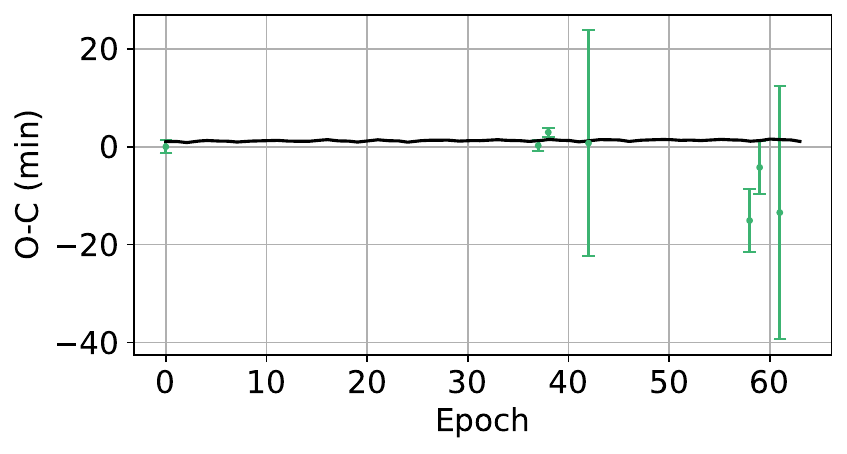}
    \includegraphics[width=0.48\textwidth]{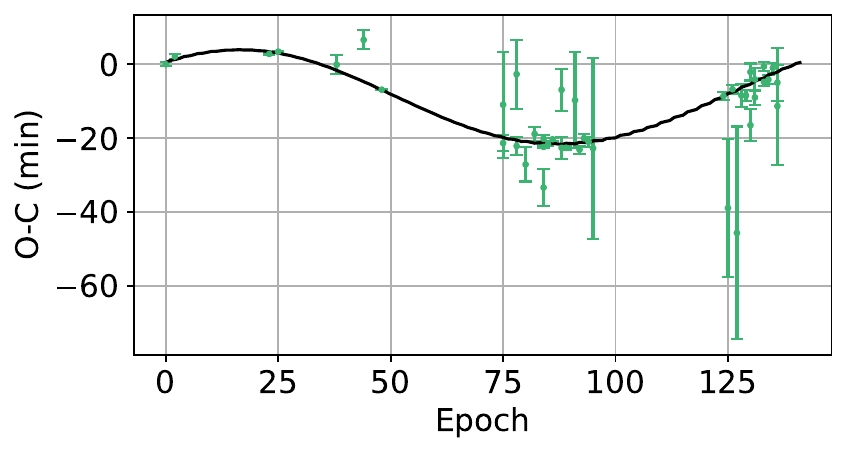}
    \includegraphics[width=0.48\textwidth]{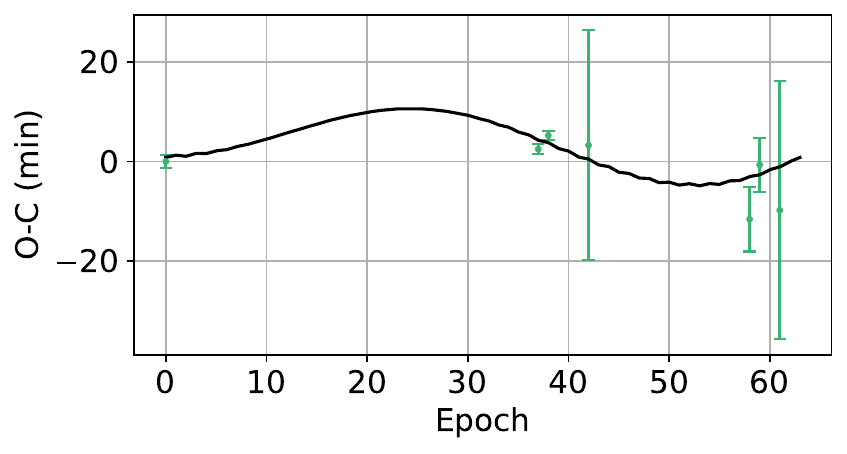}
    \includegraphics[width=0.48\textwidth]{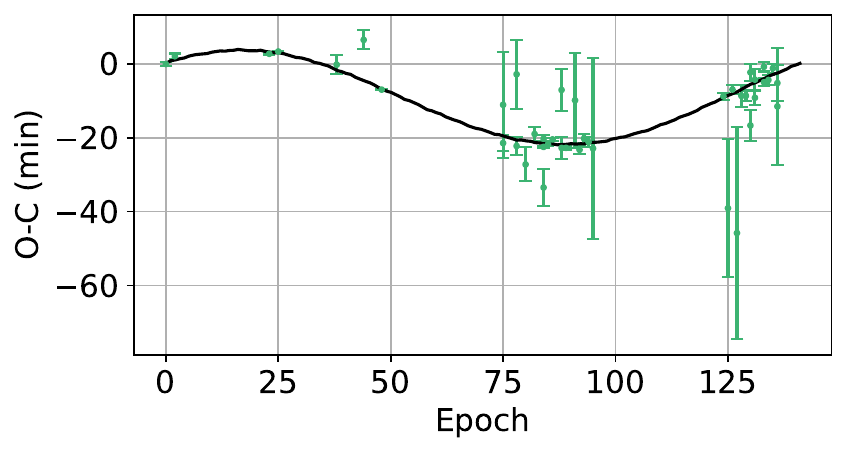}
    \includegraphics[width=0.48\textwidth]{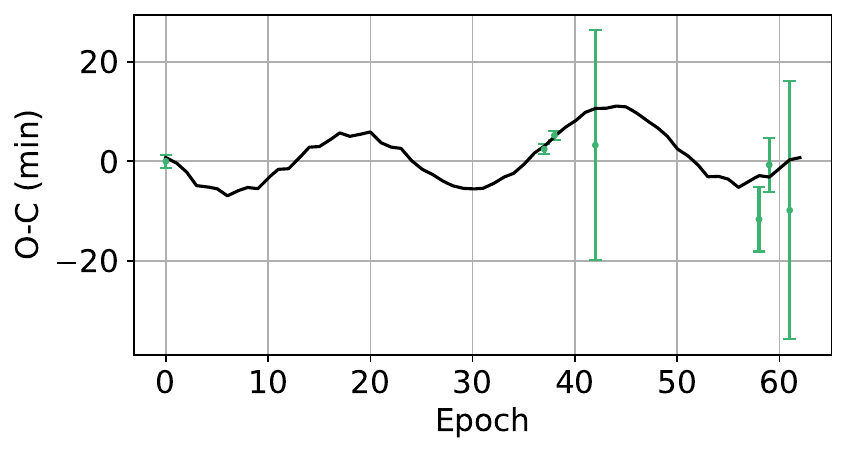}
    \caption{O--C diagram of AU Mic b ({\it left}) and AU Mic c ({\it right}), with comparison between TTVs (green) and \exostriker-generated mcmc models (black) for 11.9 days ({\it top row}), 12.6 days ({\it middle row}), and 14.1 days ({\it bottom row}).}
    \label{fig:3p_middle_exostriker_plots}
\end{figure*}

\begin{deluxetable*}{l|c|ccc|ccc}\label{tab:5-08d_exostriker_params}
    \tablecaption{\exostriker-generated best fit and MCMC modeling parameters for P$_{\rm d}$ = 5.08 days.}
    \tablehead{\multirow{2}{*}{Parameter} & \multirow{2}{*}{Unit} & \multicolumn{3}{c|}{Best Fit}  & \multicolumn{3}{c}{MCMC}       \\
                                          &                       & AU Mic b & AU Mic c & AU Mic d & AU Mic b & AU Mic c & AU Mic d }
    \startdata
K	             & m s$^{-1}$ & 3.35536   & 0.22986   & 0.50720	  & 2.87947 $\pm$ 0.34321   & 3.60029 $\pm$ 3.44133   & 0.41877 $\pm$ 0.04442    \\
\porb	         & day        & 8.46351   & 18.86127  & 5.07706	  & 8.46331 $\pm$ 0.00010   & 18.86102 $\pm$ 0.00029  & 5.07701 $\pm$ 0.00017    \\
e	             & ...        & 0.00000   & 0.00592   & 0.00016	  & 0.04872 $\pm$ 0.04662   & 0.00103 $\pm$ 0.00159   & 0.04209 $\pm$ 0.03778    \\
$\omega$	     & deg        & 98.37057  & 223.59289 & 152.78962 & 98.57816 $\pm$ 2.54783  & 223.75374 $\pm$ 1.77297 & 144.16841 $\pm$ 2.43169  \\
M$_{0}$	         & deg        & 351.61077 & 0.00025   & 268.90220 & 350.54357 $\pm$ 1.58108 & 0.24484 $\pm$ 0.20155	  & 274.43129 $\pm$ 10.64027 \\
i	             & deg        & 89.57924  & 89.43188  & 77.33666  & 89.48110 $\pm$ 0.31229  & 89.39871 $\pm$ 0.42666  & 76.88456 $\pm$ 3.04551   \\
$\Omega$	     & deg        & 23.10249  & 22.62173  & 8.17561	  & 19.11756 $\pm$ 0.53911  & 20.71402 $\pm$ 3.01682  & 2.55408 $\pm$ 1.56684    \\ \hline
$\chi^{2}$       & ...        & 114.02094 & 11.26343  & ...       & 106.79515               & 18.02865                & ...                      \\
\enddata
    \tablecomments{The K's listed here are unconstrained, but see $\S$\ref{sec:results} for discussion regarding the planets' low K's generated by \exostriker.}
\end{deluxetable*}

\begin{deluxetable*}{l|c|ccc|ccc}\label{tab:5-39d_exostriker_params}
    \tablecaption{\exostriker-generated best fit and MCMC modeling parameters for P$_{\rm d}$ = 5.39 days.}
    \tablehead{\multirow{2}{*}{Parameter} & \multirow{2}{*}{Unit} & \multicolumn{3}{c|}{Best Fit}  & \multicolumn{3}{c}{MCMC}       \\
                                          &                       & AU Mic b & AU Mic c & AU Mic d & AU Mic b & AU Mic c & AU Mic d }
    \startdata
K	             & m s$^{-1}$ & 5.16537   & 0.50637	  & 0.34978	  & 0.10817 $\pm$ 1.90115   & 2.00680 $\pm$ 1.92941   & 0.32173 $\pm$ 0.01505    \\
\porb	         & day        & 8.46341   & 18.86196  & 5.38693	  & 8.46327 $\pm$ 0.00008   & 18.85933 $\pm$ 0.00107  & 5.38888 $\pm$ 0.00188    \\
e	             & ...        & 0.00001   & 0.07780	  & 0.00299	  & 0.02897 $\pm$ 0.02697   & 0.07978 $\pm$ 0.01753   & 0.08579 $\pm$ 0.07529    \\
$\omega$	     & deg        & 98.55512  & 217.22994 & 161.43928 & 97.22742 $\pm$ 1.87410  & 216.68146 $\pm$ 1.23336 & 166.41581 $\pm$ 15.66255 \\
M$_{0}$	         & deg        & 351.43634 & 0.00001   & 265.97175 & 352.33787 $\pm$ 4.06365 & 0.36773 $\pm$ 0.32286   & 265.55255 $\pm$ 9.64814  \\
i	             & deg        & 89.57209  & 89.42620  & 61.86109  & 89.79654 $\pm$ 0.61080  & 89.68914 $\pm$ 0.70596  & 57.18937 $\pm$ 1.827605  \\
$\Omega$	     & deg        & 55.99639  & 58.92122  & 2.10453	  & 43.45119 $\pm$ 7.68065  & 81.23160 $\pm$ 27.93638 & 6.51204 $\pm$ 6.20803    \\ \hline
$\chi^{2}$       & ...        & 115.39770 & 5.10544   & ...       & 105.90689               & 10.01490                & ...                      \\
\enddata
    \tablecomments{The K's listed here are unconstrained, but see $\S$\ref{sec:results} for discussion regarding the planets' low K's generated by \exostriker.}
\end{deluxetable*}

\begin{deluxetable*}{l|c|ccc|ccc}\label{tab:5-64d_exostriker_params}
    \tablecaption{\exostriker-generated best fit and MCMC modeling parameters for P$_{\rm d}$ = 5.64 days.}
    \tablehead{\multirow{2}{*}{Parameter} & \multirow{2}{*}{Unit} & \multicolumn{3}{c|}{Best Fit}  & \multicolumn{3}{c}{MCMC}       \\
                                          &                       & AU Mic b & AU Mic c & AU Mic d & AU Mic b & AU Mic c & AU Mic d }
    \startdata
K	             & m s$^{-1}$ & 4.40058   & 0.46675   & 0.33788   & 2.56609 $\pm$ 0.74131    & 0.32794 $\pm$ 0.14726   & 0.26404 $\pm$ 0.02325    \\
\porb	         & day        & 8.46339   & 18.86185  & 5.63768   & 8.46351 $\pm$ 0.00020    & 18.86055 $\pm$ 0.00059  & 5.63897 $\pm$ 0.00207    \\
e	             & ...        & 0.00000   & 0.00677   & 0.00029   & 0.00179 $\pm$ 0.00085    & 0.03117 $\pm$ 0.02806   & 0.00997 $\pm$ 0.00676    \\
$\omega$	     & deg        & 98.50502  & 223.51766 & 166.51374 & 91.68412 $\pm$ 3.26572   & 220.60684 $\pm$ 1.08879 & 179.09135 $\pm$ 17.81811 \\
M$_{0}$	         & deg        & 351.47597 & 0.00094   & 241.03825 & 358.29595 $\pm$ 11.06134 & 0.80625 $\pm$ 0.74415   & 228.08583 $\pm$ 7.13443  \\
i	             & deg        & 89.52517  & 89.43744  & 87.60186  & 89.38133 $\pm$ 0.21181   & 89.48992 $\pm$ 0.50234  & 90.07831 $\pm$ 4.98962   \\
$\Omega$	     & deg        & 30.10388  & 28.65682  & 9.84080   & 31.92354 $\pm$ 5.43454   & 30.65984 $\pm$ 7.23061  & 11.45968 $\pm$ 7.13193   \\ \hline
$\chi^{2}$       & ...        & 113.88847 & 11.22925  & ...       & 115.86957                & 11.71456                & ...                      \\
\enddata
    \tablecomments{The K's listed here are unconstrained, but see $\S$\ref{sec:results} for discussion regarding the planets' low K's generated by \exostriker.}
\end{deluxetable*}

\begin{deluxetable*}{l|c|ccc|ccc}\label{tab:5-86d_exostriker_params}
    \tablecaption{\exostriker-generated best fit and MCMC modeling parameters for P$_{\rm d}$ = 5.86 days.}
    \tablehead{\multirow{2}{*}{Parameter} & \multirow{2}{*}{Unit} & \multicolumn{3}{c|}{Best Fit}  & \multicolumn{3}{c}{MCMC}       \\
                                          &                       & AU Mic b & AU Mic c & AU Mic d & AU Mic b & AU Mic c & AU Mic d }
    \startdata
K	             & m s$^{-1}$ & 2.16456   & 0.23630   & 0.53136   & 0.03744 $\pm$ 0.66777   & 1.01987 $\pm$ 0.81217   & 0.57447 $\pm$ 0.12867    \\
\porb	         & day        & 8.46367   & 18.86055  & 5.86086   & 8.46369 $\pm$ 0.00009   & 18.85928 $\pm$ 0.00036  & 5.86210 $\pm$ 0.00231    \\
e	             & ...        & 0.00000   & 0.00311   & 0.00003   & 0.00129 $\pm$ 0.00025   & 0.01538 $\pm$ 0.01280   & 0.00011 $\pm$ 0.00110    \\
$\omega$	     & deg        & 98.48083  & 223.82709 & 164.24881 & 102.58694 $\pm$ 7.62158 & 222.06323 $\pm$ 0.41503 & 168.01295 $\pm$ 16.83654 \\
M$_{0}$	         & deg        & 351.49999 & 0.00131   & 231.06872 & 347.37523 $\pm$ 0.39120 & 0.73499 $\pm$ 0.65918   & 216.24219 $\pm$ 3.82607  \\
i	             & deg        & 89.55965  & 89.42747  & 95.02796  & 89.64773 $\pm$ 0.46690  & 89.32042 $\pm$ 0.32416  & 99.58862 $\pm$ 6.78176   \\
$\Omega$	     & deg        & 29.19644  & 28.21852  & 15.22950  & 68.15507 $\pm$ 45.50567 & 25.26173 $\pm$ 5.59688  & 57.61649 $\pm$ 48.55583  \\ \hline
$\chi^{2}$       & ...        & 102.74262 & 11.75572  & ...       & 100.82281               & 13.43211                & ...                      \\
\enddata
    \tablecomments{The K's listed here are unconstrained, but see $\S$\ref{sec:results} for discussion regarding the planets' low K's generated by \exostriker.}
\end{deluxetable*}

\begin{deluxetable*}{l|c|ccc|ccc}\label{tab:6-20d_exostriker_params}
    \tablecaption{\exostriker-generated best fit and MCMC modeling parameters for P$_{\rm d}$ = 6.20 days.}
    \tablehead{\multirow{2}{*}{Parameter} & \multirow{2}{*}{Unit} & \multicolumn{3}{c|}{Best Fit}  & \multicolumn{3}{c}{MCMC}       \\
                                          &                       & AU Mic b & AU Mic c & AU Mic d & AU Mic b & AU Mic c & AU Mic d }
    \startdata
K	             & m s$^{-1}$ & 2.90791   & 0.46992   & 0.26161   & 2.45919 $\pm$ 0.14618   & 0.49589 $\pm$ 0.28947   & 0.47104 $\pm$ 0.24820    \\
\porb	         & day        & 8.46343   & 18.86090  & 6.20441   & 8.46362 $\pm$ 0.00029   & 18.86047 $\pm$ 0.00004  & 6.20494 $\pm$ 0.00105    \\
e	             & ...        & 0.00000   & 0.01115   & 0.00018   & 0.00452 $\pm$ 0.00323   & 0.01386 $\pm$ 0.00986   & 0.02737 $\pm$ 0.02352    \\
$\omega$	     & deg        & 98.48979  & 223.15789 & 160.36483 & 99.85587 $\pm$ 4.41672  & 222.82523 $\pm$ 1.52204 & 194.59296 $\pm$ 41.32711 \\
M$_{0}$	         & deg        & 351.49117 & 0.00036   & 238.43986 & 350.03934 $\pm$ 2.04421 & 0.09010 $\pm$ 0.04308   & 184.33273 $\pm$ 46.79969 \\
i	             & deg        & 89.55720  & 89.42914  & 94.74564  & 89.82232 $\pm$ 0.65141  & 88.96846 $\pm$ 0.01337  & 107.15764 $\pm$ 14.18751 \\
$\Omega$	     & deg        & 24.51329  & 12.84297  & 4.91602   & 13.49562 $\pm$ 8.12320  & 12.11611 $\pm$ 3.83135  & 2.67773 $\pm$ 0.61209    \\ \hline
$\chi^{2}$       & ...        & 107.43664 & 11.84116  & ...       & 99.94369                & 10.06389                & ...                      \\
\enddata
    \tablecomments{The K's listed here are unconstrained, but see $\S$\ref{sec:results} for discussion regarding the planets' low K's generated by \exostriker.}
\end{deluxetable*}

\begin{deluxetable*}{l|c|ccc|ccc}\label{tab:6-47d_exostriker_params}
    \tablecaption{\exostriker-generated best fit and MCMC modeling parameters for P$_{\rm d}$ = 6.47 days.}
    \tablehead{\multirow{2}{*}{Parameter} & \multirow{2}{*}{Unit} & \multicolumn{3}{c|}{Best Fit}  & \multicolumn{3}{c}{MCMC}       \\
                                          &                       & AU Mic b & AU Mic c & AU Mic d & AU Mic b & AU Mic c & AU Mic d }
    \startdata
K	             & m s$^{-1}$ & 5.22397   & 0.51394   & 0.30268   & 2.24280 $\pm$ 0.60448   & 0.44272 $\pm$ 0.19449   & 0.27380 $\pm$ 0.05928    \\
\porb	         & day        & 8.46343   & 18.86244  & 6.47074   & 8.46352 $\pm$ 0.00016   & 18.86056 $\pm$ 0.00033  & 6.47251 $\pm$ 0.00249    \\
e	             & ...        & 0.00000   & 0.00442   & 0.00023   & 0.02862 $\pm$ 0.02473   & 0.00453 $\pm$ 0.00215   & 0.01002 $\pm$ 0.00819    \\
$\omega$	     & deg        & 98.49871  & 223.71495 & 166.62198 & 99.63829 $\pm$ 3.59581  & 223.56221 $\pm$ 1.61921 & 178.62168 $\pm$ 17.77984 \\
M$_{0}$	         & deg        & 351.48520 & 0.00062   & 248.45972 & 349.79295 $\pm$ 0.95313 & 0.15047 $\pm$ 0.08931   & 236.71532 $\pm$ 4.90333  \\
i	             & deg        & 89.57471  & 89.46389  & 94.91569  & 89.45294 $\pm$ 0.26172  & 89.25326 $\pm$ 0.26737  & 93.57266 $\pm$ 3.26340   \\
$\Omega$	     & deg        & 27.67490  & 19.55833  & 10.23918  & 20.65656 $\pm$ 1.45446  & 17.43367 $\pm$ 3.48630  & 11.01673 $\pm$ 5.42317   \\ \hline
$\chi^{2}$       & ...        & 109.38331 & 11.23694  & ...       & 97.94533                & 13.80731                & ...                      \\
\enddata
    \tablecomments{The K's listed here are unconstrained, but see $\S$\ref{sec:results} for discussion regarding the planets' low K's generated by \exostriker.}
\end{deluxetable*}

\begin{deluxetable*}{l|c|ccc|ccc}\label{tab:11-9d_exostriker_params}
    \tablecaption{\exostriker-generated best fit and MCMC modeling parameters for P$_{\rm d}$ = 11.9 days.}
    \tablehead{\multirow{2}{*}{Parameter} & \multirow{2}{*}{Unit} & \multicolumn{3}{c|}{Best Fit}  & \multicolumn{3}{c}{MCMC}       \\
                                          &                       & AU Mic b & AU Mic c & AU Mic d & AU Mic b & AU Mic c & AU Mic d }
    \startdata
K	             & m s$^{-1}$ & 0.97504   & 0.38690	  & 0.17182   & 1.24696 $\pm$ 0.65028  & 0.02093 $\pm$ 0.13709   & 0.17048 $\pm$ 0.03388   \\
\porb	         & day        & 8.46331   & 18.85981  & 11.86306  & 8.46333 $\pm$ 0.00005  & 18.85996 $\pm$ 0.00043  & 11.86292 $\pm$ 0.00406  \\
e	             & ...        & 0.00000   & 0.00109   & 0.00000   & 0.00018 $\pm$ 0.00189  & 0.00385 $\pm$ 0.00196   & 0.00051 $\pm$ 0.00087   \\
$\omega$	     & deg        & 89.99086  & 223.99440 & 160.83877 & 89.81844 $\pm$ 1.31482 & 223.69121 $\pm$ 2.55320 & 160.94662 $\pm$ 2.61504 \\
M$_{0}$	         & deg        & 0.00005   & 0.00014   & 0.00032   & 0.17304 $\pm$ 0.06999  & 0.06776 $\pm$ 0.02942   & 0.01696 $\pm$ 0.13710   \\
i	             & deg        & 89.58985  & 89.43145  & 82.02998  & 89.50347 $\pm$ 0.29017 & 89.33758 $\pm$ 0.31993  & 81.88651 $\pm$ 0.63757  \\
$\Omega$	     & deg        & 0.00012   & 0.00000   & 0.00000   & 0.00494 $\pm$ 0.12589  & 0.01833 $\pm$ 0.09216   & 0.11655 $\pm$ 0.03333   \\ \hline
$\chi^{2}$       & ...        & 112.83347 & 11.27744  & ...       & 113.01980              & 11.41707                & ...                     \\
\enddata
    \tablecomments{The K's listed here are unconstrained, but see $\S$\ref{sec:results} for discussion regarding the planets' low K's generated by \exostriker.}
\end{deluxetable*}

\begin{deluxetable*}{l|c|ccc|ccc}\label{tab:12-6d_exostriker_params}
    \tablecaption{\exostriker-generated best fit and MCMC modeling parameters for P$_{\rm d}$ = 12.6 days.}
    \tablehead{\multirow{2}{*}{Parameter} & \multirow{2}{*}{Unit} & \multicolumn{3}{c|}{Best Fit}  & \multicolumn{3}{c}{MCMC}       \\
                                          &                       & AU Mic b & AU Mic c & AU Mic d & AU Mic b & AU Mic c & AU Mic d }
    \startdata
K	             & m s$^{-1}$ & 2.90332   & 0.48084	  & 0.18469	  & 1.25408 $\pm$ 0.39741  & 0.03573 $\pm$ 0.07120   & 0.22424 $\pm$ 0.06278   \\
\porb	         & day        & 8.46334   & 18.86145  & 12.64081  & 8.46333 $\pm$ 0.00004  & 18.86027 $\pm$ 0.00009  & 12.64921 $\pm$ 0.00874  \\
e	             & ...        & 0.00849   & 0.00278	  & 0.00341	  & 0.00646 $\pm$ 0.00045  & 0.00257 $\pm$ 0.00086   & 0.00000 $\pm$ 0.00039   \\
$\omega$	     & deg        & 89.99407  & 223.85721 & 174.65177 & 88.35089 $\pm$ 0.42306 & 223.25926 $\pm$ 2.84483 & 174.35726 $\pm$ 4.76760 \\
M$_{0}$  	     & deg        & 0.00002   & 0.00048	  & 0.00012	  & 1.65671 $\pm$ 1.52820  & 0.60281 $\pm$ 0.41896   & 2.35570 $\pm$ 2.16294   \\
i	             & deg        & 89.57125  & 89.43113  & 89.06083  & 89.58284 $\pm$ 0.38234 & 89.45028 $\pm$ 0.41961  & 89.63258 $\pm$ 1.70943  \\
$\Omega$	     & deg        & 0.00001   & 0.00000	  & 0.00000	  & 2.17332 $\pm$ 2.01326  & 0.78666 $\pm$ 0.62801   & 0.15814 $\pm$ 0.01661   \\ \hline
$\chi^{2}$       & ...        & 114.85574 & 7.30405   & ...       & 113.68872              & 7.65425                 & ...                     \\
\enddata
    \tablecomments{The K's listed here are unconstrained, but see $\S$\ref{sec:results} for discussion regarding the planets' low K's generated by \exostriker.}
\end{deluxetable*}

\begin{deluxetable*}{l|c|ccc|ccc}\label{tab:14-1d_exostriker_params}
    \tablecaption{\exostriker-generated best fit and MCMC modeling parameters for P$_{\rm d}$ = 14.1 days.}
    \tablehead{\multirow{2}{*}{Parameter} & \multirow{2}{*}{Unit} & \multicolumn{3}{c|}{Best Fit}  & \multicolumn{3}{c}{MCMC}       \\
                                          &                       & AU Mic b & AU Mic c & AU Mic d & AU Mic b & AU Mic c & AU Mic d }
    \startdata
K	             & m s$^{-1}$ & 1.94679   & 0.79913   & 0.23459   & 0.53676 $\pm$ 0.09009  & 1.45679 $\pm$ 1.21270   & 0.25084 $\pm$ 0.03547   \\
\porb	         & day        & 8.46336   & 18.85942  & 14.07157  & 8.46335 $\pm$ 0.00002  & 18.85837 $\pm$ 0.00013  & 14.07595 $\pm$ 0.00724  \\
e	             & ...        & 0.01119   & 0.00631   & 0.00109   & 0.04387 $\pm$ 0.03502  & 0.00142 $\pm$ 0.00394   & 0.00120 $\pm$ 0.00028   \\
$\omega$	     & deg        & 89.99234  & 223.54696 & 173.15636 & 89.68870 $\pm$ 0.98273 & 223.47563 $\pm$ 2.72212 & 174.56600 $\pm$ 4.52321 \\
M$_{0}$	         & deg        & 0.00001   & 0.00034   & 0.00010   & 0.33416 $\pm$ 0.23087  & 0.46562 $\pm$ 0.31050   & 2.71969 $\pm$ 2.57135   \\
i	             & deg        & 89.47507  & 89.44010  & 75.59741  & 89.63164 $\pm$ 0.44523 & 89.25623 $\pm$ 0.21991  & 74.27032 $\pm$ 0.40173  \\
$\Omega$	     & deg        & 0.00000   & 0.00000   & 0.00000   & 3.15613 $\pm$ 3.01037  & 1.28301 $\pm$ 1.15840   & 0.55842 $\pm$ 0.42973   \\ \hline
$\chi^{2}$       & ...        & 116.76747 & 3.09309   & ...       & 112.45805              & 2.94137                 & ...                     \\
\enddata
    \tablecomments{The K's listed here are unconstrained, but see $\S$\ref{sec:results} for discussion regarding the planets' low K's generated by \exostriker.}
\end{deluxetable*}

\begin{figure*}
    \centering
    \includegraphics[width=\textwidth]{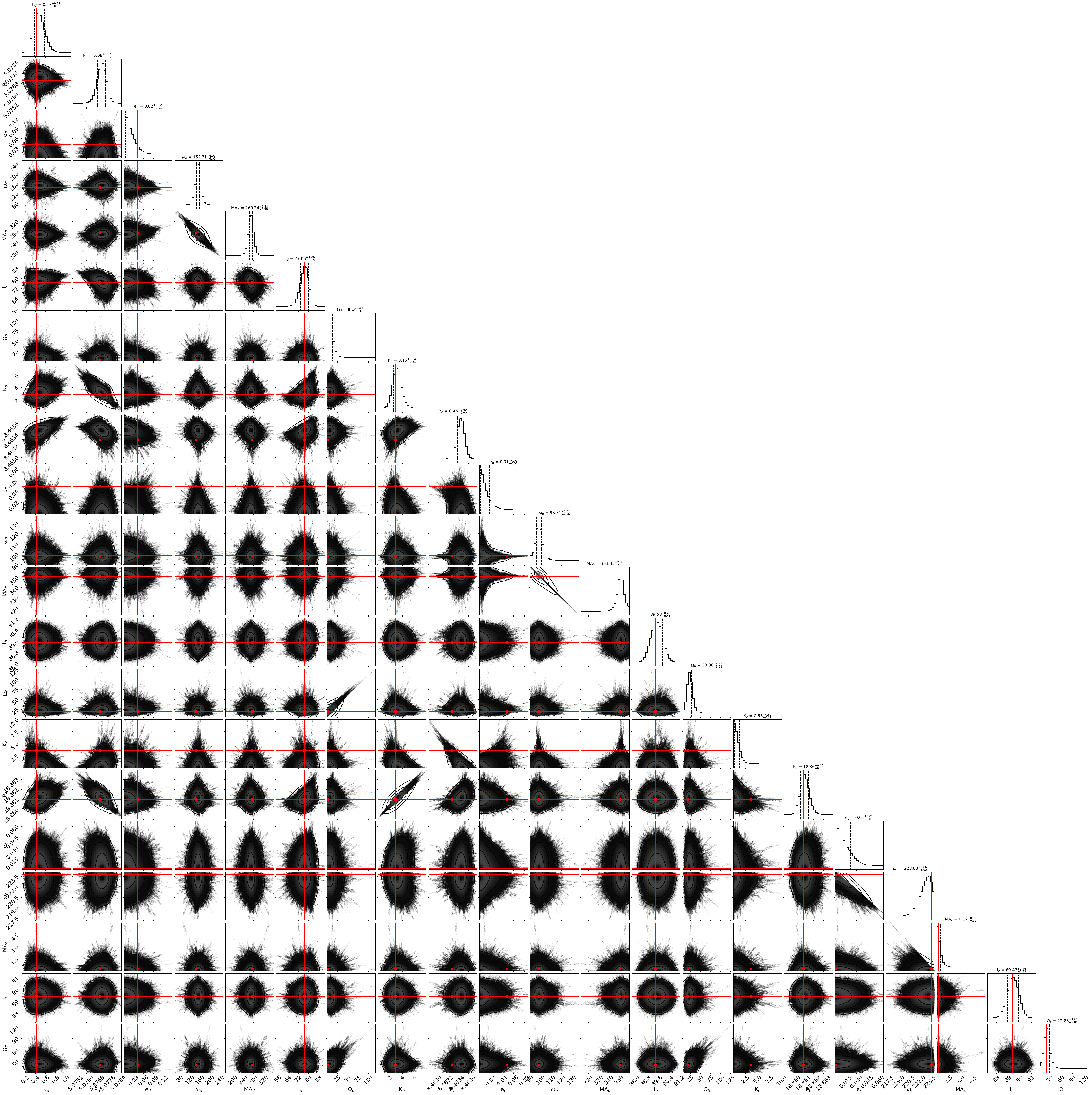}
    \caption{Corner plot of AU Mic b and c's orbital parameters from \exostriker mcmc analysis for P$_{\rm d}$ = 5.08 days.}
    \label{fig:5-08d_exostriker_cornerplot}
\end{figure*}

\begin{figure*}
    \centering
    \includegraphics[width=\textwidth]{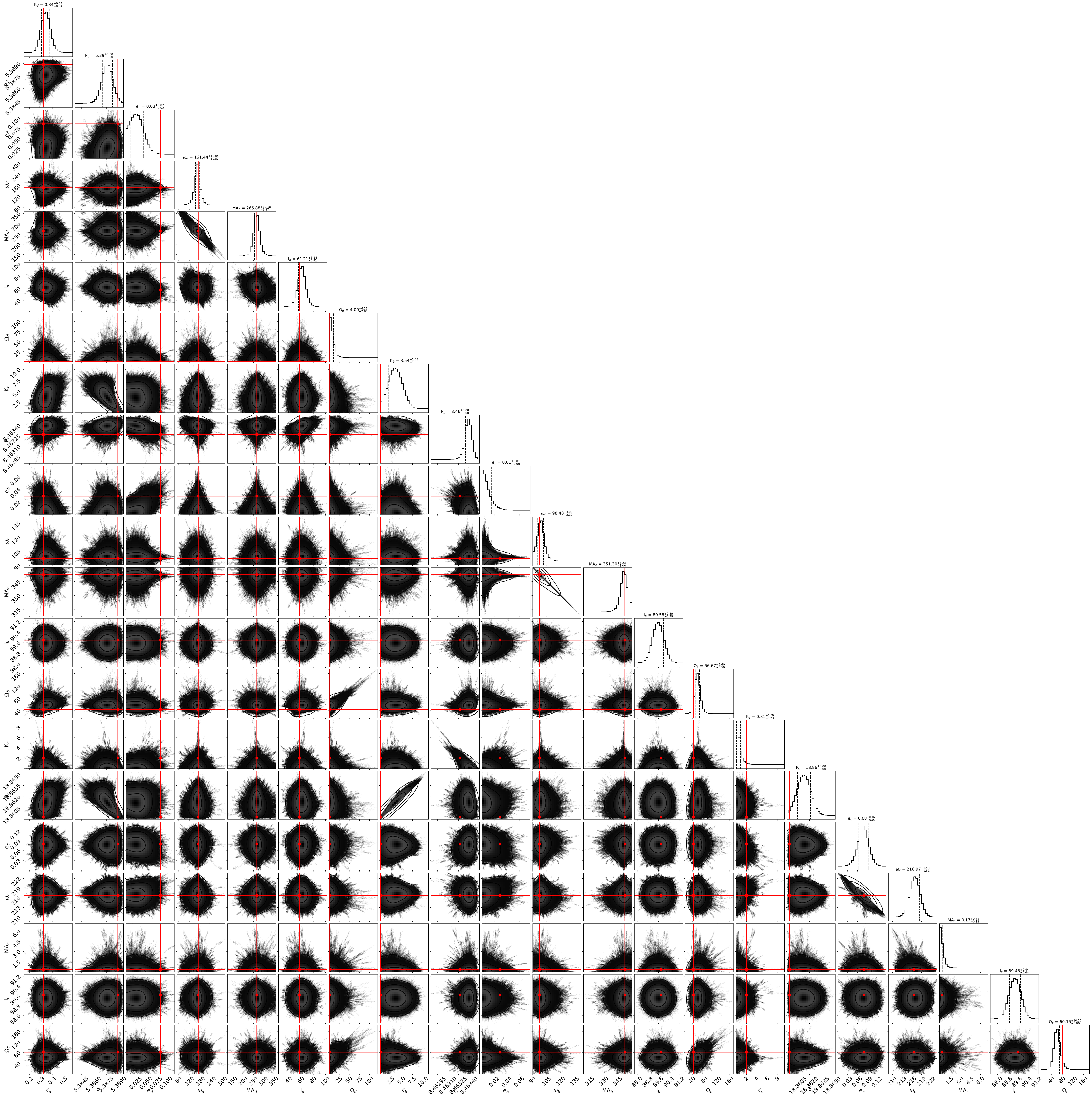}
    \caption{Corner plot of AU Mic b and c's orbital parameters from \exostriker mcmc analysis for P$_{\rm d}$ = 5.39 days.}
    \label{fig:5-39d_exostriker_cornerplot}
\end{figure*}

\begin{figure*}
    \centering
    \includegraphics[width=\textwidth]{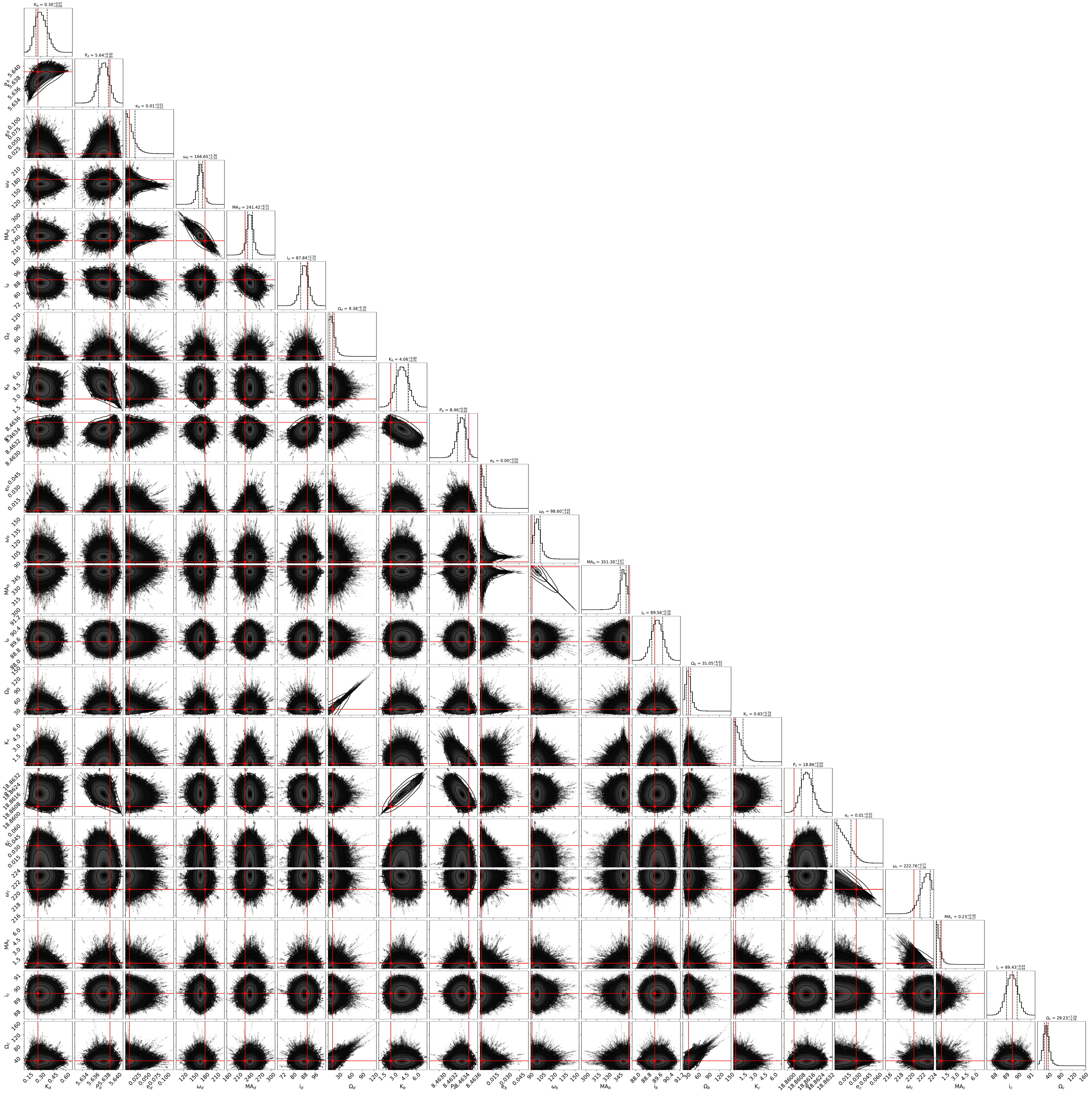}
    \caption{Corner plot of AU Mic b and c's orbital parameters from \exostriker mcmc analysis for P$_{\rm d}$ = 5.64 days.}
    \label{fig:5-64d_exostriker_cornerplot}
\end{figure*}

\begin{figure*}
    \centering
    \includegraphics[width=\textwidth]{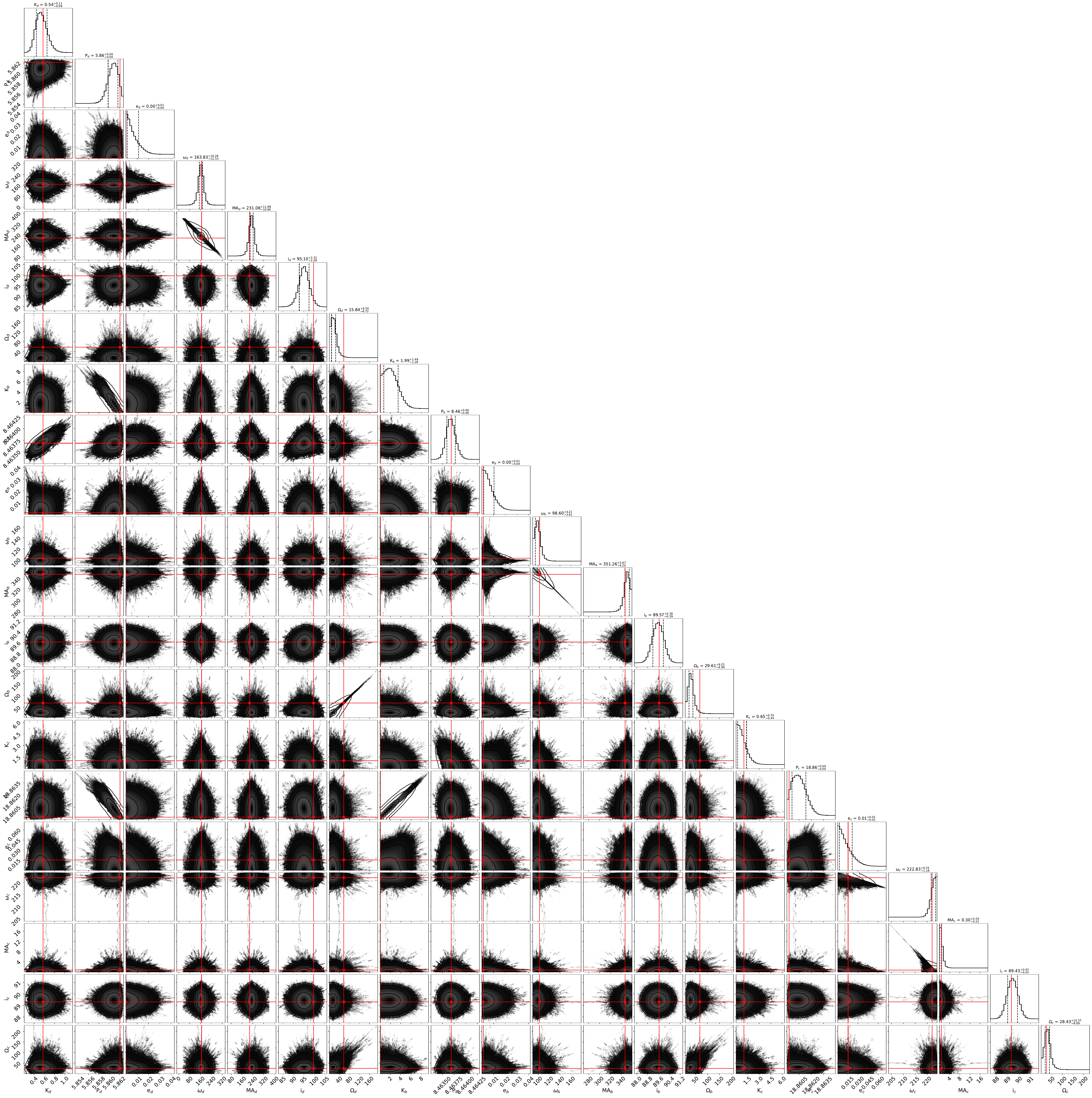}
    \caption{Corner plot of AU Mic b and c's orbital parameters from \exostriker mcmc analysis for P$_{\rm d}$ = 5.86 days.}
    \label{fig:5-86d_exostriker_cornerplot}
\end{figure*}

\begin{figure*}
    \centering
    \includegraphics[width=\textwidth]{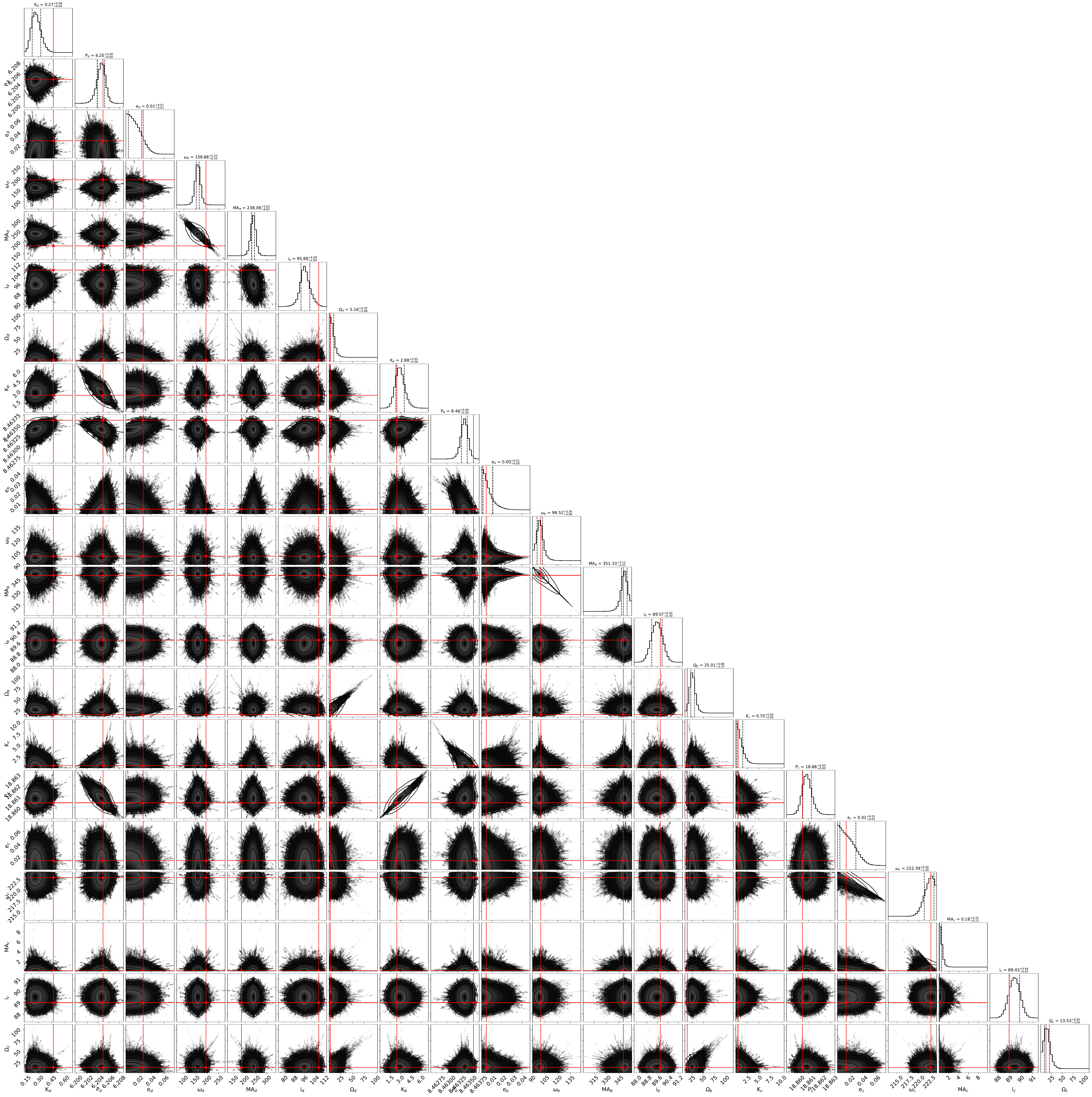}
    \caption{Corner plot of AU Mic b and c's orbital parameters from \exostriker mcmc analysis for P$_{\rm d}$ = 6.20 days.}
    \label{fig:6-20d_exostriker_cornerplot}
\end{figure*}

\begin{figure*}
    \centering
    \includegraphics[width=\textwidth]{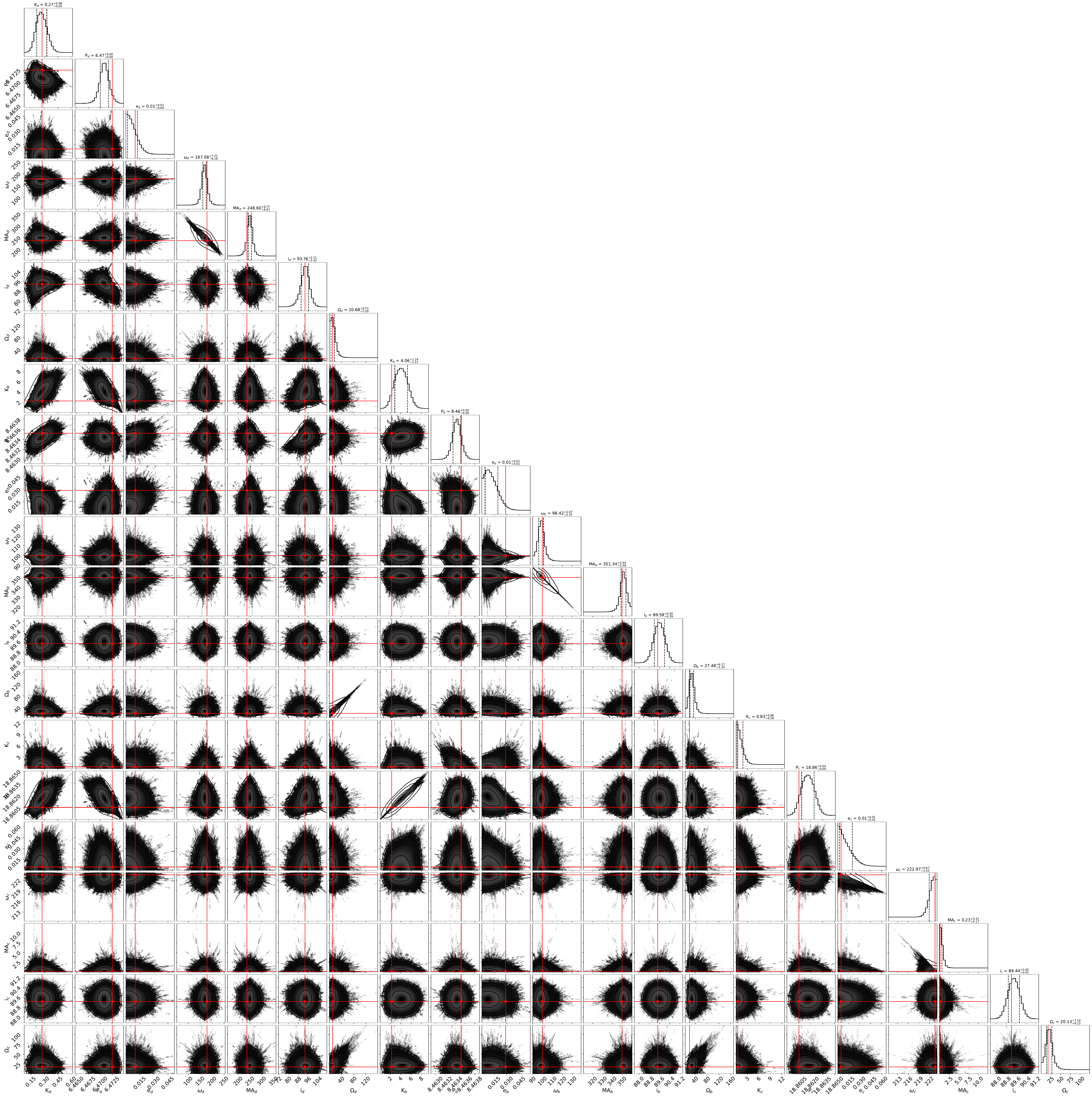}
    \caption{Corner plot of AU Mic b and c's orbital parameters from \exostriker mcmc analysis for P$_{\rm d}$ = 6.47 days.}
    \label{fig:6-47d_exostriker_cornerplot}
\end{figure*}

\begin{figure*}
    \centering
    \includegraphics[width=\textwidth]{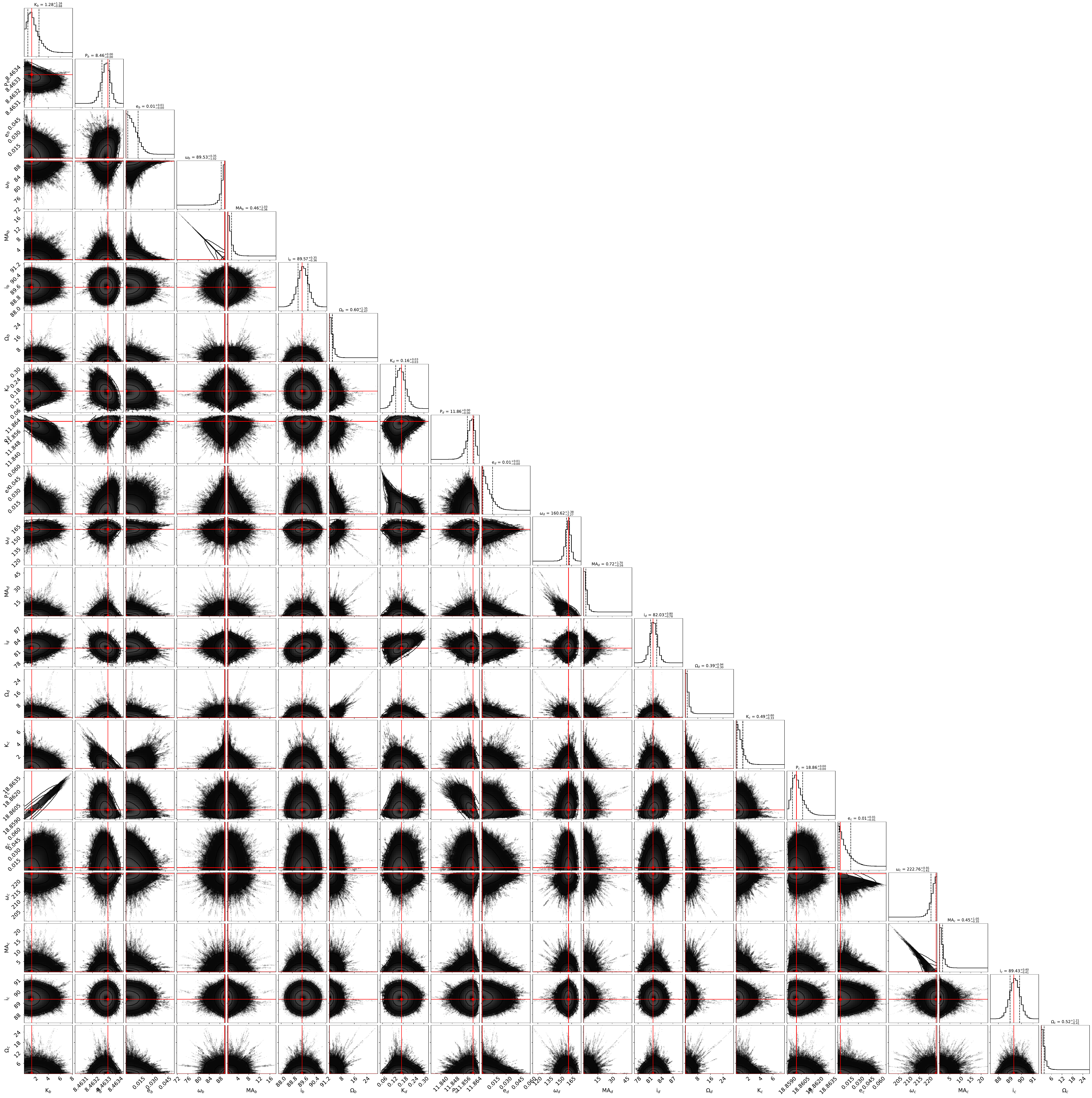}
    \caption{Corner plot of AU Mic b and c's orbital parameters from \exostriker mcmc analysis for P$_{\rm d}$ = 11.9 days.}
    \label{fig:11-9d_exostriker_cornerplot}
\end{figure*}

\begin{figure*}
    \centering
    \includegraphics[width=\textwidth]{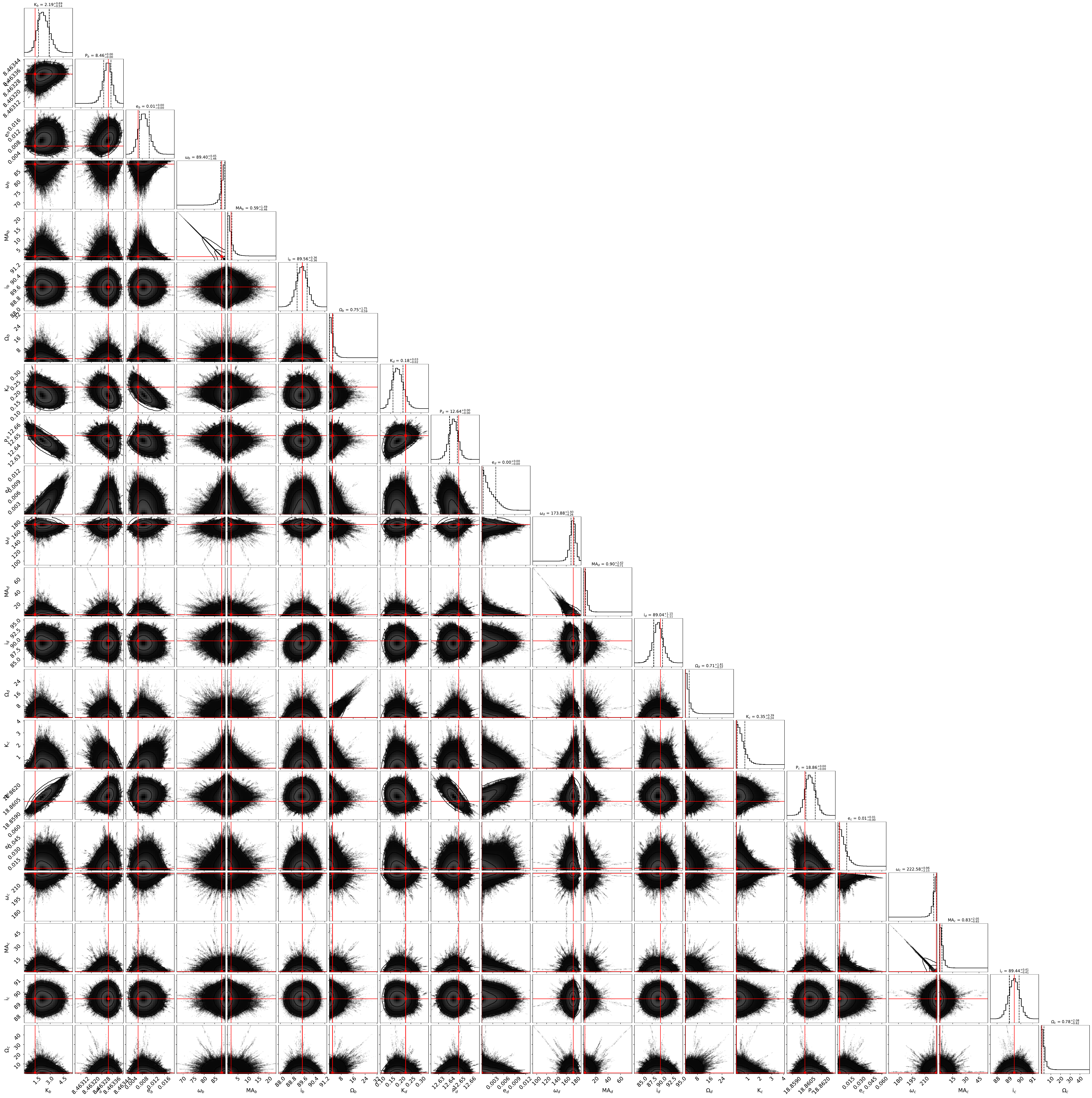}
    \caption{Corner plot of AU Mic b and c's orbital parameters from \exostriker mcmc analysis for P$_{\rm d}$ = 12.6 days.}
    \label{fig:12-6d_exostriker_cornerplot}
\end{figure*}

\begin{figure*}
    \centering
    \includegraphics[width=\textwidth]{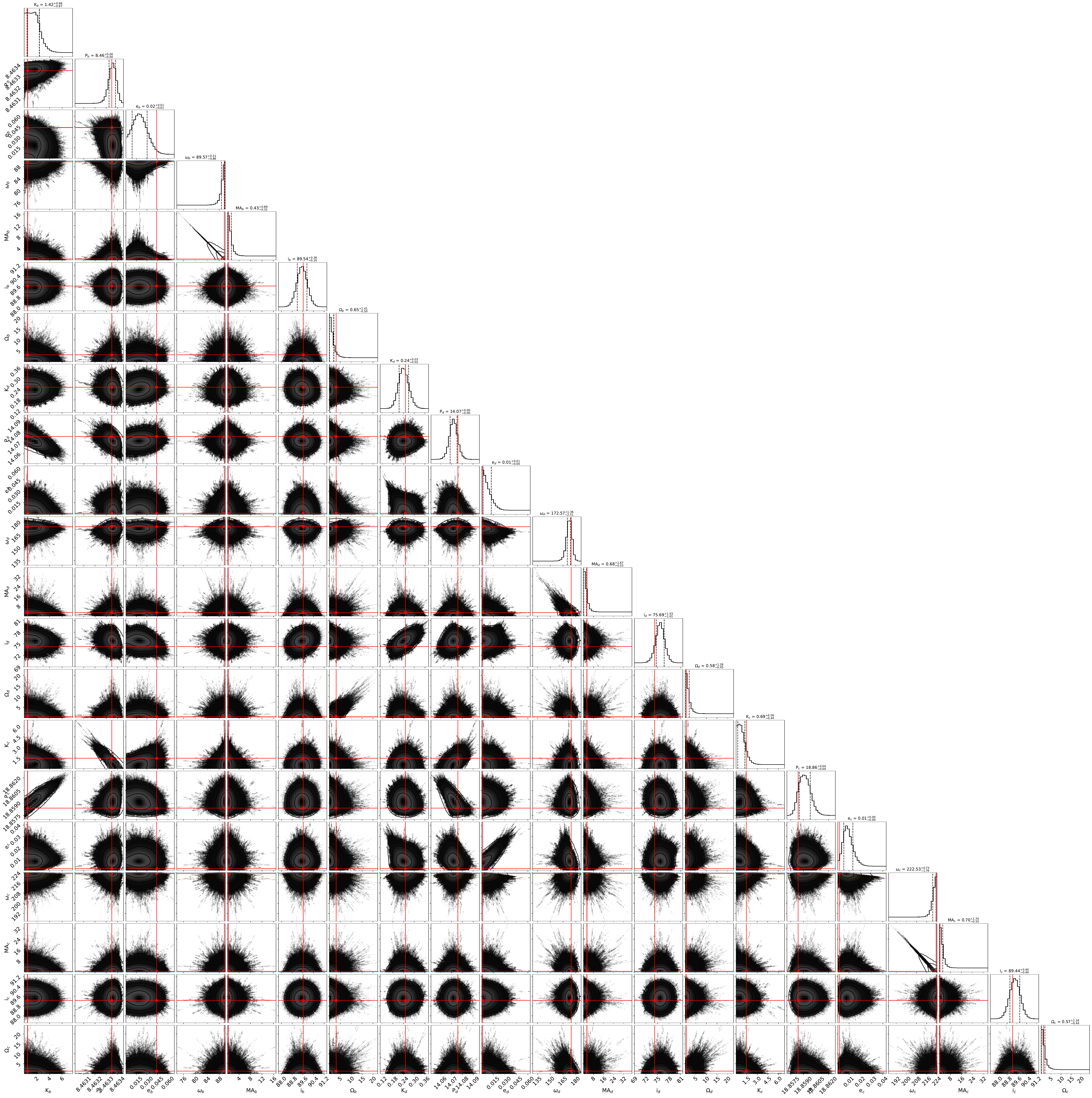}
    \caption{Corner plot of AU Mic b and c's orbital parameters from \exostriker mcmc analysis for P$_{\rm d}$ = 14.1 days.}
    \label{fig:14-1d_exostriker_cornerplot}
\end{figure*}

\begin{figure*}
    \centering
    \includegraphics[width=0.48\textwidth]{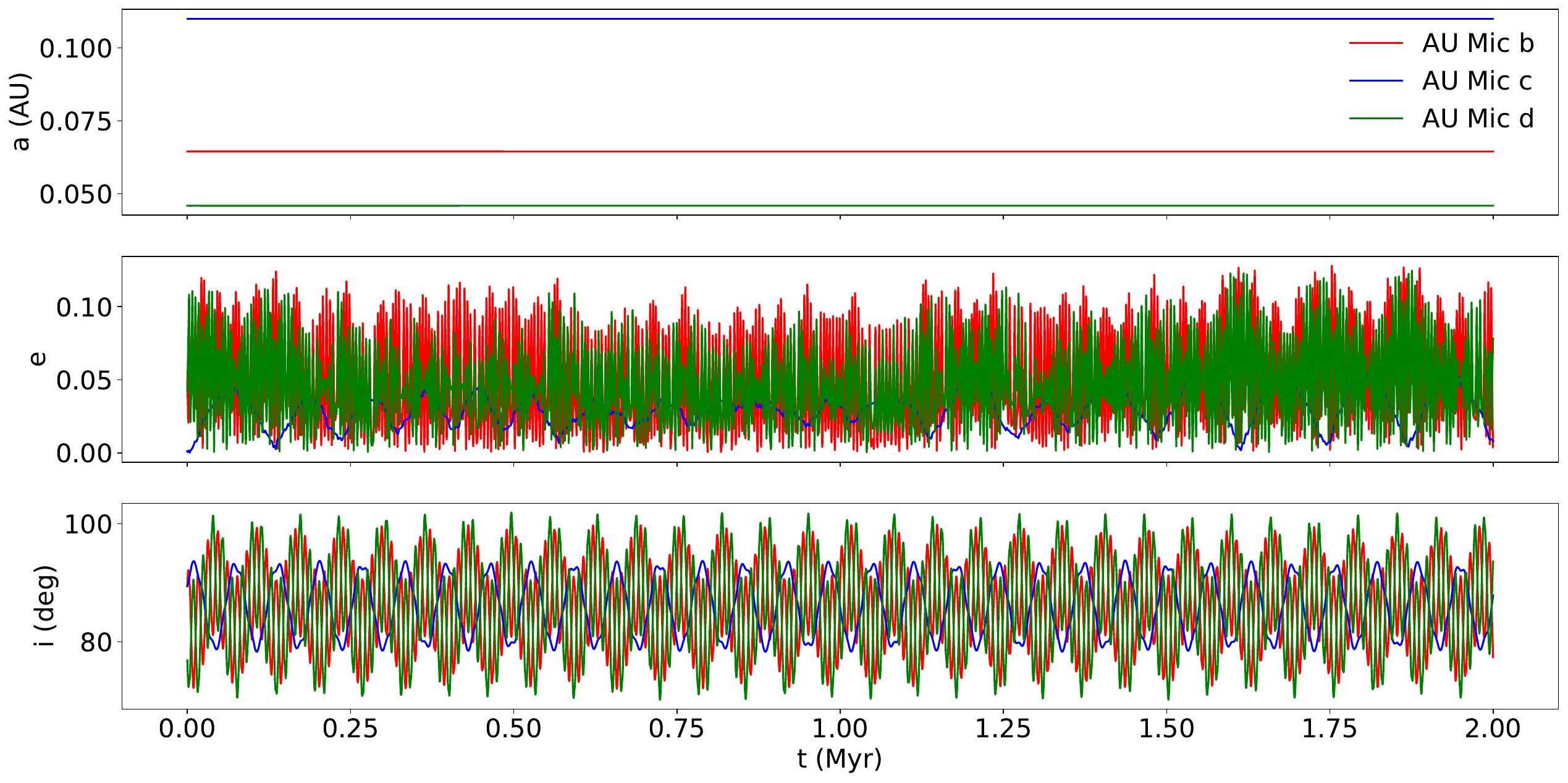}
    \includegraphics[width=0.48\textwidth]{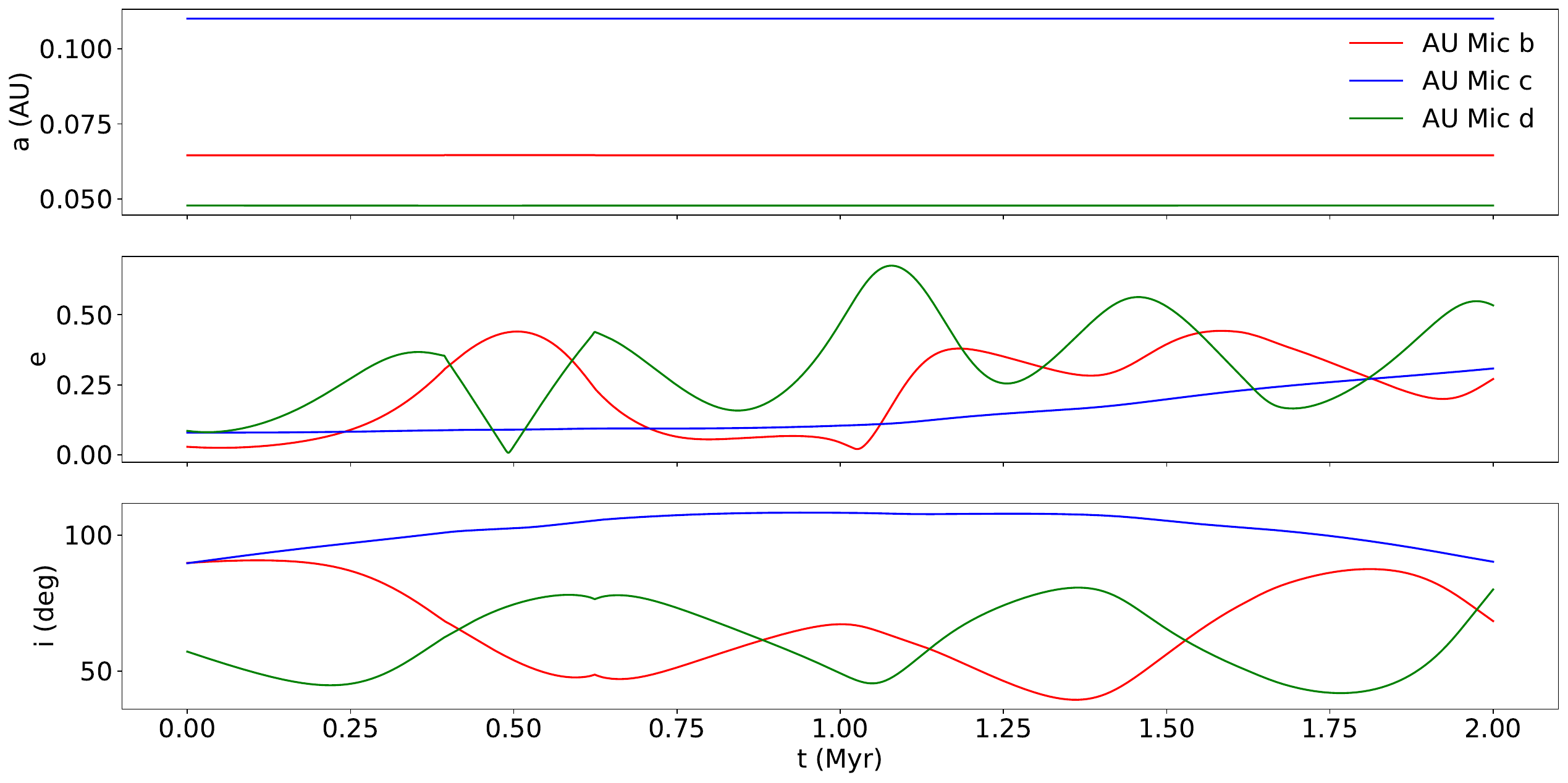}
    \includegraphics[width=0.48\textwidth]{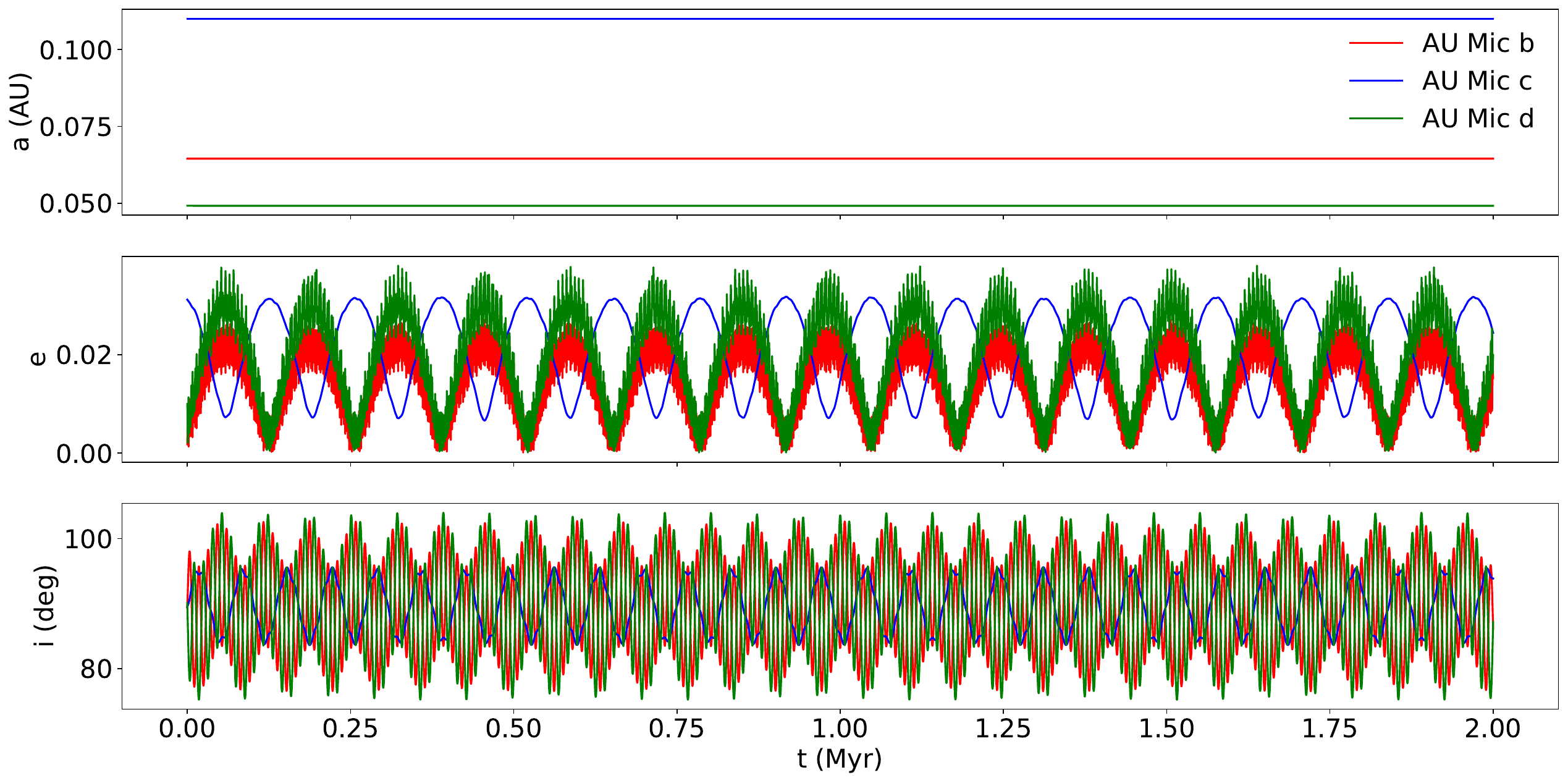}
    \includegraphics[width=0.48\textwidth]{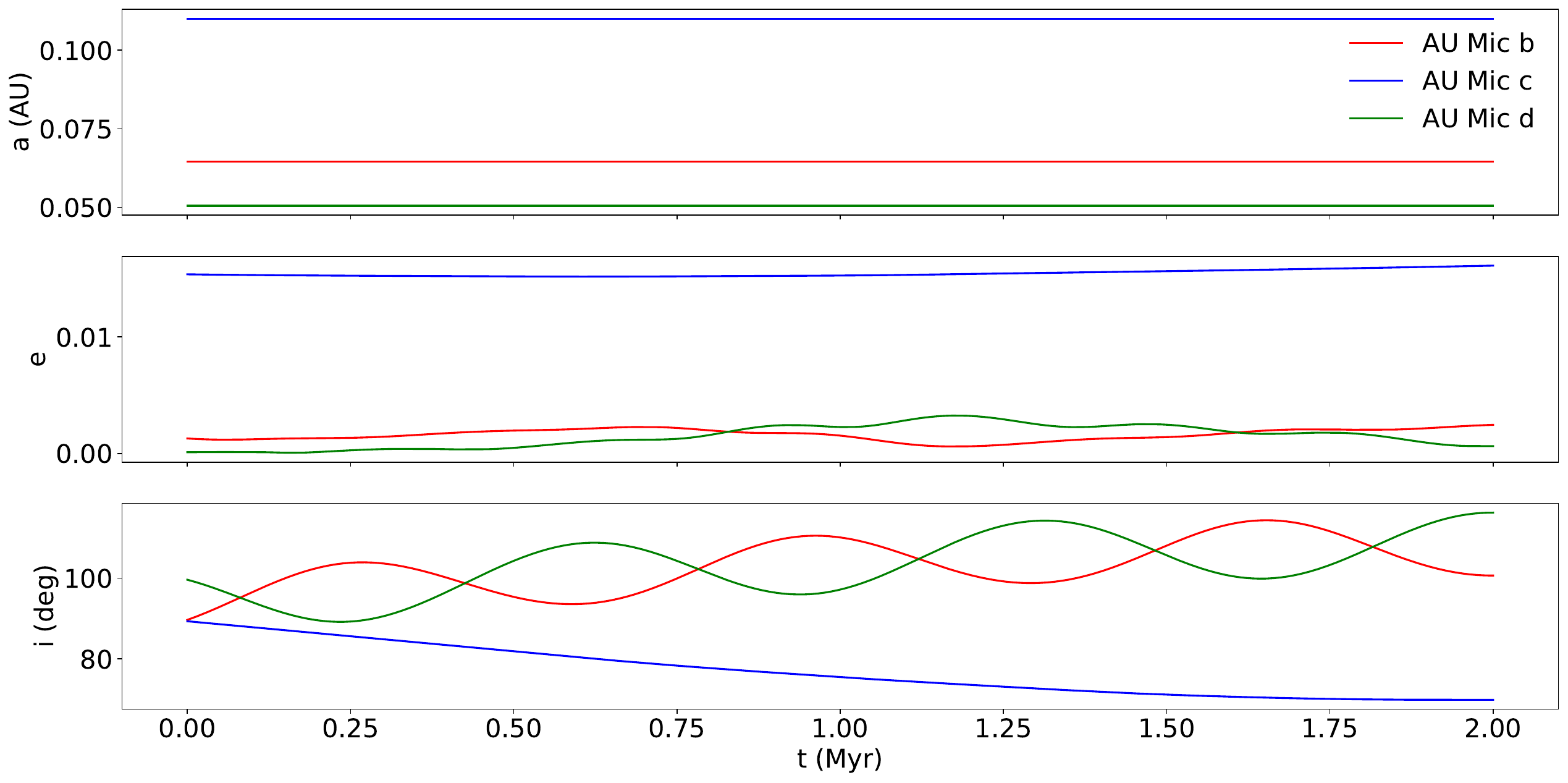}
    \includegraphics[width=0.48\textwidth]{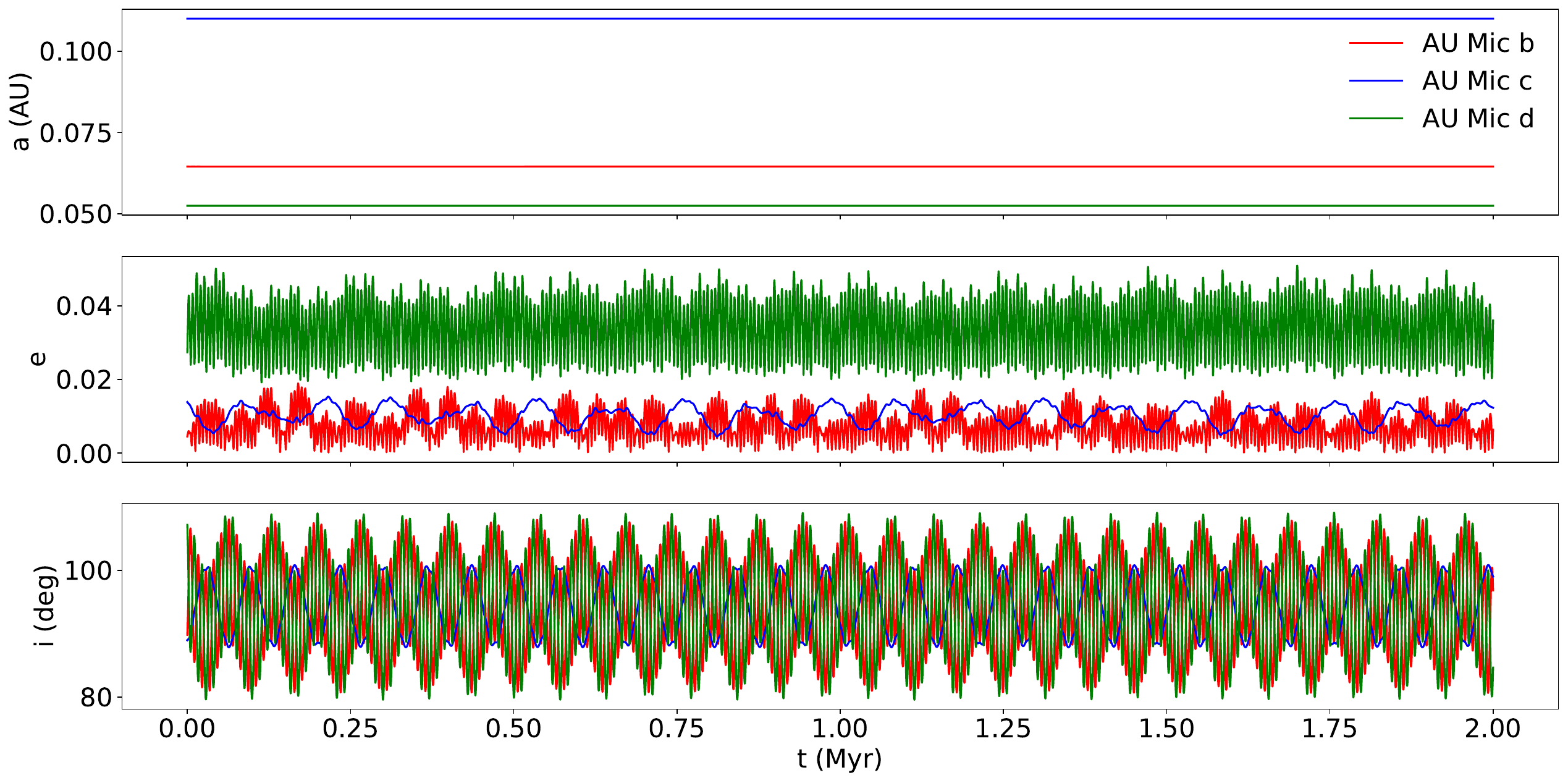}
    \includegraphics[width=0.48\textwidth]{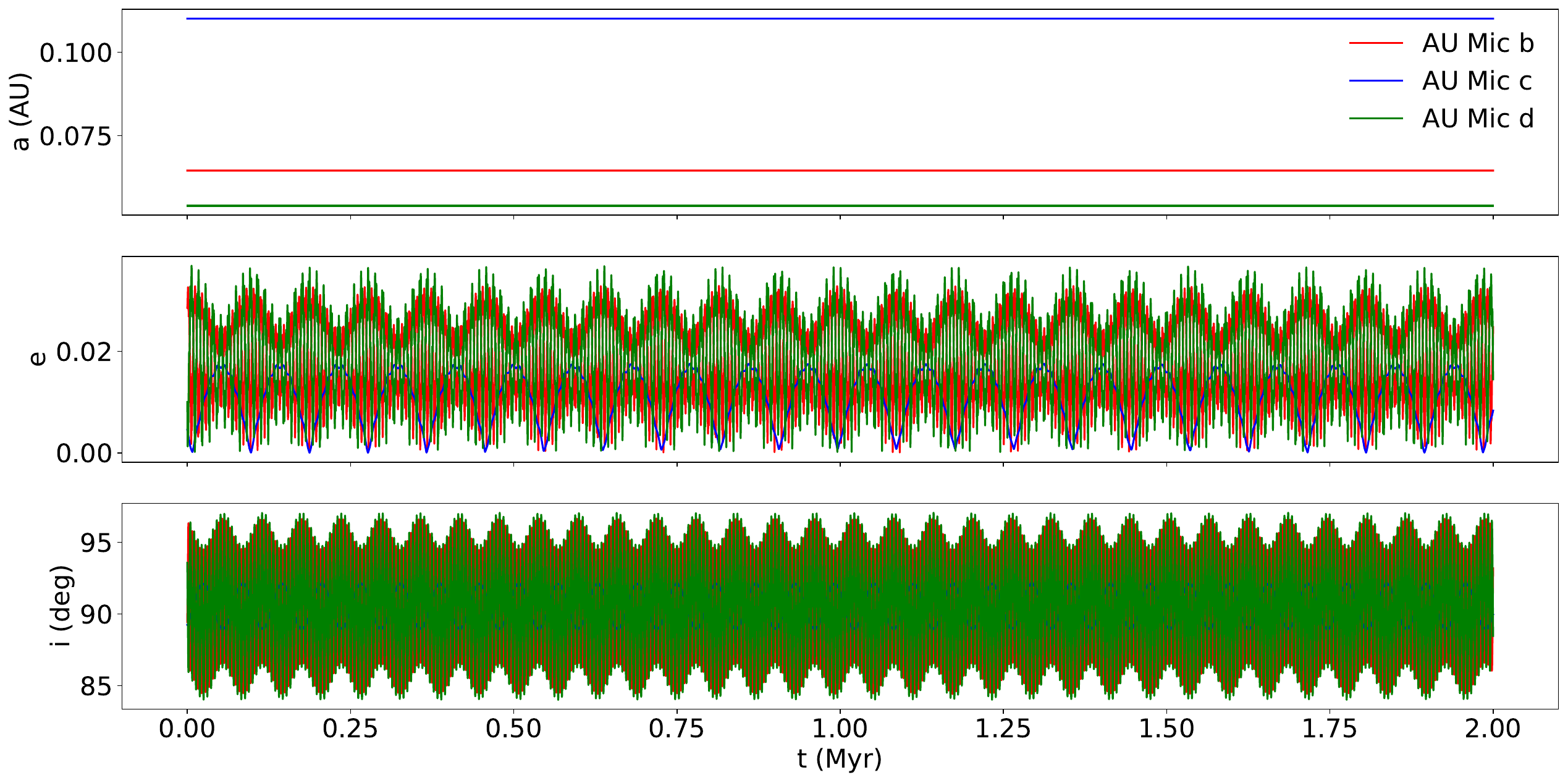}
    \includegraphics[width=0.48\textwidth]{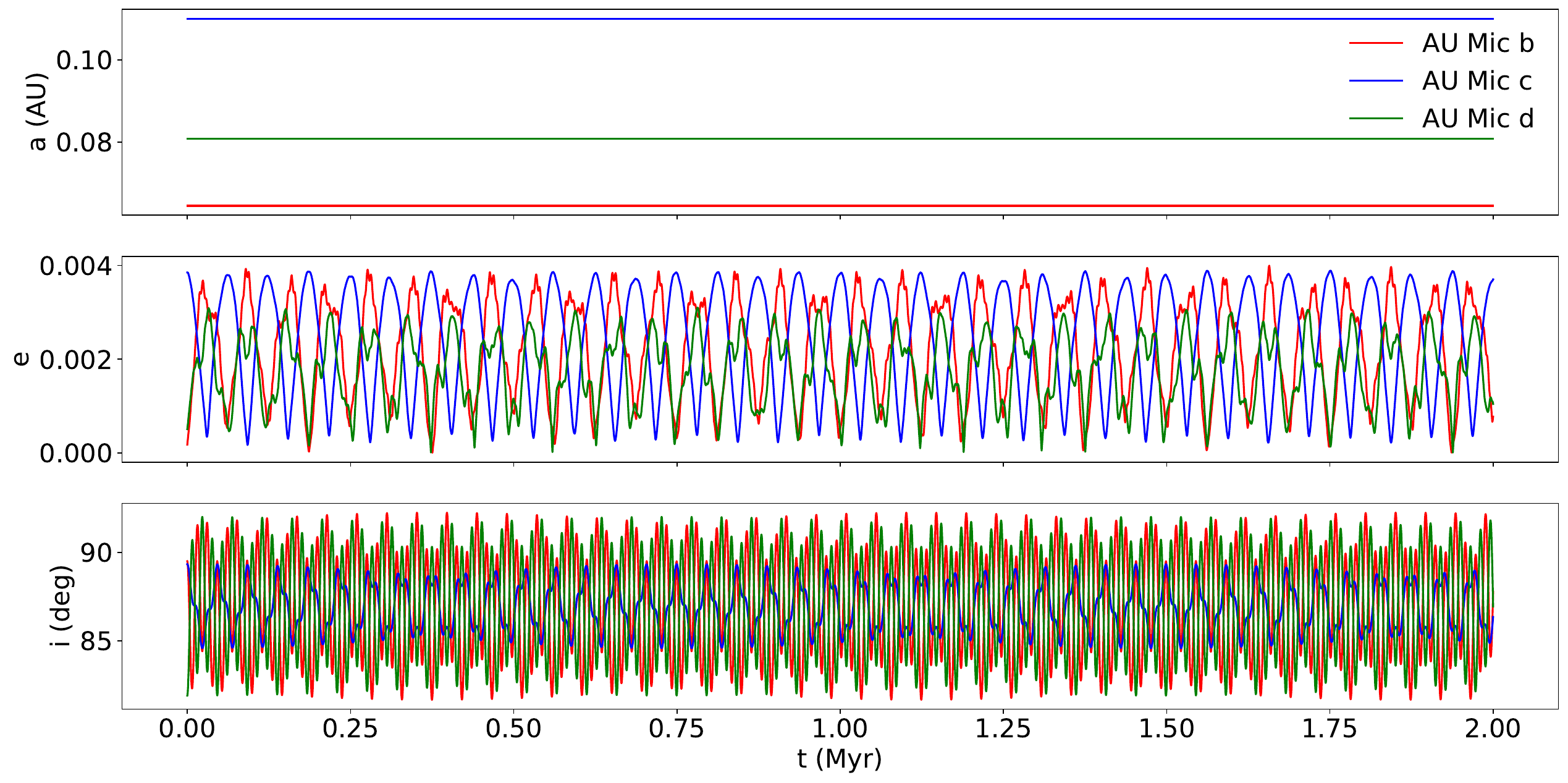}
    \includegraphics[width=0.48\textwidth]{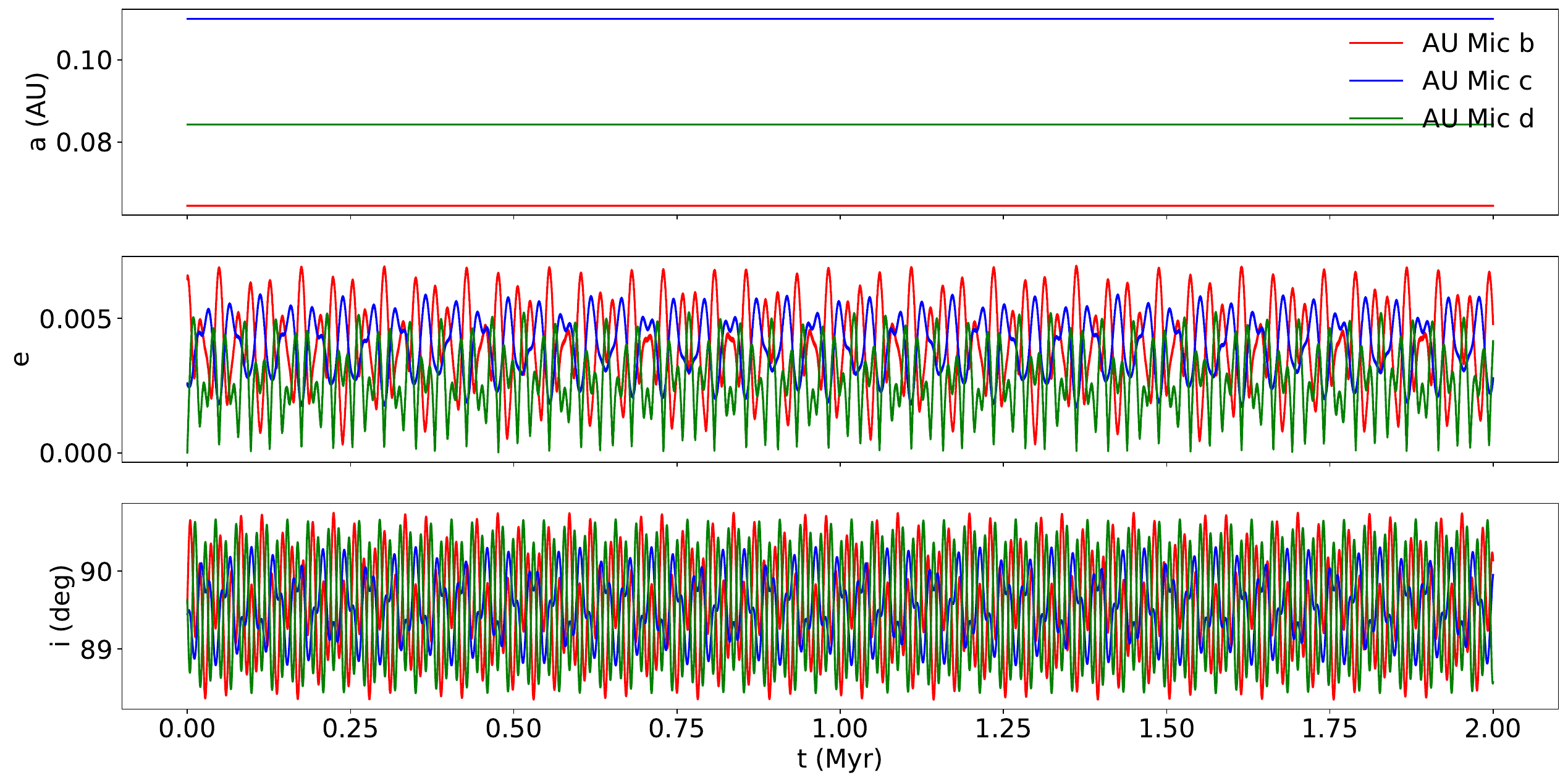}
    \includegraphics[width=0.48\textwidth]{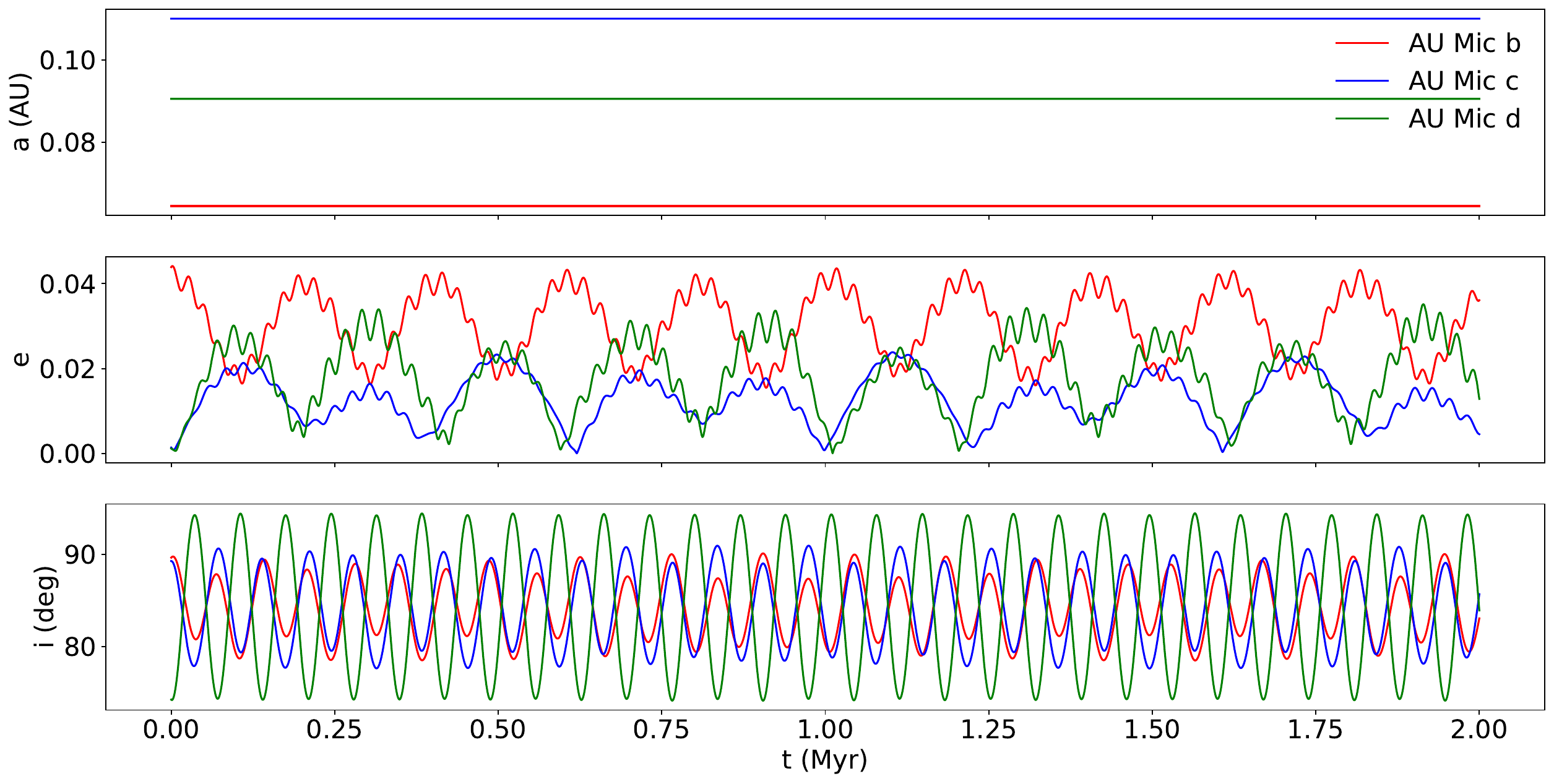}
    \caption{\rebound models of the stability of AU Mic system on timescale of 2 Myr for P$_{\rm d}$ = 5.08 days ({\it first row left}), 5.39 days ({\it first row right}), 5.64 days ({\it second row left}), 5.86 days ({\it second row right}), 6.20 days ({\it third row left}), 6.47 days ({\it third row right}), 11.9 days ({\it fourth row left}), 12.6 days ({\it fourth row right}), and 14.1 days ({\it fifth row}).}
    \label{fig:3p_others_rebound}
\end{figure*}

\begin{figure*}
    \centering
    \includegraphics[width=0.48\textwidth]{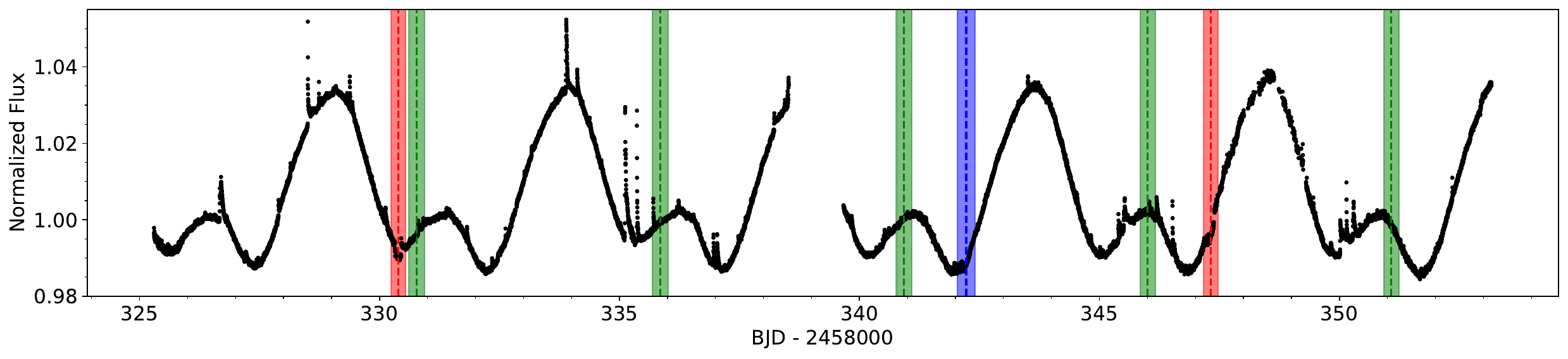}
    \includegraphics[width=0.48\textwidth]{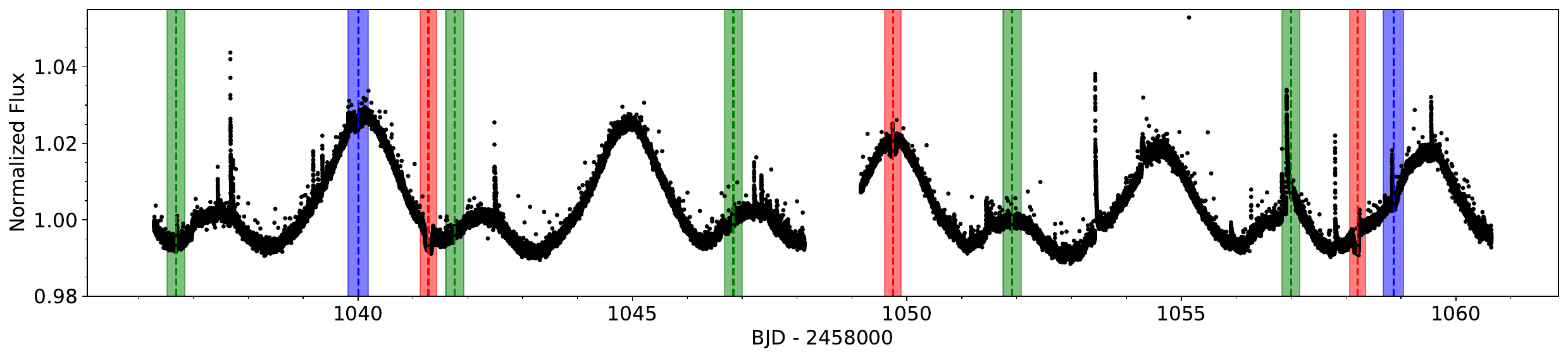}
    \includegraphics[width=0.48\textwidth]{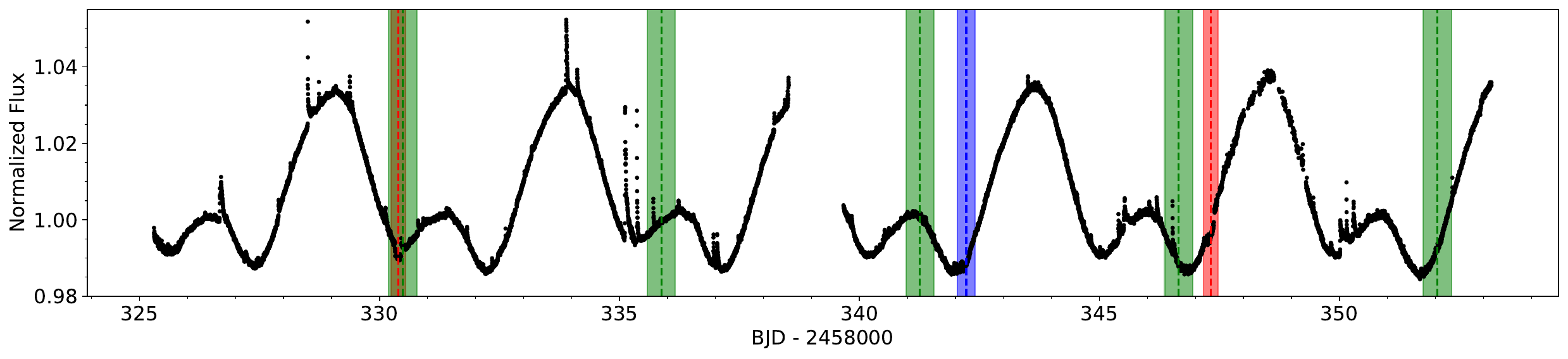}
    \includegraphics[width=0.48\textwidth]{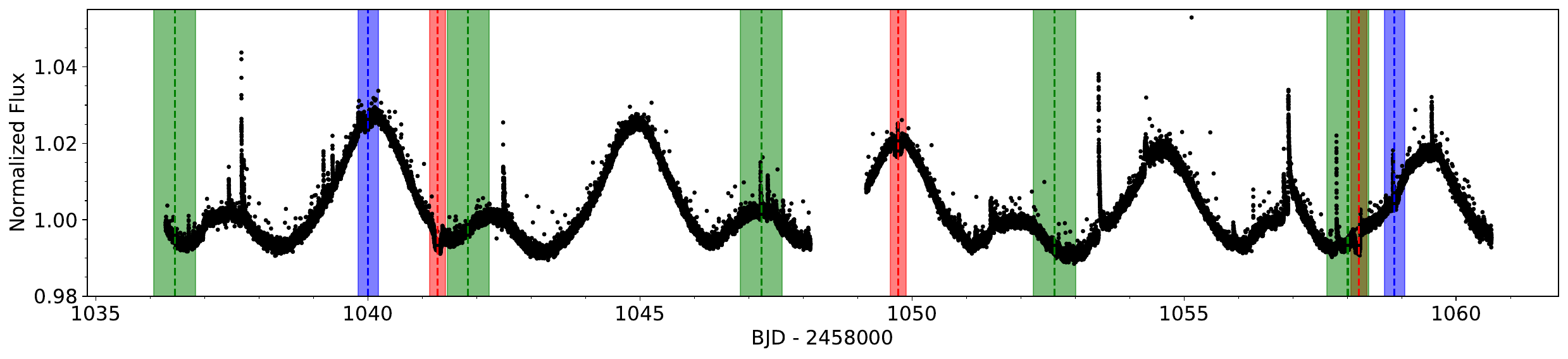}
    \includegraphics[width=0.48\textwidth]{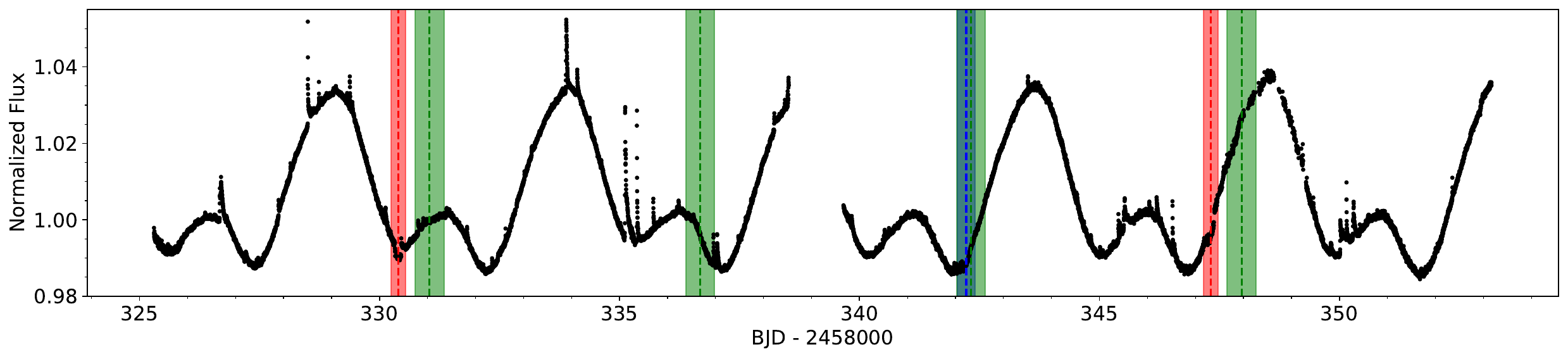}
    \includegraphics[width=0.48\textwidth]{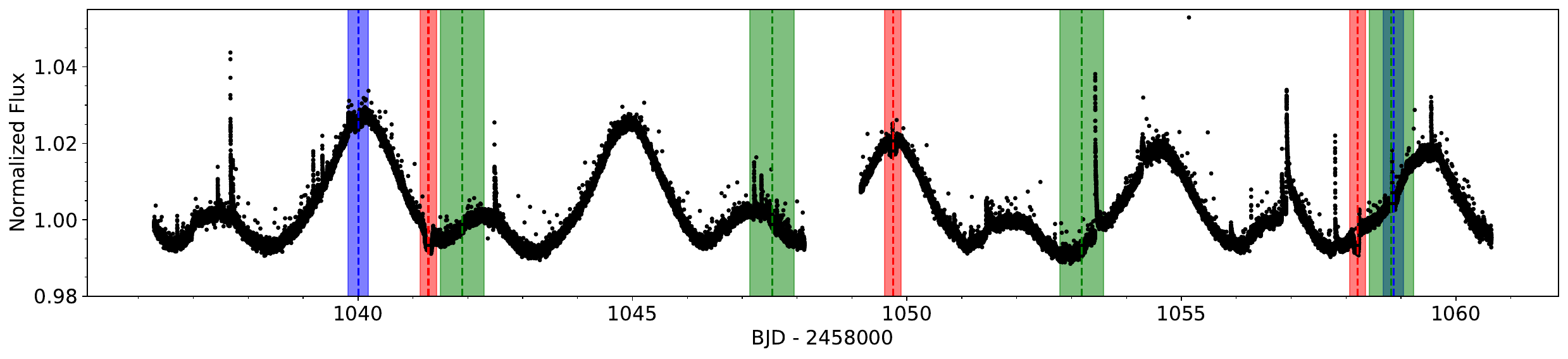}
    \includegraphics[width=0.48\textwidth]{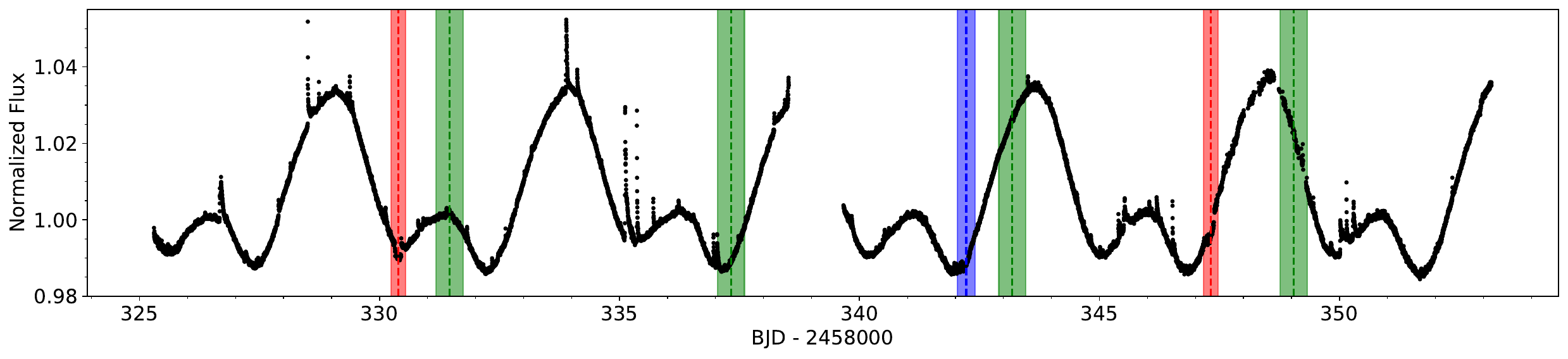}
    \includegraphics[width=0.48\textwidth]{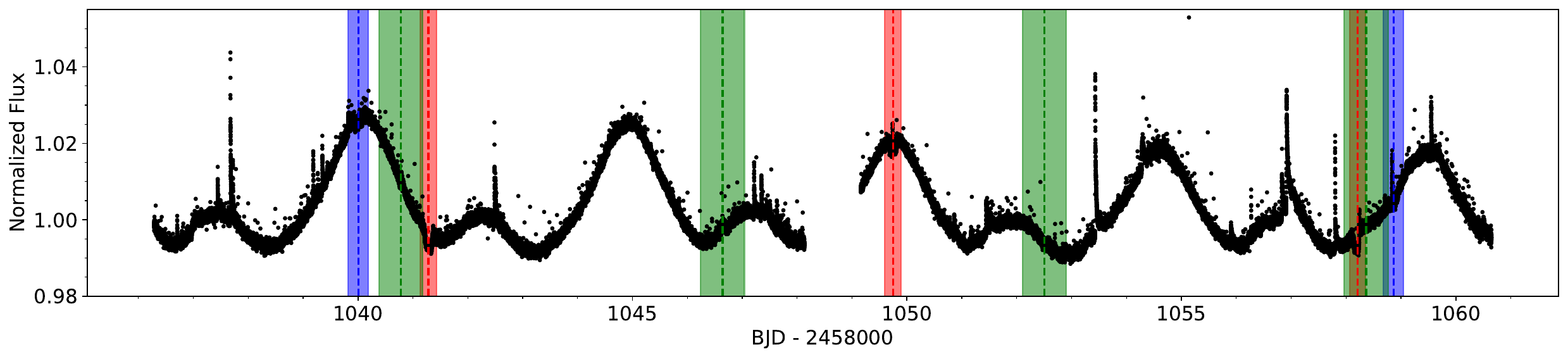}
    \includegraphics[width=0.48\textwidth]{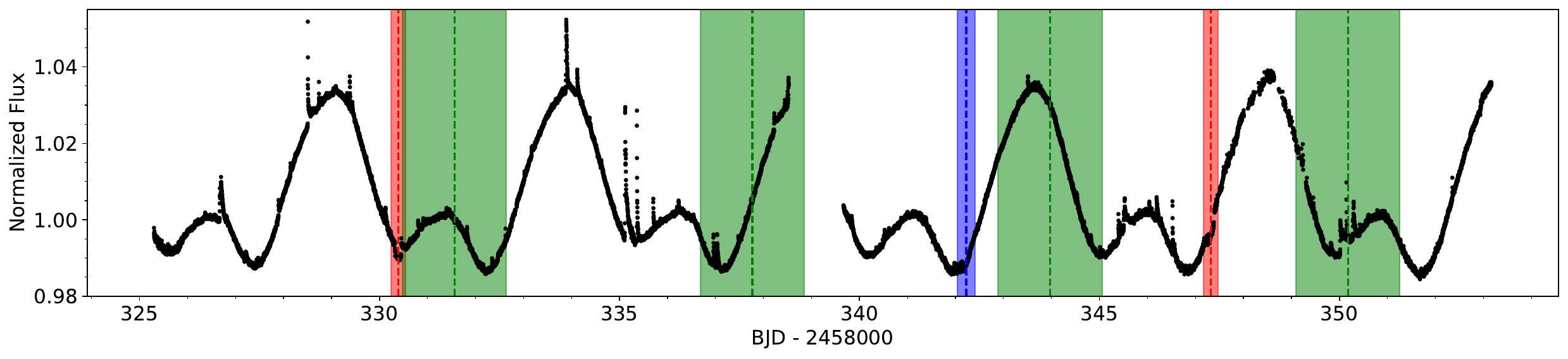}
    \includegraphics[width=0.48\textwidth]{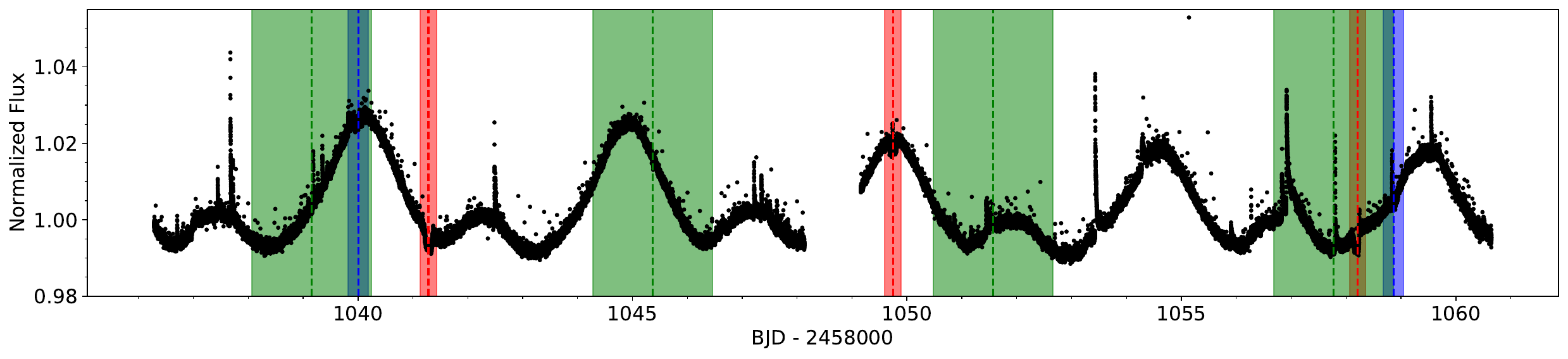}
    \includegraphics[width=0.48\textwidth]{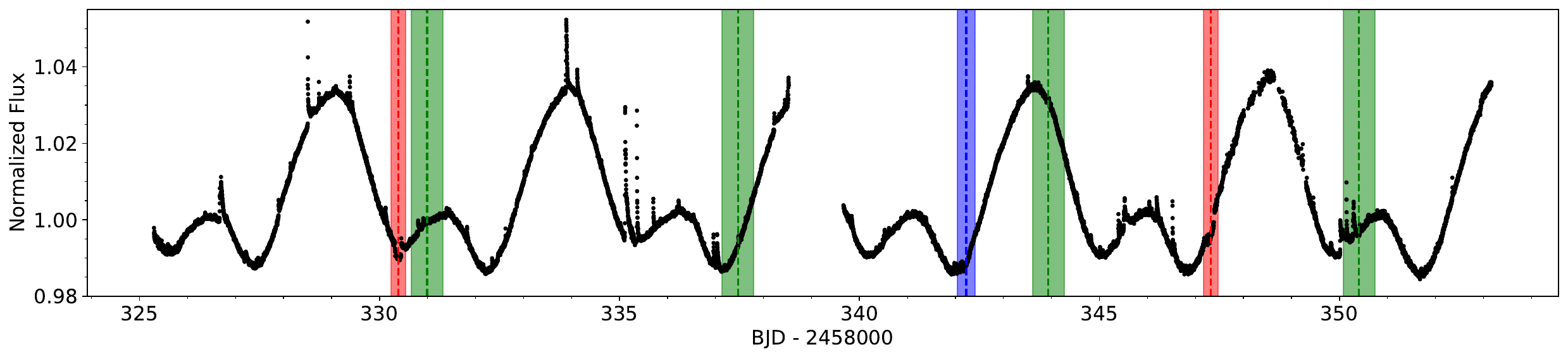}
    \includegraphics[width=0.48\textwidth]{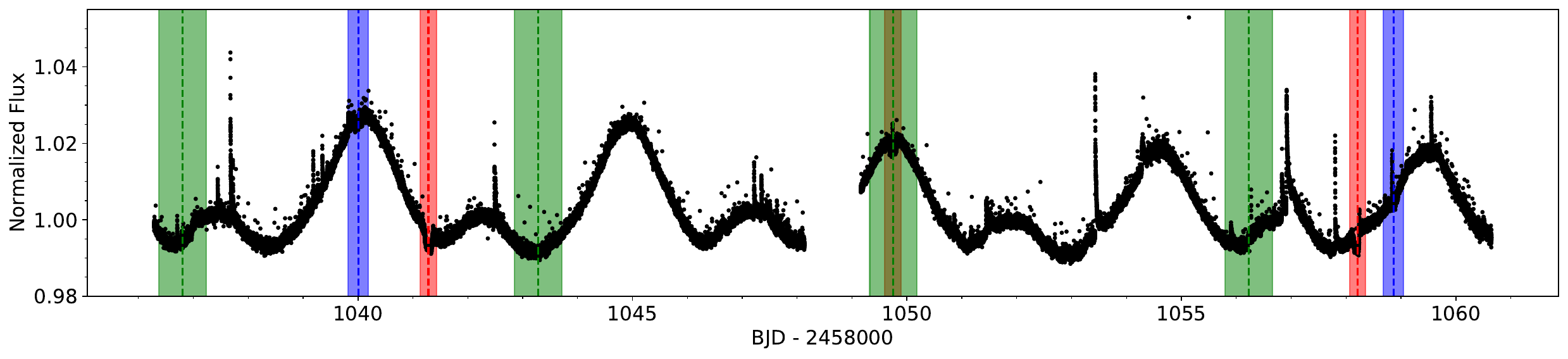}
    \includegraphics[width=0.48\textwidth]{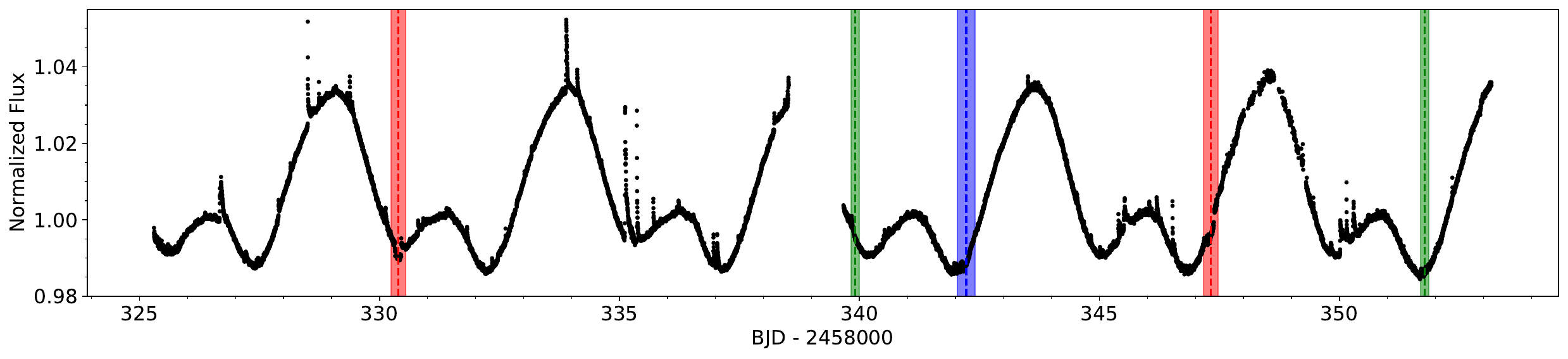}
    \includegraphics[width=0.48\textwidth]{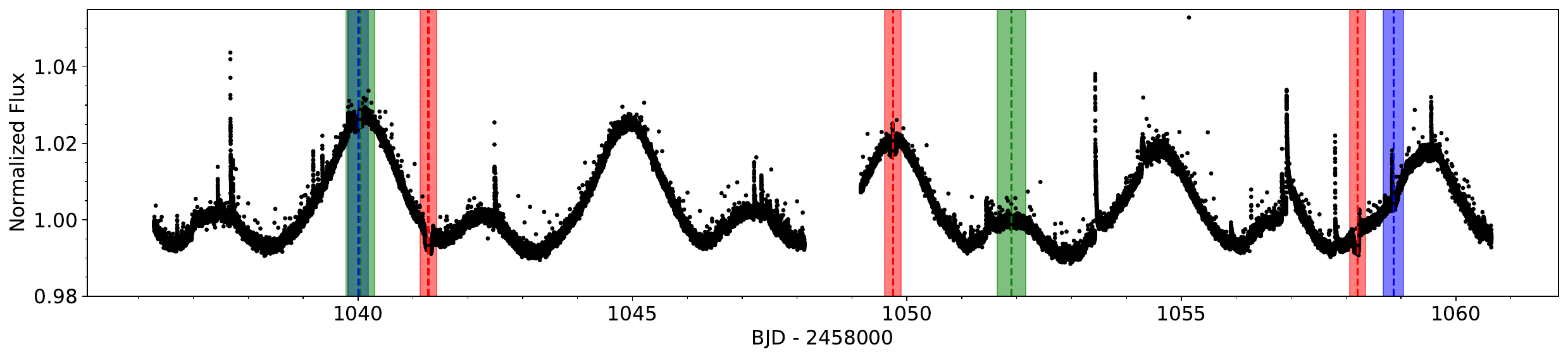}
    \includegraphics[width=0.48\textwidth]{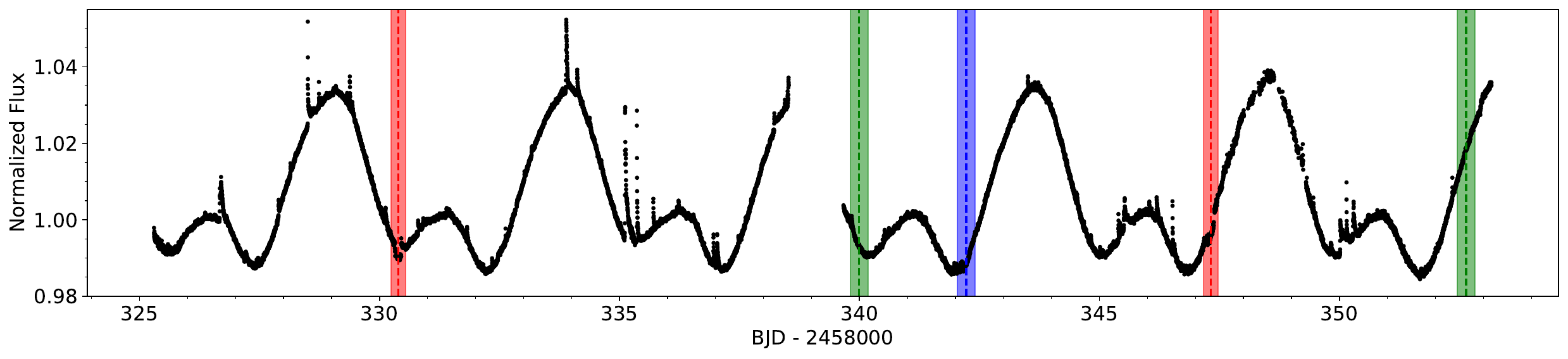}
    \includegraphics[width=0.48\textwidth]{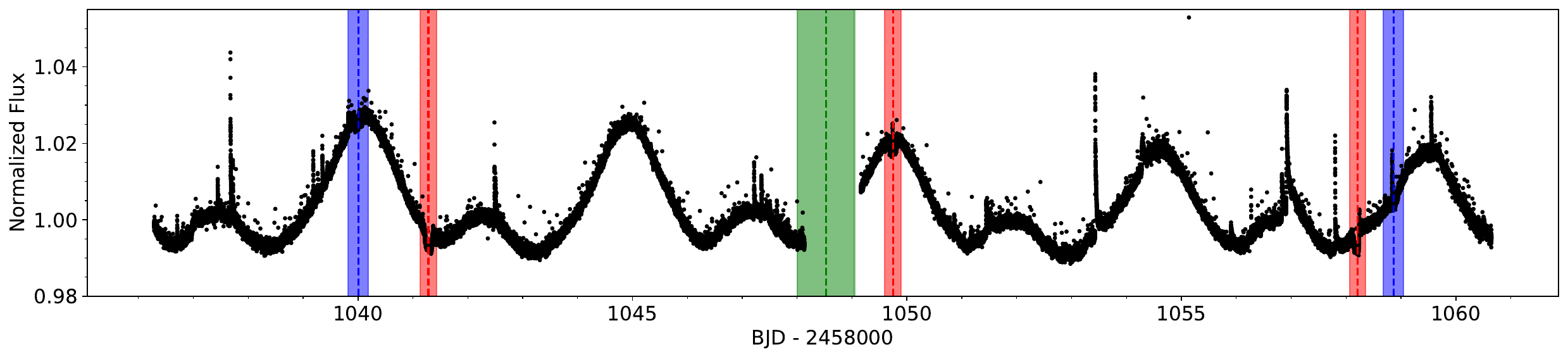}
    \includegraphics[width=0.48\textwidth]{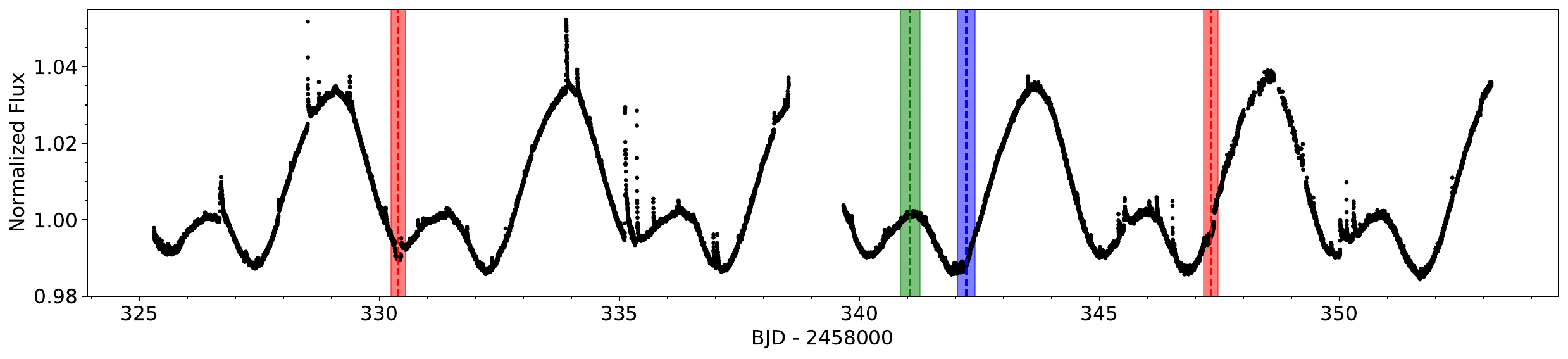}
    \includegraphics[width=0.48\textwidth]{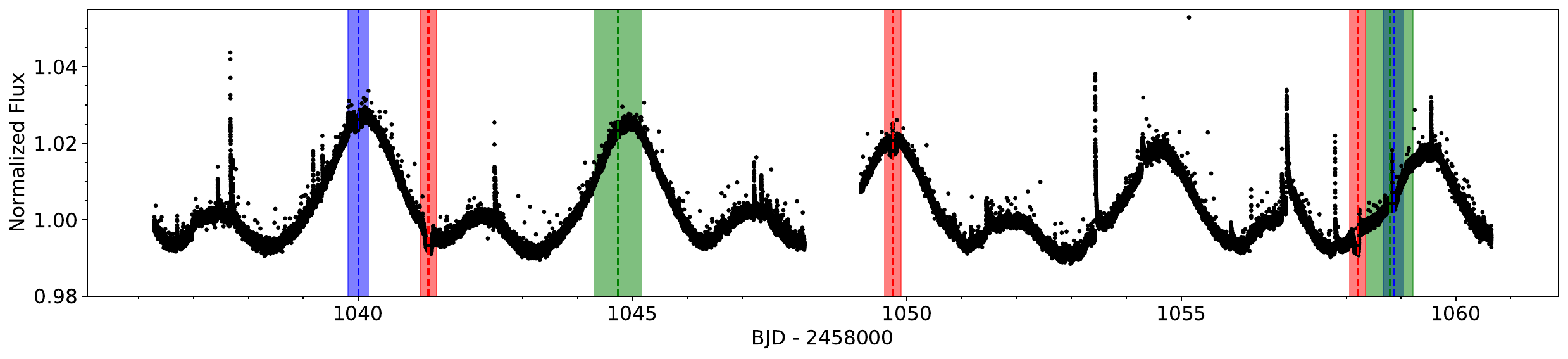}
    \caption{Raw TESS photometry of AU Mic overlaid with observed transits of AU Mic b (red) and c (blue) and range of predicted transit midpoint times for AU Mic d (green), with P$_{d}$ = 5.07 days ({\it first row}), 5.39 days ({\it second row}), 5.64 days ({\it third row}), 5.86 days ({\it fourth row}), 6.20 days ({\it fifth row}), 6.47 days ({\it sixth row}), 11.9 days ({\it seventh row}), 12.6 days ({\it eighth row}), and 14.1 days ({\it ninth row}). Left column is from cycle 1, and right column is from cycle 3. Owing to AU Mic d's estimated 160 ppm depth and AU Mic's intense stellar activity, the potential transit signatures from AU Mic d is not obvious in these plots.}
    \label{fig:d_transittimes_2}
\end{figure*}

\begin{figure*}
    \centering
    \includegraphics[width=0.43\textwidth]{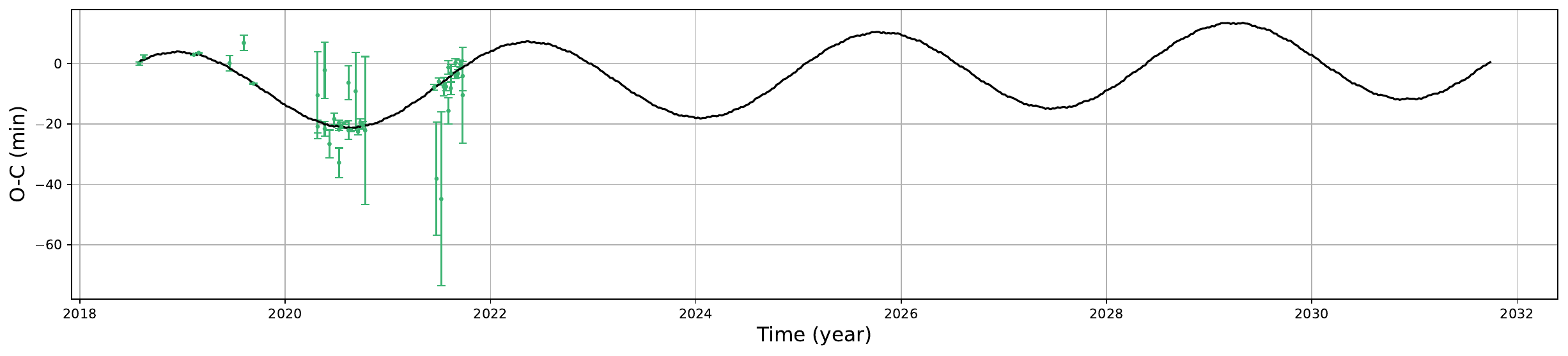}
    \includegraphics[width=0.43\textwidth]{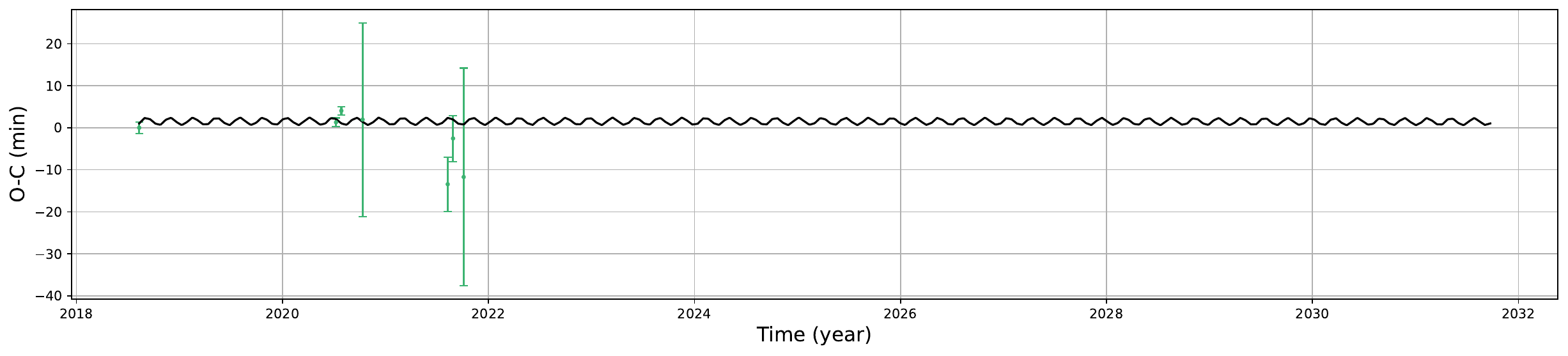}
    \includegraphics[width=0.43\textwidth]{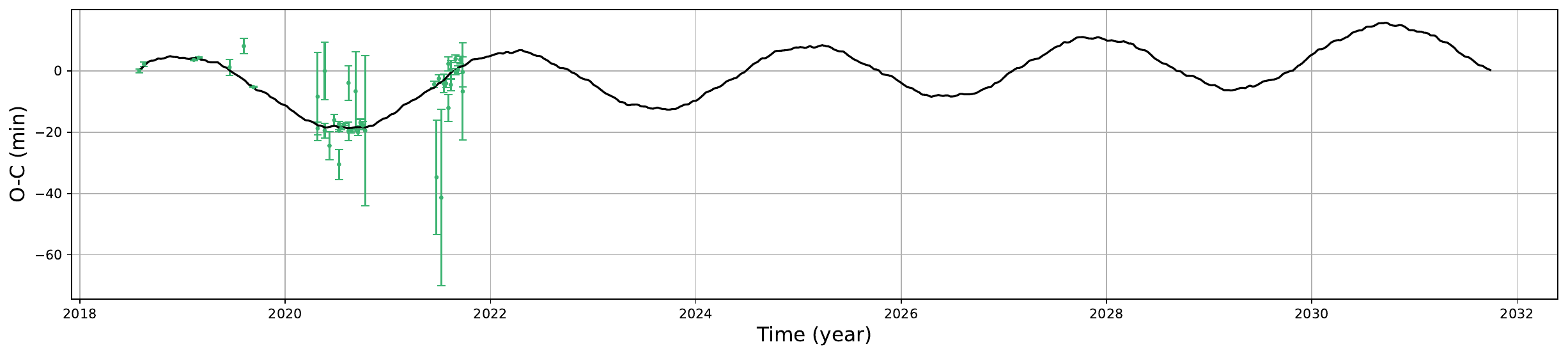}
    \includegraphics[width=0.43\textwidth]{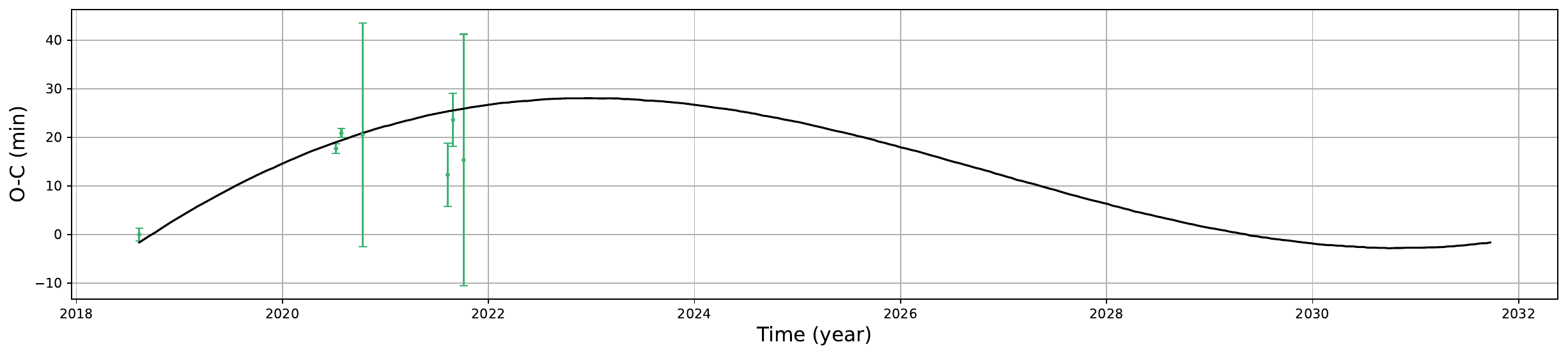}
    \includegraphics[width=0.43\textwidth]{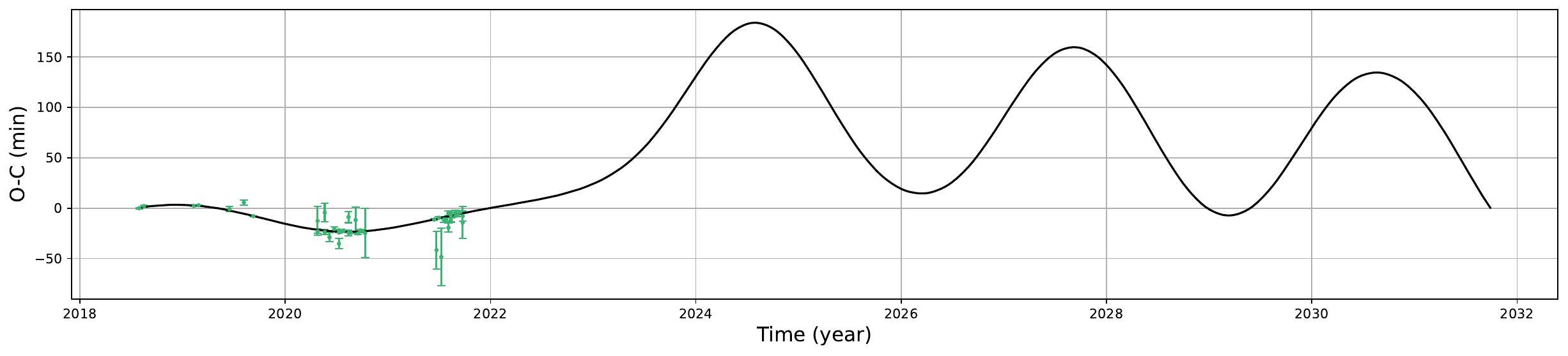}
    \includegraphics[width=0.43\textwidth]{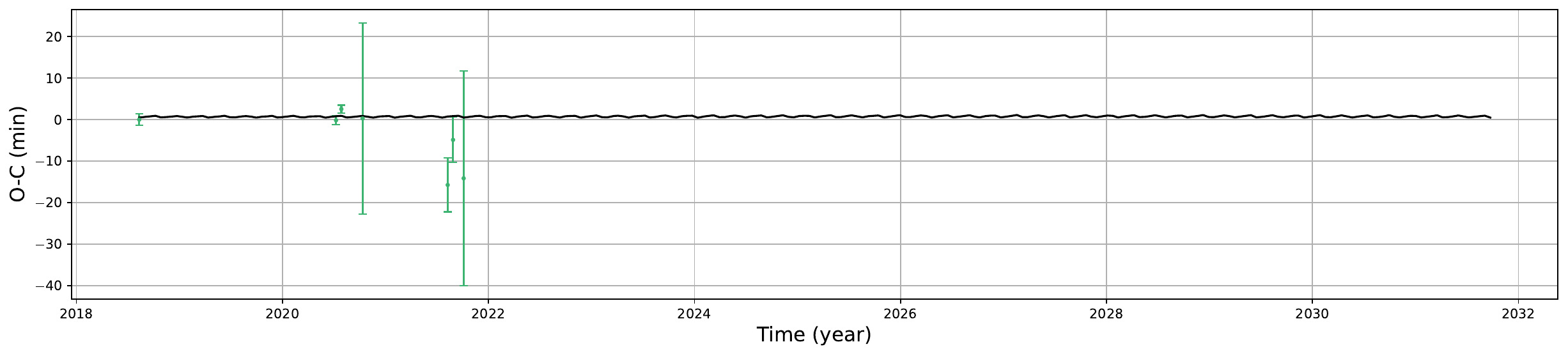}
    \includegraphics[width=0.43\textwidth]{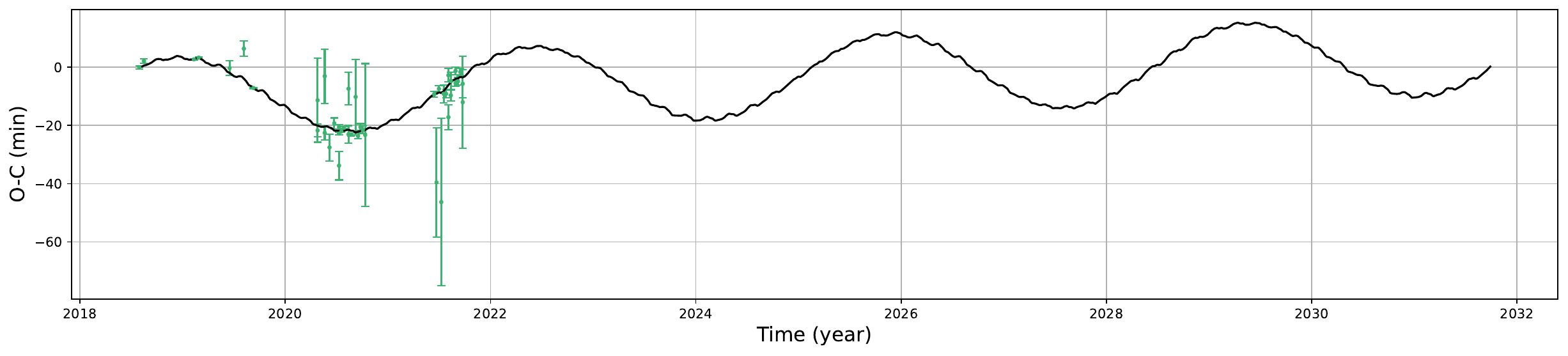}
    \includegraphics[width=0.43\textwidth]{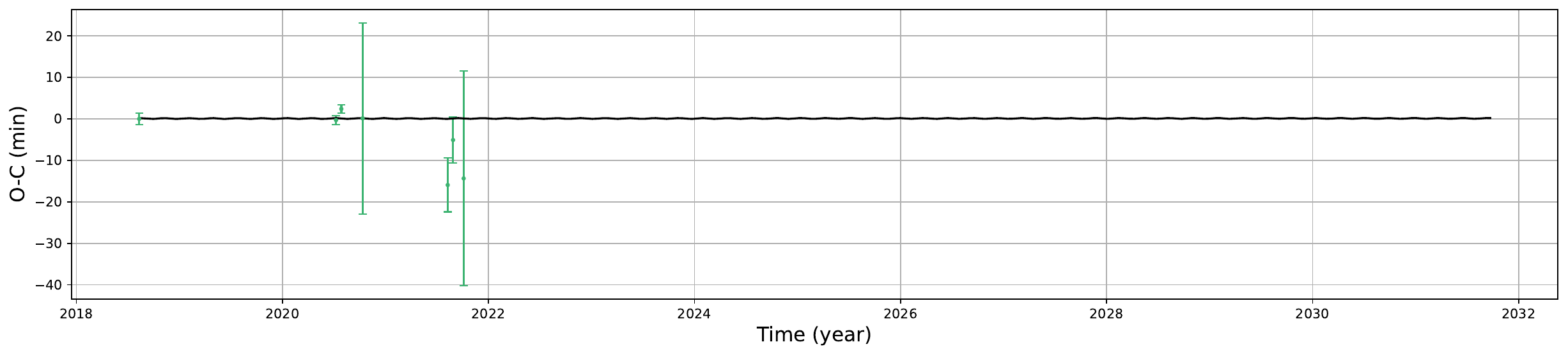}
    \includegraphics[width=0.43\textwidth]{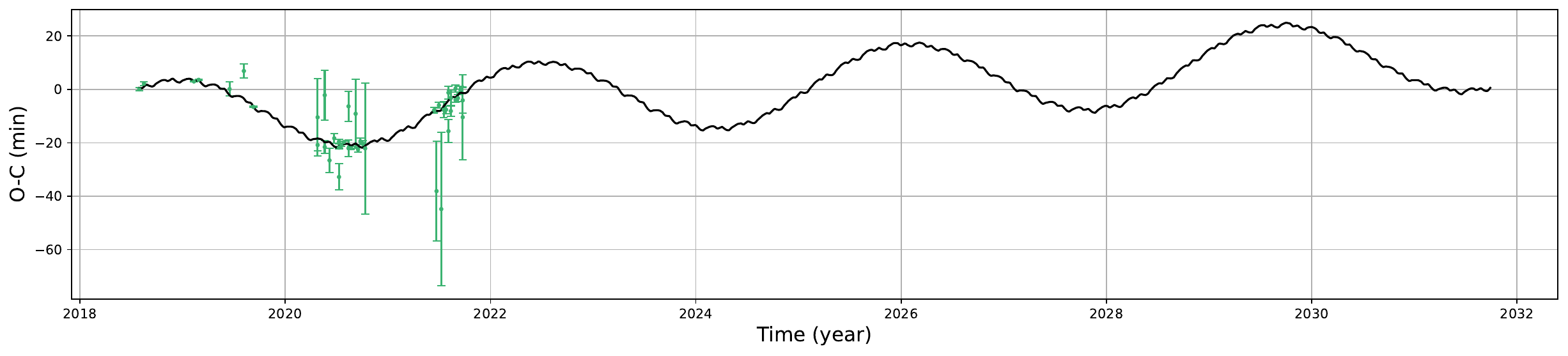}
    \includegraphics[width=0.43\textwidth]{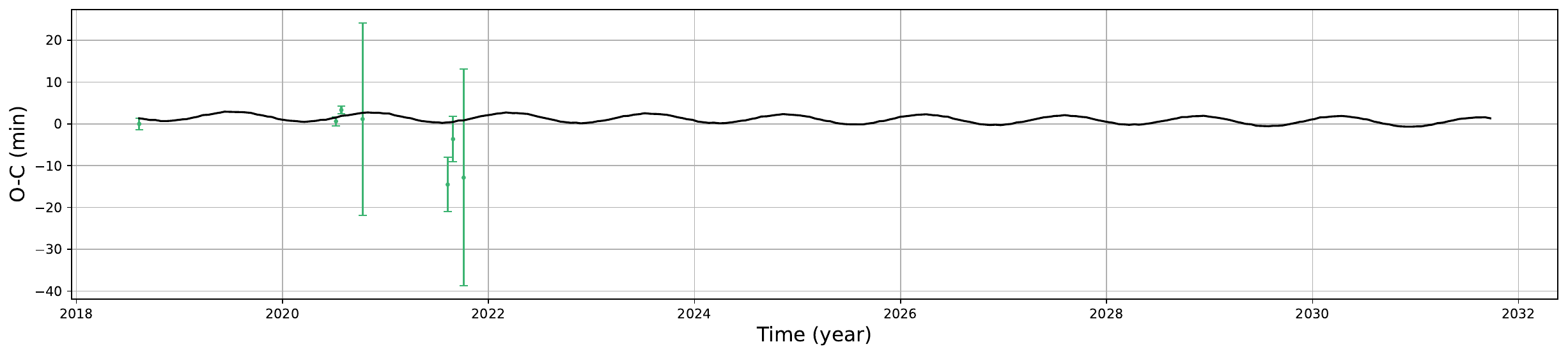}
    \includegraphics[width=0.43\textwidth]{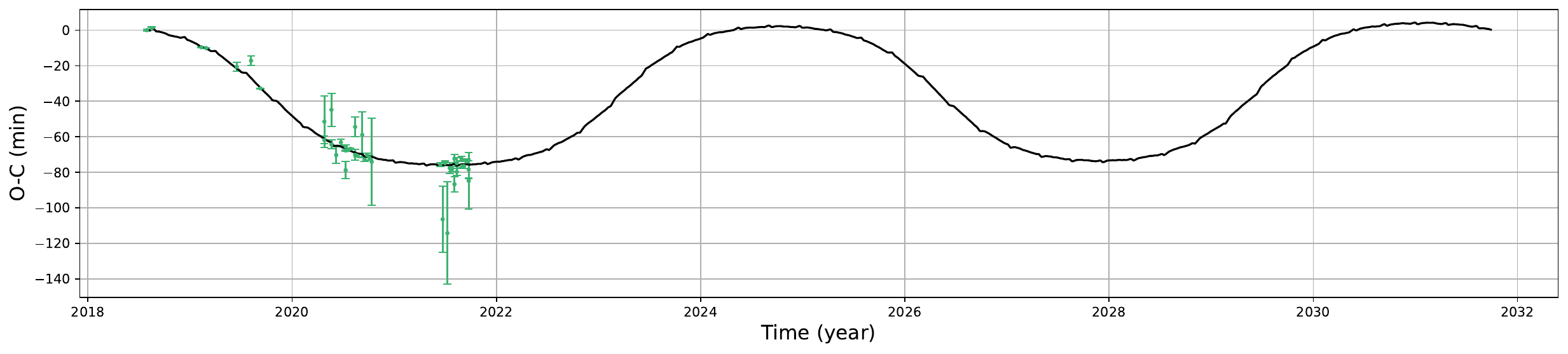}
    \includegraphics[width=0.43\textwidth]{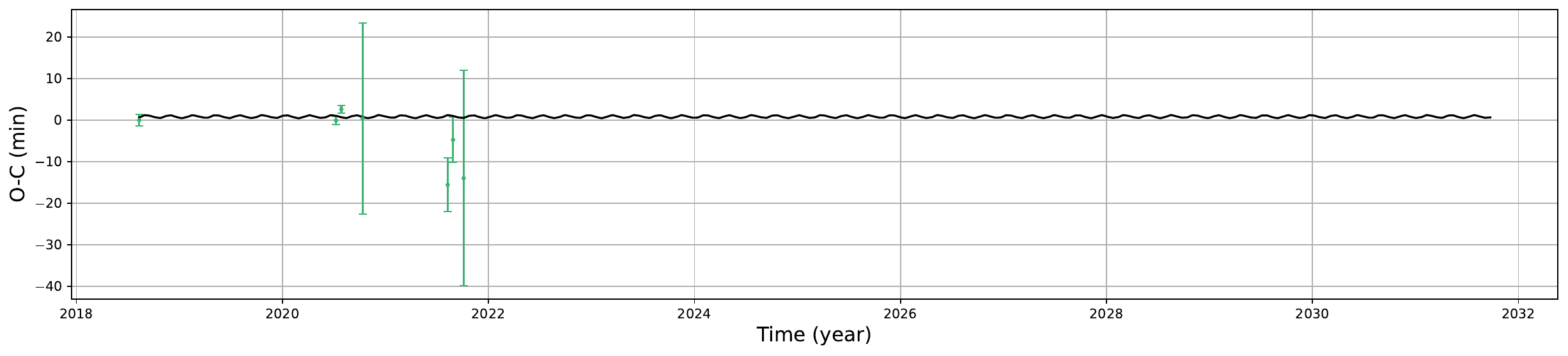}
    \includegraphics[width=0.43\textwidth]{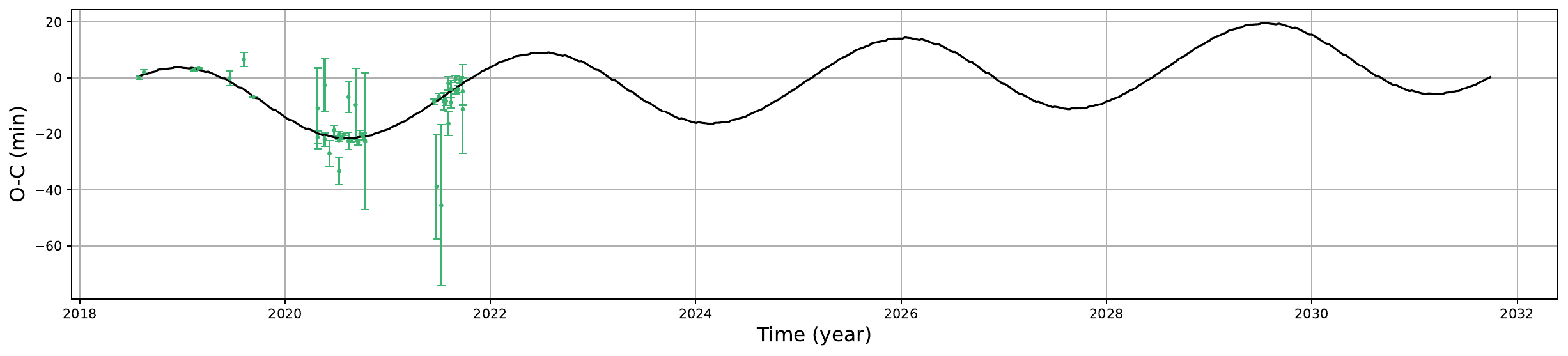}
    \includegraphics[width=0.43\textwidth]{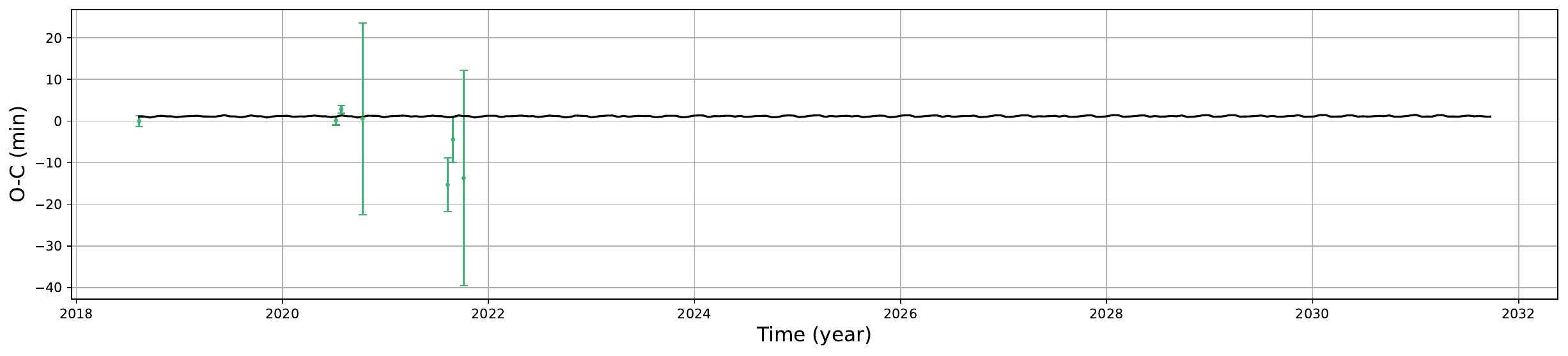}
    \includegraphics[width=0.43\textwidth]{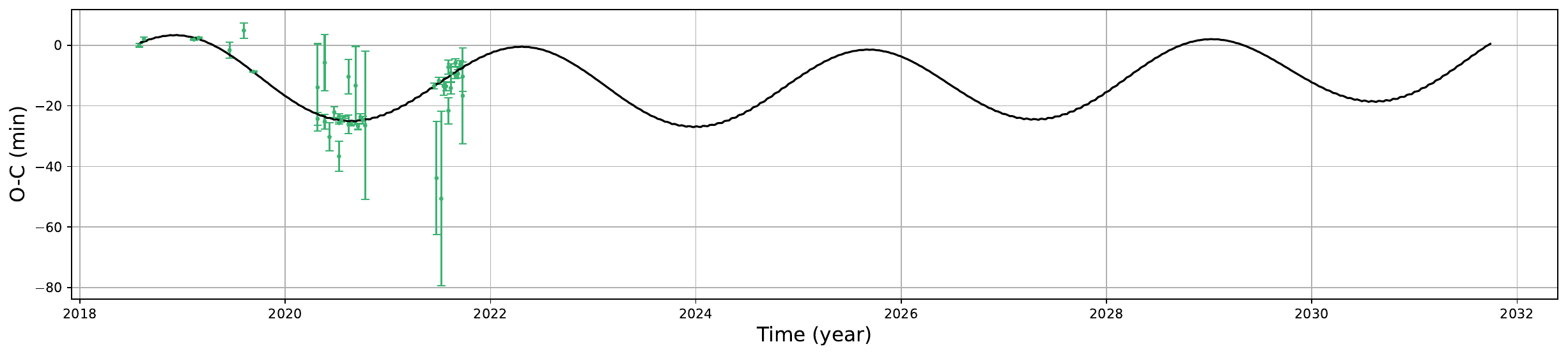}
    \includegraphics[width=0.43\textwidth]{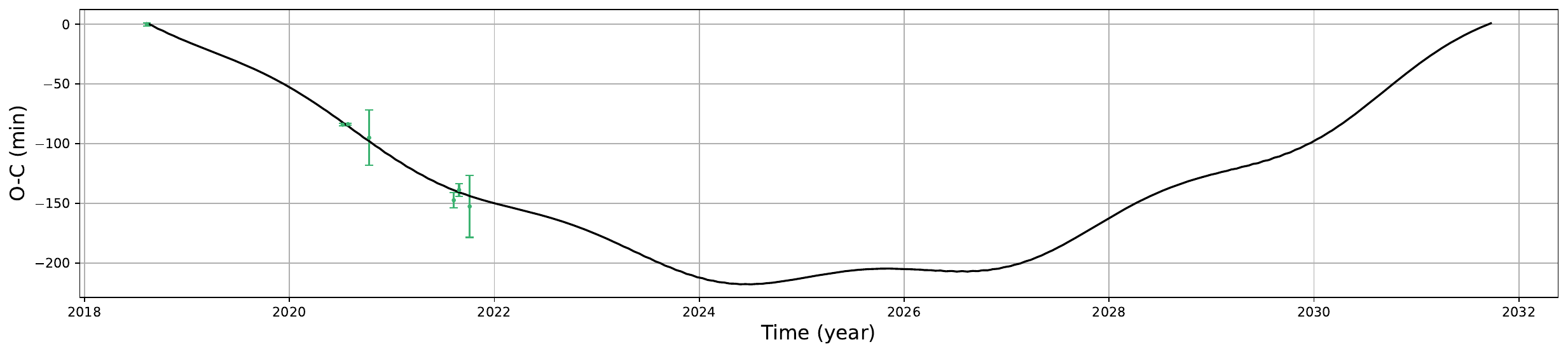}
    \includegraphics[width=0.43\textwidth]{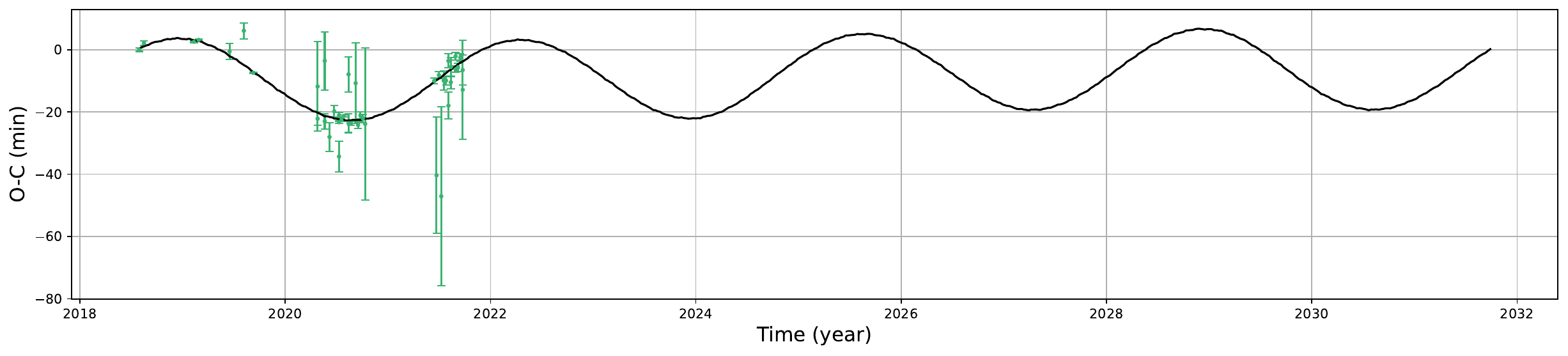}
    \includegraphics[width=0.43\textwidth]{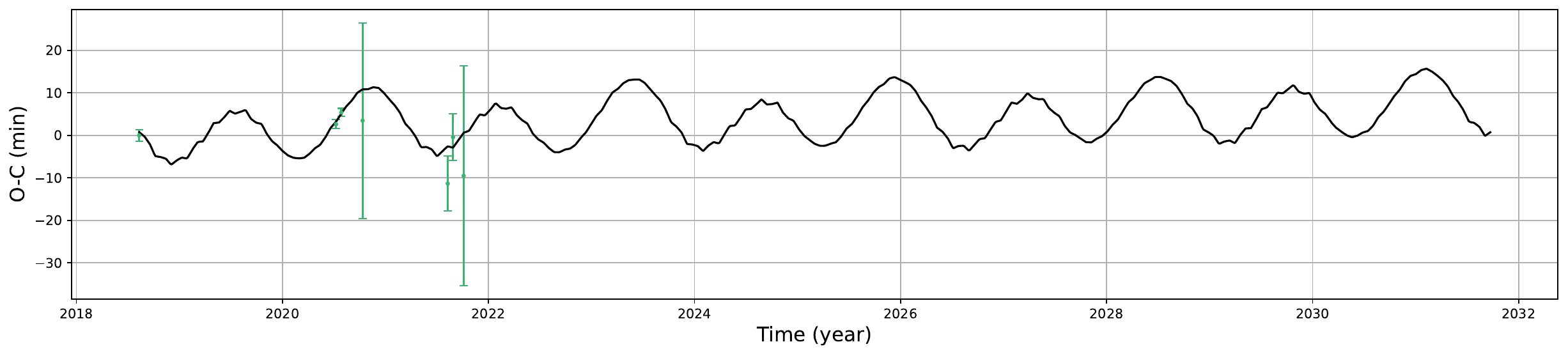}
    \caption{\exostriker-generated TTV models of AU Mic b ({\it left}) and c ({\it right}) projected over ten years since last observed transit of AU Mic system. The observed TTVs (green) is included with the model (black), with P$_{d}$ = 5.07 days ({\it first row}), 5.39 days ({\it second row}), 5.64 days ({\it third row}), 5.86 days ({\it fourth row}), 6.20 days ({\it fifth row}), 6.47 days ({\it sixth row}), 11.9 days ({\it seventh row}), 12.6 days ({\it eighth row}), and 14.1 days ({\it ninth row}).}
    \label{fig:future_ttvs_other}
\end{figure*}

\clearpage
\startlongtable
\begin{deluxetable*}{ccccc}\label{tab:aumic_b_transit_3yr_1}
    \tablecaption{List of predicted midpoint times in BJD for AU Mic b (after accounting for TTVs) over the next three year (Part I). The columns are for each 3-planet configuration.}
    \tablehead{P$_{\rm d}$ = 5.08 days & P$_{\rm d}$ = 5.39 days & P$_{\rm d}$ = 5.64 days & P$_{\rm d}$ = 5.86 days & P$_{\rm d}$ = 6.20 days}
    \startdata
2460310.79154 $\pm$ 0.02340 & 2460310.78763 $\pm$ 0.01872 & 2460310.93889 $\pm$ 0.04680 & 2460310.88002 $\pm$ 0.02106 & 2460310.86700 $\pm$ 0.06786 \\
2460319.25458 $\pm$ 0.02350 & 2460319.25095 $\pm$ 0.01880 & 2460319.40485 $\pm$ 0.04700 & 2460319.34368 $\pm$ 0.02115 & 2460319.33064 $\pm$ 0.06815 \\
2460327.71804 $\pm$ 0.02360 & 2460327.71452 $\pm$ 0.01888 & 2460327.87082 $\pm$ 0.04720 & 2460327.80741 $\pm$ 0.02124 & 2460327.79364 $\pm$ 0.06844 \\
2460336.18110 $\pm$ 0.02370 & 2460336.17825 $\pm$ 0.01896 & 2460336.33675 $\pm$ 0.04740 & 2460336.27118 $\pm$ 0.02133 & 2460336.25668 $\pm$ 0.06873 \\
2460344.64459 $\pm$ 0.02380 & 2460344.64199 $\pm$ 0.01904 & 2460344.80265 $\pm$ 0.04760 & 2460344.73512 $\pm$ 0.02142 & 2460344.72013 $\pm$ 0.06902 \\
2460353.10786 $\pm$ 0.02390 & 2460353.10580 $\pm$ 0.01912 & 2460353.26849 $\pm$ 0.04780 & 2460353.19922 $\pm$ 0.02151 & 2460353.18421 $\pm$ 0.06931 \\
2460361.57145 $\pm$ 0.02400 & 2460361.56936 $\pm$ 0.01920 & 2460361.73427 $\pm$ 0.04800 & 2460361.66304 $\pm$ 0.02160 & 2460361.64811 $\pm$ 0.06960 \\
2460370.03469 $\pm$ 0.02410 & 2460370.03278 $\pm$ 0.01928 & 2460370.19998 $\pm$ 0.04820 & 2460370.12646 $\pm$ 0.02169 & 2460370.11201 $\pm$ 0.06989 \\
\enddata
\end{deluxetable*}

\clearpage
\startlongtable
\begin{deluxetable*}{ccccc}\label{tab:aumic_b_transit_3yr_2}
    \tablecaption{List of predicted midpoint times in BJD for AU Mic b (after accounting for TTVs) over the next three year (Part II). The columns are for each 3-planet configuration.}
    \tablehead{P$_{\rm d}$ = 6.47 days & P$_{\rm d}$ = 11.9 days & P$_{\rm d}$ = 12.6 days & P$_{\rm d}$ = 12.7 days & P$_{\rm d}$ = 14.1 days}
    \startdata
2460310.84935 $\pm$ 0.03744 & 2460310.79751 $\pm$ 0.01171 & 2460310.78970 $\pm$ 0.00937 & 2460310.73989 $\pm$ 0.01405 & 2460310.79791 $\pm$ 0.00470 \\
2460319.31324 $\pm$ 0.03760 & 2460319.26066 $\pm$ 0.01176 & 2460319.25322 $\pm$ 0.00941 & 2460319.20283 $\pm$ 0.01411 & 2460319.26120 $\pm$ 0.00472 \\
2460327.77704 $\pm$ 0.03776 & 2460327.72416 $\pm$ 0.01181 & 2460327.71651 $\pm$ 0.00945 & 2460327.66589 $\pm$ 0.01417 & 2460327.72462 $\pm$ 0.00474 \\
2460336.24126 $\pm$ 0.03792 & 2460336.18739 $\pm$ 0.01186 & 2460336.17972 $\pm$ 0.00949 & 2460336.12873 $\pm$ 0.01423 & 2460336.18794 $\pm$ 0.00476 \\
2460344.70557 $\pm$ 0.03808 & 2460344.65064 $\pm$ 0.01191 & 2460344.64333 $\pm$ 0.00953 & 2460344.59174 $\pm$ 0.01429 & 2460344.65158 $\pm$ 0.00478 \\
2460353.16872 $\pm$ 0.03824 & 2460353.11392 $\pm$ 0.01196 & 2460353.10664 $\pm$ 0.00957 & 2460353.05484 $\pm$ 0.01435 & 2460353.11501 $\pm$ 0.00480 \\
2460361.63262 $\pm$ 0.03840 & 2460361.57714 $\pm$ 0.01201 & 2460361.56992 $\pm$ 0.00961 & 2460361.51776 $\pm$ 0.01441 & 2460361.57852 $\pm$ 0.00482 \\
2460370.09642 $\pm$ 0.03856 & 2460370.04046 $\pm$ 0.01206 & 2460370.03356 $\pm$ 0.00965 & 2460369.98073 $\pm$ 0.01447 & 2460370.04188 $\pm$ 0.00484 \\
\enddata
\end{deluxetable*}

\clearpage
\startlongtable
\begin{deluxetable*}{ccccc}\label{tab:aumic_c_transit_3yr_1}
    \tablecaption{List of predicted midpoint times in BJD for AU Mic c (after accounting for TTVs) over the next three year (Part I). The columns are for each 3-planet configuration.}
    \tablehead{P$_{\rm d}$ = 5.08 days & P$_{\rm d}$ = 5.39 days & P$_{\rm d}$ = 5.64 days & P$_{\rm d}$ = 5.86 days & P$_{\rm d}$ = 6.20 days}
    \startdata
2460322.63096 $\pm$ 0.03046 & 2460322.47156 $\pm$ 0.11235 & 2460322.58173 $\pm$ 0.06196 & 2460322.44772 $\pm$ 0.03781 & 2460322.57336 $\pm$ 0.00430 \\
2460341.49211 $\pm$ 0.03075 & 2460341.33078 $\pm$ 0.11342 & 2460341.44194 $\pm$ 0.06255 & 2460341.30710 $\pm$ 0.03817 & 2460341.43354 $\pm$ 0.00434 \\
2460360.35401 $\pm$ 0.03104 & 2460360.19004 $\pm$ 0.11449 & 2460360.30267 $\pm$ 0.06314 & 2460360.16643 $\pm$ 0.03853 & 2460360.29392 $\pm$ 0.00438 \\
2460379.21496 $\pm$ 0.03133 & 2460379.04925 $\pm$ 0.11556 & 2460379.16335 $\pm$ 0.06373 & 2460379.02565 $\pm$ 0.03889 & 2460379.15426 $\pm$ 0.00442 \\
2460398.07527 $\pm$ 0.03162 & 2460397.90847 $\pm$ 0.11663 & 2460398.02398 $\pm$ 0.06432 & 2460397.88484 $\pm$ 0.03925 & 2460398.01477 $\pm$ 0.00446 \\
2460416.93603 $\pm$ 0.03191 & 2460416.76769 $\pm$ 0.11770 & 2460416.88424 $\pm$ 0.06491 & 2460416.74416 $\pm$ 0.03961 & 2460416.87513 $\pm$ 0.00450 \\
2460435.79777 $\pm$ 0.03220 & 2460435.62695 $\pm$ 0.11877 & 2460435.74478 $\pm$ 0.06550 & 2460435.60353 $\pm$ 0.03997 & 2460435.73562 $\pm$ 0.00454 \\
2460454.65919 $\pm$ 0.03249 & 2460454.48616 $\pm$ 0.11984 & 2460454.60551 $\pm$ 0.06609 & 2460454.46282 $\pm$ 0.04033 & 2460454.59619 $\pm$ 0.00458 \\
\enddata
\end{deluxetable*}

\startlongtable
\begin{deluxetable*}{ccccc}\label{tab:aumic_c_transit_3yr_2}
    \tablecaption{List of predicted midpoint times in BJD for AU Mic c (after accounting for TTVs) over the next three year (Part II). The columns are for each 3-planet configuration.}
    \tablehead{P$_{\rm d}$ = 6.47 days & P$_{\rm d}$ = 11.9 days & P$_{\rm d}$ = 12.6 days & P$_{\rm d}$ = 12.7 days & P$_{\rm d}$ = 14.1 days}
    \startdata
2460322.58254 $\pm$ 0.03466 & 2460322.52001 $\pm$ 0.04516 & 2460322.40457 $\pm$ 0.00950 & 2460322.48933 $\pm$ 0.00845 & 2460322.35068 $\pm$ 0.01368 \\
2460341.44311 $\pm$ 0.03499 & 2460341.38002 $\pm$ 0.04559 & 2460341.26429 $\pm$ 0.00959 & 2460341.34919 $\pm$ 0.00853 & 2460341.20880 $\pm$ 0.01381 \\
2460360.30406 $\pm$ 0.03532 & 2460360.23997 $\pm$ 0.04602 & 2460360.12350 $\pm$ 0.00967 & 2460360.20886 $\pm$ 0.00861 & 2460360.06637 $\pm$ 0.01394 \\
2460379.16458 $\pm$ 0.03565 & 2460379.09968 $\pm$ 0.04645 & 2460378.98336 $\pm$ 0.00976 & 2460379.06903 $\pm$ 0.00869 & 2460378.92560 $\pm$ 0.01407 \\
2460398.02486 $\pm$ 0.03598 & 2460397.95978 $\pm$ 0.04688 & 2460397.84285 $\pm$ 0.00985 & 2460397.92928 $\pm$ 0.00877 & 2460397.78454 $\pm$ 0.01420 \\
2460416.88527 $\pm$ 0.03631 & 2460416.81973 $\pm$ 0.04731 & 2460416.70288 $\pm$ 0.00994 & 2460416.78973 $\pm$ 0.00885 & 2460416.64275 $\pm$ 0.01433 \\
2460435.74612 $\pm$ 0.03664 & 2460435.67976 $\pm$ 0.04774 & 2460435.56255 $\pm$ 0.01003 & 2460435.65028 $\pm$ 0.00893 & 2460435.50253 $\pm$ 0.01446 \\
2460454.60686 $\pm$ 0.03697 & 2460454.53974 $\pm$ 0.04817 & 2460454.42273 $\pm$ 0.01012 & 2460454.51091 $\pm$ 0.00901 & 2460454.36224 $\pm$ 0.01459 \\
\enddata
\end{deluxetable*}

\bibliography{main} 

\end{document}